\documentclass[3p,single-column]{elsarticle}

\usepackage{lineno,hyperref}
\journal{Journal of Computational Science}

\usepackage{numcompress}\biboptions{authoryear}

\usepackage{amsfonts}
\usepackage{multicol}
\usepackage{amsmath}
\usepackage{leftindex} % subscript in front of symbol
\usepackage{graphicx} % \scalebox
\usepackage{environ}
\usepackage{mathrsfs}
\usepackage{soul}
\usepackage{amssymb}
\usepackage{enumerate}
\usepackage{lineno,hyperref}
\usepackage{wrapfig}
\usepackage{tcolorbox}
\usepackage{amsmath}
\usepackage{tcolorbox}
\usepackage{multirow} 
\usepackage{bm}
\usepackage{stackengine}
\usepackage{amsfonts}       % blackboard math symbols
\usepackage{enumitem}
\usepackage{tabularx}
\usepackage{multirow}
\usepackage[linesnumbered,ruled,vlined]{algorithm2e}
\usepackage{algpseudocode}
\usepackage{xcolor}
\usepackage{multirow}
\usepackage{booktabs}
\usepackage{caption}
\usepackage{subcaption}
\usepackage{soul}
\usepackage{mathtools}

\makeatother

\newcommand{\bs}{\boldsymbol{s}}
\newcommand{\ba}{\boldsymbol{a}}
\newcommand{\bp}{\boldsymbol{p}}

\usepackage[utf8]{inputenc}

\newcommand{\bw}{\boldsymbol{w}}

\newcommand{\bla}{\boldsymbol{\lambda}}

\begin{document}
	
	\begin{frontmatter}
		%\title{From Digital Twins to Model-Based Reinforcement Learning } --- Miguel
		%\title{Reinforced System Assimilation: from Digital Twins to Model-Based Reinforcement Learning}
		%\title{Reinforced Model Assimilation for System Identification and Control: from digital twins to model-based reinforcement learning}
		
		\title{Reinforcement Twinning: from digital twins to\\ model-based reinforcement learning}
		
		\author[label1,label2]{Lorenzo~Schena}
		\author[label1,label3]{Pedro~Marques}
		
		\author[label1,label4]{Romain~Poletti}
		
		\author[label1]{Samuel~Ahizi}
		
		\author[label1,label5]{Jan~ Van~ den~ Berghe}
		
		\author[label1]{Miguel~ A.~ Mendez}

		%\affiliation[label1]{organization={von Karman Institute},%Department and Organization
			%            city={Rhode-St-Genése},
			%            postcode={1640}, 
			%            country={Belgium}}
		
		\affiliation[label1]{organization={von Karman Institute},%Department and Organization
			city={Rhode-St-Genése},
			postcode={1640}, 
			country={Belgium}}

		\affiliation[label2]{organization={Vrije Universiteit Brussel (VUB), Department of Mechanical Engineering},%Department and Organization
			city={Elsene, Brussels},
			postcode={1050}, 
			country={Belgium}} 
		
		\affiliation[label3]{organization={Université Libre de Bruxelles},%Department and Organization
			addressline={Av. Franklin Roosevelt 50}, 
			city={Brussels},
			postcode={1050}, 
			country={Belgium}}        
		
		\affiliation[label4]{organization={University of Ghent},%Department and Organization
			addressline={Sint-Pietersnieuwstraat 41}, 
			city={Ghent},
			postcode={9000}, 
			country={Belgium}}

		\affiliation[label5]{organization={Institute of Mechanics, Materials, and Civil Engineering (iMMC), Université Catholique de Louvain},%Department and Organization
			addressline={}, 
			city={Louvain-la-Neuve},
			postcode={1348},
			country={Belgium}}       
		
		
		
		
		\begin{abstract}
			The concept of digital twins promises to revolutionize engineering by offering new avenues for optimization, control, and predictive maintenance. We propose a novel framework for simultaneously training the digital twin of an engineering system and an associated control agent. The training of the twin combines methods from adjoint-based data assimilation and system identification, while the training of the control agent combines model-based optimal control and model-free reinforcement learning. The training of the control agent is achieved by letting it evolve independently along two paths: one driven by a model-based optimal control and another driven by reinforcement learning. The virtual environment offered by the digital twin is used as a playground for confrontation and indirect interaction. This interaction occurs as an ``expert demonstrator", where the best policy is selected for the interaction with the real environment and ``cloned" to the other if the independent training stagnates.
			We refer to this framework as Reinforcement Twinning (RT). The framework is tested on three vastly different engineering systems and control tasks, namely (1) the control of a wind turbine subject to time-varying wind speed, (2) the trajectory control of flapping-wing micro air vehicles (FWMAVs) subject to wind gusts, and (3) the mitigation of thermal loads in the management of cryogenic storage tanks. The test cases are implemented using simplified models for which the ground truth on the closure law is available. The results show that the adjoint-based training of the digital twin is remarkably sample-efficient and completed within a few iterations. Concerning the control agent training, the results show that the model-based and the model-free control training benefit from the learning experience and the complementary learning approach of each other. The encouraging results open the path towards implementing the RT framework on real systems.   
		\end{abstract}
		
		
	\end{frontmatter}
	
	% \linenumbers

	\section{Introduction}
	
	Mathematical models of engineering systems have always been fundamental for their design, simulation and control.
	Recent advances in the Internet of Things (IoT), data analytics and machine learning are promoting the notion of `dynamic' and `interactive' models, i.e. architectures that integrate and enrich the mathematical representation of a system with real-time data, evolve through the system's life-cycle, and allow for real-time interaction.
	Such kinds of models are currently referred to as \emph{digital twins}.  
	
	Although the specific definition can vary considerably in different fields  \citep{Wagner2019,Barricelli2019,chinesta_virtual_2020,rasheed_digital_2019,ammar_digital_2022,Wright2020,tekinerdogan_notion_2022,Beek2023,haghshenas2023predictive}, a digital twin is generally seen as a virtual replica of a physical object (or system) capable of simulating its characteristic, functionalities and behaviour \emph{in real-time}. 
	
	Besides the obvious need for domain-specific knowledge and engineering, constructing such a replica involves various disciplines, from sensor technology and instrumentation for real-time monitoring to virtual/augmented reality and computer graphics for interfacing and visualization, and data-driven modelling, assimilation and machine learning for model updating and control. This work focuses on the last aspects and the required intersection of disciplines, with a particular emphasis on the need for real-time interaction. Real-time predictions are essential to deploying digital twins for control or monitoring purposes and require fast models. Fast models can be built from (1) macroscopic/lumped formulations, eventually enriched by data-driven `closure' laws, (2) surrogate/template models, or (3) reduced-order models derived from high-fidelity simulations. In the context of system identifications for control purposes, the first would be referred to as `white' models; the second would be referred to as `black box models' while the third falls somewhere in between (see \cite{Schoukens2019} for the palette of grey shades in data-driven models). Regardless of the approach, the main conceptual difference between digital twinning and traditional engineering modelling is the continuous model update to synchronize the twin with the physical system and to provide a \emph{live} representation of its state from noisy and partial observations.
	
	It is thus evident that the notion of digital twin lies at the intersection between established disciplines such as system identification (see \cite{Nelles2001,Ljung2008,Nicolao2003SystemI}) and data assimilation (see \cite{Asch2016,Bocquet2023}). Both are significantly enhanced (and perhaps more intertwined) by the rapid growth and popularization of machine learning, as reviewed in the following section of this article. Both seek to combine observational data with a numerical model, and both involve an observation phase (known as training or calibration, depending on the field) in which the model is confronted with online data and a `prediction' phase in which the model is interrogated. However, these disciplines are built around different problems and use different methods.
	
	Data assimilation is mostly concerned with the \emph{state estimation} problem. It is assumed that the model is known (at least within quantifiable uncertainties), but observations are limited and uncertain. These uncertainties are particularly relevant because the system is chaotic and extremely high dimensional (e.g. atmospheric/oceanic models) and thus highly sensitive to the initial conditions from which predictions are requested. The traditional context is weather prediction and climate modelling \citep{Lorenc1986,Wang2000,Pu2018,Lahoz2010}, where the goal is to forecast the future evolution of a system from limited observations. The most common techniques are 4D variational assimilation \citep{Talagrand1987,Dimet2016,Ahmed2020} and, more recently, Ensemble Kalman methods \citep{Evensen2009,Bocquet2011,Routray2016}. Much effort is currently ongoing towards methods that combine these techniques (e.g. \citealt{Kalnay2007,Lorenc2015}.
	
	Nonlinear system identification is mostly concerned with the \emph{model identification} problem. The system is subject to external inputs and observed through noisy measurements but is deterministic. The model can be fit into structures that are derived either from first principles or from general templates such as Nonlinear AutoRegressive models with exogenous inputs (NARX, \cite{narx}) or Volterra models \citep{Schoukens2019}, and the main goal is to predict how the system responds to actuation. The traditional context is control engineering, i.e. driving the system towards a desired trajectory. Besides the stronger focus on input-output relation (and thus the focus on actuated systems), the main difference with respect to traditional data assimilation is in the deterministic nature of the system to be inferred: errors in the initial conditions are not amplified as dramatically as, for example, in weather forecasting. Therefore, the resulting uncertainties remain within the bounds of model uncertainties and the state estimation is less critical than in data assimilation. Moreover, in many actuated systems, the forced response dominates over the natural one and the initial conditions are quickly forgotten. Tools for system identification are reviewed by \cite{Nelles2001} and \cite{Suykens1996}.  
	
	Machine learning is progressively entering both communities with a wealth of general-purpose function approximations such as Artificial Neural Networks (ANNs, see \citealt{goodfellow2016deep}) or Gaussian Processes (GPr, see \citealt{CarlEdward}). More specifically, the fusion between these disciplines materializes through a combination of variational or ensemble tools from data assimilation to train dynamic models with model structures from machine learning to solve identification, forecasting or control problems. In this regard, the literature in control engineering has significantly anticipated the current developments (see neural controllers in \citealt{Suykens1996,Norgaard2000}), although their extension to broader literature has been limited by the mathematical challenges of nonlinear identification and the success of much simpler strategies based on piece-wise linearization or adaptive controllers \citep{Astrom:1994,sastry1989adaptive}.
	
	While machine learning enters the literature of data assimilation and system identification, cross-fer\-ti\-li\-za\-tion proceeds bilaterally. The need for identifying dynamic models from data is also growing in the literature on reinforcement learning, which is significantly drawing ideas from system identification and optimal control theory. Reinforcement learning is a subset of machine learning concerned with the training of an agent via trial and error to achieve a goal while acting on an environment \citep{sutton2018reinforcement,Bertsekas2019}. The framework differs from classic control theory because of its roots in sequential decision making: the environment to be controlled and the agent constitute a Markov Decision Process (MDP, \cite{Puterman1994}) rather than a dynamical system with the usual state-space representation. Model-free approaches solely rely on input-output information and function approximations of the agent's performance measures, while model-based approaches combine these with a model of the environment \citep{moerland_model-based_2022,luo_survey_2022}. The model can be used as a forecasting tool, thus playing the same role as in model predictive control, or as a playground to accelerate the learning \citep{schwenzer_review_2021}. Model-free approaches have gained burgeoning popularity thanks to their success in video games \citep{Szita2012,mnih2013playing} or natural language processing \citep{UcCetina2022}, for which the definition of a system model is cumbersome. However, these algorithms have shown significant and somewhat surprising limitations for engineering problems governed by simple PDEs \citep{Werner2023,Pino2023}. %controlla teams!
	
	This article proposes a framework combining ideas from data assimilation, system identification and reinforcement learning with the goal of: (1) training a digital twin on real-time data and (2) solving a control problem. We refer to this framework as Reinforcement Twinning (RT). The rest of the article is structured as follows. Section \ref{s2} provides a concise literature review of related works across the intersected disciplines and identifies the main novelties of the proposed framework. Section \ref{s3} presents the general framework and the key definitions, while Section \ref{s4} goes into the mathematical details of its implementation. Section \ref{s5} introduces three test cases selected for their relevance in the engineering literature and stemming from largely different fields. These are (1) the control of a wind turbine subject to time-varying wind speed, (2) the trajectory control of flapping-wing micro air vehicles (FWMAVs) subject to wind gusts, and (3) the mitigation of thermal loads in the management of cryogenic storage tanks. 
	Although all test cases are analyzed using synthetic data that might look somewhat contrived, the main focus is here to show that the model-based and the model-free training can benefit from each other. Section \ref{s6} presents the main results; section \ref{s7} closes with conclusions and perspectives.
	
	%The challenge of bridge the huge amount of data and the previous physical knowledge is found difficult to be harmonized, and more often than not, the two approaches - i.e. model-free and model-based - are seen as antipodal viewpoints.
	%The contribution of this work is to move a step forward toward the harmonization of data- and physics- based approaches, developing a unique framework which encompasses both. Starting from available models of the problem of interest, we show that it is possible to assimilate incoming data to develop a DTI tailored on the machine/asset being monitored. On top of this \textit{assimilated} system, employing classical approaches deriving from Optimal Control an optimal trajectory to keep the system in is identified. Finally, these two information are made available and blended with recent Reinforcement Learning (RL) algorithm, to cope with epistemic and measured uncertainties, leading to what we call a "Digital Twin Instance - Based Reinforcement Learning" algorithm. 
	
	\section{Related Work}\label{s2}

    This section locates the contribution of this work in the broad literature of the intersected fields. We briefly review recent developments in methods combining machine learning with data assimilation (Section \ref{s2p1}) and system identification (Section \ref{s2p2}). We then move towards the cross-fertilization in the context of model-based reinforcement learning (Section \ref{s2p3}) and focus on recent developments on the combination between model-based and model-free reinforcement learning (Section \ref{s2p4}). Finally, Section \ref{s2p5} briefly reports on the main novelties of the proposed approach.
	
\subsection{Machine Learning for Data Assimilation }\label{s2p1}
	
An extensive overview of the state of the art of data assimilation (DA) is provided by \cite{Carrassi2017} while \cite{cheng_machine_2023} give an overview of how machine learning (ML) is entering DA. \cite{Geer2021} discuss the formal links between DA and ML in a Bayesian framework while \cite{abarbanel2017machine} analyzes the link by building parallelism between time steps in a DA problem and layer labels in a deep ANN. As explored in \cite{Bocquet2011}, the formulation of hybrid DA-ML algorithms consists in introducing ML architectures either (1) to complement/correct the forecasting or the observation model or (2) to replace at least one of them. An early attempt at DA-ML hybridization within the first class was proposed by \cite{Tang2001}, who used a variational approach to train an ANN that replaces missing dynamical equations. More recent approaches within the first category are proposed by \cite{Arcucci2021} while \cite{Buizza2022} presents several approaches in both categories. In particular, \cite{Arcucci2021} implements a Recurrent Neural Network (RNN,\citealt{Madhavana}) to correct the forecasting model, while \cite{Buizza2022} use convolutional neural networks (CNNs) to improve the observations passed to a Kalman filter. Within the second class of hybrid DA-ML methods, \cite{brajard_combining_2020} and \cite{Buizza2022} use a Kalman filter to train the parameters of an ANN that acts as a forecasting model. Outside geophysical sciences, the use of ANNs as a forecasting tool in assimilation frameworks is spreading in economics \citep{Khandelwal2021} and epidemiology \citep{Nadler2020}. Finally, within the literature on data assimilation, several works have combined ML and DA strategies for model identification \citep{Bocquet2019DataAA,ayed_learning_2019}.
	
\subsection{Machine learning for System Identification and Control}\label{s2p2}
	
Recent reviews on deep learning for system identifications are proposed by \cite{Ljung2020DeepLA} and \cite{Pillonetto2023} while earlier reviews are provided by \cite{Ljung2008}, \cite{Nicolao2003SystemI}. System identification has been historically influenced by developments in machine learning and adopted the use of ANNs for system modelling already in the 90's \citep{Chen1990,Zhang1991SystemIU,Sjoeberg1994}. Early works (see also \citealt{Suykens1996} and \citealt{Norgaard2000}) focused on the use of feedforward neural networks (FNNs) or recurrent neural networks (RNNs); these can be seen as special variants of nonlinear autoregressive and nonlinear state-space models. The popularization of ML has significantly enlarged the zoology of ANNs architectures used as model structures for state-spacem odeling. The most popular examples include variants of the RNN, such as long-short-term memory (LSTM) networks \citep{Hochreiter1997} and Echo State Networks (ESN, \cite{Jaeger2004}), as well as convolutional neural networks (CNNs,\cite{LeCun1989}) and the latest developments in neural ODEs (also known as ODE-net,\cite{Chen2018}). 
	
	Important contributions within the first class of approaches are the works by \cite{Ljung2020DeepLA}, \cite{Gonzalez2018} and \cite{bucci2018model} who use LSTM networks \citep{Hochreiter1997}. These are variants of RNNs, which are very popular in natural language processing and use three gates (forget/input/output) to preserve information over longer time steps, thus allowing for better handling of long-term dependencies. \cite{Canaday2020} used ESNs, which gained popularity for forecasting in stochastic systems. These are variants of RNNs implementing the reservoir computing formalism, i.e. ANNs in which weights and biases are selected randomly. \cite{Andersson2019} used CNNs, which gained popularity in image recognition and allow for identifying patterns at different scales, while \cite{ayed_learning_2019} and \cite{Rahman2022} use ODE-nets, recently proposed as a new paradigm for neural state-space modelling. ODE-nets treat the layers of an ANN as intermediate time steps of a multistep ODE solver and are commonly trained using the adjoint method, a ubiquitous tool in data assimilation. An overview of the challenges of learning models of dynamical systems from data has been recently presented by \cite{Bucci2023}, who also propose a ``gradual'' approach which seeks to build models of gradually increasing complexity.

	\subsection{Model Based Reinforcement Learning and Optimal Control}\label{s2p3}
		
	Recent reviews of model-based reinforcement learning (MBRL) are proposed by \cite{chatzilygeroudis2019survey}, \cite{luo2022survey} and \cite{moerland2022modelbased}. MBRL uses a model of the environment to guide the agent, with the goal of increasing its sample efficiency (i.e. reducing the number of trials required to learn a task). A model allows the agent to sample the environment with arbitrary state-action values. In contrast, a model-free approach can only sample visited states, eventually stored in a replay buffer to allow multiple re-uses \citep{haarnoja2018soft, lillicrap2019continuous}. In the reinforcement learning terminology, a model allows the agent to \emph{plan}, giving the possibility to \emph{imagine} the consequences of actions and, more generally, update policy and/or state-action value predictors without interacting with the environment.
	
	MBRL methods can be classified based on how the model is built and how it is used. Concerning the approaches for model constructions, a large arsenal of function approximations has been deployed in the literature. ANNs \citep{HUNT19921083,kurutach2018model,nagabandi2017neural}, Gaussian Processes (GPr,\citet{pilco,approx_gaussians}), time-varying linear models \citep{real_time_oc} are some examples. Concerning model usage, the simplest approach consists of using it solely to sample the state-action space, while more advanced approaches use the model in the training steps \citep{janner2021trust, gu2016continuous}. One of the seminal MBRL approaches is the Dyna algorithm proposed by \cite{sutton_dyna}. This uses experience (i.e. trajectories in the state-action space and associated rewards) collected from both environment and model. More sophisticated approaches use the model to ``look ahead'', i.e. predicting the future evolution of the system under a given policy. A notable example is the Monte Carlo Tree Search in the AlphaGO \citep{Silver2016,Silver2018} algorithm that reached superhuman performances in the game of Go. Moving from Markov Decision Processes to dynamical systems, the boundaries between MBRL and Model Predictive Control (MPC) become blurred, particularly when the latter uses black-box models (see \citealt{hedengren_nonlinear_2014,Rawlings2000} and \citealt{nagabandi2017neural} or \citealt{Weber2017}).
	
	Several authors have proposed methods to leverage physical insights in the models for MBRL, using physics-informed neural networks \citep{raissi2019physics} as \cite{Liu2021} or Deep Lagrangian Networks (DeLaN \citealt{lutter2019deep}) as \cite{ramesh2023physicsinformed}. A more extreme approach is proposed by \cite{lutter2020differentiable}, who uses `white-box' models (eventually enhanced as proposed by \citealt{lutter2020differentiable}) in MBRL. This work shows a traditional physics-based model trained via machine learning methods. The opposite case is provided by \cite{liu2023adjointbased}, who uses an ANN agent trained with adjoint-based optimization. An overview of methods to integrate physics-based models and machine learning is provided by \cite{osti_1478744} and \cite{Willard2020}. Regardless of its structure, a differentiable model of the environment can be trained using optimal control theory \citep{stengel1994optimal}.

\subsection{The Hybridization of Model-Based and Model-Free Control}\label{s2p4}
The idea of combining model-based and model-free control methods is gaining momentum in the literature of robotics and control and is generally referred to as hybrid learning \citep{pinosky2022}. Hybrid learning seeks to combine the model-based methods' sample efficiency with the model-free methods' robustness (see \citet{dulacarnold2019challenges}). Hybrid approaches differ in how the model-based and model-free policies are chosen or merged. An early idea proposed by \cite{hybrid1997} consisted of keeping these separate and independent, with procedures known as Hierarchical Mixtures of Experts (HME) \citep{jordanmixture1994, tham1995} used to identify the best of the two at a given state. More recently, a stronger interaction was proposed by \cite{freed2024unifying}, who used a model-based policy to bound a model-free policy, with the goal of guaranteeing the stability of the controller even during the exploration. A similar approach is developed in the context of continuous reinforcement learning \citep{yıldız2021continuoustime, lutter2021continuoustime,doya2000reinforcement} by \cite{pinosky2022}.
Alternatively, if models are cast in a locally linear or linear-quadratic form, iterative optimal control methods \citep{tassaSynthesisStabilizationComplex2012, vandenbergIteratedLQRSmoothing2014} have been used to generate a pool of ``expert" local policies from which a model-free learner must generalize and improve upon \citep{levine2013guided, levine2014learning, chebotar2017combining, kahn2017plato}, or to warm-start a model-free policy search step \citep{qu2020combining}. 
\citep{abbeel_using_2006} proposed an approach decoupling the policy evaluation and policy improvement steps, thereby developing a hybrid method in which one step is carried out in a model-free approach and the other proceeds in a model-based approach. 

An alternative path is to obtain a hybridization in the value function space. Value functions aim to predict each state's value and quantify how well a controller can perform from a given state of the system; value function approximators can thus be used to guide optimal controllers as proposed by \citep{pong2020temporal}. Models can also be used to generate roll-outs to enhance model-free value estimation methods \citep{heess2015learning}, potentially weighting the collected data according to the trust in the model at hand \citep{feinberg2018modelbased}. Other combinations propose the use of control-theoretic tools to provide guarantees on model-free methods. An example can be found in \cite{berkenkamp2017safe}, in which results on Lyapunov stability theory are used to identify safe regions in the state-space \citep{richards2018lyapunov} in which the controller is constrained to operate. In addition, if considerations on robustness are also made, this effectively leads to an $H_\infty$ model-free RL approach \citep{hanH_InftyModelfree2020}.

\subsection{Novelties of the Proposed Approach}\label{s2p5}

The RT framework proposed in this work is a hybrid learning approach in which the model-based part is based on optimal control and adaptive modelling using methods from data assimilation. The first novelty within the landscape of model-based methods is the use of physics-based models with closure laws adapted in real-time: the derived model allows for real-time interaction and adapts to the system evolution, becoming a digital twin of the real system. Although the model is here trained using an ensemble adjoint-based approach to compute the gradient of the cost function driving the model derivation, variants of the algorithm using Ensemble Kalman filtering are possible and currently being explored. Departing from traditional data assimilation, we assume that all states are observable and the underlying dynamic is deterministic. Moreover, since the main role of the model is to provide input-output relations in an actuated system, the approach is closer to system identification than traditional data assimilation. 

The second novelty within the land-scale of hybrid learning is in the strong interaction between the model-based and the model-free control agents, which alternate the role of disciple and master. A system similar to HME decides which policy is promoted to master and used on the real system. However, both are continuously trained in parallel, using the digital twin as a playground and sharing the collected expertise. We show that these agents can learn from each other via different mechanisms.

%, in which model-based and model-free directly interact with each other, alternating the role of disciple and master. The model-based search is grounded in an adaptive, white box (physics-based) model trained with methods from data assimilation.
%This model allows for real-time interaction and can be seen as a digital twin because it is trained in real time to reproduce the performance of a specific system, \textcolor{blue}{in a specific condition}. Concerning the training, we use an ensemble adjoint-based approach to compute the gradient of the cost function driving the model derivation. This gives remarkable robustness to noise. Variants of the algorithm using Ensemble Kalman filtering are currently being explored.

%However, contrary to traditional data assimilation, we assume that all states are observable and the underlying dynamic is deterministic. Moreover, since the main role of the model is to provide input-output relations in an actuated system, the approach is closer to system identification than traditional data assimilation. Finally, contrary to MPC, we assume that not all inputs to the system can be predicted; hence, the model cannot `look ahead'. However, the ensemble approach to handle the unpredictable inputs allows for online planning. Extensions to an MPC formalism are left to future work.

	\section{Definitions and General Formulation }\label{s3}
	
\begin{figure*}
		\centering
		\includegraphics[width=0.95\linewidth]{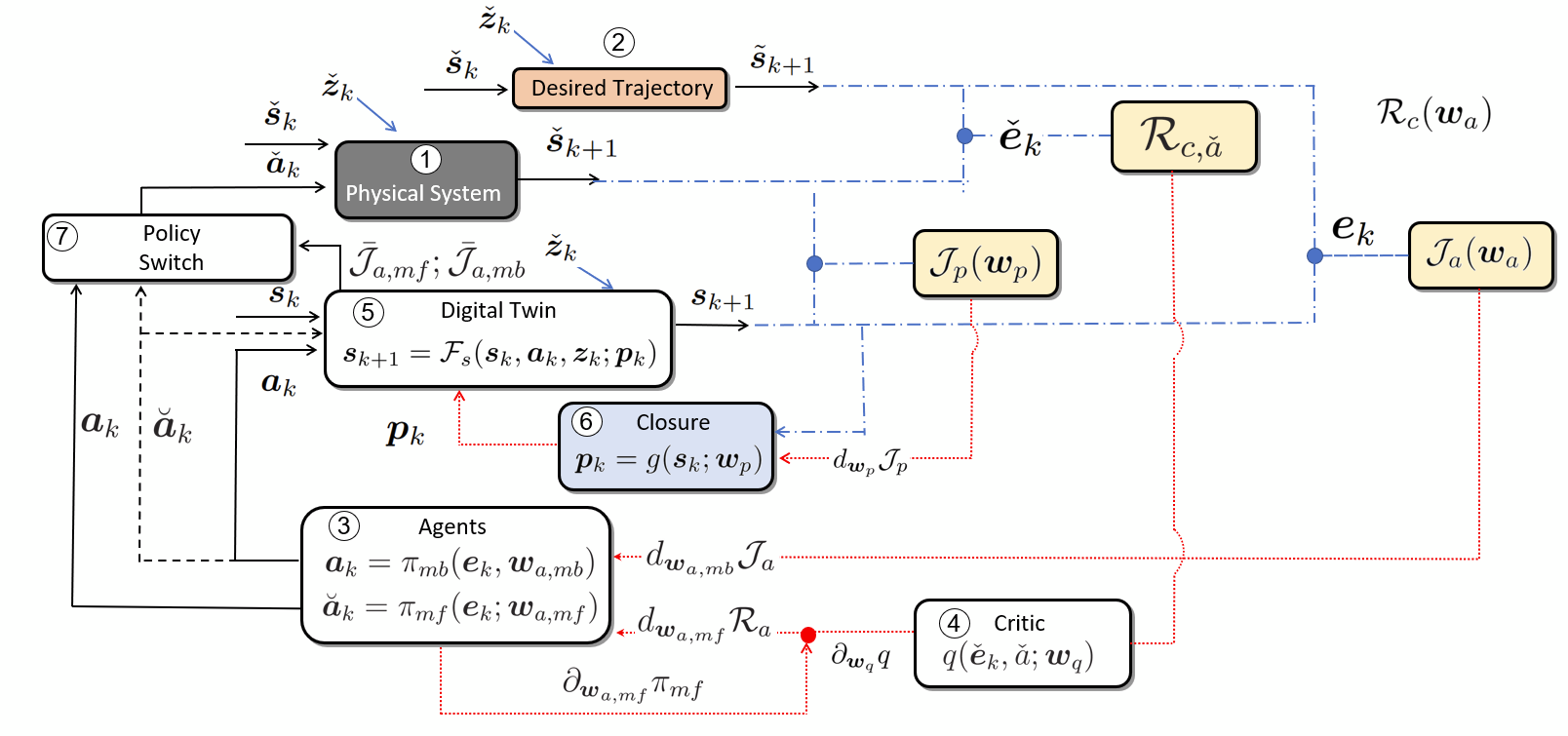}
		\caption{Structure of the proposed approach to blend assimilation, model-based and model-free control. We use one control agent, acting on the physical system and its digital twin, and we combine the learning from a model-free and a model-based approach. The dashed blue lines are used to track the distance measures, while the red dotted lines track the updates of the parameters involved in the training. The full algorithmic implementation is provided in Algorithm \ref{alg:rsa}.}
		\label{Fig1}
	\end{figure*}

	The main components of the RT approach implemented in this work are illustrated with the aid of the schematic in Figure \ref{Fig1}. All components are numbered following the presentation below.
	
	We consider a dynamical system (component 1) with states $\check{\bm{s}}\in\mathbb{R}^{n_s}$ evolving according to \emph{unknown} dynamics. We treat this system as a black box and consider it the \emph{real} environment. We assume that these states can be sampled (measured) at uniform time intervals $t_k=k\Delta t$ over an observation time $T_o$ to collect the sequence $\check{S}=[\check{\bm{s}}_1,\dots \check{\bm{s}}_{n_t}]$ with $n_t=T_o/\Delta t+1$. All variables are sampled in the same way, and we use subscripts to identify the sample at specific time steps.
	
	The real system evolves under the influence of control actions $\check{\bm{a}}_k$,
	provided by an agent, and exogenous (uncontrollable) inputs $\check{\bm{z}}_k=\check{\bm{d}}_k+\check{\bm{\epsilon}}_k\in\mathbb{R}^{n_z}$. We assume that these consist of a `large' scale ($\check{\bm{d}}_k$) and a `small scale' ($\check{\bm{\varepsilon}}_k$) contribution, but neither of the two can be forecasted. These are treated as random processes with different integral time scales. 
	With no loss of generality, we assume that observations and actions occur simultaneously. Therefore, in the reinforcement learning terminology, the collection of $n_t$ observations/action pairs within the observation time $T_o$ defines an \emph{episode}.
	
	The actions seek to keep the system along a desired trajectory $\tilde{S}=[\tilde{\bm{s}}_1,\dots,\tilde{\bm{s}}_{n_t}]$ (component 2) and we introduce the \emph{real error} $\check{\bm{e}}_k=\check{e}(\check{\bs}_k,\tilde{\bs}_k)\in\mathbb{R}^{n_s}$ to measure the discrepancy between the current and the target states. The control agent (component 3) acts according to a \emph{policy} which takes the form of a parametric function $\check{\ba}_k=\pi(\check{\bm{e}}_k;\bm{w}_a)$ in the weights $\bm{w}_a\in\mathbb{R}^{n_{\pi}}$. The control problem involves identifying the weights $\bm{w}_a$ through an iterative ``learning process" while the agent interacts with the system. This process is carried out by evolving two policies in parallel: the best one is promoted to ``live" policy and interacts with the real system while the other becomes ``idle" and continues to be trained in the background.
	
	One of these policies evolves according to model-free strategy. It is denoted as $\pi_{mf}$ and is defined by the set of parameters $\bm{w}_{a, mf}$. The other evolves according to a model-based strategy. It is denoted as $\pi_{mb}$ and is defined by the set of parameters $\bm{w}_{a, mb}$. The decision of which of these becomes ``live" or ``idle '' is taken by policy switch (element 7 in Figure \ref{Fig1}) that selects either $\bm{w}_a=\bm{w}_{a, mf}$ or $\bm{w}_a=\bm{w}_{a, mb}$ depending on which of the two is producing the best results on the virtual environment and the digital twin (component 5).
	
	The working principle of this policy switch is described in section \ref{s4p3}. What follows is a description of the model-free and the model-based training, treated as independent strategies.

	\begin{itemize}
		
		\item The model-free policy ($\pi_{mf}$) training relies on classic model-free reinforcement learning. The optimal weights $\bw_{a, mf}$ are defined according to a reward function, which measures how well the system is kept close to the desired trajectory. This evaluation is made \textit{online}, updating the policy after each episode. The optimal weights maximize this function, which is written as the summation of instantaneous rewards $r_k=r(\check{\bm{e}}_k,\check{\ba}_k)\in\mathbb{R}$, possibly discounted by a factor $\gamma$:
		
		\begin{equation}
			\label{R_c}
			\mathcal{R}_c(\bm{a}, \bm{e})=\sum^{n_t-1}_{k=0} \gamma^{k} r_k(\check{\bm{e}}_k,\check{\ba}_k)\,.
		\end{equation} 
		
		The definition of the instantaneous reward function is problem-dependent and discussed in the following sections. As described in section \ref{s4p1}, the model-free loop implemented in this work is an actor-critic algorithm, which relies on a parametric approximation of the state-action value function $Q(\check{\bm{e}},\check{\ba};\bw_q)$ in the parameters $\bm{w}_q$. This estimates the expected reward for an action-state pair and is essential in computing the update of the policy parameters $\bm{w}_a$. This parametric function is referred to as `critic' (component 4) and is essential in computing the gradient $d_{\bw_{a, mf}} \mathcal{R}_c$ driving the training of the agent from the model-free side (see connections in dashed red in Figure \ref{Fig1}). We highlight that the use of the error state $\bm{e}$ rather than the state $\bm{s}$ in the reward and $Q$ functions is here justified by our focus on tracking problems.

		\begin{comment}
			$$\bm{p}_k=g(\bs_k;\bw_p)$$
			
			$$\check{e}_k$$
			
			$$e_k$$
			
			$$d_{\bw_a} \mathcal{J}_a$$
			
			$$d_{\bw_a} \mathcal{R}_a$$
			
			$$d_{\bw_p} \mathcal{J}_p$$
			
			$$\partial_{\bw_a} \pi$$
			$$\partial_{\bw_q} q$$
			
			$\bm{s}_{k+1}=\mathcal{F}_s(\bm{s}_k,\bm{a}_k,\bm{z}_k;\bm{p}_k)$    
			
			$\bm{a}_{k}=\pi(\bm{e}_k;\bm{w}_a)$    
			
			$\check{\bm{a}}_{k}=\pi(\check{\bm{e}}_k;\bm{w}_a)$    
			
			$$\tilde{\bm{s}}_k$$
			
			$$\bm{a}_{k}=\pi(\bm{e}_k;\bm{w}_a)$$    
			
			$$\check{\bm{a}}_{k}=\pi(\check{\bm{e}}_k;\bm{w}_a)$$

			$$q(\check{\bm{e}},\check{\bm{a}};\bm{w}_q)$$
			
		\end{comment}
		
		\item The model-based policy ($\pi_{mb}$) optimization loop combines system identification and optimal control. System identification seeks to continuously adapt and improve the digital twin (component 5). This is a model of the system which evolves the `virtual states' $\bs_k$ to $\bs_{k+1}$, subject to the same exogenous inputs $\check{\bm{z}}_k$ as the physical system and under the `virtual actions' $\bm{a}_k$. This solver advances the digital twin in real time and restarts at the end of each episode.
		
		The model relies on unknown parameters $\bm{p}_k\in\mathbb{R}^{n_p}$, provided by a closure parametric function $\bm{p}_k=g(\bs_k;\bw_p)$ (component 6) in the weights $\bw_p\in\mathbb{R}^{n_g}$. This closure function embeds unknown physics or terms that would render the prediction overly expensive. As we shall see in Section \ref{s5}, these parameters can be classic empirical coefficients (e.g. aerodynamic coefficients or heat/mass transfer coefficients) of a carefully crafted physics-based model of the system. The forward step of the system is driven by an ODE solver, which defines the function $\bm{s}_{k+1}=\mathcal{F}_s(\bm{s}_{k},\bm{a}_{k},\bm{z}_{k};\bm{p}_{k})$.
		
		The system identification consists in finding the optimal weights $\bw_p$ according to a cost function $\mathcal{J}_p(\bw_p)$ which measures how close the virtual trajectory ${S}=[{\bs}_1,\dots {\bs}_{n_t}]$ resembles the real one. The optimal weights $\bw_p$ minimize this function, defined as
		
		\begin{equation}
			\label{J_p}
			\mathcal{J}_p(\bw_p)=\mathbb{E}_{\sim S(\bm{z},\bm{w}_p)}\biggl \{\int^{T_o}_{0} \mathcal{L}_p \big(\check{\bs}(t_k),\bs(t;\bw_p)\big) \; \mathrm{dt}\,\biggr \},
		\end{equation} with $\mathcal{L}_p: \mathbb{R}^{n_s\times n_s}\rightarrow \mathbb{R}$ the Lagrangian function for the identification problem and the expectation $\mathbb{E}_{\sim S(\bm{z},\bm{w}_p)}\{\}$ over possible virtual trajectories accounting for the stochastic exogenous inputs. More specifically, starting from one sample signal $\bm{z}_k$, with $k=0,n_t-1$, we build a set of $N_z$ realizations of exogenous inputs as described in section \ref{s4p2} and we denote as $S(\bm{z},\bm{w}_p)$ the set of trajectories generated for a given sample of the exogenous input $\bm{z}$ and a given set of weights $\bw_p$. Two reasons motivate the choice for the integral formulation in \eqref{J_p} even if only a discrete set of samples $\hat{\bs}(t_k)$ is available. First, the integral is particularly handy in deriving the adjoint problem driving the computation of the gradient $d\mathcal{J}/d\bm{p}$, as presented in section \ref{s4p2}. Second, the choice keeps the approach independent from the specific numerical solver to advance the virtual states $\bs(t)$. We return to both points in section \ref{s4p3}. 
		We here stress that the closure model $g(\bs;\bw_g)$ makes the digital twin predictive only for a given sample of the exogenous disturbance signal and within $t\in[0,T_o]$, i.e. no forecasting is considered. 
		
		% These are maybe details for later ? TODO
		% we use large regression to approximate its large-scale contribution $\check{\bm{d}}_k$ and model the small-scale contribution $\check{\varepsilon}_k$ as a Gaussian process with zero mean and prescribed covariance. We 
		
		The model-based control seeks to find the optimal policy to keep the digital twin along the desired trajectory. We introduce the \emph{virtual error} $\bm{e}(t_k)=e(\bm{s}_k,\tilde{\bm{s}}_k)$ to measure the distance between the current virtual state and the target state and $\bm{a}_{mb} = \pi_{mb}(\bm{e};\bw_{a, mb})$ the \emph{virtual actions} taken on the digital twin. The performances of the controller are measured by the cost function $\mathcal{J}_a(\bm{w}_{a, mb})$. For convenience, we define this cost function similarly to the one in \eqref{J_p}, i.e.: 
		
		\begin{equation}
			\label{J_a}
			\mathcal{J}_a(\bw_{a,mb})=\mathbb{E}_{\sim S(\bm{z},\bw_{a,mb}) }\biggl \{\int^{T_o}_0 \mathcal{L}_a \big(\tilde{\bs}(t),\bs(t;\bw_{a,mb})\big) \; \mathrm{dt} \,\biggr \}\,,
		\end{equation} with $\mathcal{L}_a: \mathbb{R}^{n_s}\rightarrow \mathbb{R}$ the Lagrangian function for the model-based control problem. The same notation as in \eqref{J_p} applies. The optimal control weights for the model-based approach are those that minimize this function. The process of computing a policy update using the virtual environment (model-based loop) without interacting with the real system is here referred to as \textit{offline (policy) planning}.
		
		It is worth stressing that the weights $\bm{w}_{a,mb}$ minimizing $\mathcal{J}_a$ in \eqref{J_a} do not necessarily maximize $\mathcal{R}_c$ in \eqref{R_c}. These problems will not have the same solution unless the virtual environment closely follows the real one. The way these loops interact in the agent's training is described in the following section, along with more details on the computation of the gradient $d\mathcal{J}_p/d\bw_p $ driving the identification and the gradient $d\mathcal{J}_a/d\bw_{a, mb}$ driving the agent's training from the model-based side (see connections in dashed red in Figure \ref{Fig1}).
		
	\end{itemize}
	
	\begin{comment}
		$$\Delta \bm{w}_p$$
		$$\Delta_f \bm{w}_a$$
		$$\Delta_b \bm{w}_a$$
		
	\end{comment}
	
	%The expectation operator in \eqref{R_c} accounts for the impact of the stochastic contribution to the exogenous inputs. In the proposed framework, this is the only source of randomness as the system is treated as deterministic.
	
	% \mathbb{E}_{\sim \check{S}}

	% I put these for the figures
	
	%$$\bm{a}_k=\pi({e}_k,\check{\bm{z}}_k;\bm{w}_a)$$
	
	%$$\check{\bm{a}}_k=\pi(\check{e}_k,\check{\bm{z}}_k;\bm{w}_a)$$

	%It is worth noticing that the Lagrangian function here acts as a kernel function to measure the similarity between the real and virtual trajectories, \textcolor{blue}{potentially accounting for the fact that states are not available at the same time steps}. The integral formulation becomes particularly handy in deriving the adjoint problem driving the assimilation as presented in section \ref{sec_4_2}. \textcolor{blue}{To prevent model-bias, $M$ assimilation processes are generated, starting from $M$ different initial conditions.}

	\section{Mathematical Tools and Algorithms}\label{s4}
	
	We here detail the building blocks of the proposed approach. Only the general architecture is discussed, leaving customizations to specific environments to Section \ref{s5}. The model-free paradigm (connection of components 1-2-3-4) is illustrated in section \ref{s4p1}. This leverages standard Deep Reinforcement Learning (DRL) methodologies. The model-based paradigm (connections of components 1-2-3-5-6) is illustrated in section \ref{s4p2}. This combines adjoint-based nonlinear system identification and adjoint-based optimal control. The model-free and model-based loops could be used independently, and we first introduce their working principle independently. Then, section \ref{s4p3} details how these are connected and presents the proposed algorithm.

	\subsection{The Model-Free loop: (1)-(2)-(3)-(4)}\label{s4p1} %maybe try to have consistency between figure and steps indicated here / in the algorithm
	
	We consider a classic actor-critic reinforcement learning formalism (\citealt{sutton2018reinforcement}, \citealt{bhatnagar_natural_2009}). Although more recent variants exist, we here focus on the classic Deep Deterministic Policy Gradient (DDPG) algorithm proposed by \cite{lillicrap2019continuous} with the prioritized experience replay strategy by \cite{Schaul2015}. This algorithm offers a good compromise between performance and simplicity, and it has been proven successful in several flow control tasks \citep{ bucci2019control, Pino2023}.
	
	The DDPG trains two functions while interacting with the system. The first, referred to as \emph{actor}, aims to approximate the policy function $\pi_{mf}: \check{\bm{e}} \rightarrow \check{\ba}$, which provides the action to be taken at a certain state. This policy is deterministic, but we add noise generated by a random process to explore the action space. 
	
	The second, referred to as \emph{critic}, aims to approximate the state action value function \textcolor{blue}{$Q:(\check{\bm{e}},\check{\ba})\rightarrow \mathbb{R}$}; this provides the expected reward for taking action $\check{\ba}$ while the system is in the error state $\check{\bm{e}}$ and then following a certain policy. 
 
We use $Q$ to denote the (unknown) true $Q$ value and $q(\check{\bm{e}},\check{\ba};\bw_q)$ to denote the approximation from the $Q-$network. In line with the previous section and Figure \ref{Fig1}, the parameters $\bw_{a, mf}$ denote the weights and biases of the actor-network while $\bw_q$ denotes the weights and biases for the critic network. These are identified through an online optimization process, which is based on stochastic, momentum-accelerated gradient-based optimization as in classic deep learning (see \cite{goodfellow2016deep}). The gradient driving the optimization of the policy reads (see \cite{silver_deterministic_nodate}): 
	
	\begin{equation}
		\label{eq:policy_grad}
		\frac{d\mathcal{R}_c}{d\bw_a} = \mathbb{E}_{\sim \check{S}}\biggl \{ \frac{d q(\check{\bm{e}},\check{\ba};\bw_q)}{d\ba} \frac{d \pi(\check{\bm{e}};\bw_a)}{d\bw_a} \biggr\}\,,
	\end{equation} where $\mathcal{R}_c$ is the reward function in \eqref{R_c}.
	
	The expectation is evaluated on an ensemble of trajectories or transitions and is readily available using back-propagation on the actor and the critic networks. Its accuracy strongly relies on the accuracy of the critic network. The recursive nature of the state value action function allows for measuring the performances of the critic from the Bellman equation \citep{sutton2018reinforcement}. For the true $Q$ function, this reads:
	
	\begin{equation}
		\label{eq:Q_critic1}
		Q(\check{\bm{e}}_{k}, \check{\ba}_{k}) = \mathbb{E}_{\sim \bs_{k+1}} \biggl \{ r_k + \gamma Q (\check{\bm{e}}_{k+1}, \ba^{\pi}_{k+1})\biggr\}\,.
	\end{equation} where $r_k=r(\check{\bm{e}}_k, \check{\ba}_k)$. The expectation over possible future states accounts for the stochastic nature of the exogenous inputs, and $\bm{a}^{\pi}$ denote actions according to the current policy $\pi$ (i.e. $\check{\bm{a}}^{\pi}_k=\pi(\check{\bm{e}}_k;\bm{w}_a)$). The same recursive form should then exist for the q network. Therefore, defining 
	
	\begin{equation}
		\label{eq:y_t_critic}
		y_k = r_k + \gamma \; q(\check{\bm{e}}_{k+1}, \pi(\check{\bm{e}}_{k+1});\bw_q)
	\end{equation} the expected state-action value at step $k$, and as $\delta_k=y_k-q(\check{\bm{e}}_k,\check{\bm{a}}_k)$ the Temporal Difference (TD) error, a cost function for training the critic network can be written as 
	
	\begin{equation}
		\label{eq:Q_critic}
		\mathcal{J}_q(\bw_q)= \mathbb{E}_{\sim \bs_k, \bs_{k+1} } \bigl \{\delta_k^2 \bigr \}\,,
	\end{equation} where the expectation $\mathbb{E}_{\sim \bs_k, \bs_{k+1} }$ is computed over a set of transitions from one state to the following (regardless of their position in the system trajectory).
	
	The gradient of \eqref{eq:Q_critic} with respect to $\bw_q$ can be easily computed using back-propagation on the critic network. However, the training of the critic network is notoriously unstable \citep{Mnih2013,Mnih2015}. The two classic approaches to stabilize the learning consist in (1) using an under-relaxation in the update of the weights and (2) keeping track of a large number of transitions by using a replay buffer. The implementation of both methods is discussed in Section \ref{s4p3}. 
	
	Finally, the sampling from the buffer is carried out in the form of batches to evaluate the gradient $d \mathcal{J}_q / d\bw_q$. The sampling prioritizes transitions with the largest TD-error, $\delta_k$, \citep{schaul2016prioritized} building a non-zero probability distribution for each sample proportional to it, i.e.  $p_i = |\delta_i| + \mathrm{tol}$, with $\mathrm{tol}$ a small offset. This yields to the sampling probability of the $i-$th transition, $P(i)$ to be expressed as:
	
	\begin{equation}
		\label{eq:td_error}
		P(i) = \frac{p_i^{\alpha^*}}{\sum_k p_k^{\alpha^{*}}}\;,
	\end{equation}
	in which $\alpha^*$ sets the prioritization influence, with $\alpha^*=0$ implying uniform sampling.  
	
	\subsection{The Model-Based Loop (1)- (2)- (3)-(5)-(6) }\label{s4p2}
	
	The assimilation and model-based control problem in the loop (1)-(2)-(3)-(5)-(6) share the same mathematical framework. The dynamical system describing the evolution of the digital twin is written as
	
	\begin{equation}
		\label{sys}
		\begin{cases}
			\dot{\bs}&=f(\bs,\ba,\bm{z},\bp)\\
			\bm{s}(0)&=\bm{s}_0\\
			\ba&=\pi_{mb}(\bm{e},\bw_{a, mb})\, ;\, \bp=g(\bm{s},\bw_{p})\\
		\end{cases}\,,
	\end{equation} where $f:\mathbb{R}^{n_s\times n_a \times n_p\times n_z}\rightarrow \mathbb{R}^{n_s}$ is the flow map of the dynamical system and is here assumed to be known. 
	
	This function could be derived from first principles, as described in section \ref{s5} for the selected test cases, or using a general-purpose function approximator such as ANNs. The distinction is irrelevant to the method presented in this section, and we note in passing that using an ANN to approximate $f$ in combination with the training techniques described in this section leads to neural ODEs \citep{Chen2018}.
	
	The problem of \emph{training} the agent and the closure law consists in finding the optimal parameters $\bw_a$ and $\bw_p$, respectively, using continuously provided data. The training data consists of a target trajectory $\tilde{\bs}(t)$ for the control problem and measurements of the states $\check{\bs}(t)$ for the identification problem.
	The performances of the assimilating agent are measured according to \eqref{J_p} and \eqref{J_a}, respectively (see Figure \ref{Fig1}).
	
	These functions are interconnected because they depend on the states, which depend on both the closure variables and the actions (thus both on $\bw_a$ and $\bw_p$). However, we here consider these independently: $\bw_p$ should be identified to minimize $\mathcal{J}_p$ regardless of the performance of the controlling agent, and $\bw_a$ should be identified to minimize $\mathcal{J}_a$ regardless of the performances of the assimilating agent. This is why only the relevant functional dependency is considered in equations \eqref{J_p} and \eqref{J_a}.
	
	To maximize the sample-efficiency training strategy, the optimization of these functions is carried out using gradient-based optimization along with efficient computation of the gradients $d \mathcal{J}_p/d\bw_p$ and $d\mathcal{J}_a/d\bw_a$ using the adjoint method \citep{Errico,Bradley,Cao2003}. This method bypasses the need for computing the sensitivities $d\bs/d\bm{p}$ and $d\bs/d{\ba}$. Moreover, because of the similarity in the cost functions \eqref{J_p} and \eqref{J_a}, the adjoint problems share key terms. Concerning the gradient driving the system identification, the adjoint-based evaluation gives

	\begin{equation}
		\label{Exp_dpdJ}
		\frac{d\mathcal{J}_p}{d\bm{p}}=
		\mathbb{E}_{\sim S(\bm{z},\bm{w}_p)}\bigg\{\int^{T_o}_0 \bla^T_p(t)\frac{df}{d\bp} \frac{dg}{d\bm{w}_p} \; \mathrm{dt}\biggr \}\,,
	\end{equation} where the expectation operator is the same used for the cost function definition in \eqref{J_a}, the gradient $dg/d\bw_p$ is readily available from the closure definition and $\bla_p\in\mathbb{R}^{n_s}$ is the vector of Lagrange multipliers (co-states) computed from the terminal value problem:
	
	\begin{equation}
		\label{lambda_p}
		\begin{dcases}
			\dot{\bla}_p&=-\Bigg(\frac{df}{d \bs}\Bigg)^T \bla_p -\Bigg(\frac{d\mathcal{L}_p}{d \bs}\Bigg)^T\\
			\bla_p(T_o)&=\bm{0}\,.
		\end{dcases}\,
	\end{equation} 
	
	Similarly, the computation of the gradient $d\mathcal{J}_a/d\bp$ driving the training of the controlling agent is: 
	
	\begin{equation}
		\label{Exp_dadJ}
		\frac{d\mathcal{J}_a}{d\ba} =\mathbb{E}_{\sim S(\bm{z},\bm{w}_a)}\bigg\{\int^{T_o}_0 \bla^T_a(t)\frac{df}{d\ba}\,\frac{d\pi}{d\bm{w}_a}\; \mathrm{dt}\biggr \}\,,
	\end{equation} where the gradient $d\pi/d\bm{w}_a$ is available from the policy definition (e.g. via backpropagation if this is an ANN) and the evolution of the Lagrange multiplier $\bla_a (t)\in\mathbb{R}^{n_s}$ is determined by the terminal problem:
	
	\begin{equation}
		\label{lambda_a}
		\begin{dcases}
			\dot{\bla}_a&=-\Bigg(\frac{df}{d\bs}\Bigg)^T \bla_a -\Bigg(\frac{d\mathcal{L}_a}{d\bs} \Bigg)^T\\
			\bla_a(T_o)&=\bm{0}\,.
		\end{dcases}
	\end{equation} 
	
	In both problems, one has one adjoint state evolution $\bla_p(t)$ and $\bla_a(t)$ for each of the $N_z$ trajectories used to evaluate the expectation in \eqref{Exp_dpdJ}. The methodology to compute the population of trajectories is described in the remainder of this section. 
	
	Given a sample of the temporal evolution of the exogenous inputs $\bm{z}_k=\bm{z}(t_k)$, with $k=[0,n_t-1]$, we build an analytic approximation of its large-scale component $\bm{d}_k=\bm{d}(t_k)$ using Support Vector Regression (SVR, \cite{Smola2004}). Let $\boldsymbol{d}_k[j]\in\mathbb{R}$ denote the $j$-th component of the exogenous input sampled at time step $k$, with $j = [0, n_z - 1]$. We seek to derive a continuous function $\tilde{\bm{d}}[j](t)$ such that $\tilde{\bm{d}}[j](t_k)\approx \boldsymbol{d}_k[j]$. This is a regression problem that we tackle with support vector regression. Hence, the continuous function is written as 
	
	\begin{equation}
		\label{svr}
		\tilde{\bm{d}}[j](t;\boldsymbol{\alpha}[j])=\sum^{n_t-1}_{k=0} \kappa(t_k,t)\alpha[j]_k
	\end{equation} with $t\in\mathbb{R}$, $\kappa(t_1,t_2)=\exp(-\gamma ||t_1-t_2||^2_2)$ a Gaussian kernel and $\gamma$ a user defined scale parameter. We stress that $t\in[0, T_o]$ since no forecasting attempt is made. The vector of coefficients $\boldsymbol{\alpha}[j]=[\alpha[j]_1,\dots \alpha[j]_{n_t}]$ is computed by minimizing the cost function (see \citealt{Chang2011}) 
	
	\begin{equation}
		J(\boldsymbol{\alpha}[j])=\frac{1}{2} ||\boldsymbol{\alpha}[j]||^2_2+ C \biggl(\sum^{n_t-1}_{k=0} \varepsilon_j+\varepsilon^*_j \biggr)\,,
	\end{equation} subject to $\bm{z}[j](t_k)-\tilde{\bm{d}}[j](t_k,\boldsymbol{\alpha})<\epsilon+\varepsilon_j$, $\bm{z}[j](t_k)-\tilde{\bm{d}}[j](t_k,\boldsymbol{\alpha})>\epsilon+\varepsilon_j$ and $\varepsilon_j \varepsilon^*_j>0$ for all $j$ in $[1,n_t]$. The coefficients $C$ and $\epsilon$ are user-defined parameters controlling the regularization of the regression.
	
\begin{figure}
		\centering
		\includegraphics[width=.65\linewidth]{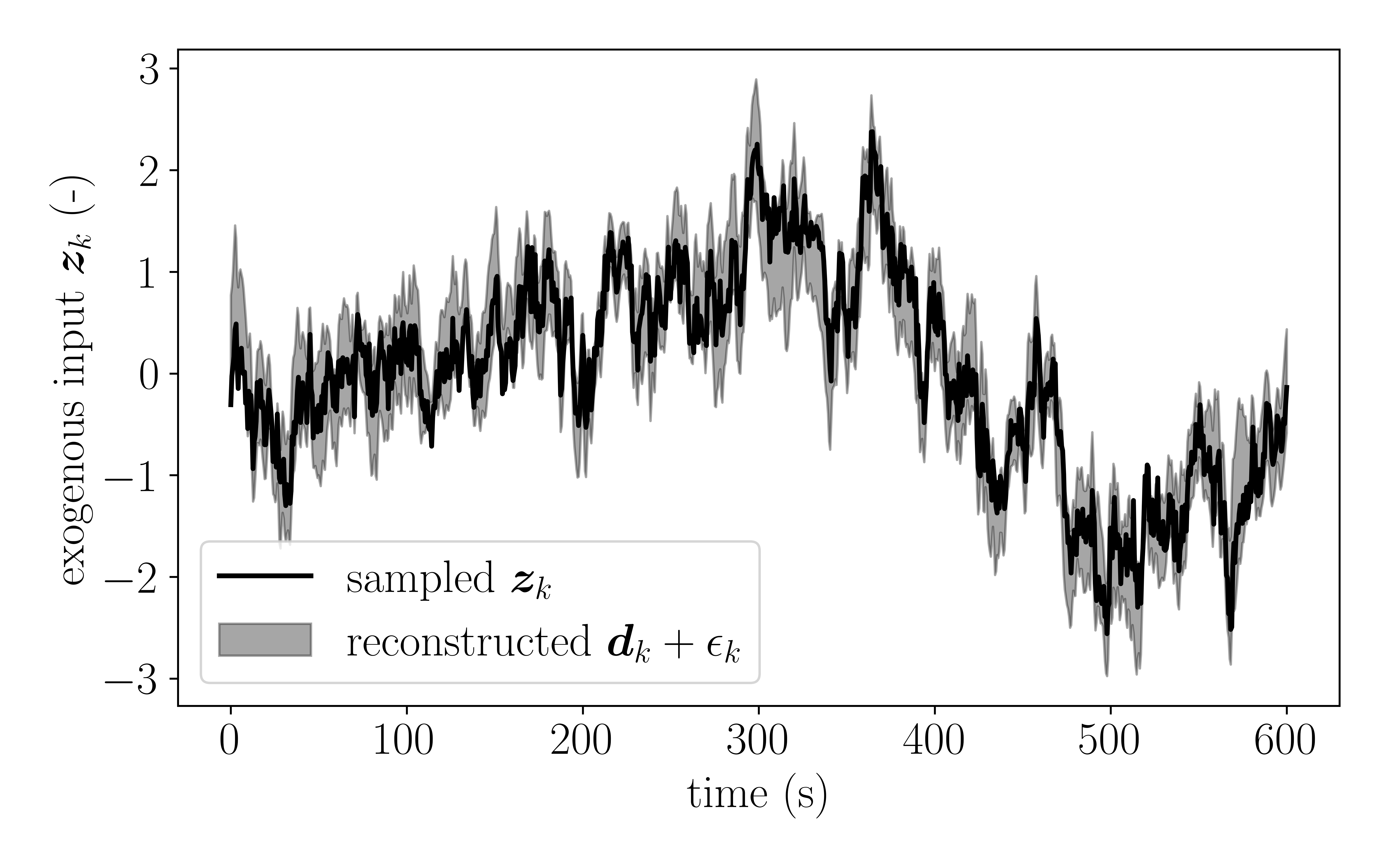}
		\caption{Example of SVR regression of sampled exogenous input. A filtered exogenous input is passed as an input of the SVR (continuous line) which outputs $N_z=10$ possible realizations.}
		\label{fig:svr_example}
	\end{figure}

	The regression model generates $N_z$ samples of possible exogenous disturbances. These are modelled as a set of $n_z$ Gaussian processes with mean $\tilde{\bm{d}}[j](t;\boldsymbol{\alpha}[j])$ and covariance matrices $\boldsymbol{\Sigma}_\varepsilon[j]$. We sample these processes $N_z$ times to obtain the exogenous disturbances in the virtual trajectories. An example of SVR regression is shown in Figure \ref{fig:svr_example} together with $N_z=10$ random samples of the modelled exogenous disturbance.
	
	\subsection{The proposed RT algorithm}\label{s4p3}
	
	The proposed algorithm is illustrated in \ref{alg:rsa}. It consists of six steps, repeated over $n_e$ episodes. These are described in the following.

	\begin{algorithm}[h!]
		\caption{Reinforcement Twinning (RT) Algorithm (Part 1)}
		\label{alg:rsa}
		\SetAlgoLined
		Initialize model-free $\pi_{mf}(\bw_{a, mf}^{(0)})$ and model-based policies $\pi_{mb}(\bw_{a, mb}^{(0)})$ and set live policy $\pi=\pi_{mf}$\\
		Initialize critic network with weights $\bw^{(0)}_q$ and $\bw^{(0)}_q$ and targets $\bw_q'$\\
		Initialize closure function's weights $\bw^{(0)}_p$\\
		Initialize long term memory buffer $\mathcal{D}_{\mathrm{L}}$ and short term memory $\mathcal{D}_{\mathrm{S}}$\\
		Initialize Step 6 counters and thresholds $C_{\pi}$, $P_{\pi}$, $C_{w}$, $P_{w}$, $T_J$ and $T_w$ \\
		\For{$i = 1, \dots, n_e$}{
			\Comment{\textcolor{blue}{Step 1: Run a real episode}} \textcolor{white}{I cheat to have the comment where I need it, because left comment c} 
			\For{$k = 1, ..., n_t$}{
				Sample $\xi(i)$; if $\xi(i)\neq0$, sample a random action $\bm{n}_k$ from $\mathcal{N}_n(\mu_a,\Sigma_a)$  \\
				Select action according to policy and exploration noise, $\check{\ba}_{k} = \pi(\check{\bm{e}}_{k} ; \bw^{(i)}_a) + \xi(i)\bm{n}_k$\\
				Act on real system, then collect reward $r_k$, states $\check{\bm{s}}_{k+1}$ and tracking error $\check{\bm{e}}_{k+1}$ \\
				Store error transitions $(\check{\bm{e}}_k, \check{\ba}_k, r_k, \check{\bm{e}}_{k+1})$ and associated TD errors $\delta_k$ in $\mathcal{D}_{\mathrm{L}}$\\ 
				Store state transitions $(\check{\bm{s}}_k,\check{\bm{z}}_k, \check{\ba}_k, {\bm{\check{s}}_{k+1}})$ in $\mathcal{D}_{\mathrm{S}}$\\ 
			}
			\Comment{\textcolor{blue}{Step 2: Update Critic and compute the model-free policy update}\textcolor{white}{a smaller cheating to have the s}} 
			\For{$n=1,\dots,n_Q$}{
				Sample $\textsc{min}(n_b,n_t)$ transitions from $\mathcal{D}_{\mathrm{L}}$ and compute $\mathcal{J}_q$ and $d\mathcal{J}_q/dq$ from \eqref{eq:Q_critic}\\
				Compute $q(\check{\bm{e}},\check{\bm{a}}_k)$ and the update $\Delta \bm{w}^{(i)}_q$, then update $\bm{w}^{(i)}_q\leftarrow\bm{w}^{(i)}_q+\Delta \bm{w}^{(i)}_q$ \\
				Use target for $\tau$ under-relaxation:  $\bm{w}^{(i)}_q\leftarrow \tau \bm{w}^{(i)}_q + (1-\tau) \bm{w}'_q$ and update target $\bm{w}'_q=\bm{w}^{(i)}_q$  \\ 
				Update the TD errors in the buffer $\mathcal{D}_{\mathrm{L}}$ for sampled transitions \\ 
			}
			\For{$n=1, \dots, n_A$}{
				Sample $\textsc{min}(n_b,n_t)$ transitions from $\mathcal{D}_{\mathrm{L}}$ \\
				Compute $\mathcal{R}_c(\bm{w}_a)$ and new set of weights $\bw_{a, mf}^{(n)} \leftarrow \bw_{a, mf}^{(n-1)} + \Delta_f \bm{w}_{a, mf}$ from $d\mathcal{R}_c /d\bw_{a, mf} $ in \eqref{eq:policy_grad}\\
				Update model-free actor policy: $\pi_{mf}^{(n)} = \pi_{mf}(\bw_{a, mf}^{(n)})$ \\
			}
			\Comment{\textcolor{blue}{Step 3: Prepare inner-episode regressors}\textcolor{white}{llllllllllllllllllllllllllllllllllllllllllllllllllllllllllllllllllllllllllleeellllll}} \\ 
			Use $n_S$ series of $\bm{z_k}$ from $\mathcal{D}_{\mathrm{S}}$ use \eqref{svr} to build $\mathcal{N}^{(l)}_z(\tilde{\bm{d}}\big[j\big](t;\boldsymbol{\alpha}[j]),\Sigma[j])$ with $j\in[1,n_z]$, $l\in[1,n_S]$\\
			\Comment{\textcolor{blue}{Step 4: Optimize digital twin on the $n_s\times N_Z$ virtual episodes with real actions }\textcolor{white}{more and co lll}} \\ 
			\For{$n=1,\dots,n_G$}{
				\For{$l=1,\dots n_S,\; \quad \mbox{with}$ $\quad g(\bm{s},\bm{w}_p^{(i)})$}{
					Forward $N_z$ virtual trajectories with $\bm{s}_0=\check{\bm{s}}^{(l)}_0$, $\bm{a}_k$ from $\check{\bm{a}}^{(l)}_k$, $\bm{z}\sim\mathcal{N}^{(l)}_z()$\\
					Compute $\mathcal{J}_p(\bm{w}_p^{(i)})$ in \eqref{J_p} solve $N_z$ adjoint problems for $\bla_p(t_k)$ for each member \\
					Compute $d\mathcal{J}_p/d\bp $ from the ensemble (see eq. \ref{Exp_dpdJ}) and the associated updates $\Delta \bm{w}^{(l)}_p$ \\
				}
				Compute the update $\bm{w}^{(i)}_p\leftarrow \bm{w}^{(i)}_p+\Delta \bm{w}^{(i)}_p$ \\ 
			}
			Save current optimal set of $\bw_p = \bw_p^*$ \\
			$\dots$ \\
		}
	\end{algorithm}
	
	\begin{algorithm}[h!]
		\setcounter{AlgoLine}{36}
		\SetAlgoLined
		\SetKwBlock{Begin}{\dots}{end}
		\Comment{\textcolor{blue}{Step 5: Run $N_z$ virtual episodes with current policy and compute model-based policy update }}
		\Begin{
			\For{$n=1, \dots n_{mb}\;\quad \mbox{with}$ $\quad g(\bm{s},\bm{w}_p^{*})$}{
				Forward $N_z$ virtual trajectories with $\bm{s}_0=\check{\bm{s}}^{(l)}_0$, $\bm{a}_k=\pi_{mb}(\bm{s}_k;\bm{w}^{(i)}_a)$, $\bm{z}\sim\mathcal{N}^{(l)}_z()$\\
				Compute $\mathcal{J}_a(\bm{w}_a^{(i)})$ in \eqref{J_a} and solve $N_z$ adjoint problems to get $\bla_a(t_k)$ for each member \\
				Compute $d \mathcal{J}_a /d\ba$ from the ensemble (see eq. \ref{Exp_dadJ})\\
				Compute the new set of weights $\bw_{a, mb}^{(n)} \leftarrow \bw_{a, mb}^{(n-1)} + \Delta \bm{w}^{(l)}_a$\\ 
				Update model-based actor policy $\pi_{mb}^{(n)} = \pi_{mb}(\bw_{a, mb}^{(n)})$ \\
			}
			Save current optimal set of $\bw_{a, mb} = \bw_{a, mb}^*$ \\
			\Comment{\textcolor{blue}{Step 6: Policy switching and parametrization cloning}\textcolor{white}{llllllllllllllllllllllllllllllllllllllllllllllllllllllllllllllllll}} \\ 
			Forward $N_z$ virtual trajectories with $\bm{s}_0=\check{\bm{s}}^{(l)}_0$, $\bm{a}_k = \pi_{\mathrm{live}}(\bm{s}_k,\bw_{a, \mathrm{live}})$, $\bm{z}\sim\mathcal{N}^{(l)}_z()$. Get \, $\bar{\mathcal{J}}_{a, \mathrm{live}}  $ \\ 
			Forward $N_z$ virtual trajectories with $\bm{s}_0=\check{\bm{s}}^{(l)}_0$, $\bm{a}_k = \pi_{\mathrm{idle}}(\bm{s}_k,\bw_{a, \mathrm{idle}})$, $\bm{z}\sim\mathcal{N}^{(l)}_z()$. Get \, $\bar{\mathcal{J}}_{a, \mathrm{idle}} $ \\ 
			%\pm \sigma \mathcal{J}_{a, mf}
			Policy switching and weight cloning based on the decision tree (Figure \ref{decisionTree}) \\
		}
		\caption{Reinforcement Twinning (RT) Algorithm (Part 2)}
	\end{algorithm}

	\begin{itemize}  
		\item \textbf{Step 1: Interaction with the real environment (lines 7-13)}. A sequence of $n_t$ interactions with the real system takes place within an observation time $T_o$. 
		These episodes are denoted as \emph{real} as opposed to the \emph{virtual} ones carried out on the digital twin. These interactions may have an exploratory phase, carried out by adding a random component ($\bm{n}_k$) sampled from a random process. We here use a Gaussian process with mean $\bm{\mu}_a$ and covariance matrix $\bm{\Sigma}_a$ (line 6). The balance between exploration and exploitation is controlled by the sequence $\xi(i)$, which tends to zero as exploration ends. This sequence could be linked to the performance of the digital twin assimilation, but we leave it as a user-defined sequence for the purposes of this work. 
		During this phase, transitions are stored in the long-term memory buffer $\mathcal{D}_{\mathrm{L}}$ and the short-term memory buffer $\mathcal{D}_{\mathrm{S}}$. The first is used to update the critic network and the model-free policy and contains transitions ranked by their TD error $\delta_k$ (see eq. \ref{eq:Q_critic}), which is updated as the training of the critic progresses. The second is used for training the digital twin and contains transitions ordered in time and arranged in trajectories. Therefore, $\mathcal{D}_{\mathrm{L}}$ contains $n_L$ (random) transitions while $\mathcal{D}_{\mathrm{S}}$ collects $n_S$ trajectories. The TD error defines both the sampling and the cleaning of the $\mathcal{D}_{\mathrm{L}}$ buffer. The triangular distribution in \eqref{eq:td_error} is used both for sampling and cleaning: the transitions with the lowest probability of being sampled are also the ones continuously replaced by the new ones.
		On the other hand, no specific criteria are considered for the sampling/cleaning of the buffer $\mathcal{D}_{\mathrm{S}}$, which are simply sampled and stored in chronological order. 
		
		\item \textbf{ Step 2: Critic Update and model-free policy update (lines 14-24)}. An optimization with $n_Q$ iterations is carried out on the critic network, followed by an optimization with $n_A$ iterations for the policy. Both use batches of $n_b$ transitions sampled from $\mathcal{D}_{\mathrm{L}}$. 
		Concerning the training of the critic network (lines 14-19), this loss function and its gradient are computed from \eqref{eq:Q_critic}. This step is essentially a training of the critic network in a supervised learning formalism. Similarly to what was proposed by \cite{lillicrap2019continuous} for the original DDPG, the update of the critic is modulated by an under-relaxation $\tau$  using a target network (line 17). At every $n_Q$ iteration, the optimized critic re-evaluates the TD errors of the sampled transitions in $\mathcal{D}_{\mathrm{L}}$ (line 18). Limiting the TD updates for the current batch of trajectories allows for a pseudo-random update of the most critical transitions, according to the critic. In addition, this approach is also computationally cheaper than re-evaluating all the stored transitions in the buffer, usually in the order $\mathcal{O}(10^6)$. Successively, the critic is then used to train the policy in lines 20-24.
		In this second stage, the gradient with respect to the cost function is computed using eq \eqref{eq:policy_grad}. Note that we use the notation $\Delta \bm{w}$ to denote the optimization update for the weights $\bm{w}$, such that the updating reads $\bm{w}\leftarrow \bm{w}+\Delta \bm{w}$. Therefore, the update in a simple gradient descent over a cost function $J(\bm{w})$, for example, reads $\Delta \bm{w}=-\eta \; dJ(\bm{w})/d\bm{w}$. This makes the notation independent of the choice of the optimizer. Differently from the original DDPG algorithm in \cite{lillicrap2019continuous}, the updates of the critic and the policy networks are carried out via two separate loops rather than in a single one. In our numerical experiments, we found this approach to lead to more stable model-free policy updates.
		
		\item \textbf{ Step 3: Fit random processes for exogenous inputs (line 26)}. In principle, both the forward and backward integration of the ODEs driving the digital twin could be carried out using time steps that differ from the ones controlling the interactions with the real system. This would require some interpolation on the exogenous disturbances as well as on the states and the actions. The data storage between forward/backward evaluations could leverage modern checkpointing techniques (see \citealt{Zhang2023}). However, keeping the focus on the first proof of concept of the reinforcement twinning idea, we here assume that the numerical integration (both forward and backward) is carried out using the same time stepping from the interaction with the environment (step 1).
		
		Nevertheless, we use the SVR regression on the exogenous input (see \ref{svr}) to construct the set of random processes (here treated as Gaussian processes) from which the ensemble of possible realizations of the (stochastic) exogenous input will be sampled. These are used for ensemble averaging of the gradient over possible trajectories. For each of the trajectory $l\in [1,n_S]$ available in the short-term memory buffer $\mathcal{D}_{\mathrm{S}}$, we denote as $\bm{z}^{(l)}\sim\mathcal{N}^{(l)}_z()$ a sample of the Gaussian processes generating the evolution of possible exogenous inputs. 
		
		\item \textbf{ Step 4: Optimize digital twin on the real actions (lines 27-36).} The goal of this step is to find the closure parameters $\bm{w}_p$ such that the digital twin's prediction matches with the $n_S$ trajectories of the real system, stored in $\mathcal{D}_{\mathrm{S}}$. Consequently, the loss function $\mathcal{J}_p$ in \eqref{J_p} is minimized using a gradient-based optimization loop. This consists of $n_G$ iterations and computes the updates $\Delta \bm{w}_p$ from the gradient evaluated along the $n_S$ trajectories available in $\mathcal{D}_{\mathrm{S}}$. Again, fixing the numerical time stepping to the one of the real interaction avoids the need for action interpolation but might become problematic for stiff problems. For each trajectory (loop in lines 27-33), a total of $N_z$ possible realizations of the exogenous input are considered (sampled from the $\mathcal{N}^{(l)}_z$ processes built in the previous step), and the gradient is computed using the adjoint method (see eq. \ref{Exp_dpdJ}). The associated updates are then averaged (line 32), and the optimization step is carried out (line 34). Each optimization step requires a total of $n_S \times N_Z$ forward and backward virtual episodes. It is worth highlighting that the ensemble averaging at this step would be essential in cases for which (1) the system is chaotic and highly sensitive to random disturbances (hence producing "largely different" trajectories for "similar" inputs) or (2) the policy becomes an exploratory/identification policy rather than a control one (hence the goal is not to keep a trajectory but explore as much as possible the state-action space). On the other hand, if the action is solely driven by the goal of trajectory tracking control and the system is inherently a noise-rejection system, the interest in computing many trajectories and then averaging the associated gradients is lost. The presented test cases fall within this second category; future work will consider cases under the first category. 
		
		\item \textbf{ Step 5: Model-based policy update. (lines 37-43)}. This step is structurally similar to the previous: it computes the ensemble-averaged adjoint-based gradient (line 41) of a cost function ($\mathcal{J}_a$) and performs a number ($n_{mb}$) of gradient-based optimization steps (line 43). In this case, however, the interactions with the digital twin follow the model-based policy $\pi_{mb}$ instead of the actions stored in $\mathcal{D}_{\mathrm{S}}$. This allows to evaluate the current set of model-based policy weights $\bm{w}_{a,mb}$. The response of the digital twin is governed by the closure law defined by the current best set of weights, i.e. $\bm{p}=g(\bs, \bw_p^*)$. 
		
		\item \textbf{ Step 6: Policy switching and parametrization cloning. (lines 44-47)}.  This step decides which of the policies $\pi_{mb}$ or $\pi_{mf}$ becomes ``live" and which one becomes ``idle". The first is deployed in the next interaction with the real system, while the second continues to be trained with different mechanisms. The decision in this step is based on the policy performances in the virtual environment, evaluated and averaged over $N$ episodes with randomized initial conditions and disturbances. The flow chart for this step is illustrated in Figure \ref{decisionTree}. Defining as $\bar{\mathcal{J}}_{a, \mathrm{live}}$ and $\bar{\mathcal{J}}_{a, \mathrm{idle}}$ the averaged cost functions for the ``live" and ``idle" policies, we introduce a tolerance factor $T_J\in\mathbb{R}$, a threshold $P_\pi\in\mathbb{N}$, and a counter $C_\pi\in \mathbb{N}$ to avoid overly impulsive decisions. If the live policy performs worse than the idle one more than the tolerance factor ($\bar{\mathcal{J}}_{a, \mathrm{live}} /  \bar{\mathcal{J}}_{a, \mathrm{idle}} < T_J$), the counter $C_\pi$ gets incremented. If $C_\pi$ exceeds the threshold $P_\pi$, the ``idle'' policy takes over and becomes ``live''. On the other hand, if the opposite is true, we check that the idle policy is not stuck in a local minima, by analyzing its weight variation:
  
		\begin{equation}
			\label{eq:dweights_}
			d\bw_{a} = \frac{\| \bw_{a}^{i} -\bw_{a}^{i-1}\|}{\| \bw_{a}^{i} \|} \,,
		\end{equation}
	
       where $i$ indicates the current number of the episode ($i \in [1,n_e]$). To this end, we introduce a tolerance $T_w$, a counter $C_w$ and a threshold $P_w$ and repeat the previous decision process: if $d\bw_a$ falls below $T_w$, $C_w$ increases. If $C_w$ goes beyond $P_w$, it is reasonable to assume that the policy optimization step is trapped in a local minimum, and we replace the ``idle" parametrization by cloning the ``live" one. 

	\end{itemize}
	
	\noindent
	We close with a few general remarks on the proposed algorithm.  
	
	\paragraph{Remark 1} The $n_S$ hyperparameter in Steps 4 and 5 defines how much the digital twin should consider past experiences. In this work, we set $n_S$=1, thus focusing on the last trajectory solely, while having a variable $N_z$ to promote robustness against randomness on the exogenous inputs and to take into account possible sampling errors which may occur in real-world scenarios. 
	
	\paragraph{Remark 2} The idle policy continues its training through different mechanisms. No special considerations are needed if the idle policy is the model-based policy: this continues to be trained on the digital twin regardless of whether it is active on the real system or not. On the other hand, if the idle policy is model-free one, the learning occurs indirectly through two mechanisms that establish a ``demonstration'' from the model-based to the model-free: (1) the storage of ``expert" transitions in the memory buffer $\mathcal{D}_{\mathrm{L}}$ and (2) the continuous training of the value function Q. The off-policy nature of the reinforcement learning algorithm here implemented allows the agent to learn from policies that are exploratory or simply suboptimal according to the current Q function. In this case, when the model-free policy is idle, and the Q function continues to be updated by observing the model-based policy in action on the real system-- a sort of ``demonstration" mechanism is established. The fact that the model-free could learn by the demonstration and eventually overtake the lead as a live policy is arguably the most significant contribution of this work.

	\paragraph{Remark 3} The model-free update (Step 2), the model-based update (Steps 3,4,5), and the individual evaluations occurring in (Step 6) are independent processes and can be parallelized, greatly reducing computational time.
	
	% put a clear page to have a better impagination next 
	\begin{figure*}[h!]
		\centering
		\includegraphics[width=0.45\textwidth]{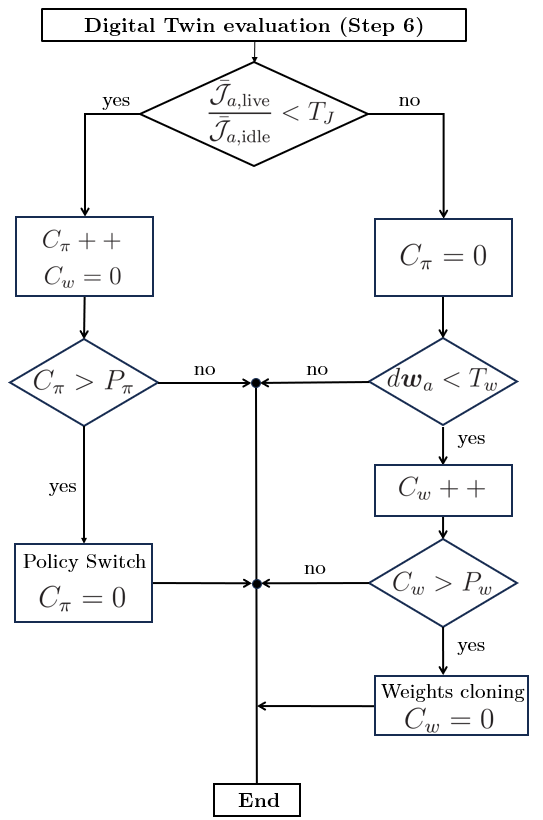}
		\caption{Decision tree in step 6 of the Reinforcement Twinning algorithm (see \ref{alg:rsa}. This step defines 1) which of the model-free or model-based policy becomes live and idle and 2) whether one of the two is consistently under-performing and is thus replaced by a clone of the other. }
		\label{decisionTree}
	\end{figure*}

	\section{Selected Test Cases}\label{s5}
	
	We present the three selected test cases: (1) the control of a wind turbine subject to time-varying wind speed (Section \ref{sec:testcase_1}), (2) the trajectory control of flapping-wing micro air vehicles (FWMAVs) subject to wind gusts (Section \ref{sec:testcase_2}) and (3) the mitigation of thermal loads in the management of cryogenic storage tanks (Section \ref{sec:testcase_3}). These are very simplified versions of the real problems, and the reader is referred to the specialized literature for more details.

	\subsection{Wind Turbine Control}\label{sec:testcase_1}
	
	\paragraph{\textbf{Test case description and control problem}} 
	
	We consider the problem of regulating the rotor's angular speed $\omega_r$ and the extracted power $P_e$ of a variable speed variable pitch horizontal turbine of radius $R$, subject to time-varying wind velocity $u_\infty$ (see \citealt{Laks2009,Pao2009}). 
	
\begin{figure}[h!]
		\centering 
		\begin{subfigure}[b]{0.32\linewidth}
			\centering
			\includegraphics[width=\linewidth]{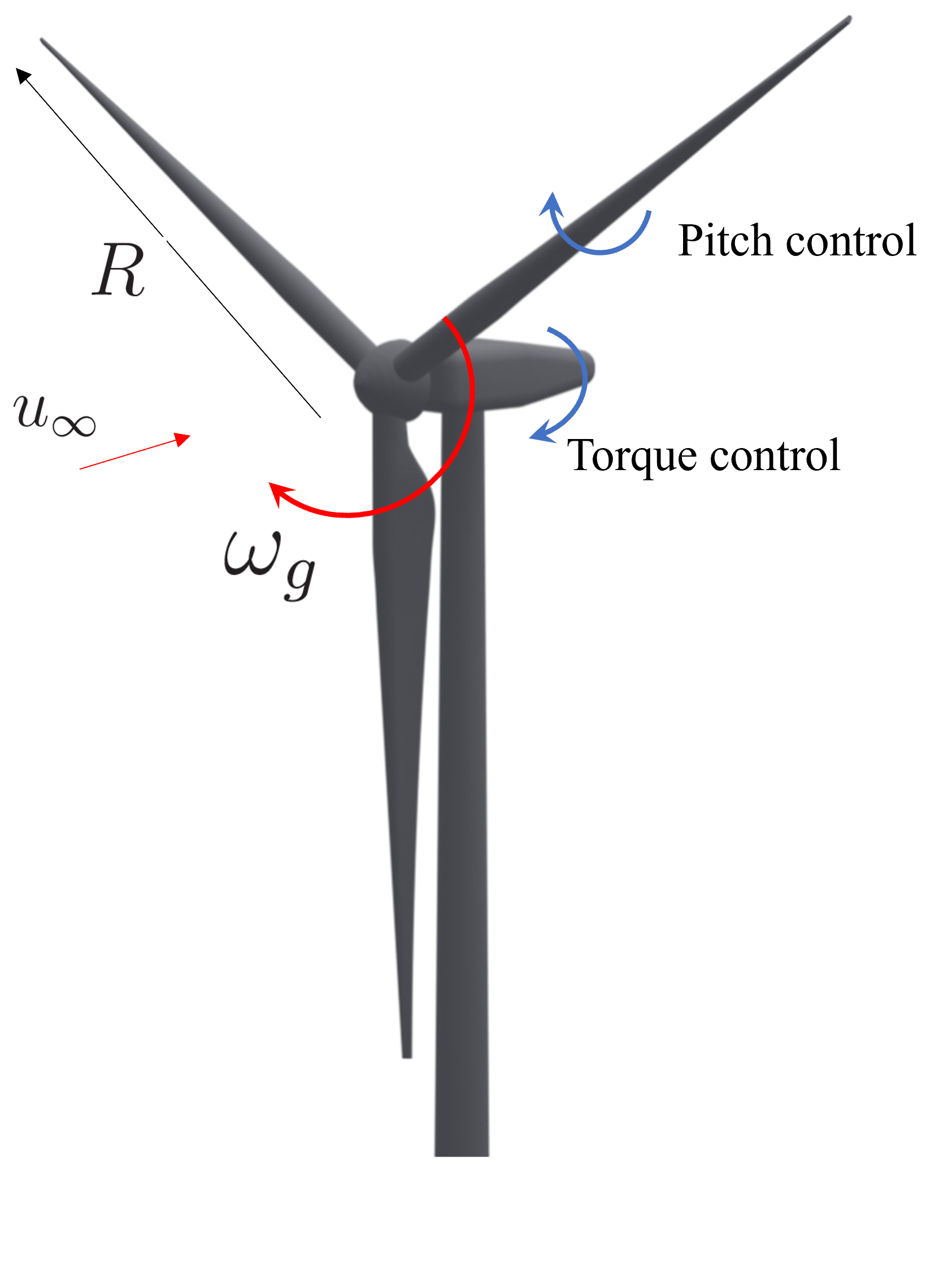}
			\caption{}
			\label{fig:wind_turbine_sketch}
		\end{subfigure}
		\hspace{2cm}
		\begin{subfigure}[b]{0.45\linewidth}
			\centering
			\includegraphics[width=\linewidth]{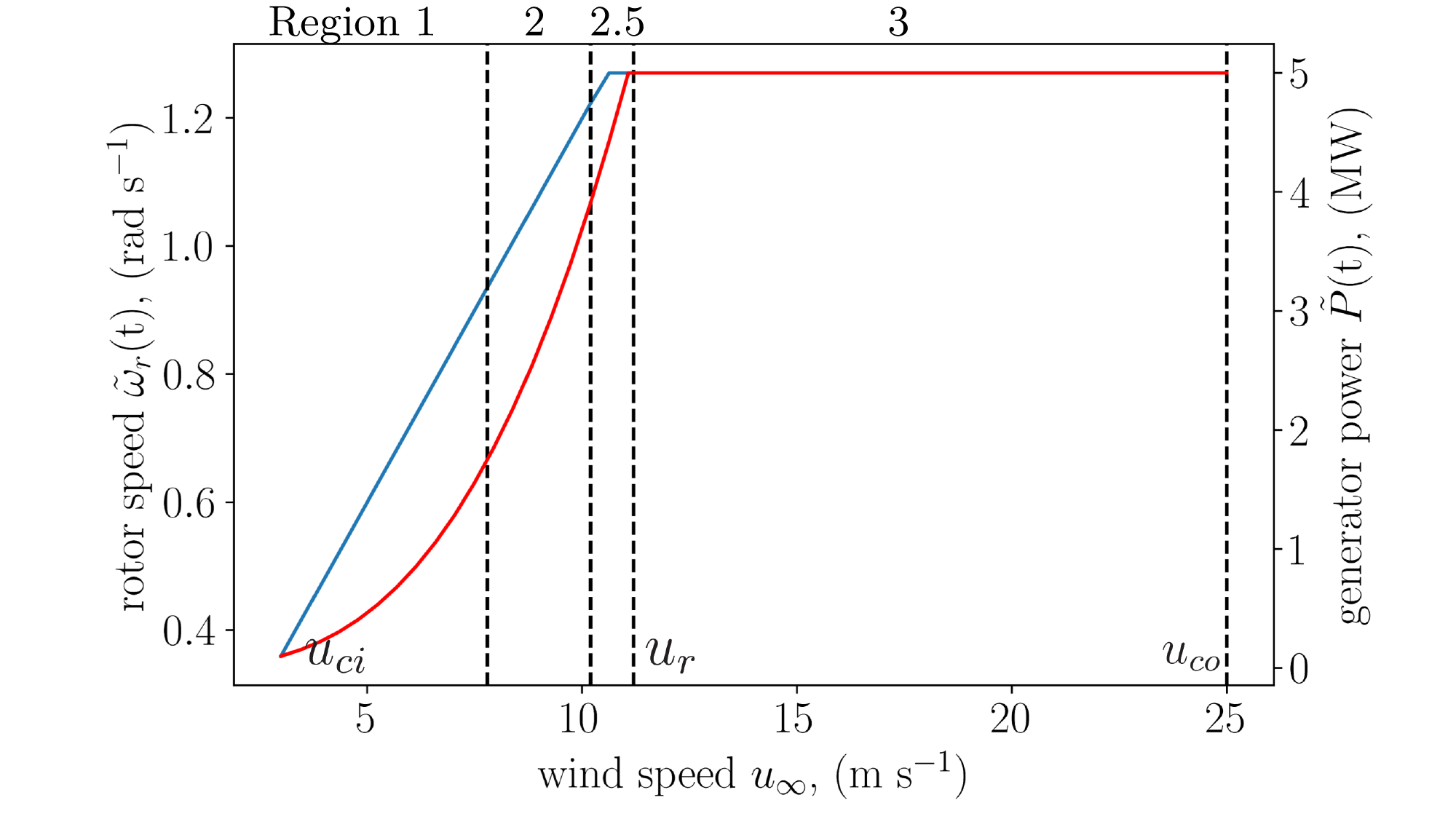}
			\caption{}
			\label{fig:ideal_power_curvee}
		\end{subfigure}
		\caption{Sketch of the main parameters involved in wind turbine control (\ref{fig:wind_turbine_sketch}) and reference power curve (\ref{fig:ideal_power_curvee}), $\tilde{P}$ (in red), and rotor speed $\tilde{\omega}_r$ (in blue).}
		\label{fig:ideal_power_curve}
	\end{figure}

	% Explain the control pilosphy.
	The operation (and thus the control objectives) of a wind turbine depends on the wind speed in relation to three regions, identified by three velocities: cut-in ($u_{ci}$), rated ($u_r$) and cut-out ($u_{co}$) wind speeds. These are shown in Figure \ref{fig:ideal_power_curve} for the NREL 5-MW turbine considered in this test case (details in the following subsection).
	In region 1 ($u_\infty<u_{ci}$), the wind speed is too low to justify the turbine operation and no power is produced. In region 2 ($u_{ci}<u_\infty<u_r$), the goal is to maximize the energy production. This is usually achieved by pitching the blades at the optimal value $\beta^*$ and acting on the generator torque to keep the optimal rotational speed. In region 3 ($u_\infty>u_r$), the turbine is in above-rated conditions, and the goal is to keep the power extraction at its nominal value $P_N$, thus reducing the aerodynamic load acting on the rotor. This is usually achieved by keeping the nominal generator torque and acting on $\beta$. Finally, the wind turbine is shut down if $u_\infty > u_{co}$, preventing mechanical and electrical overloading.
	
	% Explain the test case that you are considering
	Particularly critical is the transition between regions 2 and 3 (region 2.5 in Fig.\ref{fig:ideal_power_curve},) where gusts or lulls could result in conflicting requirements between torque and pitch controllers. This region is typically handled via set point smoothing, mitigating transients when switching from one control logic to another \citep{bianchi2007wind, abbas2020update}. 
	% Then the control problem:
	\par In this test case, we combine a Maximum Power Point Tracking (MPPT) controller in below-rated conditions \citep{bhowmik_performance_1999, howlader_parameter_2010} with a regulation control in above-rated conditions. The controller can act on both the generator torque and the pitch angle, receiving as sole input the generator speed $\omega_g$, as typically done in multi-megawatt wind turbines \citep{jonkman2009definition}. The current wind speed $u_\infty$ acts as exogenous input, which we assume can be provided by real-time measurements. We thus collect a sequence $u_\infty(t_k)$ with $t_k=k\Delta t$. To avoid high-frequency excitation of the system, we filter the generator speed signal using an exponential smoothing \citep{Zheng2022} with the recursive relations:
	
	\begin{equation}
		\label{etilde}
		\hat{y}(t_k)=(1-\hat{\alpha}) y(t_k) + \hat{\alpha}\cdot y(t_{k-1})\,,
	\end{equation} with $\alpha$ being the low-pass filter coefficient computed as $\hat{\alpha} =\exp(-2\pi f_{c} \Delta t)$, with $f_{c}$ the minimum cut-off frequency, $y$ the generic raw signal and $\hat{y}$ its low-pass filtered version.
	
	We employ this filter for the generator rotational speed, $\omega_g$ and the measured wind speed, $u_\infty$. From the latter, it is possible to define the target rotational speed $\tilde{\omega}_g$ and extracted power $\tilde{P}$ from the reference curves in Figure \ref{fig:ideal_power_curve}. In Region 2, these two variables are linked by the operating conditions, while in Region 3, these are defined by mechanical constraints. In the below-rated conditions, the power is maximized if the rotor operates at its optimal tip-speed-ratio $\lambda_{r, *}= \omega_r R / u_\infty$. Thus, one can define the reference generator speed relying on the measured wind speed as
 
	\begin{equation}
		\label{eq:omega_ref}
		\tilde{\omega}_g = N_g \frac{\lambda_{r, *} \hat{u}_{\infty}}{R}\,,
	\end{equation} where $N_g$ is the gearbox ratio. This variable is constrained in the range $\omega_{g, min} \le \tilde{\omega}_g \le \omega_{g, \mathrm{rated}}$, taking into account the machine specifications.
	Therefore, given $\tilde{\omega}_g$ the target rotational velocity, we define the tracking errors as ${e}_1(t)=(\check{\omega}_g - \tilde{\omega}_{g})/\gamma_1$, and its integral over the past rotor revolution, ${e}_2(t) = (\int_{t_i - T_r}^{t_i} e_1(t')\; \mathrm{dt}')/\gamma_2$, where $\boldsymbol{\gamma} = [\gamma_1, \gamma_2]$ ensure that both inputs lie in [-1, 1], and consider these as the inputs of the control logic. We then define the actuation vector as
	
	\begin{equation}
		\bm{a}(\bm{e}(t),\bw_a)= \begin{bmatrix}\tau_g \\\beta  \end{bmatrix} 
		=
		\begin{cases}
			\mbox{clip}\bigl(w_{a1} e_1(t) + w_{a2}e_2(t),\tau_{g,min},\tau_{g,max})\\%, \dot{\tau}_{g, max}
			\mbox{clip}\bigl(w_{a3} e_1(t) +w_{a4}e_2(t),\beta_{min},\beta_{max})\\ %, \dot{\beta}_{max})
		\end{cases}
	\end{equation} where $\mbox{clip}(x,x_1,x_2)$ is a smooth step returning $x_1$ if $x<x_1$, $x$ if $x_1<x<x_2$ and $x_2$ if $x>x_2$. In this work, we define it as 
 
	\begin{equation}
		\label{clip}
		\mbox{clip}(x,x_1,x_2)=\frac{1}{2}(\mbox{tanh}(x)+1)(x_2-x_1)+x_1\,,
	\end{equation} though any differentiable smooth step fits the purposes of bounding the action space while ensuring differentiability. This allows to limit the actuation rate of change according to the actuators' specifications, $\dot{\tau}_{g, max}$ and $\dot{\beta}_{max})$. We further assume that the controller can be informed by measurements of the turbine's rotational velocity $\check{\omega}$ and the generator torque $\check{\tau}_g$. It is thus possible to define a tracking error on the power as $e_P=\check{\tau}_g \cdot \omega(t;\bw_a) - \tilde{P}$ and use this signal to measure the controller's performance. The reference power $\tilde{P}$ follows the $K\omega^2$-law \citep{bossanyi_design_2000} in under-rated conditions, setting
	
	\begin{equation}
		\label{eq:tau_g_squared}
		\tau_g (t) = K \omega_g(t)^2 \quad \quad K = \frac{\pi \rho R^5 C_{p, max}}{2 \lambda_{r, *}^3 N_g^3} \,,
	\end{equation}	with
	
	\begin{equation}
		\label{eq:P_ref}
		\tilde{P} = K \omega_g(t)^3 \,.
	\end{equation}
	
	The target power production, also accounting for Region 3, is this  $\tilde{P} = \textsc{min}(P_N, K \omega_g(t)^3)$. Combining these two references, the cost function driving the model-based optimal control law is
	
	\begin{equation}
		\label{eq:control_functional_mb}
		\mathcal{J}_{a} = \frac{\alpha_1}{2 \; T_o} \int_0^{T_o} e_1(t)^2 \; dt + \frac{\alpha_2}{2 \; T_o} \int_0^{T_o} e_P(t)^2 \; dt \,,
	\end{equation} while for the model-free counterpart, we define the reward as 
	
	\begin{equation}
		\label{eq:control_functional_mf}
		\mathcal{R}_a = - \frac{\alpha_1}{2 \; n_t} \sum^{n_t-1}_{k=0} e_1^2(t_k) - \frac{\alpha_2}{2 \; n_t} \sum^{n_t-1}_{k=0} e^2_P(t_k).
	\end{equation}
	
	In both equations, $\alpha_1$ and $\alpha_2$ ensure that the two components of the cost function have comparable orders of magnitude. The choice of including both the velocity \emph{and} the power extraction is justified by the fact that these parameters are independent, in the sense that one could extract the same amount of power with different generator velocities by acting on the torque, possibly violating mechanical constraints.
	
	\paragraph{\textbf{Selected conditions and environment simulator}} 
	
\begin{figure*}
		\centering
		\begin{subfigure}[b]{0.48\linewidth}
			% \centering
			\includegraphics[width=\linewidth]{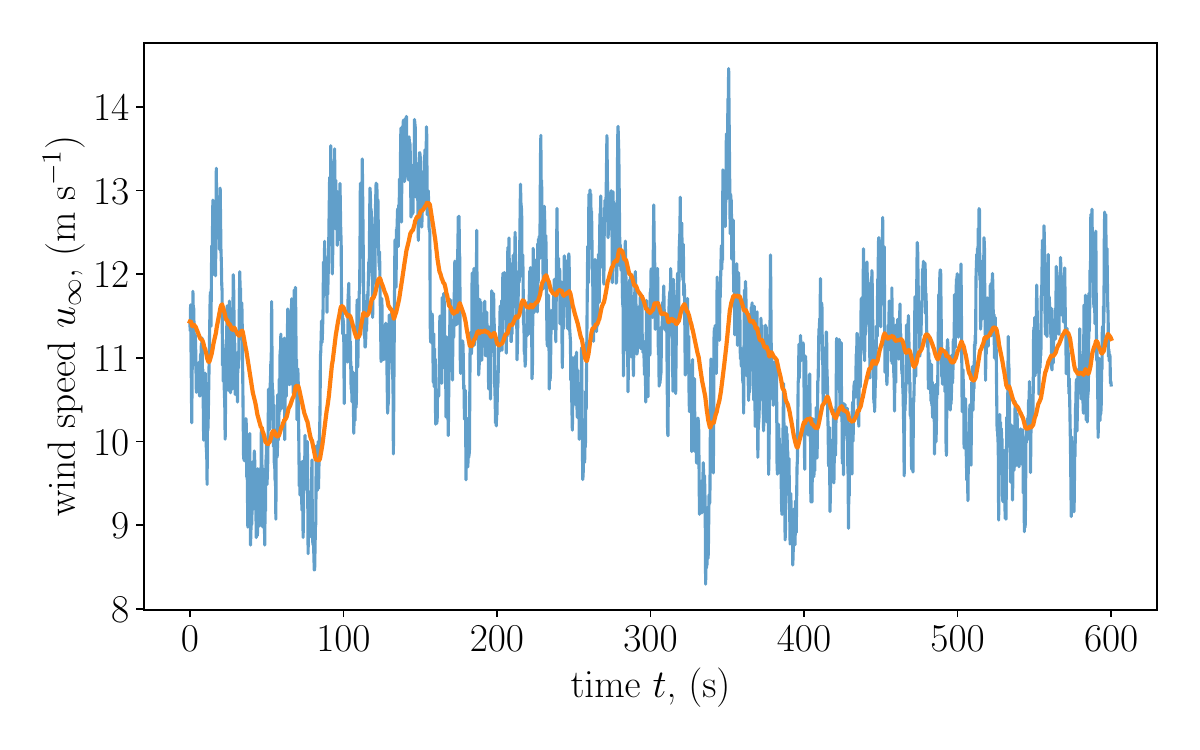}
			%\caption{Sample wind speed}
			\caption{}
			\label{fig:example_wind}
		\end{subfigure}
		\hfill
		\begin{subfigure}[b]{0.42\linewidth}
			% \centering
			\includegraphics[width=\linewidth]{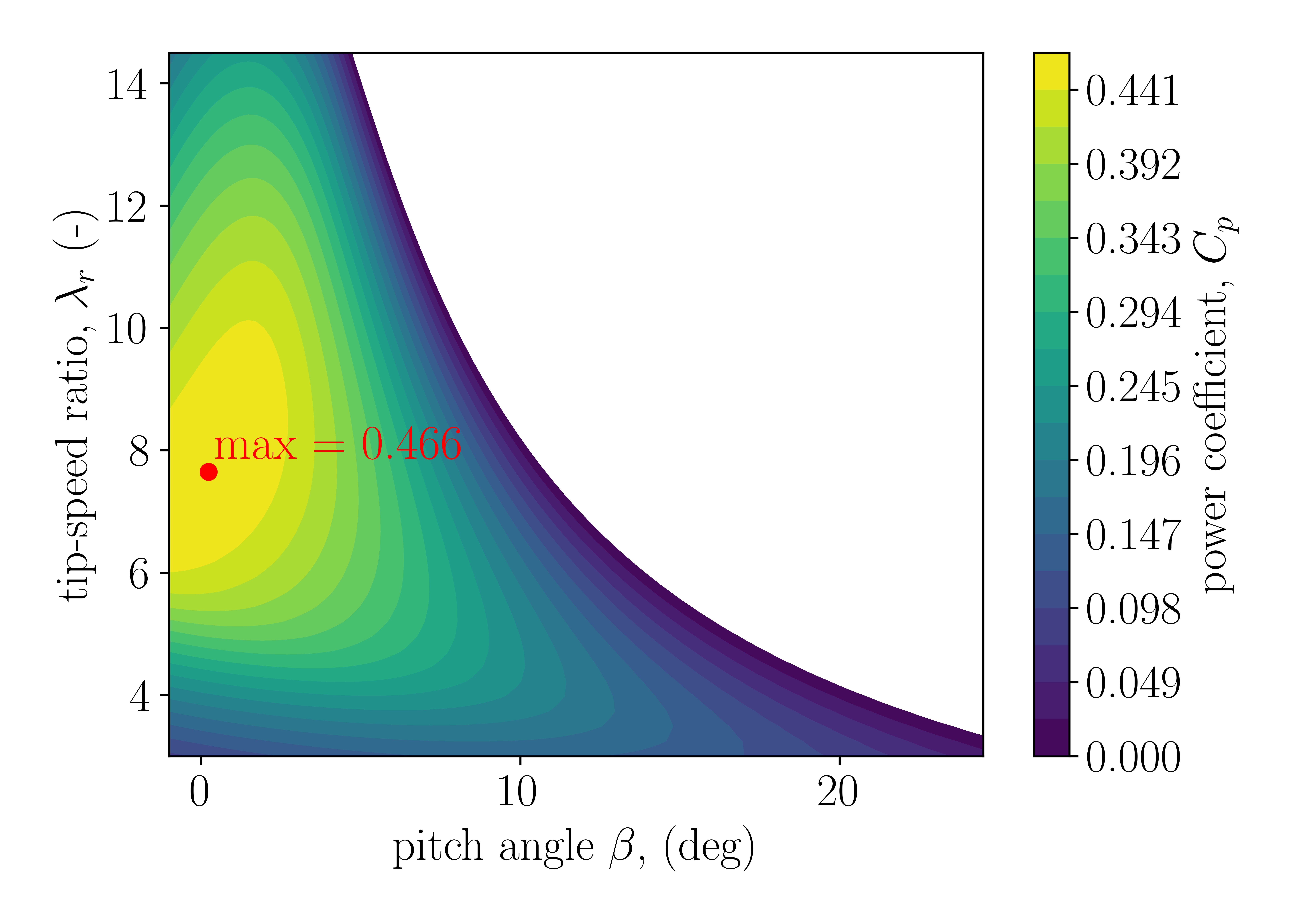}
			% \caption{Power coefficient of the NREL 5 MW}
			\caption{}
			\label{fig:cp_coeff_nrel}
		\end{subfigure}
		\caption{ Fig \ref{fig:example_wind}): sample of the wind speed (exogenous input) considered for the test case 1, unfiltered (in blue) and filtered (orange). Fig \ref{fig:cp_coeff_nrel} shows the contour of the power coefficient $C_p(\lambda_r,\beta)$ for the NREL 5MW turbine considered for control purposes. The controller's objective is to keep the turbine near the maximum ($C_p=0.46$) acting on the motor torque or the blade pitching. }
		\label{fig:overview}
	\end{figure*}

	Disregarding aeroelastic effects on both the blades and the tower, the simplest model of the rotor dynamics is provided by the angular momentum balance, which leads to a first-order system for the turbine's rotor's angular speed $\omega_r$: 
	
	\begin{equation}
		\label{eq:first_order_wt}
		\dot{\omega}_r = \frac{1}{J} \big(\tau_a - \tau_g N_g) \,,
	\end{equation} where $J$ is the rotor's moment of inertia, $\tau_a$ is the aerodynamic torque, $\tau_g$ is the opposing torque due to the generator and the gear train, and $N_g$ the gearbox ratio. The gearbox efficiency is assumed to be unitary. The aerodynamic torque is linked to the wind velocity as 
	
	\begin{equation}
		\label{eq:tau_a}
		\tau_a = \frac{1}{2} \rho A_r \; \frac{C_p( \lambda_r, \beta)}{\omega_r} \; u_\infty^3\,,
	\end{equation} where $\rho$ is the air density, $A_r=\pi R^2$ the turbine's swept area, and $C_p=P_e/(1/2 \rho A u^3_\infty)$ is the power coefficient, that is the ratio between the extracted power $P_e$ and the available one. For a given turbine design, this coefficient is a function of the tip-speed ratio $\lambda_r$ and the blade pitching angle $\beta$. A predictive model can thus be obtained by combining \eqref{eq:first_order_wt} and \eqref{eq:tau_a} and a pre-computed $C_p( \lambda_r, \beta)$ map. This map is usually obtained via Blade Elementum Momentum Theory (BEMT, \citealt{moriarty2005aerodyn}), and we do the same for simulating the real environment in this test case. The resulting map for the selected turbine is shown in Figure \ref{fig:cp_coeff_nrel}.
	
	We consider the NREL 5-MW reference offshore wind turbine \citep{jonkman2009definition}, commonly used as a reference in the literature \citep{de2022influenceNREL_turb, de2022dynamicNREL, coquelet2022reinforcement}. This turbine has $R=63$ m, $J=115 926$ kg m$^2$ and $\omega_{r, \mathrm{rated}} = 1.27$ rad s$^{-1}$ for $u_\infty\ge 11.4$ m s$^{-1}$. The gearbox ratio is $N_g=97$. The actuation bounds of the pitch angle and torque actuators are respectively $(\beta_{\mathrm{min}}, \beta_{\mathrm{max}})=(0, 90)$ deg and $(\tau_{\mathrm{min}}, \tau_{\mathrm{max}})=(0, 43093)$ Nm. Their maximum rate of change are $\dot{\beta}_{max}=8$ deg s$^{-1}$ and $\dot{\tau}_{g, max}=15 000$ N m s$^{-1}$, respectively. From these, we define the cut-off frequency as the harmonic associated with the highest rate of change, obtaining $f_{c,\tau}=0.05$ Hz and $f_{c,\beta}=0.04$ Hz. We set the filter smoothing coefficient according to the slowest pitch actuator, finding $\hat{\alpha}=0.98$. 
	
	We take $T_o = 600$ s for the observation time and assume a sampling of $f_s = 4$ Hz for all the simulated instrumentation. We simulate the incoming wind speed time series using the stochastic, turbulent wind simulator TurbSim \citep{jonkman2006turbsim} considering a wind velocity with average $u_\infty = 11$ m s$^{-1}$ and turbulence intensity TI$= 16\%$, thus oscillating between below and above rated. An example of velocity time series is shown in Figure  \ref{fig:example_wind} together with the associated filtered signal.
	
	\paragraph{\textbf{Digital Twin definition}} 
	
	The wind energy community has put significant effort into the development of digital twins of onshore and offshore wind turbines, with applications ranging from load estimations \citep{Pimenta_2020, wes-2023-50} to real-time condition monitoring \citep{Olatunji2021OverviewOD, Fahim2022MachineLD}. In this work, we solely focus on the rotor's aerodynamic performance and the twinning process seeks to identify a closure law for the power coefficient from data, starting from \eqref{eq:first_order_wt} and \eqref{eq:tau_a}. In fact, the BEMT, commonly used to obtain the $C_p$ curve, assumes that the flow is steady, and this is questionable in the presence of large fluctuations in the wind speed \citep{leishman2002challenges}. Moreover, the coefficient can change over time as a result of turbine wear (e.g. due to blade erosion) or in the presence of extreme conditions (e.g. ice, dirt or bug buildup) \citep{STAFFELL2014775}. This is why any strategy based on a pre-computed $C_p$ curve is likely sub-optimal \citep{johnson2004,johnson2006} and the idea of continuous updates of this closure law from operational data has merit. Although this exercise is purely illustrative, the approach can be readily extended to more complex and multi-disciplinary model formulations. 
	
	The digital twin employs the power coefficient parametrization formulated by \cite{saint-drenan_parametric_2020}
	
	\begin{equation}
		\label{eq:c_p_correlation}
		\begin{cases}
			\begin{split}
				C_p(\lambda_r, \beta) = &c_1 \Big(\frac{c_2}{\xi} - c_3 \beta - c_4 \xi \beta - c_5 \beta^{c_{12}} - c_6\Big) e^{-\frac{c_7}{\xi}} + c_8 \xi
			\end{split}
			\\
			\xi^{-1} = (\lambda_r + c_9 \beta)^{-1} - c_{10} (\beta^3 + 1)^{-1}
		\end{cases}
	\end{equation} with the parameters $\bm{w}_p=[c_1,c_2,\dots c_{12}]$ to be identified. This model is known to be valid up to a maximum pitch excursion of $\beta = 30$ deg. Therefore, we limit the maximum allowable pitch command to this value in our numerical experiments.
	
	The assimilation optimizes for $\bm{w}_p$, aiming to minimize the error between the predicted dynamics $\omega_g(t)$ and the real system behaviour $\check{\omega}_g(t)$. This translates into minimizing the cost function
	
	\begin{equation}
		\label{eq:assimilation_functional}
		\mathcal{J}_{p} = \frac{1}{2} \int_0^T \Big(\omega_g(t ; \bw_p) - \check{\omega}_g(t)\Big)^2 \; \mathrm{dt} \,.
	\end{equation}

	\subsection{Trajectory control of flapping wings micro air vehicles}\label{sec:testcase_2}
	
	\paragraph{\textbf{Test case description and control problem}} 
	
	This test case considers the control of a Flapping-Wing Micro Air Vehicle (FWMAV) which aims to move from one position to another in the shortest possible time. FWMAVs are sometimes considered better alternatives to fixed-wing and rotary-wing micro aerial vehicles because of their higher agility and manoeuvrability \citep{Haider2020}. The design and control of these drones seek to mimic the remarkable flight performances of hummingbirds, which can undertake complex manoeuvres with an impressive response time (see, for example, \citealt{Cheng2016} and \citealt{Ortega2018}).
	
	%, for example, analyzed the flight of a hummingbird escaping from a potential threat and reported drastic alterations to the magnitude and direction of flight forces coupled with body pitch and roll rotations.
	
\begin{figure*}[!ht]\center
		\begin{subfigure}{0.45\textwidth}
			\centering
			\includegraphics[width=0.7\textwidth]{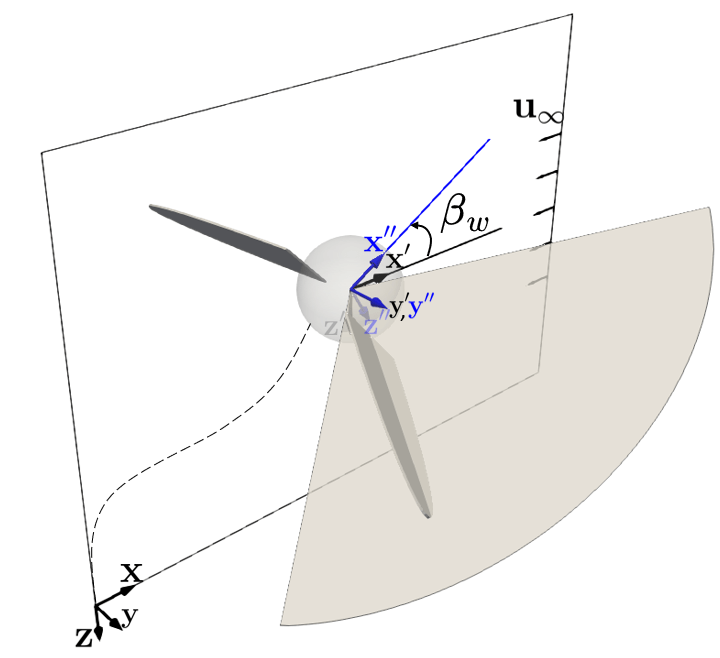}
			\caption{ }
			\label{drone_img_fig1} 
		\end{subfigure}
		\hspace{2mm}
		\begin{subfigure}{0.5\textwidth}
			\centering
			\includegraphics[width=0.95\textwidth]{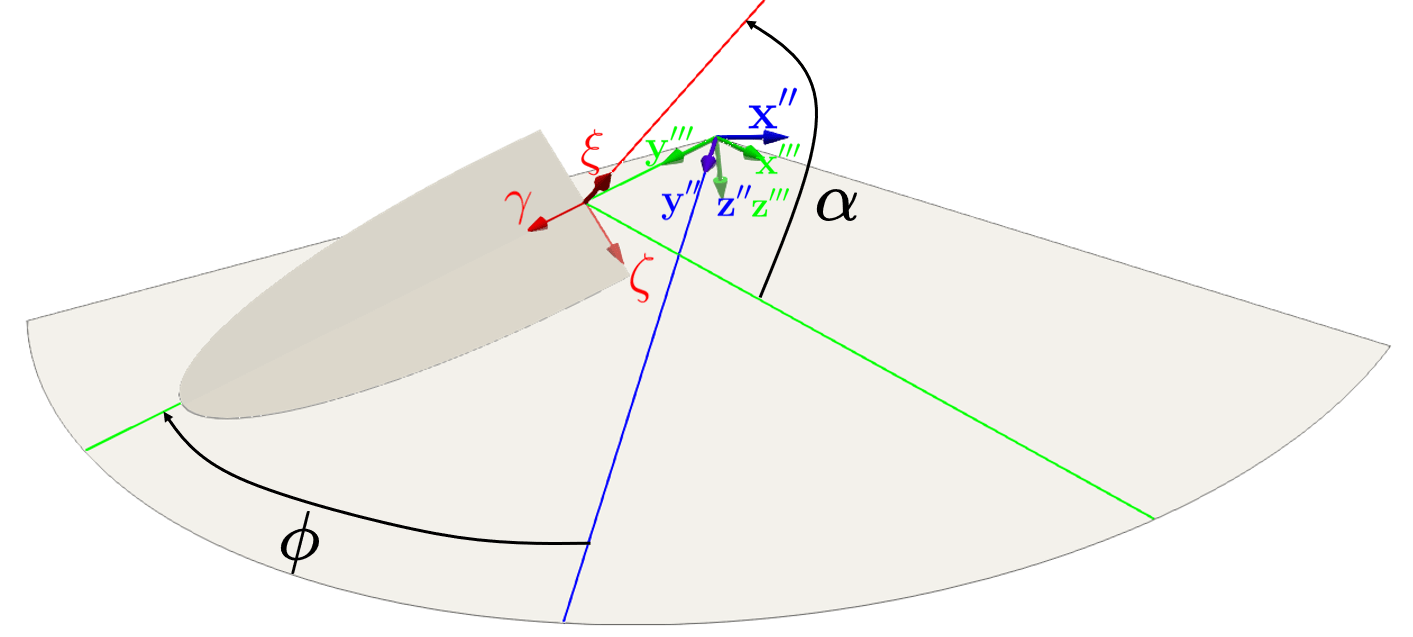}
			\caption{ }
			\label{drone_img_fig2}  
		\end{subfigure}
		\caption{(a) Schematic of the flapping drone going from the initial position $\bm{\tilde{x}_0}$ to the final position $\bm{\tilde{x}_f}$. ($x'y'z'$) is the stroke frame that is tilted by $\beta_w$ from the inertial frame ($xyz$) (b) Zoom on the stroke plane that holds three frames (stroke frame $x'y'z'$, flapping plane $x''y''z''$ and the wing-fixed frame  $(\xi,\gamma,\zeta)$) which defines the flapping angle $\phi$ and the pitching angle $\alpha$. The wing frame is offset on the wing root only for the clarity of the figure. }
		\label{drone_img_fig12}   
	\end{figure*}

	Reproducing these performances requires advanced knowledge of the flapping wing's aerodynamics in the low Reynolds number regime, as well as the fluid-structure interaction resulting from the wing's flexibility. These remain significant challenges for advanced numerical simulations \citep{Fei2019,Xue2023}. The proposed test case shows how simplified models of the wing's aerodynamics could be derived in real-time while an FWAVs seeks to achieve its control target. The considered problem is illustrated in Figure \ref{drone_img_fig1}. A drone initially at $\bm{x}_0 = [x_0,z_0]$ m and $\bm{\dot{x}}_0 = 0$ m/s is requested to move to $\bm{\tilde{x}_f} = [\tilde{x}_f,\tilde{z}_f]$ m as fast as possible, and then hover in that position (i.e. $\dot{\bm{\tilde{x}}}_f=[0,0]$) by continuously adapting its flapping wing motion. We challenge the controller by introducing gusty wind disturbances $\bm{u}_{\infty}$, %\textcolor{blue}{with the subscript $i$ distinguishing the definition with respect to the inertial reference frame}. 
	
	%! Put everything in plural
	% Body kinematics
	To simplify the body's kinematics, we consider only 2D trajectories characterized by the position vector $\bm{x}(t)$ defined in the ground frame ($x,y,z$). 
	We also attach a moving frame ($x', y', z'$) to the body with no rotational freedom.  
	% Wing kinematics
	Starting from the body frame, three Euler angles  ($\beta_w,\phi,\alpha$) and three reference frames characterize the wing kinematics (Figure \ref{drone_img_fig2}). The wings always flap symmetrically so that the position of one wing defines the other one.
	
	The body frame is first pitched by a stroke plane angle $\beta_w(t)$ along $y'$ to define the stroke plane frame ($x'',y'',z''$). The wing tips are constrained to flap in the stroke plane ($x'',y''$) as it flaps along the normal $z''$. 
	The flapping frame ($x''',y''',z'''$) follows this rotational motion and defines the flapping angle $\phi(t)$ between the wing symmetry axis $y'''$ and the lateral direction $y''$ (Figure \ref{drone_img_fig2}).
	The last frame is the wing-fixed frame $(\xi,\gamma,\zeta)$ that results from the pitching rotation along the wing symmetry axis. The pitching angle $\alpha(t)$ is then defined between the chord-normal direction $\xi$ and the stroke plane axis $x'''$. The reader is referred to \cite{Whitney2010} and \cite{Cai2021} for a complete treatment of the flapping wing kinematics. 
	%The 2D motion in the fixed frame $(x,z)$ avoids the need for other two angles between body and wing frames.
	We select an harmonic parametrization for $\phi$ and $\alpha$:
	
	\begin{eqnarray}
		\label{drone_eq_phi_and_alpha}
		\phi (t) = A_{\phi}\cos(2 \pi  ft) \quad \mbox{and}\quad 
		\alpha (t) = A_{\alpha}\sin(2 \pi  ft), 
	\end{eqnarray} 
	where the beginning of the stroke ($t=0$) is when the flapping amplitude is maximal ($\phi (t) = A_{\phi}$) and the chord plane is perpendicular to the stroke plane ($\alpha (t) = 0^o$).
	
	We assume that the flapping frequency $f$ and the pitching amplitude $A_{\alpha}$ are fixed and set the flapping amplitude $A_\phi$ as a control parameter together with the stroke plane angle $\beta_w$, i.e. $\bm{a}=[A_\phi,\beta_w]^T$. We link the action vector to the state error using a 
	Proportional–Derivative (PD) controller with an offset and capped outputs:
	
	\begin{equation}
		\bm{a}(\bm{e};\bw_a)= 
		\begin{cases}
			\mbox{clip}\bigl(w_{a1}+w_{a2} e_{x} +w_{a3} e_{z}+w_{a4} \dot{e}_{x}+w_{a5} \dot{e}_{z} ,A_{\phi,min},A_{\phi,max})\\
			\mbox{clip}\bigl( w_{a6}+w_{a7} e_{x} +w_{a8} e_{z}+w_{a9} \dot{e}_{x}+w_{a10} \dot{e}_{z},\beta_{w,min},\beta_{w,max})\\
		\end{cases}
	\end{equation} where $\bm{e}=[e_{x},e_{z},\dot{e}_{x},\dot{e}_{z}]$, with $e_x=x_f-x$, $e_z=z_f-z$ and dots denoting differentiation in time. The control action parameters are $\bm{w}_a\in\mathbb{R}^{10}$ and the function $\mbox{clip}(x,x_1,x_2)$ is the smoothing operation introduced in \eqref{clip}. The bias terms $w_{a1},w_{a6}$ helps stabilize the drone in hovering conditions once the target position is reached since this operation requires a continuous effort even when the state error is zero.
	
	The reward function driving the model-free controller is written as 
	
	\begin{equation} \label{drone_eq_rewardDisc}
		\mathcal{R}_c = -\frac{1}{2n_t}\sum_{k=1}^{n_t} ||\bm{\check{e}}_k||_2^2 +   \alpha_1|| \bm{\dot{\check{e}}}_k\circ \bm{h}(\bm{x}_k) ||^2_2 \,.
	\end{equation} 
	
	\noindent
	where $\circ$ denotes the Hadamard (entry by entry) product 
	and $\bm{h}(\bm{x}_k)$ is a vector of two Gaussian functions with mean $\bm{x}_f$ and standard deviation $\sigma=0.71$:
	
	\begin{equation}
		\bm{h}(\bm{x}_k)= \frac{\bm{x}^2_f}{2} \mbox{exp}\biggl[-\frac{1}{2}\biggl(\frac{\bm{x}_k - \bm{x}_f}{\sigma}\biggr)^2\biggr]\,.
	\end{equation}
	
	%     \begin{equation}
		%  \bm{h}(\bm{x}_k)=\sqrt{
			% \frac{1}{2}\tanh\biggl [c_1\bigl(\bm{x}_k-(1-c_2)\bm{x}_f\bigr)\biggr]-\frac{1}{2}\tanh\biggl [c_1\bigl(\bm{x}_k-(1+c_2)\bm{x}_f\bigr)\biggr]}\,,
		%     \end{equation} with $c_1=4$ $c_2=1/8$ two hyper-parameters.

	This function is zero far from the target position and unitary once the drone reaches it, so the role of the second term in \eqref{drone_eq_rewardDisc} is to penalize large velocities once the drone approaches the goal. The term $\alpha_1\in\mathbb{R}$ in \eqref{drone_eq_rewardDisc} weights the importance of the penalty. The cost function driving the model-based controller reproduces the same approach in a continuous domain and reads:
	
	\begin{equation} \label{drone_eq_rewardCont}
		\mathcal{J}_a = \frac{1}{2T_0}\int_0^{T_0}||\bm{e}(t)||_2^2+ \alpha_1 || \bm{\dot{e}}(t) \circ h(\bm{x}(t_k))||^2_2 dt\,.
	\end{equation}

	\paragraph{\textbf{Selected conditions and environment simulator}} 
	We consider a drone with mass $m_b=3$ g and a constant flapping frequency of $f=20$ Hz. Given the initial position $\bm{x}_o=[0,0]$ m, we set the target position at $\bm{x}_f=[5,5]$ m. We assume that the manoeuvre should take less than $n_t=100$ flapping cycles, hence we consider an observation time of $T_o=5$ s. This defines the observation time for the twinning.
	We fix the pitching amplitude to $A_\alpha=45$ deg and let the flapping amplitude vary in the range $A_\phi\in[50;88]$ deg. These values are comparable to what is observed in hummingbird's flight, although the harmonic parameterization in \eqref{drone_eq_phi_and_alpha} oversimplifies their actual wing dynamics \citep{Kruyt2014}. The stroke plane angle is bounded in the range $\beta_w\in[-30,30]$ deg.
	
	We assume that the drone's wings are semi-elliptical and rigid with a span $R=0.05$ m and a mean chord $\overline{c}= 0.01$ m. The wing roots are offset of $R_0=0.0225$ m from the body barycenter, which is also the centre of rotation of the wings. 
	We decouple their motion from the body such that the wing inertia does not influence the body motion \citep{Taha2012}. This hypothesis simplifies the body dynamics, which becomes only a function of the aerodynamic forces produced by the wings and the gravitational force. These are assumed to apply on the barycenter of the body, which is treated as a material point. Therefore, restricting the dynamics to the $(x,z)$ plane, the force balance gives
 
	\begin{align}\label{drone_eq_x}
		m_b \ddot{x} &= F_{x,w}  -D_b \cos(\beta_b) \\
		m_b \ddot{z} &= F_{z,w}  - m_bg - D_b  \,\sin(\beta_b)  \label{drone_eq_y}
	\end{align} where the subscript $w$ refers to forces produced by the wing and the subscript $b$ identifies body quantities. Hence the aerodynamic forces produced by the flapping wings are $\bm{F}_w=[F_{x,w},F_{z,w}]^T$, $D_b$ is the magnitude of the drag force exerted on the body and $\beta_b$ is the angle between the drag force and the $x$ axis. Considering the velocity of the drone $\dot{\bm{x}}=[\dot{x},\dot{z}]^T$ and the incoming horizontal wind disturbance (both in the ground frame), this angle is defined as $\beta_b=\tan^{-1}({\dot{z}}/{(\dot{x} + u_{\infty}}))$. The magnitude of the drag force is computed as 
	%The drone's body does not generate lift.
	
	\begin{equation} \label{drone_eq_bodyDrag}
		D_b =\frac{1}{2}\rho S_b C_{D,b} U_b^2, 
	\end{equation}
	
	\noindent
	where $\rho$ is the air density, $S_b=0.0005$ m$^2$ is the body's cross-sectional area, $C_{D,b}$ is the body's drag coefficient, here taken as unitary, and $U_b$ is the modulus of the relative velocity between the drone body and the wind, i.e. 
	$U_b=\sqrt{(\dot{x} + u_{\infty})^2 + \dot{z}^2}$.
	The wind disturbance $u_{\infty}$ results from a stationary Gaussian process with mean value $\overline{u}_{\infty}=1$ m/s, in opposite direction to the $x$ versor, and covariance matrix defined by a Gaussian kernel $\kappa(u_{\infty}(t),u_{\infty}(t+\tau))=\exp(-\tau^2/(2\sigma^2_f))$ with $\sigma_f=3\mbox{s}$. 
	
	To compute the wing forces in \eqref{drone_eq_x} and \eqref{drone_eq_y} we follow the semi-empirical quasi-steady approach of \cite{Lee2016}, based on the work by \cite{Dickinson1999}. This is a common approach in control-oriented investigations (see e.g \cite{Cai2021,Fei2019}) because it provides a reasonable compromise between accuracy and computational cost, at least for the case of smooth flapping kinematics and near hovering conditions (see \cite{Lee2016}).
	
\begin{figure*}[!ht]\center
		\begin{subfigure}{0.6\textwidth}
			\centering
			\includegraphics[width=\textwidth]{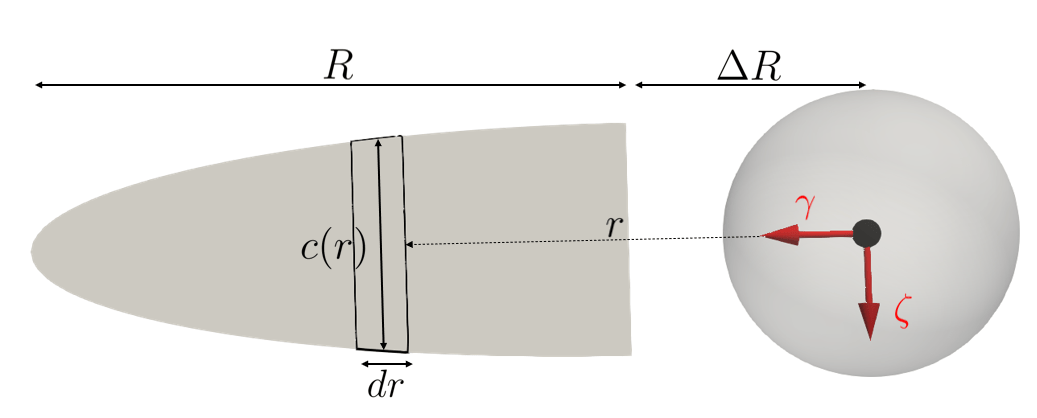}
			\caption{ }
			\label{drone_img_fig3} 
		\end{subfigure}
		\hspace{2mm}
		\begin{subfigure}{0.3\textwidth}
			\centering
			\includegraphics[width=\textwidth]{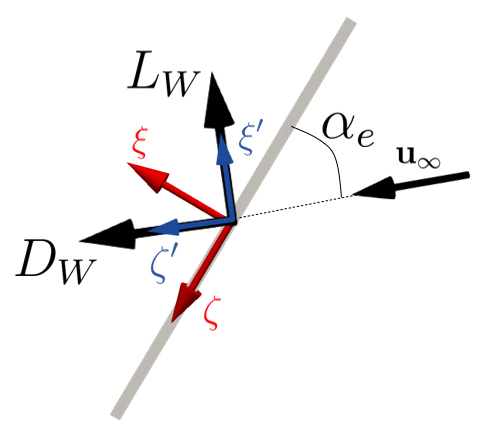}
			\caption{ }
			\label{drone_img_fig4}  
		\end{subfigure}
		\caption{(a)  Main dimensions of the semi-elliptical wing. An element part of the BET is also highlighted (b) Effective angle of attack $\alpha_e$, lift $L_w$ and drag $D_w$ shown with the wind frame $(\xi',\gamma',\zeta')$ and wing-fixed frame $(\xi,\gamma,\zeta)$ on a chord section of the wing}
		\label{drone_img_fig34}   
	\end{figure*}

	% Flow :'(
	The approach is based on BEMT, which partitions the wing into infinitesimal elements (see Figure \ref{drone_img_fig3}). Each element produces an infinitesimal aerodynamic force that must be integrated along the span to get the total lift and drag. 
 For smooth flapping motion \eqref{drone_eq_phi_and_alpha}, the forces due to the flapping dominate all the other physical phenomena (\cite{Lee2016}) so that the lift $L_w$ and drag $D_w$ expressions are
	%to get the total lift $L_w$ and drag $D_w$ of the wing. 
	
	\begin{equation} \label{drone_eq_lift}
		L_w(t)    = \frac{1}{2}\rho C_{L,w}(\alpha) \int^{R}_{\Delta R} U_w^2(t,r)  c(r)dr 
	\end{equation}
	\begin{equation} \label{drone_eq_drag}
		D_w(t) = \frac{1}{2}\rho C_{D,w}(\alpha) \int^{R}_{\Delta R} U_w^2(t,r)   c(r)dr, 
	\end{equation}
	
	\noindent
	where $C_{L,w}$ and $C_{D,w}$ are the lift and drag coefficient of the wing and $U_w$ is the modulus of the relative velocity between the wing displacement, the body displacement and the wind. This velocity depends on the span-wise coordinate of the wing section $r$ (see Figure \ref{drone_img_fig4}) and is computed as
	
	\begin{equation} \label{drone_eq_Uw}
		\bm{U}_w(r,t,\bm{u}) = [\bm{U}_{w,x},\bm{U}_{w,y},\bm{U}_{w,z}] =  \Biggl( {\bm{R}}_2 (\alpha)\begin{bmatrix}
			0 \\
			0 \\
			\dot{\phi} 
		\end{bmatrix} \Biggr) \times \begin{bmatrix}
			0 \\
			r \\
			0 
		\end{bmatrix}  +   {\bm{R}}_2(\alpha)  {\bm{R}}_3(\phi) 
		{\bm{R}}_2(\beta_w)
		\Biggl(\begin{bmatrix}
			\dot{x} \\
			\dot{y}  \\
			0 
		\end{bmatrix}  + \begin{bmatrix}
			U_{wi} \\
			0  \\
			0 
		\end{bmatrix} \Biggr)
	\end{equation}
	
	The first term is the (linear) flapping velocity of the wing, while the second term gathers the velocity of the body and the wind. Both terms are projected in the wing frame thanks to three rotation matrices:
	
	\begin{subequations}
		
		\begin{equation}
			{\bm{R}}_2 (\alpha) = \begin{bmatrix}
				\cos(\alpha) & 0 & \sin(\alpha)\\
				0 & 1 & 0\\
				-\sin(\alpha) & 0 & \cos(\alpha) 
			\end{bmatrix} 
			\hspace{1cm}
			{\bm{R}}_3(\phi) = \begin{bmatrix}
				\cos(\phi) &  \sin(\phi) & 0\\
				-\sin(\phi) & \cos(\phi) & 0\\
				0 & 0 & 1
			\end{bmatrix}
		\end{equation}
		
		\begin{equation}
			{\bm{R}}_2(\beta_w) = \begin{bmatrix}
				\cos(\beta_w) & 0 & \sin(\beta_w)\\
				0 & 1 & 0\\
				-\sin(\beta_w) & 0 & \cos(\beta_w) 
			\end{bmatrix}
		\end{equation}
		
	\end{subequations} 
	
	\noindent
	where ${\bm{R}}_2(\beta_w)$ transform velocities from the 
	$(x',y',z')$ frame to the $(x'',y'',z'')$ frame, ${\bm{R}}_3(\phi)$ from  $(x'',y'',z'')$ to $(x''',y''',z''')$ and ${\bm{R}}_2(\alpha)$ from $(x''',y''',z''')$ to $(\xi,\gamma,\zeta)$. 
	
	The force coefficients $C_{L,w}$ and $C_{D,w}$ in equations \eqref{drone_eq_lift} and \eqref{drone_eq_drag} are modeled similarly to \cite{Lee2016} and \cite{Sane2001}:
	
	\begin{align} \label{drone_eq_Clift}
		C_{L,w} &= a\sin(2\alpha_e) \\
		C_{D,w} &= b + c\;(1-\cos(2\alpha_e)) \label{drone_eq_Cdrag},
	\end{align}
	
	\noindent
	where $\bm{w}_p=[a,b,c]$ are the closure parameters for this problem.
	This parametrization approximates the influence of the Leading Edge Vortex (LEV), generated at a high pitching angle due to flow separation and stably attached on the suction side (see \cite{Sane2003}). The corresponding force coefficients depend on the effective angle of attack $\alpha_e = \cos^{-1}(U_{w,z}$/$||\bm{U_w}||_2)$, defined between the chord direction $\zeta$ and the relative velocity of the wing $\bm{U_w}$. \\
	
	%In short, when the wing flaps at a high pitching angle, the flow separates from the leading edge and forms a bounded vortex that remains stably attached to the upper surface of the wing. This LEV generates a low-pressure region, and so a net lift and drag (see \citep{Sane2003}).

	Finally, the forces in equations \eqref{drone_eq_lift} and \eqref{drone_eq_drag} were computed in the reference wind frame $(\xi',\gamma',\zeta')$, which rotates the wing frame by $\alpha_e$ along $\gamma$. The drag $D_w$ is then aligned with $\xi'$ while the lift is perpendicular to the drag (Figure \ref{drone_img_fig4}). These forces are successively transformed into the wing frame and then into the body frame to retrieve $F_{x,w}$ and $F_{z,w}$ in the dynamic equations \eqref{drone_eq_x} and \eqref{drone_eq_y}.
	
	In this test case, the real environment is simulated using $\bm{w_p}=[a,b,c]=[1.71,0.043,1.595]$ in \eqref{drone_eq_lift} and \eqref{drone_eq_drag}, as approximated from \citep{Lee2016} for semi-elliptical wings. For the digital twin, these are the parameters to be identified while interacting with the environment to minimize the cost function:
	
	\begin{equation}
		\label{drone_eq_Jp}
		\mathcal{J}_{p} = \frac{1}{2} \int_0^T \Big(\bm{s}(t ; \bw_p) - \bm{\check{s}}(t)\Big)^2 \; \mathrm{dt} \,,
	\end{equation}
	with $\boldsymbol{s}(t)=[x(t), z(t)]$ the drone position at time $t$.
	
	\subsection{Thermal Management of Cryogenic Storage}\label{sec:testcase_3}
	
	\paragraph{\textbf{Test case description and control problem}} 
	We consider the thermal management of a cryogenic tank. This is a fundamental challenge in long-duration space missions, which require storing large amounts of cryogenic propellant for long periods. These fluids must be stored in the liquid phase to maximize the volumetric energy density. However, their storage temperatures are extremely low (e.g., -250 $^{\circ}$C for liquid hydrogen), and their latent heat of evaporation is much smaller than that of non-cryogenic propellants. These combined factors challenge the storage because evaporation due to thermal loads produces a continuous pressure rise over time \citep{Salzman1996, Motil2007, chai_cryogenic_2014}. 
	
\begin{figure}[h!]
		\centering
		\begin{subfigure}[b]{0.38\textwidth}
			% \centering
			\includegraphics[width=\linewidth]{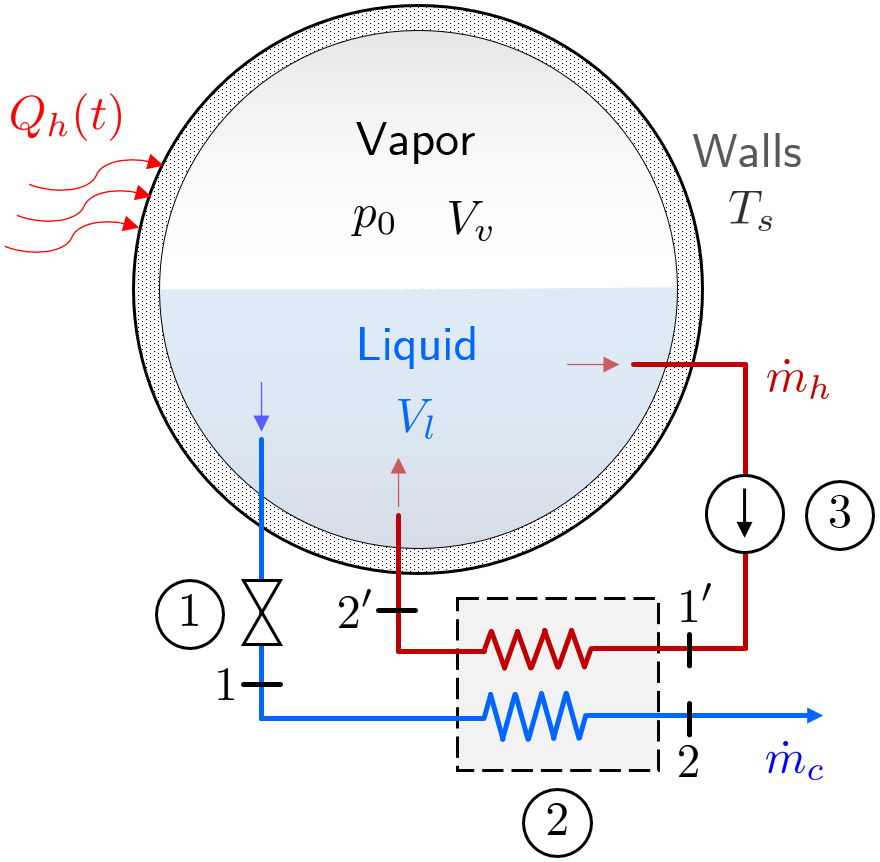}
			\caption{}
			\label{fig:sketch}
		\end{subfigure}
		\hspace{9mm}
		\begin{subfigure}[b]{0.42\textwidth}
			% \centering
			\includegraphics[width=\linewidth]{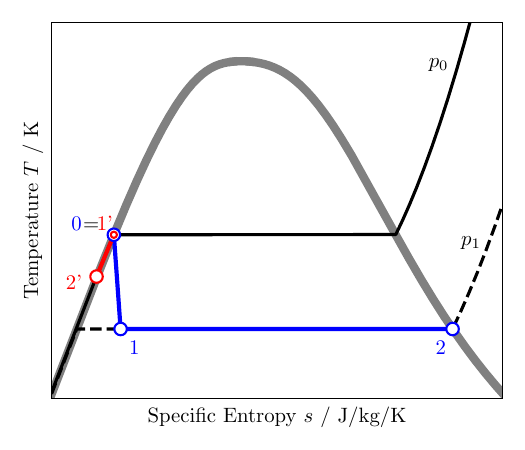}
			\caption{}
			\label{TS_D}
		\end{subfigure}
		\caption{Schematic of the thermal management problem considered in this work. A cryogenic tank of volume $V_t$, containing a volume $V_l$ of saturated liquid and $V_t-V_l$ saturated vapour, is subject to a heat load $Q_h (t)$. The figure on the right traces the state of the system on the temperature-specific entropy diagram considering the simplified model used to simulate the environment.}
		\label{fig:caseII}
	\end{figure}

	The efficient management of this problem requires a combination of advanced insulation strategies (see \citealt{Mer2016, Jiang2021}) and venting control strategies \citep{Lin1991_TVS}. In this test case, we consider a classic active Thermodynamic Venting System (TVS), similar to the one presented in \cite{Lin1991_TVS} and recently characterized in \cite{Imai2020,Qin2021}. Figure \ref{fig:sketch} shows the schematics of the TVS configuration.
	
	We consider a tank of volume $V_t$ filled with a volume of liquid $V_l(t)$, operating at a nominal pressure $p_0$ while subject to a time-varying heat load $Q_h(t)$. The TVS is composed of a vented branch and an injection loop. In the vented branch (blue line in Fig.\ref{fig:sketch}), a mass flow rate $\dot{m}_c$ is extracted from the tank, expanded through a Joule-Thomson valve (1) and used as a cold source in a heat exchanger (2) before being expelled. In the injection loop, a mass flow rate $\dot{m}_h$ is circulated by a pump (3) into the hot side of the heat exchanger and reinjected into the tank as a subcooled liquid. This injection can be carried out via a submerged jet or a spray bar \citep{Hastings2005,Wang2017_TVS,Hastings2003_TVS}, but this distinction is unnecessary for the illustrative purposes of this work. An extensive literature review of TVS approaches is presented by \cite{Barsi_THESIS}, who also introduced and tested two simplified approaches for the thermodynamic modelling of the problem (see \citealt{Barsi2013} and \citealt{Barsi2013a}). Figure \ref{TS_D} maps the (ideal) key state of the fluids at the inlet and outlet of each component in the $T-s$ (temperature vs. specific entropy) diagram. The state of the liquid and the vapour in the tank are denoted with subscripts ($l$) and ($v$), the states ($1$) and ($2$) are the inlet and outlet of the heat exchanger on the vented line, while states ($1'$) and ($2'$) are their counterparts on the injection side. 
	
	The control task is to keep the tank pressure below the maximum admissible one while venting the least amount of liquid. The simplest control problem, implemented in this work, consists of employing the pressure difference $\Delta p_V= p_0-p_1$ as the sole control parameter. 
	% The simplest control problem could be set by having only the pressure difference $\Delta p_V= p_0-p_1$ as a control parameter. 
	% This could be linked to both mass flow rates $\dot{m}_v$ and $\dot{m}_h$ in the J-T valve (1) and the pump (3) by limiting the range of operating conditions of the heat exchanger and imposing that only latent heat is to be recovered from the venting line. Therefore, defining as $\Delta T(p_1)=T_l-T_1(p_1)$ the temperature drop produced by the J-T device and $\Delta h_c(p_1)=h_2 - h_1 = h_{v,sat}(p_1) - h_{l,sat}(p_0) $ the specific latent heat remaining on the vented line, one has 
	The $\Delta p_V$ is linked to the mass flow rates in the J-T valve (1) and the pump (3) by limiting the operating range of the heat exchanger and by imposing that only latent heat is to be recovered from the venting line. Therefore, defining as $\Delta T_l(p_1)=T_l-T_1(p_1)$ the temperature drop produced by the J-T device and $\Delta h_c(p_1)=h_2 - h_1 = h_{v,\text{sat}}(p_1) - h_{l,\text{sat}}(p_0) $ the specific latent heat remaining on the vented line, one has 
	
	\begin{equation}
		\label{mh}
		\dot{m}_h(p_1)=\frac{\Delta h_c(p_1)}{c_{p,l}\Delta T_l(p_1)}\;\dot{m}_c(p_1)\,.
	\end{equation}
	
	Furthermore, to limit the control action exclusively to $\Delta p_V$, we assume that the pump (3) is equipped with flow regulation to follow \eqref{mh} and we link the vented flow rate $\dot{m}_c$ to $\Delta p_V$ and the J-T valve's characteristic through 
	
	\begin{equation}
		\dot{m}_c(\Delta p_v)=\sqrt{\Delta p_V/K_{JT}},
	\end{equation}
	
	\noindent
	with $K_{JT}$ the valve's pressure drop constant. 
	Considering the maximum pressure drop along the JT-valve $\Delta p_\text{max}$, the desired operational pressure $\tilde{p}$ and defining $e(t)=p(t)-\tilde{p}$ the tracking error, we consider a simple policy parametrization of the form
	
	% \begin{equation}
		% \Delta p_V=a(\mathbf{s},\bw_a)=\begin{cases}
			% 0 \quad \text{(TVS is off)} & \text{ if } p(t)\leq p_{min}, \\
			% \mbox{clip}\bigl(w_{a1} e +w_{a2} e^2\bigr,\Delta p_{min},\Delta p_{max})  & \text{ if } p(t)>p_{max}, \\
			% \end{cases}
		% \end{equation} 

	\begin{equation}
		\Delta p_V=a(e,\bw_a)= \frac{1}{2} \; \Delta p_\text{max}\left( \tanh( w_{a1} e +  w_{a2} e^3 + w_{a3}) + 1 \right).
	\end{equation} 
	
	We treat $\Delta p_{max}$ as a pre-defined constant and use $\bm{w}_a=[w_{a1},w_{a2},w_{a3}]$ as policy parameters. $w_{a3}$ acts as an offset in the policy, controlling the threshold pressure above which the valve opens, while $w_{a1}$ controls the slope of the hyperbolic tangent, i.e. the sensitivity of the policy to the tracking error. The cost function driving the model-based controller and the model-free agent are 
	
	\begin{equation}
		\label{eq:cryo_J_a}
		\mathcal{J}_a=\frac{1}{T_o}\int^{T_o}_0 \alpha_1 e^{2}(t) + \alpha_2 \, \dot{m}_c \mathrm{dt} 
		\quad \mbox{and} \quad
		\mathcal{R}_c=-\frac{1}{n_t}\sum^{n_t-1}_{0} \alpha_1 e(t_k)^2 + \alpha_2 \dot{m}_c(t_k)\,,
	\end{equation} 
	\noindent
	with the $\alpha_1,\alpha_2\in\mathbb{R}$ used to non-dimensionalize the inputs and weight the penalty on the mass consumption versus the primary objective of keeping the pressure constant. 
	
	To set up the data-driven identification to this control law, we assume that the tank is instrumented with a pressure sensor sampling $\check{p}(t)$ and a flow rate sensor on the vented line to measure $\check{m}_v(t)$.

	\begin{figure*}
		\centering
		\begin{subfigure}[b]{0.45\textwidth}
			% \centering
			\includegraphics[width=\linewidth]{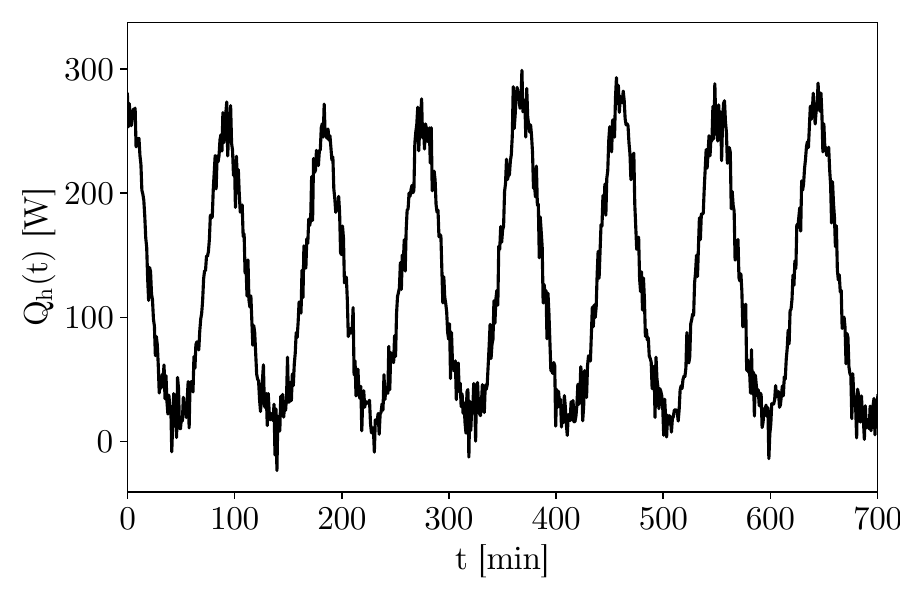}
			\caption{}
			\label{fig:Load}
		\end{subfigure}
		\hfill
		\begin{subfigure}[b]{0.45\textwidth}
			% \centering
			\includegraphics[width=\linewidth]{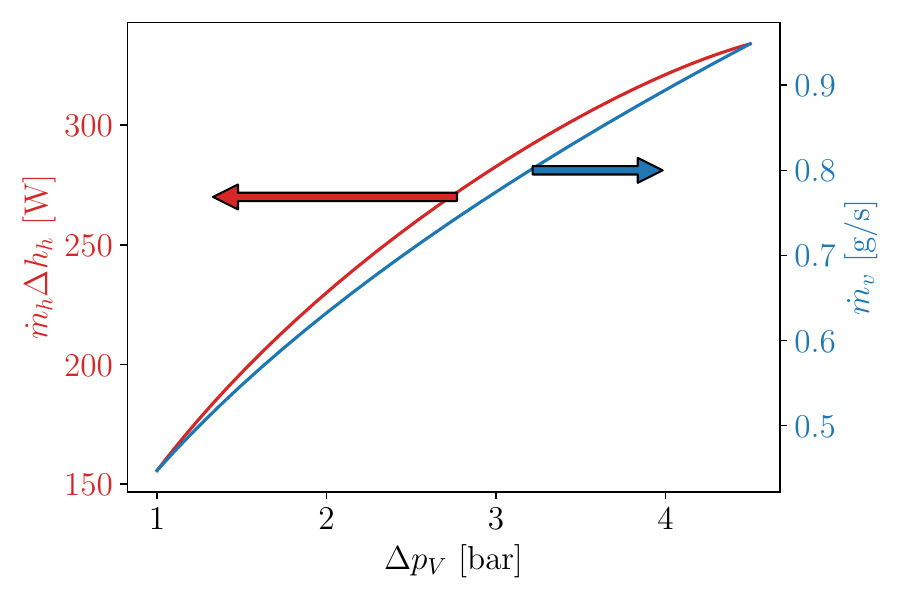}
			\caption{}
			\label{Thermal_C}
		\end{subfigure}
		\caption{ Sample profile of the thermal load (exogenous input) considered in test case 2 (Fig a) and impact of the control parameter ($p_1$) on the vented flow rate and subcooling of the injection line. }
	\end{figure*}
	
	\paragraph{\textbf{Selected conditions and environment simulator}} 
	We consider a cryogenic tank with volume $V_t=60$ m$^3$ with a desired operational pressure $\Tilde{p} = 500$ kPa filled with liquid hydrogen. 
	At the start of each episode, the tank is $70\%$ full and pressurized to an initial value sampled from a normal distribution with mean $\overline{p}_0=500$ kPa and standard deviation $\sigma_{p_0} = 25$ kPa.
	% The corresponding saturation temperature is $T=27.2$ K.
	The thermal load applied to the tank fluctuates from a minimum of roughly $50$ W to a maximum of $300$ W with a period of $T_h\approx 1.5$ h. This cyclic load mimics the large thermal fluctuations a tank could face in orbit as it is periodically exposed to direct sun radiation. The chosen period is approximately the time it takes the International Space Station (ISS) to complete an orbit around Earth. The load profile is shown in Figure \ref{fig:Load}. This load is constructed as a pseudo-random signal made of Gaussian-like peaks randomly varying within $[0.9,1.1]$ times the nominal peak, lifted by the minimal load and polluted by uniform noise with a standard deviation equal to $10\%$ the nominal peak. 
	
	The simulation of the real environment and the digital twin are carried out using the same simplified model.
	In particular, we employ the homogeneous model presented in \cite{Barsi_THESIS}, adapted to account for the venting line similarly to \cite{Mer2016a}. This approach assumes that the liquid and the vapour are in saturation conditions. This is an oversimplification because the liquid is typically subcooled and the vapor is usually superheated. We refer to \cite{Marques2023} and \cite{Barsi2013a} for more advanced models and to \cite{panzarella_validity_2003} and \cite{panzarella_pressure_2004} for an in-depth discussion on the limits of thermodynamic models. Focusing only on the storage problem (no out-flow except venting), the conservation of internal energy of the solid, and the mass and enthalpy of the saturated mixture read
	
	\begin{subequations}
		\begin{equation}
			\frac{d}{dt}\big[m_s c_s T_s\bigr]=Q_h(t) - Q_{s\rightarrow f}\,,
		\end{equation}
		\begin{equation}\label{eqa}
			\frac{d}{dt}\big[\rho_v (V_t-V_l(t))+\rho_l V_l(t)\bigr]=-\dot{m}_c\,,
		\end{equation}    
		\begin{equation}\label{eqb}
			\frac{d}{dt}\big[\rho_v h_v (V_t-V_l(t))+\rho_l h_l V_l(t)\bigr]=Q_{s\rightarrow f}+V_t\frac{dp}{dt}-\dot{m}_c h_l +\dot{m}_h \Delta h_h\,,
		\end{equation}
	\end{subequations} where $\rho$ is the density, $h$ is the specific enthalpy, and $\Delta h_h=h_{2'}-h_{1'}$ is the variation of specific enthalpy on the injection loop provided by the heat exchanger. All properties are at saturation conditions, and the subscripts $l$ and $v$ refer to the liquid and the vapour phases, respectively. We consider the mass $m_s$ and the wall's specific heat capacity $c_s$ constant. Denoting as $H=\rho h$ the (volumetric) enthalpies and using the chain rule to have $d(\cdot)/dt=(d(\cdot)/dp)\;(dp/dt)$, these can be written as 
	
	\begin{subequations}
		\begin{equation}
			m_s c_s\frac{dT_s}{dt}=Q_h(t) - Q_{s\rightarrow f}\,,
		\end{equation}
		\begin{equation}
			\frac{d V_l}{dt}=\frac{1}{(\rho_l-\rho_v)}\Biggl[\Biggl(\frac{d \rho_v}{dp }-\frac{d\rho_l}{dp} \Biggr) \frac{dp}{dt} V_l(t)-\dot{m}_c-\Biggl(\frac{d \rho_v}{d p}\Biggr)\frac{d p}{d t}V_t\Biggr],
		\end{equation}    
		\begin{equation}
			\frac{dp}{dt}\biggl\{\biggl(\frac{d H_l}{dp}-\frac{d H_v}{dp}\biggr)V_l(t)+\biggl(\frac{d H_v}{dp}-1\biggr)V_t\biggr\}= \biggl(H_g - H_l\biggr)\frac{dV_l}{dt} + Q_{s\rightarrow f}-\dot{m}_c h_l +\dot{m}_h \Delta h_h\,.
		\end{equation}
	\end{subequations}
	
	The heat transferred from the wall to the fluid mixture is modelled using Newton's cooling law $Q_{s\rightarrow f} = U_s (T_s - T)$, where $U_s$ is the overall heat transfer coefficient. Taking all liquid and vapour properties from thermophysical property libraries, this is a closed set of implicit differential equations for $\bm{s}=[T_s, V_l, p]$. The fluid properties are retrieved through the open-source library CoolProp \citep{REFPROP10}, which is based on multiparameter Helmholtz-energy-explicit-type formulations and also provides accurate derivatives.
	
	For a given design of the evaporator, one can link the subcooling in the injection line (term $\dot{m}_h\Delta h_h$ in \eqref{eqb}) to $p_1$ using standard methods for heat exchanger design (see \citealt{book_HT}). This can be written as 
	$\dot{m}_h\Delta h_h= \varepsilon c_{p,l}\Delta T(p_1) $, with $\varepsilon=1-e^{-{NTU}}$ the efficiency of the heat exchanger, $NTU=U_{HX}A/(\dot{m}_h c_p)$ the number of transfer units and $U_{HX}$ the overall heat transfer coefficient. In this test case we assume that the heat exchanger is a shell-and-tube design with 200 tubes of $D=5$ mm diameter and $L=2$ m length for a total exchange area of $A=6.2$ m$^2$. Taking a wall/fluid heat transfer coefficient of $U_s=50$ W/(m$^2$K), a global heat transfer coefficient of $U_{HX}=75$ W/(m$^2$K) and a valve with $K_{JT}=500$ GPa$\cdot$s$^2/$kg$^2$ gives the thermal power exchange and the vented flow rate in Figure \ref{Thermal_C}.
	
	Model predictions are thus possible once these three parameters are provided. We consider these the model closure for the digital twin, hence $\bm{p}=\bm{w}_p=[U_s,U_{HX},K_{JT}]$. These are to be inferred from real-time data by minimizing the following cost function. 
	
	\begin{equation}
		\mathcal{J}_p=\frac{1}{T_o}\int^{T_o}_0 \alpha_3 (\check{T}_s(t)-T_s(t))^2 + \alpha_4 (\check{V}_l(t)-V_l(t))^2 + \alpha_5 (\check{p}_l(t)-p_l(t))^2 dt \,,
	\end{equation} 
	\noindent
	where the parameters $\alpha_3,\alpha_4,\alpha_5\in\mathbb{R}$ are used to give comparable weight to these terms despite the largely different numerical values. To simplify the calibration of this digital twin model, in addition to the pressure sensor, we assume that a liquid level indicator is available, allowing for monitoring the liquid level $\check{V}_l(t)$, as well as temperature sensors to acquire $\check{T}_s(t)$.

	\section{Results}\label{s6}
	
	This section presents the performances of the RT algorithm in controlling and assimilating a wind turbine subject to time-varying wind speed (T1, Sect. \ref{sec:testcase_1}), a flapping wing micro air vehicle (FWMAV) against head wind (T2, Sect. \ref{sec:testcase_2}) and a cryogenic tank exposed to fluctuating thermal load (T3, Sect. \ref{sec:testcase_3}). The hyper-parameters selected for each test case are listed in Table \ref{tab:table_hpo}. We use five different random seeds for our simulations to extract performance statistics across different initial states of the optimizers. For both T1 and T2 we take the gradients ensemble over $N_z=$5 realizations, whereas in T3 we considered $N_z=$1, given the lower level of noise in the exogenous disturbance. Finally, for all test cases, we set $n_S=$1, as mentioned in Section \ref{s4p3}.
	
	Section \ref{6_1} presents the overall learning performances for both the twinning and the controller for all test cases, while the following sections report on the digital twins predictions and the identified control laws for each test case.

	%For the overall limited level of noise in the exogenous disturbances considered in this work, the ensemble strategy for the adjoint-based assimilation and model-based control was not deployed. Hence $N_z=1$ and $n_s=1$ in algorithm \ref{alg:rsa}.

	% RT algorithm in controlling a wind turbine, a flapping wing micro air vehicle (FWMAV) and a cryogenic tank. In each of these, we adapted the algorithm with the hyper-parameters as listed in Table \ref{tab:table_hpo}.
	
	\begin{table}[ht]
		\centering
		\caption{RMA Experiments: test cases configurations and hyper parameters settings.}
		\label{tab:table_hpo}
		\begin{tabular}{c|ccc|ccccc|cccc}
			& $T_o$ (s) & dt (s) & $n_e$ & $N_z$ & $n_Q$ & $n_A$ &  $n_{G}$ &  $n_{mb}$ & $P_\pi$ & $P_w$ & $T_J$ & $T_w$ \\
			\hline
			T1 (Sect. \ref{sec:testcase_1}) & 600 & 5$\cdot 10^{-1}$ & 20 & 5& 500 & 1000  & 15 & 10 & 2  & 3 & 1 & 0.5\\
			T2 (Sect. \ref{sec:testcase_2}) & 5 & 5$\cdot 10^{-4}$ & 20 & 5 & 1000 & 1000  & 15 & 5 & 4  & 4  & 1 & 0.5\\
			T3 (Sect. \ref{sec:testcase_3}) & 4.32 $\cdot 10^5$ & 30 & 20 & 1 &1000   & 1000  & 30  & 5  & 4 & 20 & 1 & 0.5\\
		\end{tabular}
	\end{table}

	\begin{figure}[h!]
		\begin{subfigure}[b]{0.45\textwidth}
			\includegraphics[width=\textwidth]{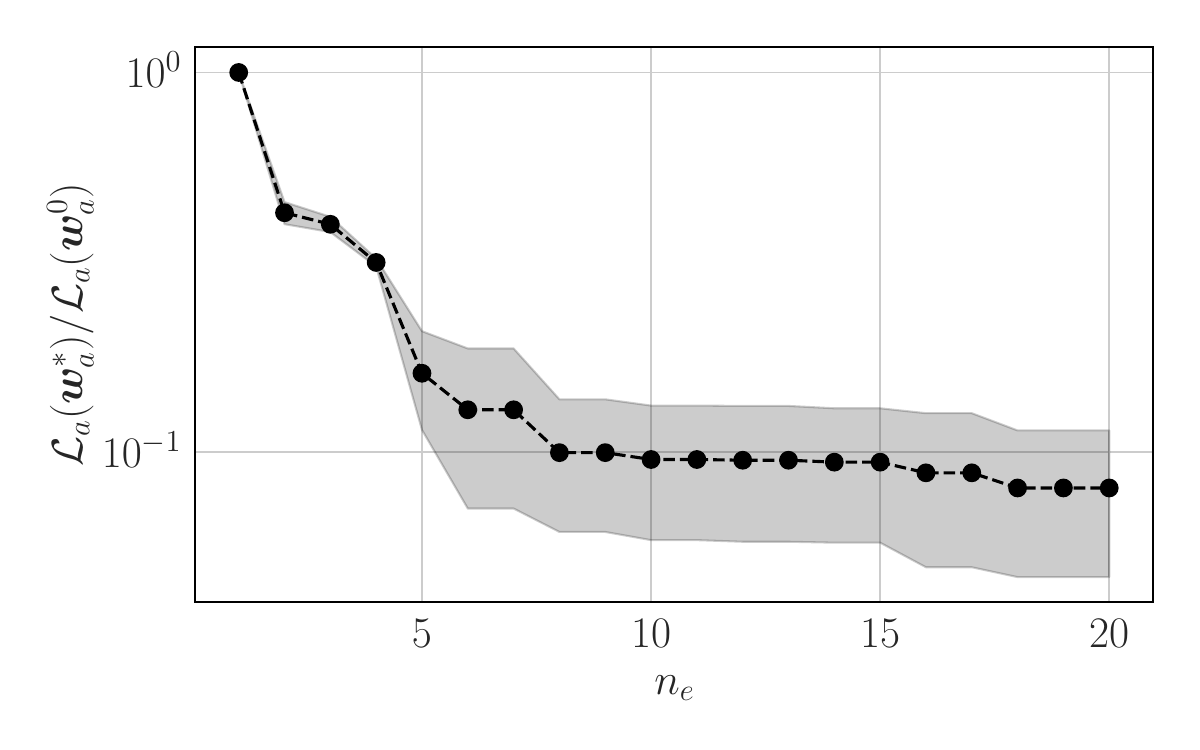}
			\caption{T1: control performances}
			\label{fig:10a}
		\end{subfigure}
		\hfill 
		\begin{subfigure}[b]{0.45\textwidth}
			\includegraphics[width=\textwidth]{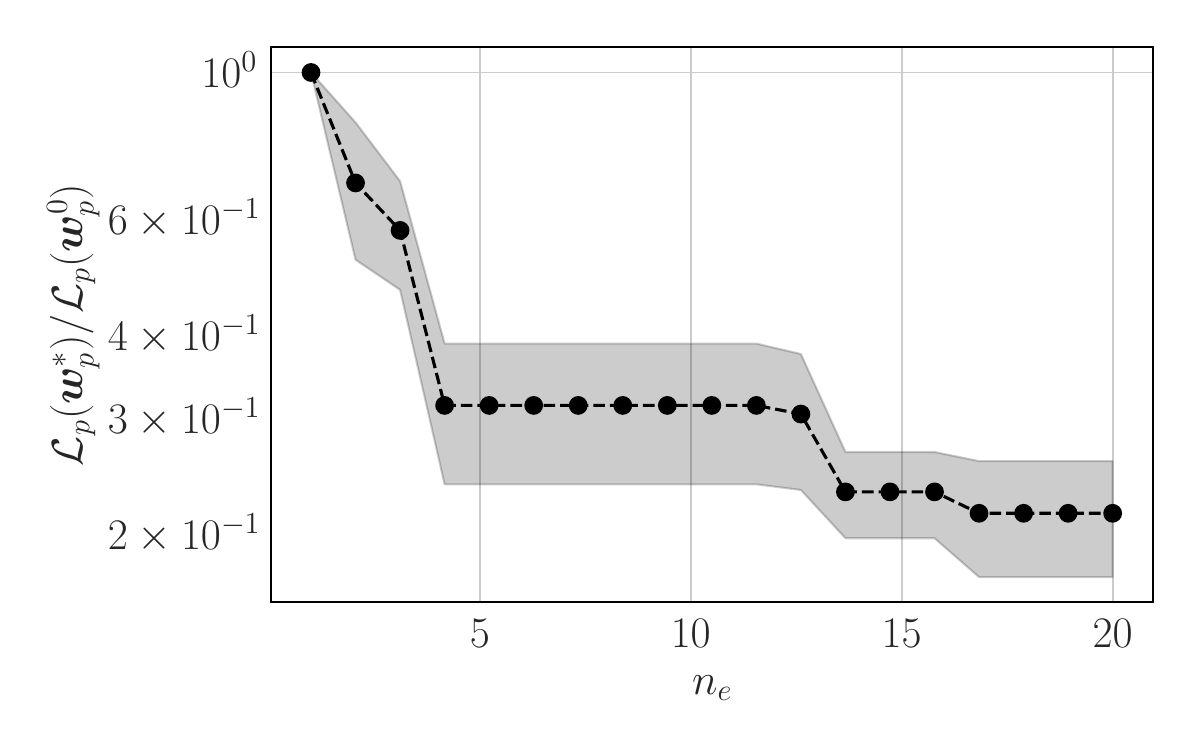}
			\caption{T1: assimilation performances}
			\label{fig:10b}
		\end{subfigure}
		
		\begin{subfigure}[b]{0.45\textwidth}
			\includegraphics[width=\linewidth]{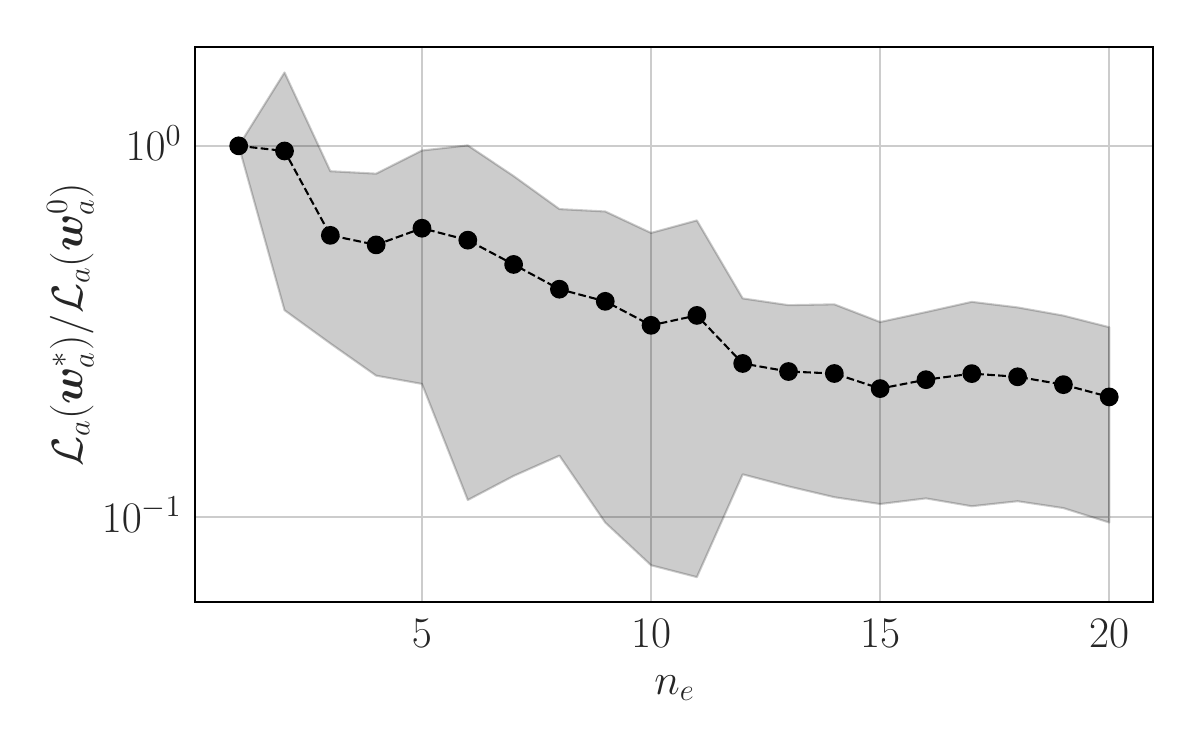}
			\caption{T2: control performances}
			\label{fig:10c}
		\end{subfigure}
		\hfill 
		\begin{subfigure}[b]{0.45\textwidth}
			\includegraphics[width=\linewidth]{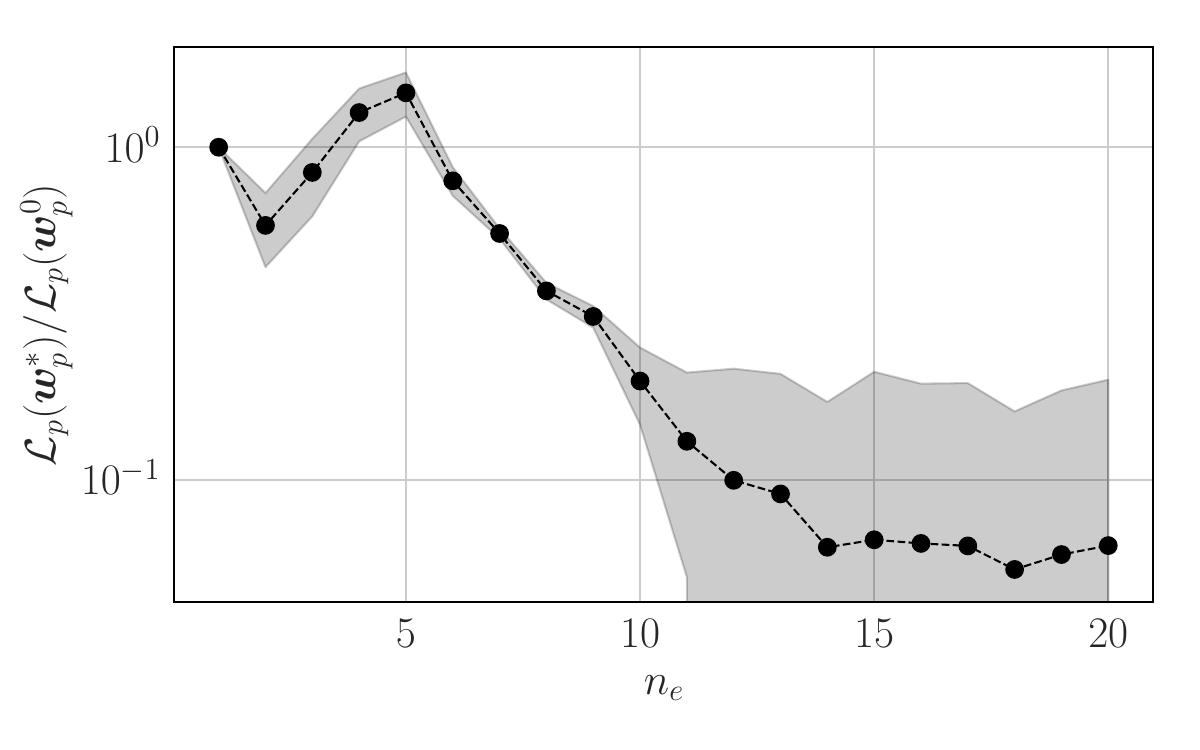}
			\caption{T2: assimilation performances }
			\label{fig:10d}
		\end{subfigure}
		\begin{subfigure}[b]{0.45\textwidth}
			\includegraphics[width=\linewidth]{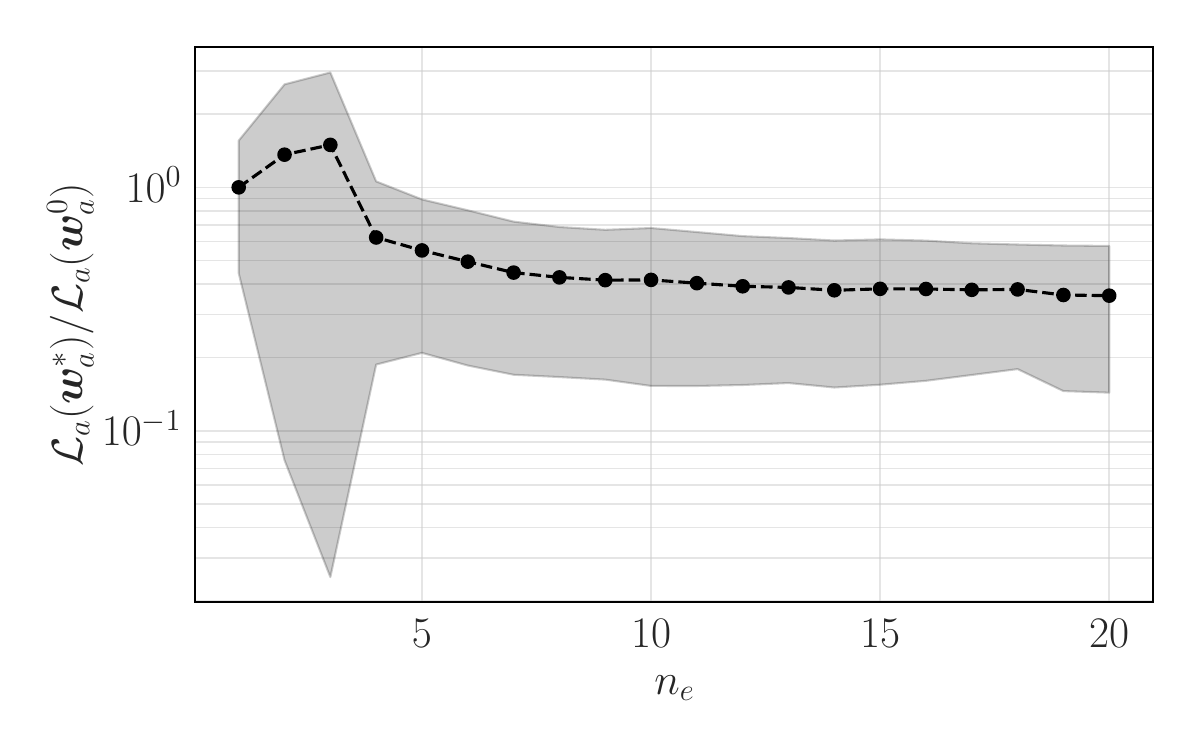}
			\caption{T3: control performances}
			\label{fig:10e}
		\end{subfigure}
		\hfill
		\begin{subfigure}[b]{0.45\textwidth}
			\includegraphics[width=\linewidth]{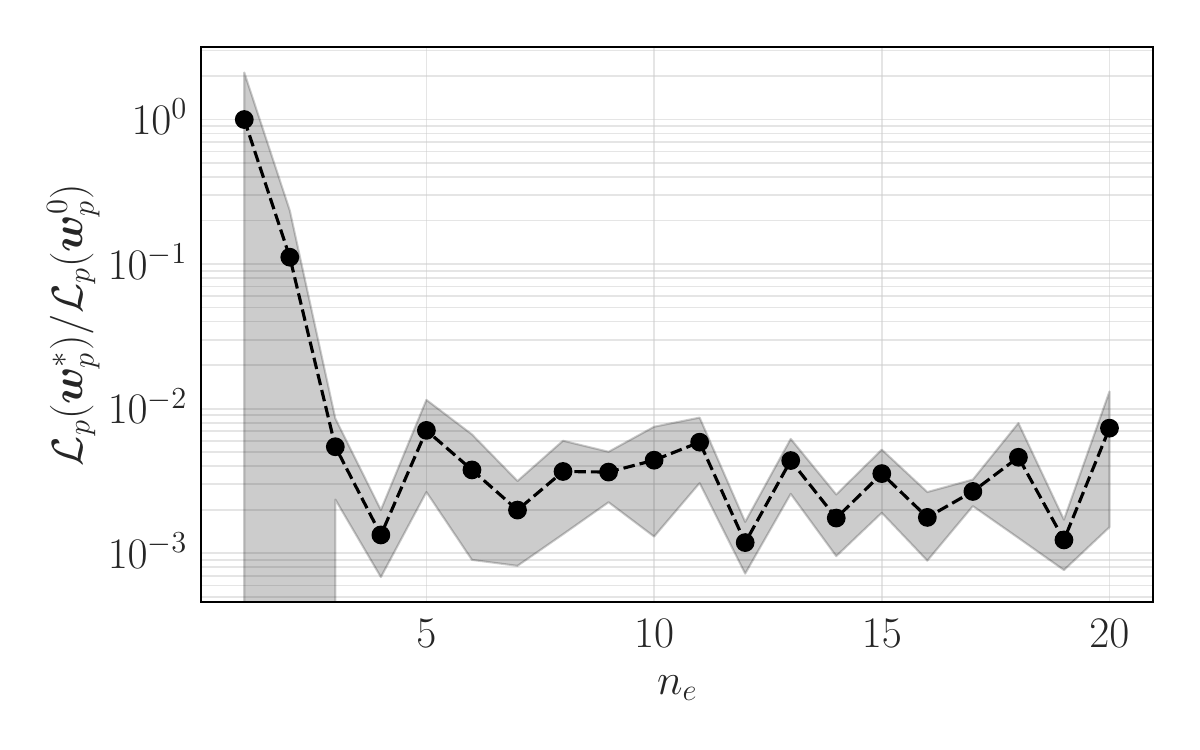}
			\caption{T3: assimilation performances}
			\label{fig:10f}
		\end{subfigure}
		
		\caption{Control $\mathcal{L}_a(\bw_a)$, and assimilation performances, $\mathcal{L}_p(\bw_p)$. Both are normalized by the initial cost, $\mathcal{L}_a(\bw_a^0)$ and $\mathcal{L}_p(\bw_p^0)$, respectively.}
		\label{fig:10}
	\end{figure}

	\begin{table}[ht]
		\centering
		\caption{RMA Learning performances, normalized by the mean initial cost.}
		\label{tab:table_learning}
		\begin{tabular}{c|c|cc}
			& $\mathcal{L}(\bw_p^*) / \mathcal{L}(\bw_p^0)$& $\mathcal{L}(\bw_{a, mb}^*) / \mathcal{L}(\bw_{a, mb}^0)$ & $\mathcal{L}(\bw_{a, mf}^*) / \mathcal{L}(\bw_{a, mf}^0)$ \\
			\hline
			T1 (Sec. \ref{sec:testcase_1}) & (2.1 $\pm $ 0.8) $\cdot 10^{-1}$ & \textbf{(0.8 $\pm$ 0.5) $\cdot 10^{-1}$} & (1.3 $\pm$ 1.2) $\cdot 10^{-1}$\\
			T2 (Sec. \ref{sec:testcase_2}) & (3.2 $\pm$ 4.9) $\cdot 10^{-2}$ & (4.18 $\pm$ 0.01) $\cdot 10^{-1}$ & \textbf{(2.01 $\pm$ 0.45) $\cdot 10^{-1}$} \\
			T3 (Sec. \ref{sec:testcase_3}) & (3.15 $\pm$ 0.9) $\cdot 10^{-4}$  &\textbf{(4.47 $\pm$ 2.68) $\cdot 10^{-1}$} & (4.80 $\pm$ 3.15) $\cdot 10^{-1}$\\
		\end{tabular}
	\end{table}

	\subsection{Learning Performances}\label{6_1}
	
	The learning performances for all the three cases are reported in Figure \ref{fig:10} and their numerical values tabulated in Table \ref{tab:table_learning}. The Figure shows the mean control or assimilation performances (first or second column, respectively) with a dashed line, and a confidence interval with a standard deviation above and below the mean performances.
	The proposed RT algorithm learns to control effectively the system in a limited number of iterations for all test cases, showing good sample efficiency. Indeed, the optimized policy performs up to ten times better than the initial one, as shown by the average control performance ($\mathcal{L}_a$). The large standard deviation at the beginning of the training of T2 and T3 can be explained by the initial exploration phase of the model-free loop. However, the standard deviation gets roughly constant as the policy optimization converges to a specific region of the search space, as shown in Figures \ref{fig:10c} and \ref{fig:10d}. Analysing the adjoint-based assimilation performances, Figures \ref{fig:10b}, \ref{fig:10d} and \ref{fig:10f} show that the methodology is able to tailor the digital twin on the current system specifics within five to ten iterations. At the end of the training process, the assimilation reduces the initial cost $\mathcal{L}^{(0)}_p$ from the initial guess $\bm{w}_p$ by a factor $10$, $100$ and $10000$ for T1, T2 and T3, respectively.
	
\begin{figure}
		\centering
		\begin{subfigure}[b]{0.47\linewidth}
			\centering
			\includegraphics[width=\linewidth]{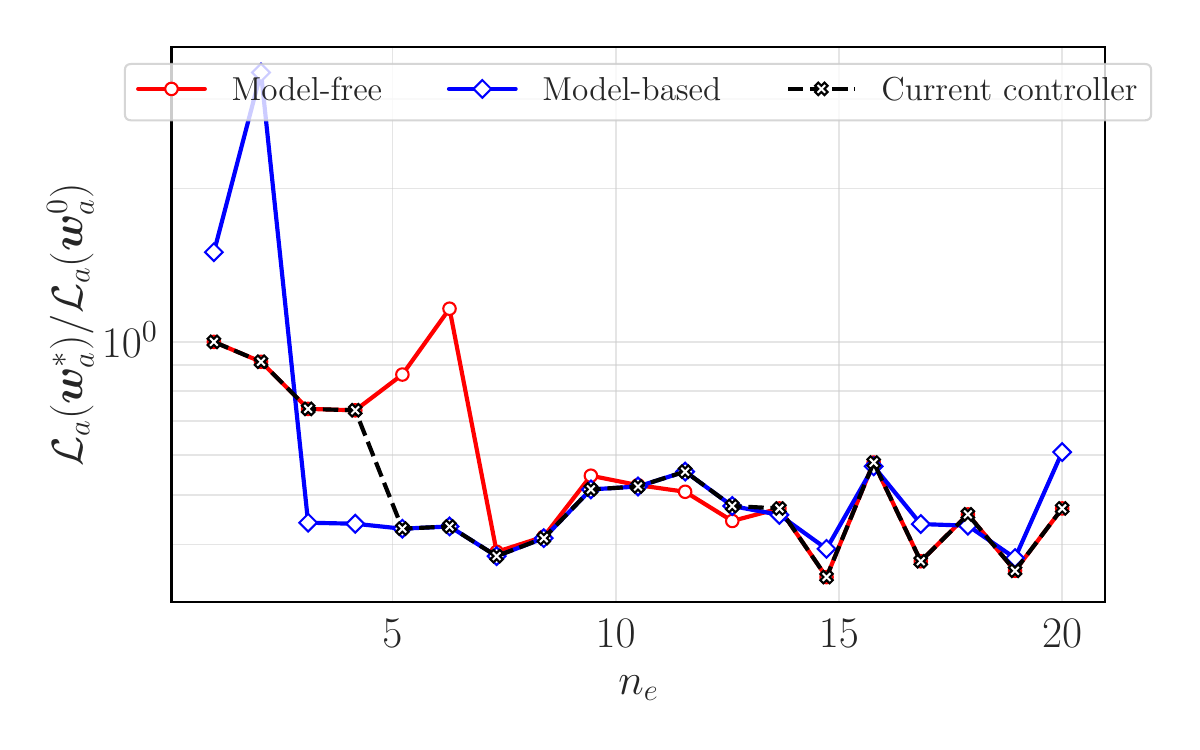}
			\caption{T1}
			\label{fig:11a}
		\end{subfigure}
		\centering
		\begin{subfigure}[b]{0.47\linewidth}
			\centering
			\includegraphics[width=\linewidth]{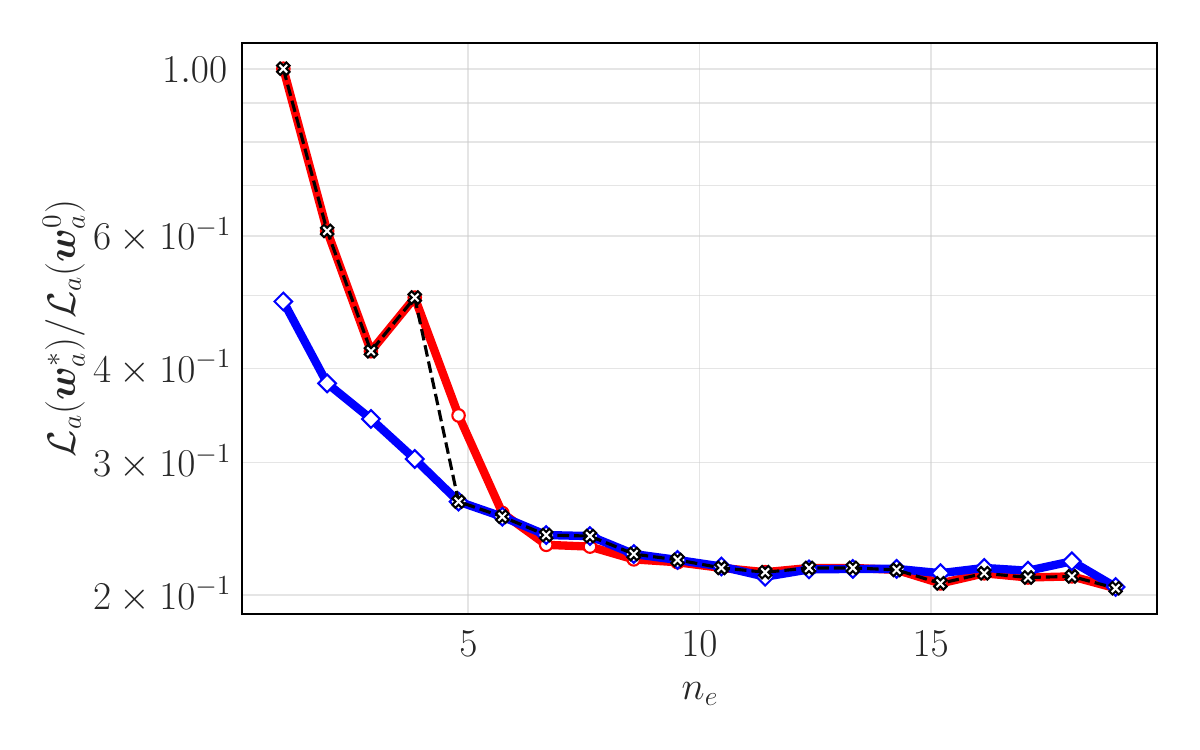}
			\caption{T2}
			\label{fig:11b}
		\end{subfigure}
		\centering
		\begin{subfigure}[b]{0.47\linewidth}
			\centering
			\includegraphics[width=\linewidth]{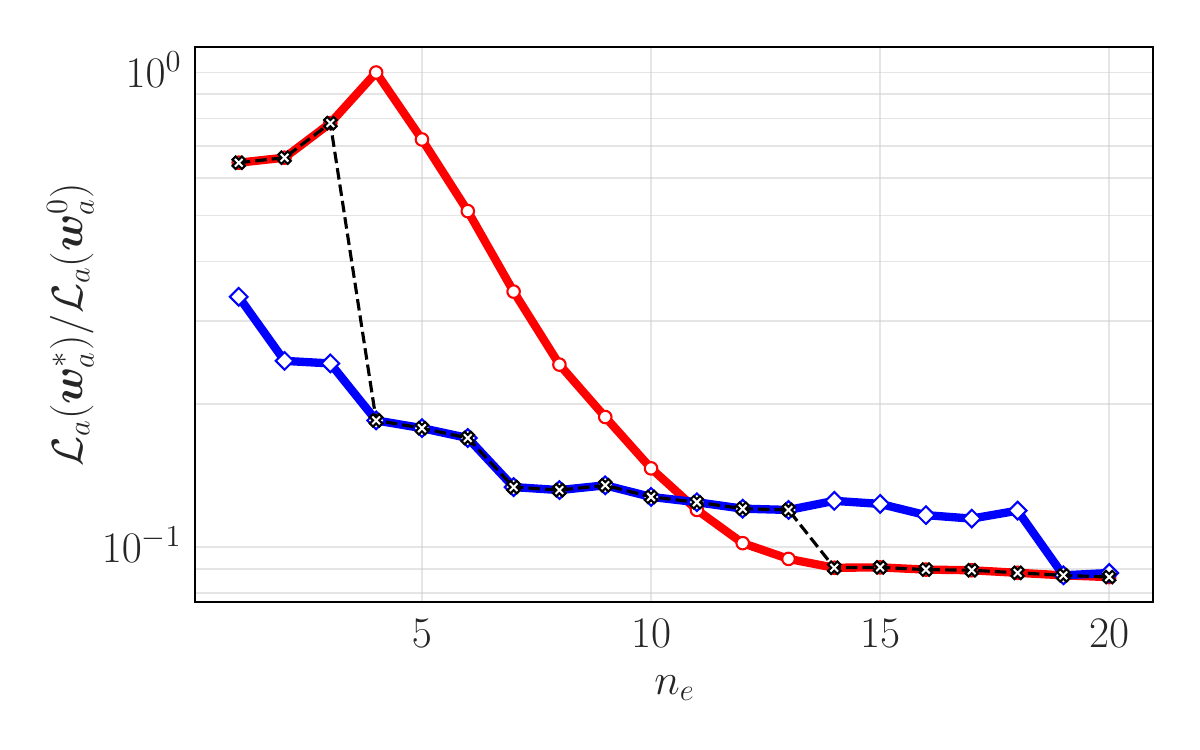}
			\caption{T3}
			\label{fig:11c}
		\end{subfigure}
		\caption{Policy switch detail. Switch between ``live" and ``idle" policies, during interactions with the real system.}
		\label{fig:11}
	\end{figure}

	This, in turn, reflects on the model-based policy optimization (Step 5). In fact, as the digital twin predictions of the system response improve, the quality of the policy optimized ``offline" increases. This also affects the policy switch (Step 6), since the performances on the twin become more representative of the real responses. The effect of Step 6 is illustrated in Fig. \ref{fig:11}. All experiments begin with the model-free policy driving the interactions with the real environment, i.e. as ``live policy", while the model-based is initially set as ``idle". As shown in Figure \ref{fig:11}, the switch between model-free and model-based policies occurs at $n_e=7$ in T1 and $n_e=4$ for both T2 and T3, as $C_\pi>P_\pi$ in the policy switch in Figure \ref{decisionTree}. After the switch, the model-free control performance registers a steeper improvement despite remaining idle. This is due to a combination of factors: (1) the model-based acts as an ``expert" for the model-free, providing high quality samples recorded in $\mathcal{D}_\mathcal{L}$ from which the model-free loop can learn; and (2) the critic function approximation gets better at predicting the value associated to the recorded transitions. Interestingly, in both T1 and T3 (Figure \ref{fig:11a}, \ref{fig:11c}), the model-free learns a better policy by imitating the model-based counterpart, eventually ``surpassing the master" and earning again the status of a ``live" policy again within a few episodes. 
	
	Comparing the performances of the model-based and model-free loops, the model based performs slightly better than the model based in T1, but worst in T2 and equally in T3, where they converge to the same policy. The different performances in the different cases highlights the benefits of having two approaches to tackle the same problem and the merit in their combination.

	\subsection{Performances on Wind Turbine Control}\label{subsect:res_wt}

\begin{figure}
		
		\begin{subfigure}[b]{0.45\textwidth}
			\includegraphics[width=\textwidth]{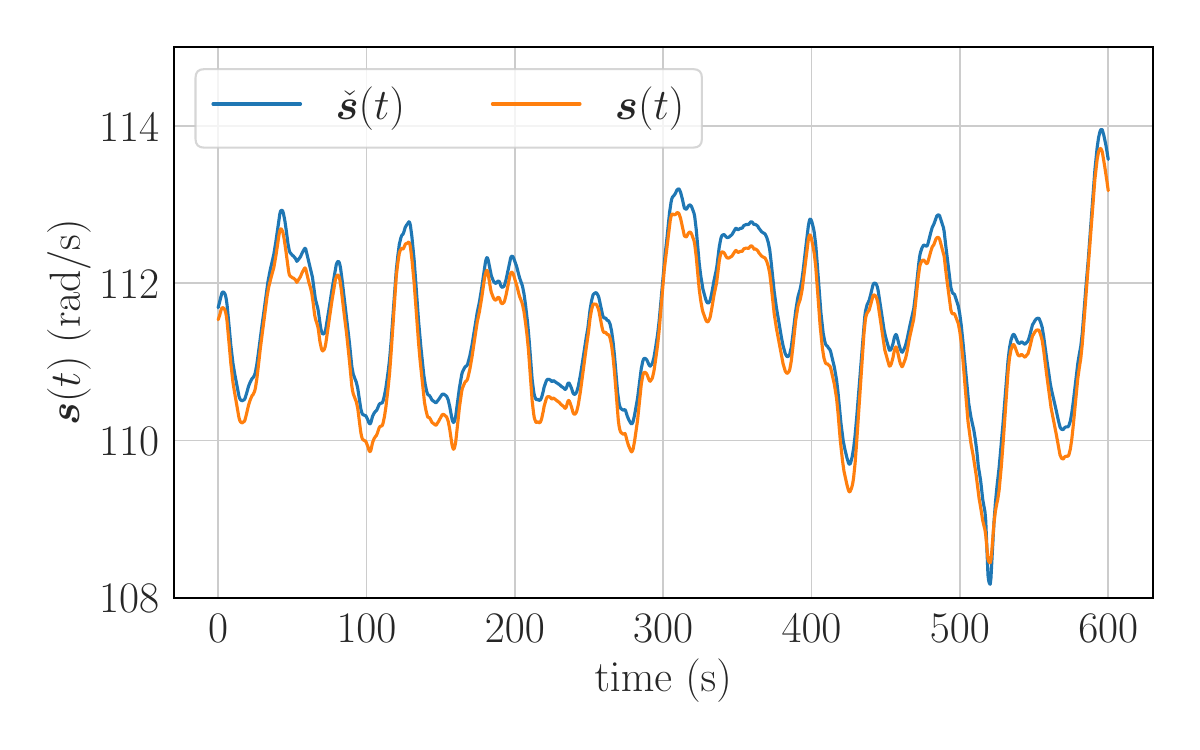}
			\caption{}
			\label{fig:12a}
		\end{subfigure}
		\hspace{2cm}
		\begin{subfigure}[b]{0.28\textwidth}
			\includegraphics[width=\textwidth]{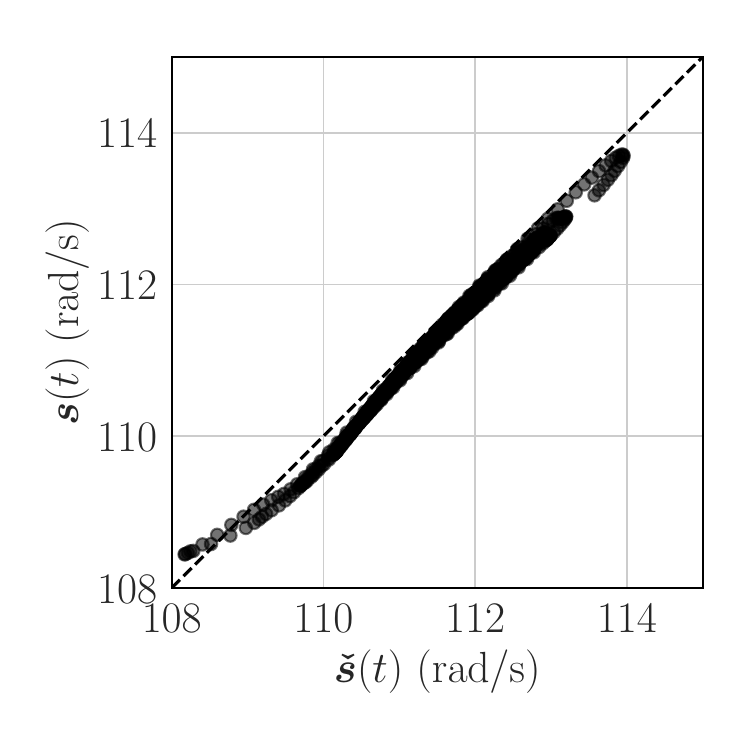}
			\caption{}
			\label{fig:12b}
		\end{subfigure}
		\caption{T1 - Digital Twin agreement with real system. Evaluation of the optimal $\bw_p = \bw_p^*$ with randomized initial conditions and disturbances. Time-series (a) and scatter plot (b), plotting real state on the x-axis and predicted one on the y-axis.}
		\label{fig:12}
	\end{figure}

	The digital twin performances on T1 are shown in Figure \ref{fig:12}. Figure \ref{fig:12a} shows the average prediction versus the average system evolution as a function of time, while Figure \ref{fig:12b} plots one versus the other. We conclude that the digital twin successfully tracks the dynamics of the real system albeit with a systematic under-estimation of roughly 1 (rad /s) or 9.55 (RPM), especially at low rotational speeds. Whether this bias is inherently linked with the chosen parametrization or an insufficient number of samples requires further investigation. Differently from T2 and T3, the coefficients to be identified are unknown \textit{a priori}, as the initial coefficients $\bw_p^0$, derived in the work of \cite{saint-drenan_parametric_2020}, describe a power curve in steady wind conditions. Therefore, the average assimilation loss $\mathcal{L}_p \approx \mathcal{O}(10^{-1})$ highlights the capability of the algorithm to adapt online, possibly taking into consideration changes in the operative conditions of the system itself.
	
	%Nevertheless, the violin plot (figure on the right) concerning this evaluation confirms the good performance of the digital twin, with a clear peak at $\mathcal{J}_p \approx 100$. Arguably, the spreading at higher costs may be due to the low-speed conditions described above. 
	
	% %A fictitious time-series showing the mean over the one hundred simulations of the real state, $\check{\bs}(t)$, and \textit{assimilated}, or predicted, state, $\bs(t)$, represented in Figure \ref{fig:wt_ass}, left, presents an overall good agreement between the two. As strengthened by the scatter plot in the middle, the digital twin tends to underestimate the effective generator speed in lower regimes. In absolute value, the agreement is sufficiently good, showing a deviation between the two of roughly $\Delta \bs \lessapprox 1$ (rad / s) or 9.55 (RPM). Whether this behaviour is inherently linked with the parametrization chosen or due to an insufficient number of samples is up to debate, and will be inquired in future works. Nevertheless, the violin plot concerning this evaluation confirms the good performance of the digital twin, with a clear peak at $\mathcal{J}_p \approx 1E2$. Arguably, the spreading at higher costs may be due to the low-speed conditions described above. 
	
\begin{figure}[h!]
		\centering
		\begin{subfigure}[b]{0.45\textwidth}
			\centering
			\includegraphics[width=\textwidth]{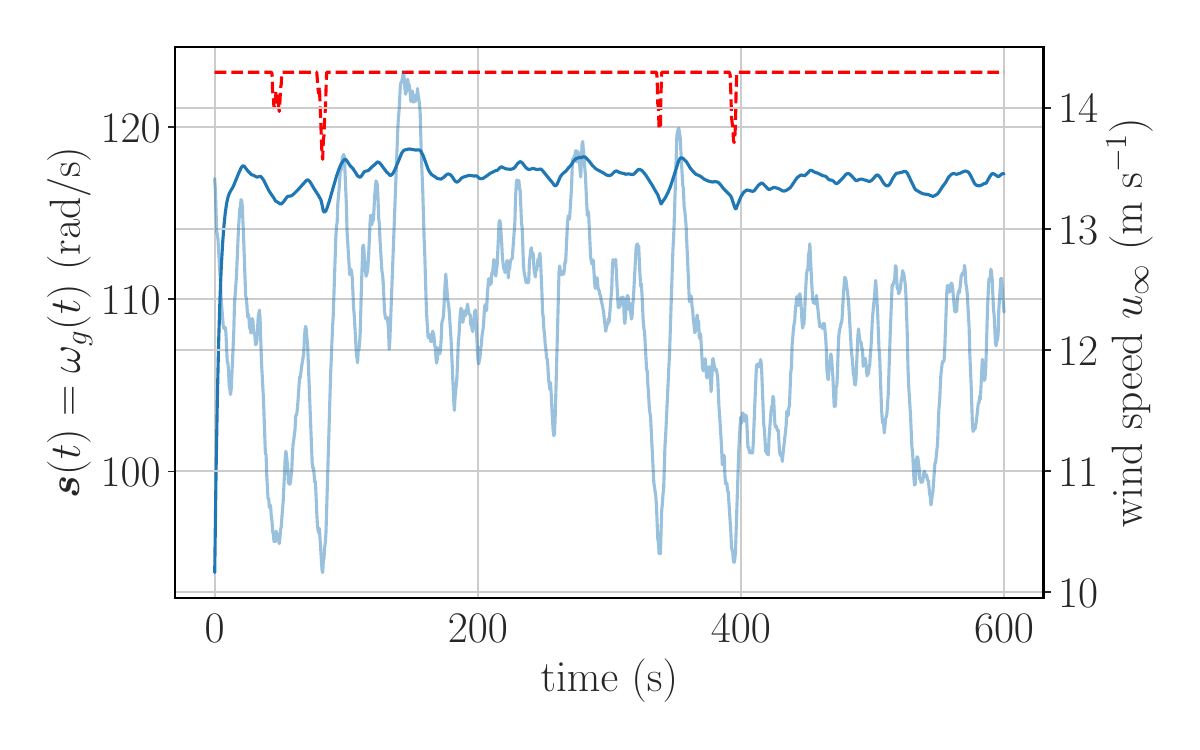}
			\caption{}
			\label{fig:00a}
		\end{subfigure}
		\begin{subfigure}[b]{0.45\textwidth}
			\centering
			\includegraphics[width=\textwidth]{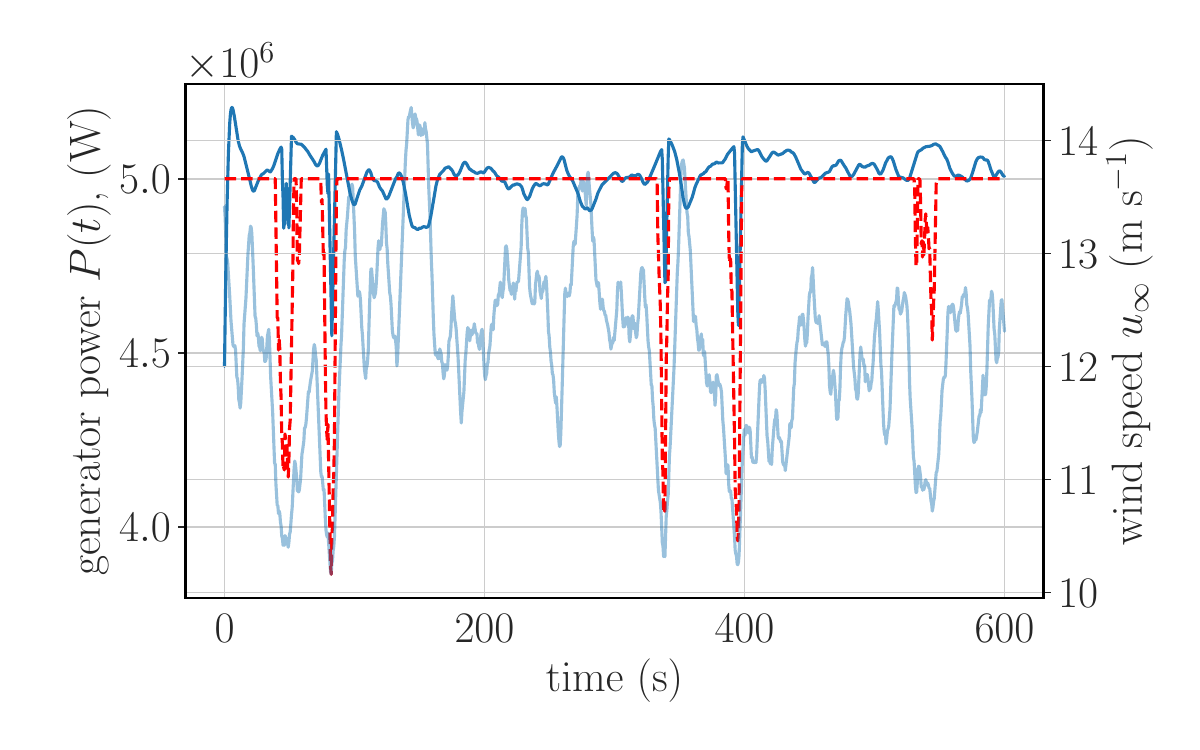}
			\caption{}
			\label{fig:00b}
		\end{subfigure}
		\hfill 
		\begin{subfigure}[b]{0.45\textwidth}
			\centering
			\includegraphics[width=\textwidth]{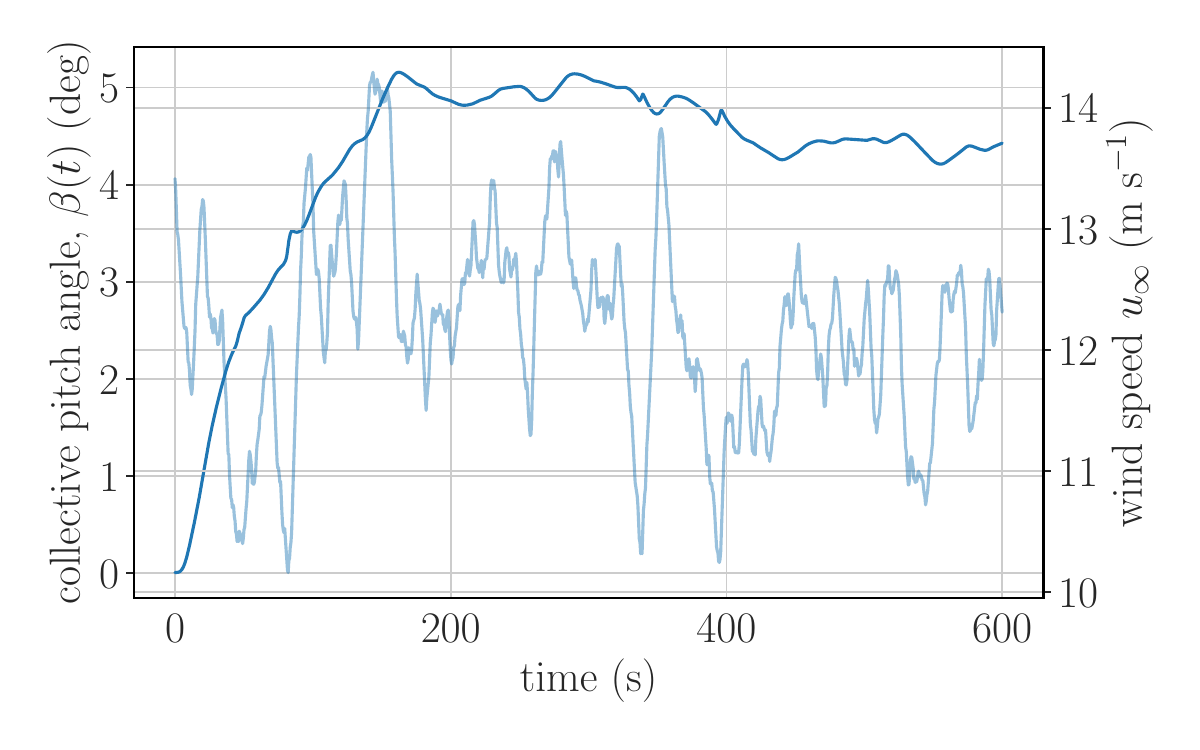}
			\caption{}
			\label{fig:00c}
		\end{subfigure}
		\begin{subfigure}[b]{0.45\textwidth}
			\centering
			\includegraphics[width=\textwidth]{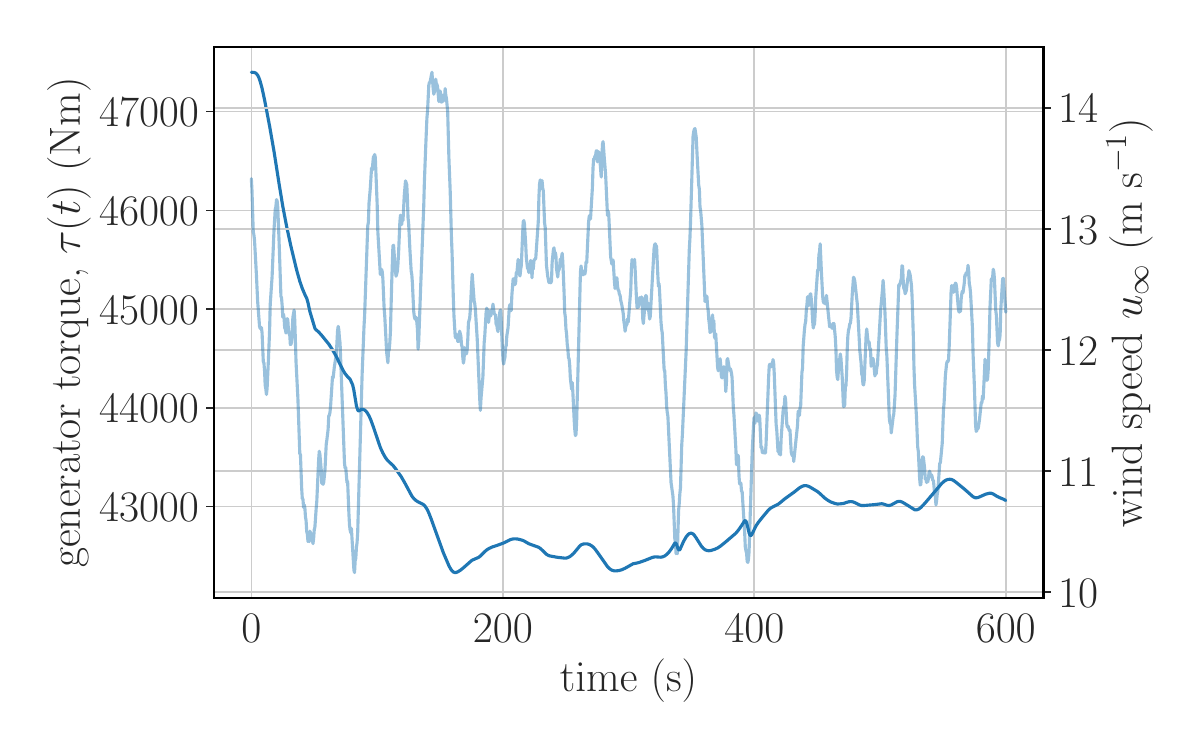}
			\caption{}
			\label{fig:00d}
		\end{subfigure}
		\caption{T1 - Control performances. System evolution against set-points (dashed red line) in (a) and (b), filtered actuation in (c) and (d).}
		\label{fig:00}
	\end{figure}

	The control performances are summarized in Figure \ref{fig:00}. The controller is able to keep the system close to the desired set-points, as shown in Figures \ref{fig:00a} and \ref{fig:00b}. This is especially true for the power tracking, always kept very close to $\tilde{P}$ despite the noticeably varying wind speeds (Figure \ref{fig:00b}). On the other hand, the controller consistently keeps the generator speed ($\omega_g$) below-rated conditions by approximately 8 rad/s. This lower-than-rated rotational speed reflects on the actuated torque, in Figure \ref{fig:00d}, which is always in the middle between the allowed action space. In addition, the very fast actuation rate on the torque ($\dot{\tau}$ = 15 kN/s) allows to rapidly adapt in case of abrupt changes in the wind speed, while fine tuning the actuation with the pitch angle $\beta$. The latter follows consistently the wind speed, which is desirable. At higher wind speeds than rated ($u_\infty > 11.4$ m/s) it consistently plays on the blades' angle to reduce the aerodynamic torque to keep the system close to desired conditions. This aerodynamic de-rating of the rotor is visible in Figure \ref{fig:14b} which shows a peak of $C_p$ at rated conditions ($u_\infty \approx 11$ (m/s)) with a $C_p = 0.42$, steadily decreasing for higher winds.
	
	We note that the interplay between the two set-points on power and generator speed could be the modified fine-tuning the normalization coefficients $\alpha_1$ and $\alpha_2$, as commonly done in multi-objective optimization problems.

	\begin{figure}[b]
		\centering
		\includegraphics[width=0.7\textwidth]{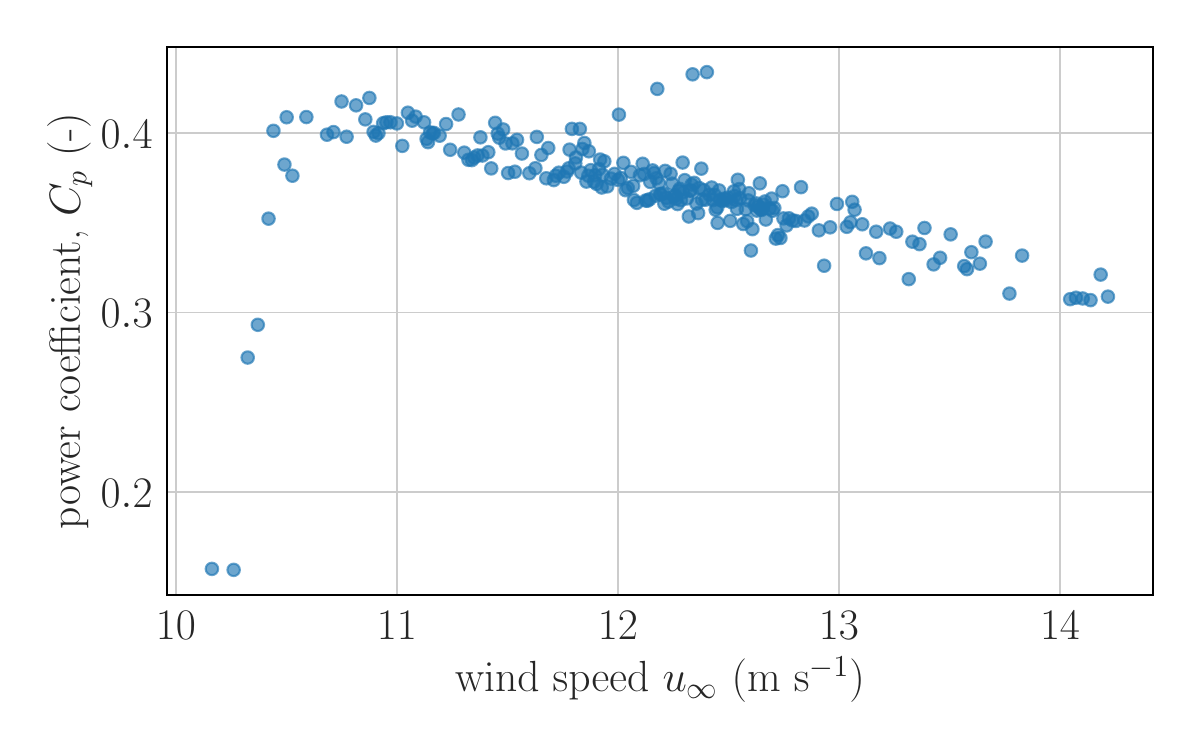}
		\caption{T1 - Power coefficient ($C_p$) against operative wind speeds.}
		\label{fig:14b}
	\end{figure}

\begin{figure}[h!]
		\centering
		\begin{subfigure}[b]{0.30\textwidth}
			\centering
			\includegraphics[width=\textwidth]{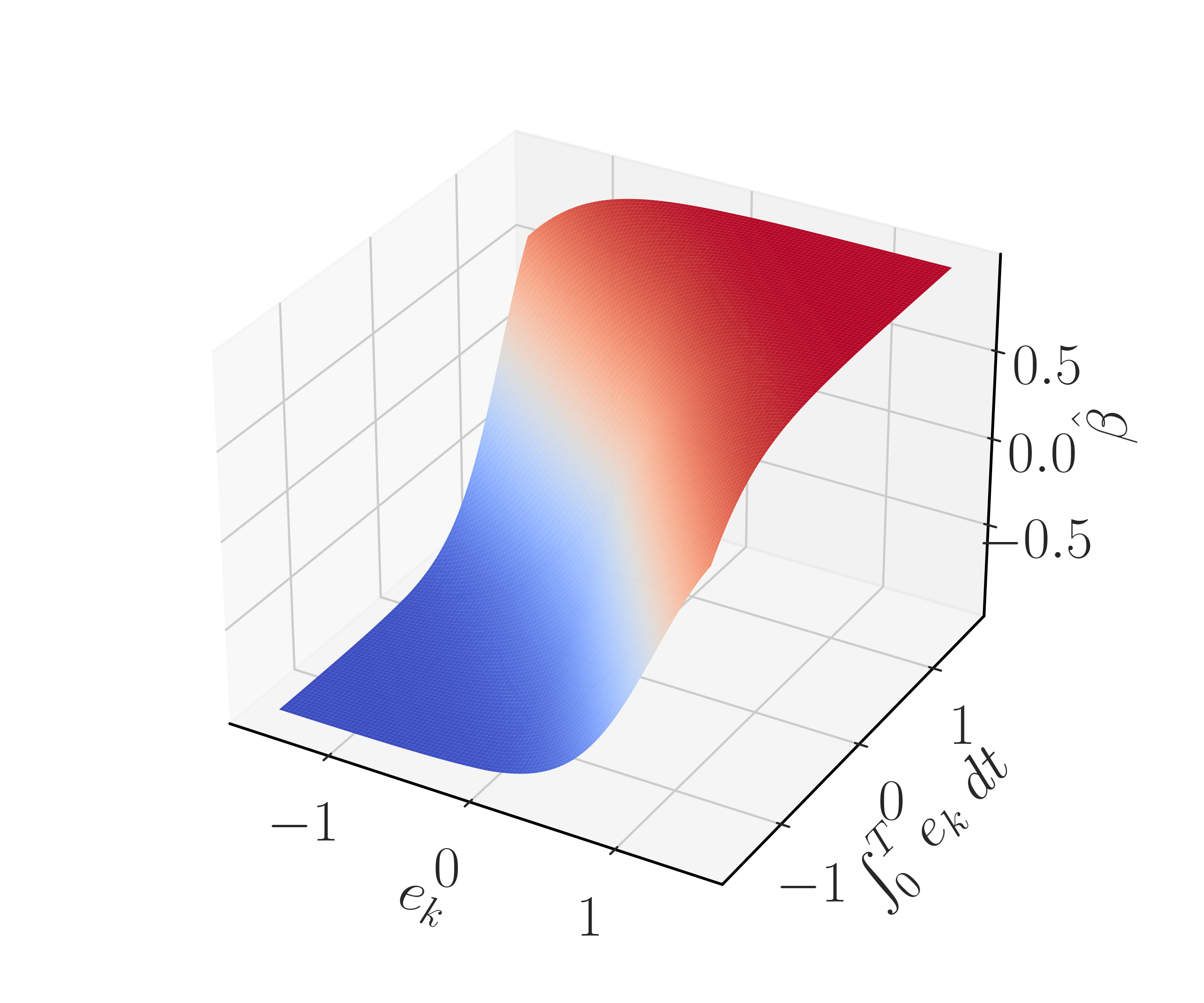}
			\caption{$\beta = \pi_{mb} (e_k, \int e_k)$}
			\label{fig:14a}
		\end{subfigure}
		\hspace{2cm}
		\begin{subfigure}[b]{0.30\textwidth}
			\centering
			\includegraphics[width=\textwidth]{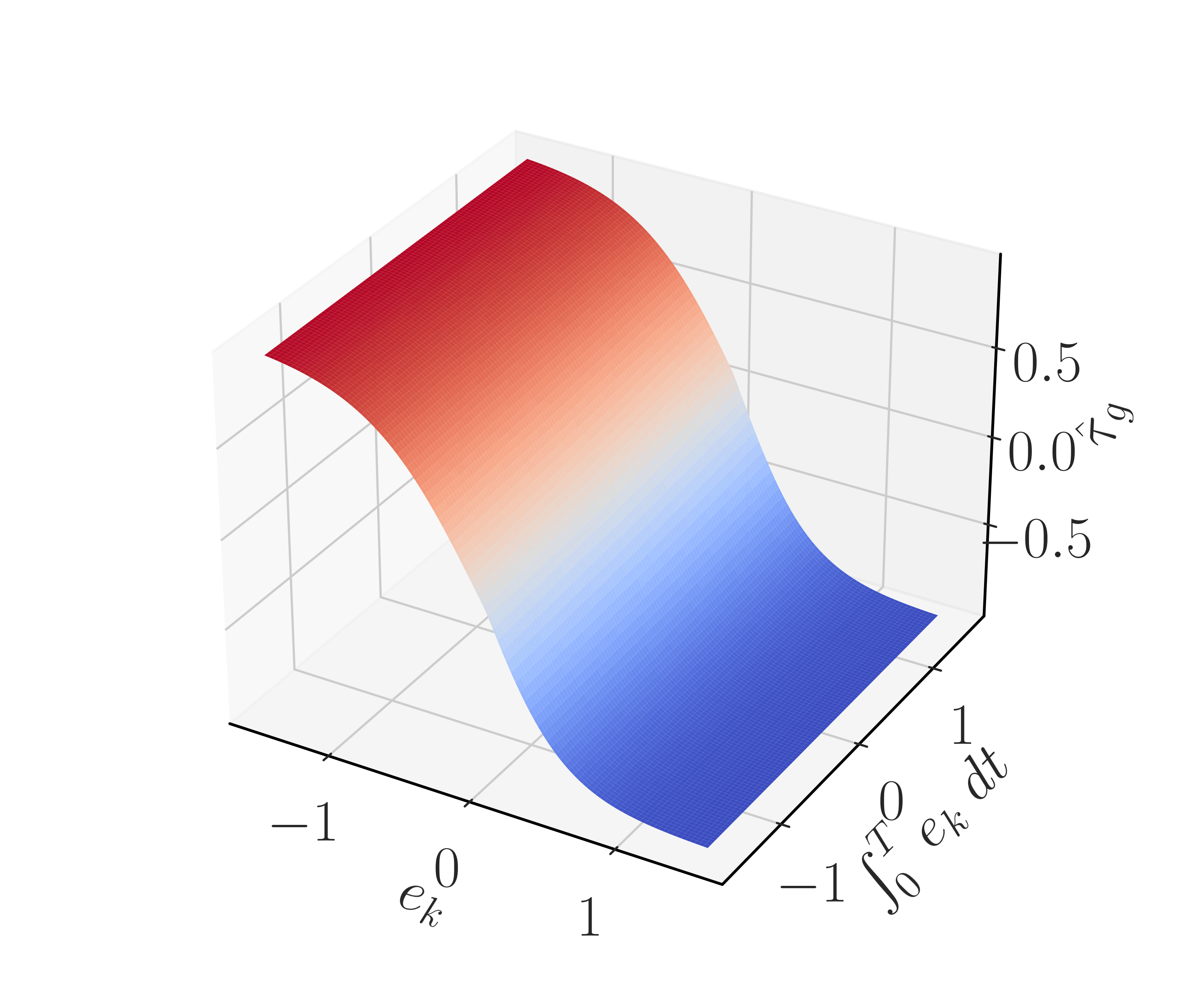}
			\caption{$\tau = \pi_{mb} (e_k, \int e_k)$}
			\label{fig:14b}
		\end{subfigure}
		\hfill 
		\begin{subfigure}[b]{0.30\textwidth}
			\centering
			\includegraphics[width=\textwidth]{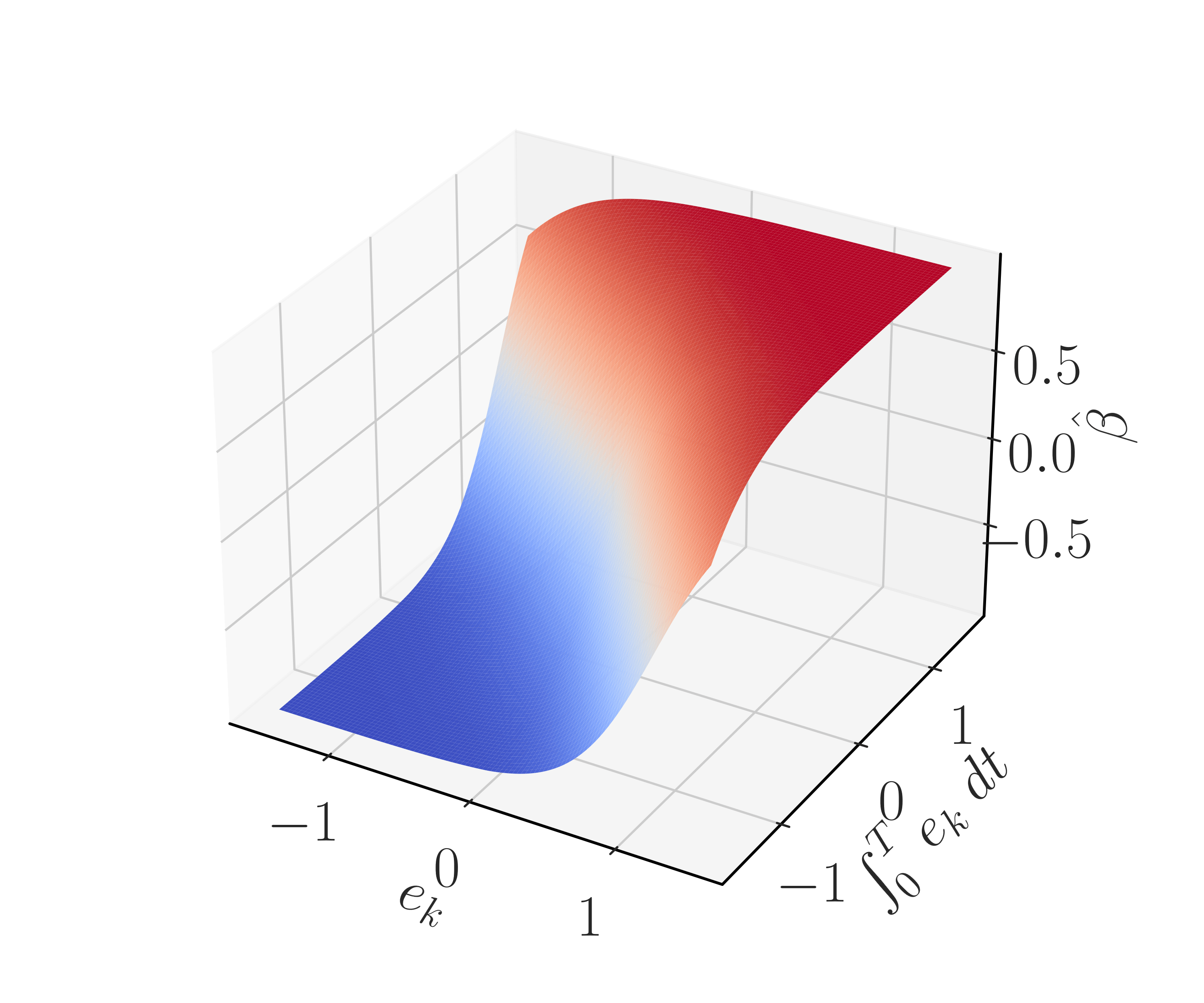}
			\caption{$\beta = \pi_{mf} (e_k, \int e_k)$}
			\label{fig:14c}
		\end{subfigure}
		\hspace{2cm}
		\begin{subfigure}[b]{0.30\textwidth}
			\centering
			\includegraphics[width=\textwidth]{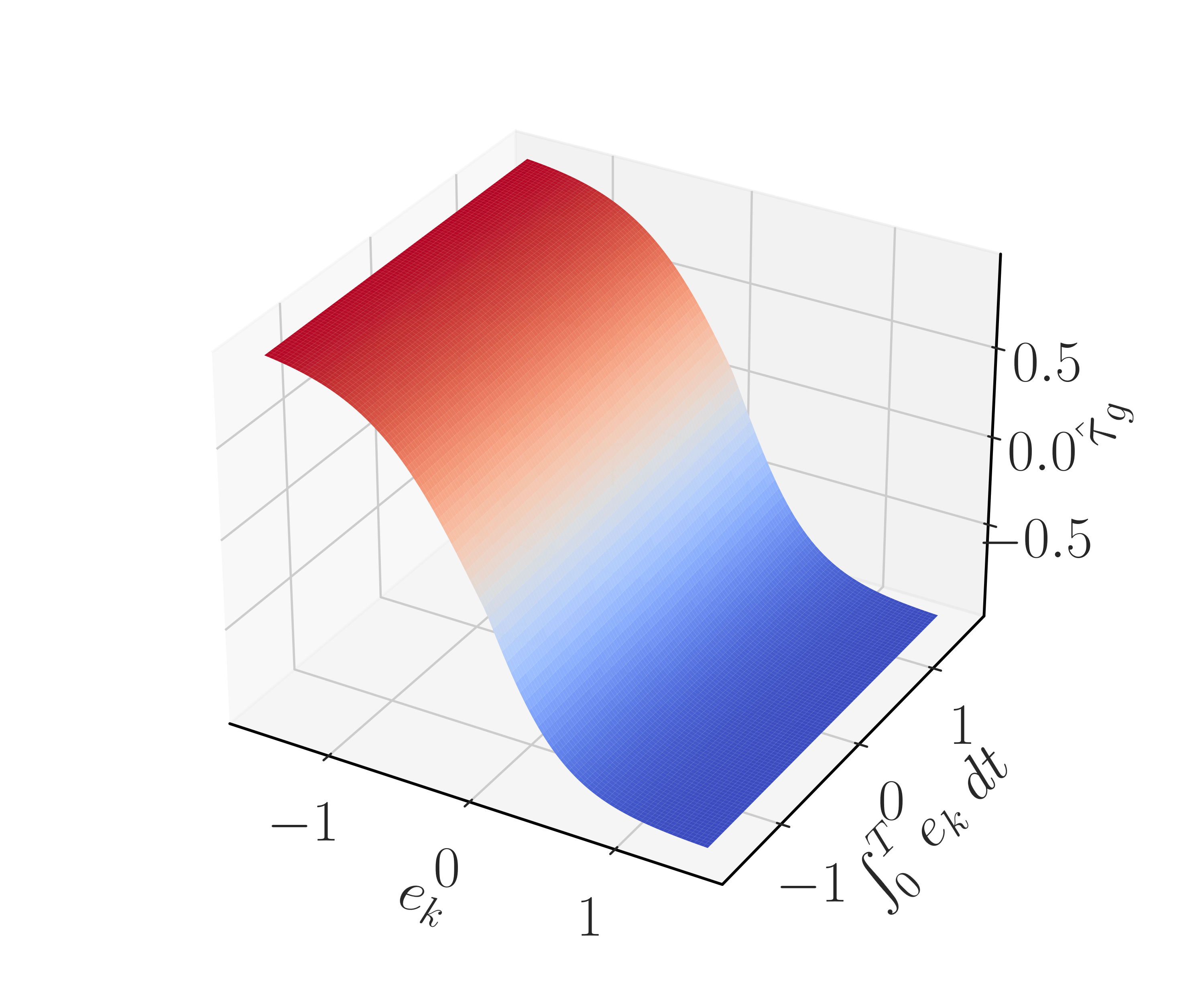}
			\caption{$\tau = \pi_{mf} (e_k, \int e_k)$}
			\label{fig:14d}
		\end{subfigure}
		\caption{T1 - Normalized policy contours. Model-based (a - b) and model-free (c - d) actors, as a function of the error $e_k(t) = \omega_g(t) - \tilde{\omega}_g(t)$ and its integral over the past revolution. }
		\label{fig:13}
	\end{figure}

	The model-based and model-free policies are compared in Figure \ref{fig:13}. The two optimization loops converge to essentially the same parametrizations, leading to the same policy functions, as also supported by the very similar mean cost functional $\mathcal{L}_a$ listed in Table \ref{tab:table_learning}. The only (minor) difference appears in the threshold of torque actuation. In both cases, when the error on the rotational speed goes negative ($e_i = \omega_g - \tilde{\omega}_g < 0$), the controller reduces the pitch angle (Figures \ref{fig:14a},\ref{fig:14c}) to increase the aerodynamic efficiency and, consequently, the aerodynamic torque, with the goal of increasing the rotational speed to get closer to the set point. The opposite is true when the generator is rotating too fast: the controller increases the angle of attack of the blades to slow the generator down. The integral of the error over the past revolution, $\int e$, essentially decides on the slopes of the control surface. Interestingly, looking at the torque actuation shown in Figures \ref{fig:14b} and \ref{fig:14d}, it seems that this control actuation is almost insensitive to this control input, relying on the scalar error solely. This control surface is directly connected with the objective of keeping the power around the rated value of $P = 5$ MW. In fact, as the error becomes negative, the controller increases the torque to keep the power at the end of the generator constant ($P_g = \omega_g \cdot \tau_g$). On the other hand, when the error increases, it reduces the resistive torque, relying on the pitch actuator to reduce the speed while keeping the power constant.

\begin{figure}[h!]
		\centering
		\begin{subfigure}[b]{0.45\textwidth}
			\centering
			\includegraphics[width=\textwidth]{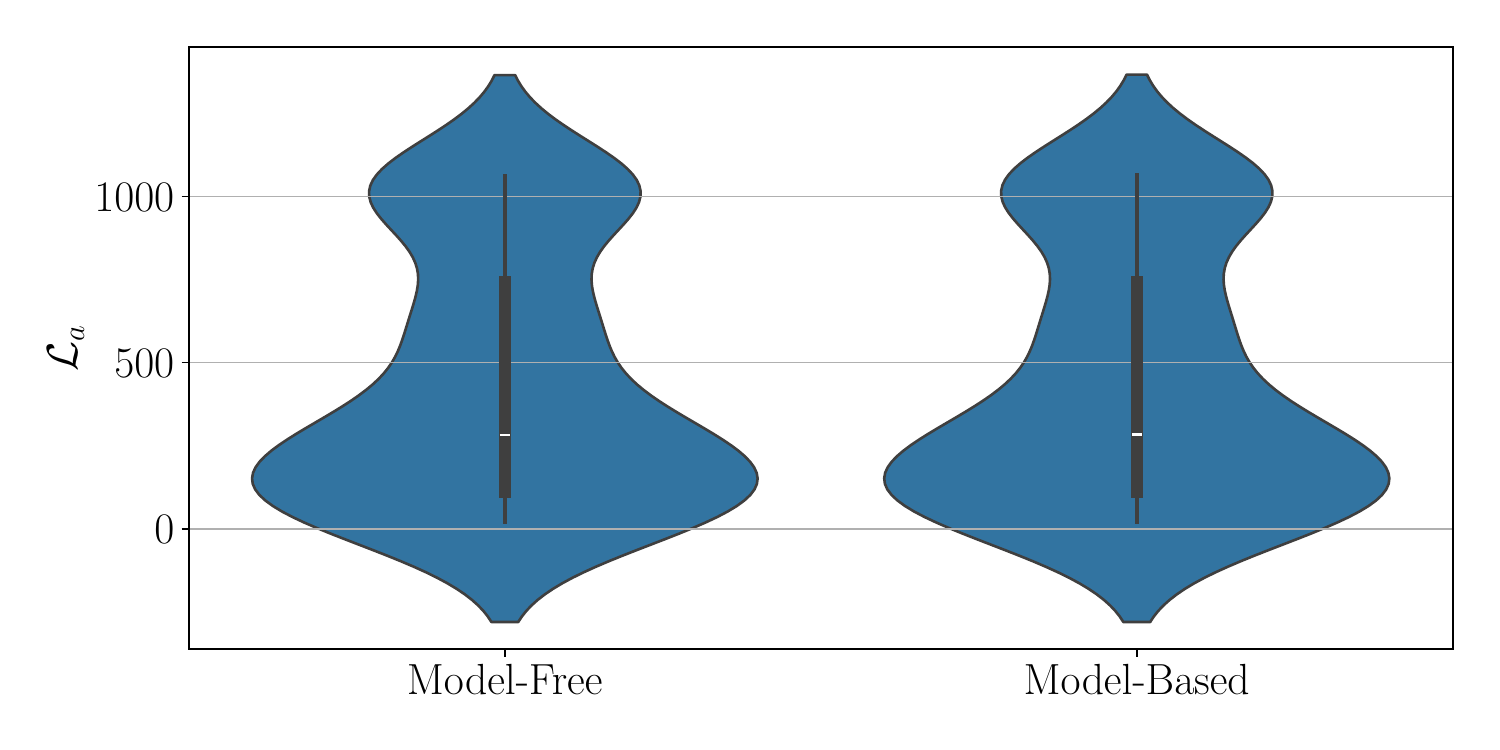}
			\caption{}
			\label{fig:15b}
		\end{subfigure}
		\hspace{2cm}
		\begin{subfigure}[b]{0.23\textwidth}
			\centering
			\includegraphics[width=\textwidth]{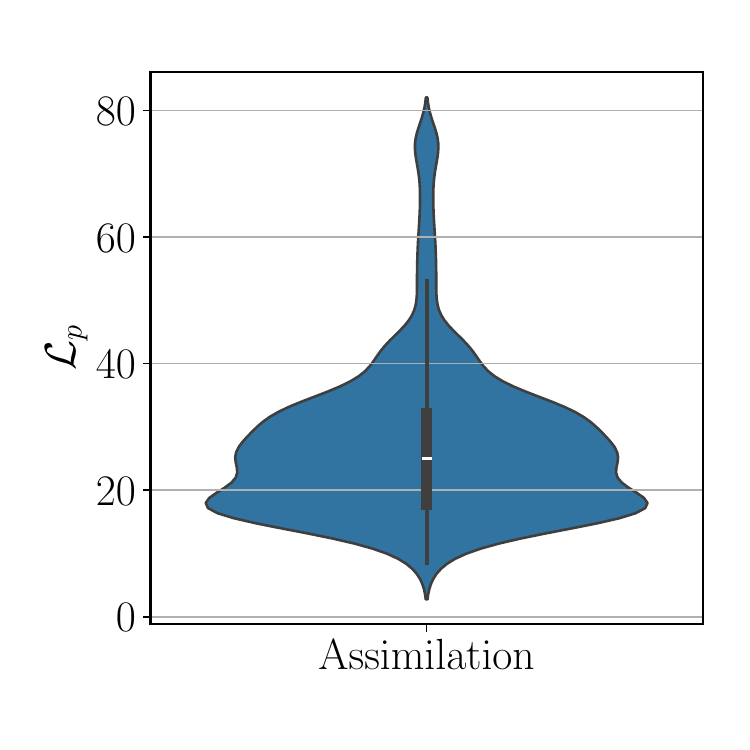}
			\caption{}
			\label{fig:15a}
		\end{subfigure}
		\caption{T1 - Robustness analysis of the control policies (a) and assimilated closure (b). }
		\label{fig:15}
	\end{figure}

	We conclude this analysis by investigating the robustness of the 
	identified models and the control laws. Figure \ref{fig:15} shows the distribution of the average  (a)  control cost function for both model based and model free and (b) assimilation, obtained by running one-hundred simulations on randomized initial conditions and wind profiles (unseen during the training phase) and using the identified model parameters $\bw_p^*$ and policy parameters $\bw_{a, mb}^*$, $\bw_{a, mf}^*$.
	The control performances shows a multi-modal distribution in Figure \ref{fig:15a}, with the strongest peak in the region of better performances ( $\mathcal{L}_a < \mathcal{O}(10^2)$). 
	This shows that at least one local minima exists  within the range of investigate conditions and the training can eventually fall into the sub-optimal one depending on the combination of initial condition and wind profiles. A possible solution to mitigate the sensitivity to the training condition could be to extend the training to a broader range of wind profiles and/or initial conditions.
	
	The assimilations performances, on the other hand, are reliably on-track with respect to what experienced during the learning phase. The digital twin tracks the real system with very small deviation ($\mathcal{L}_p<\mathcal{O}(10^2)$) for most experienced operative conditions. 
	
	\subsection{Performances on a FWMA Trajectory Control}

\begin{figure*}[!ht]\center
		\begin{subfigure}[b]{0.5\textwidth}
			\centering
			\includegraphics[width=\linewidth]{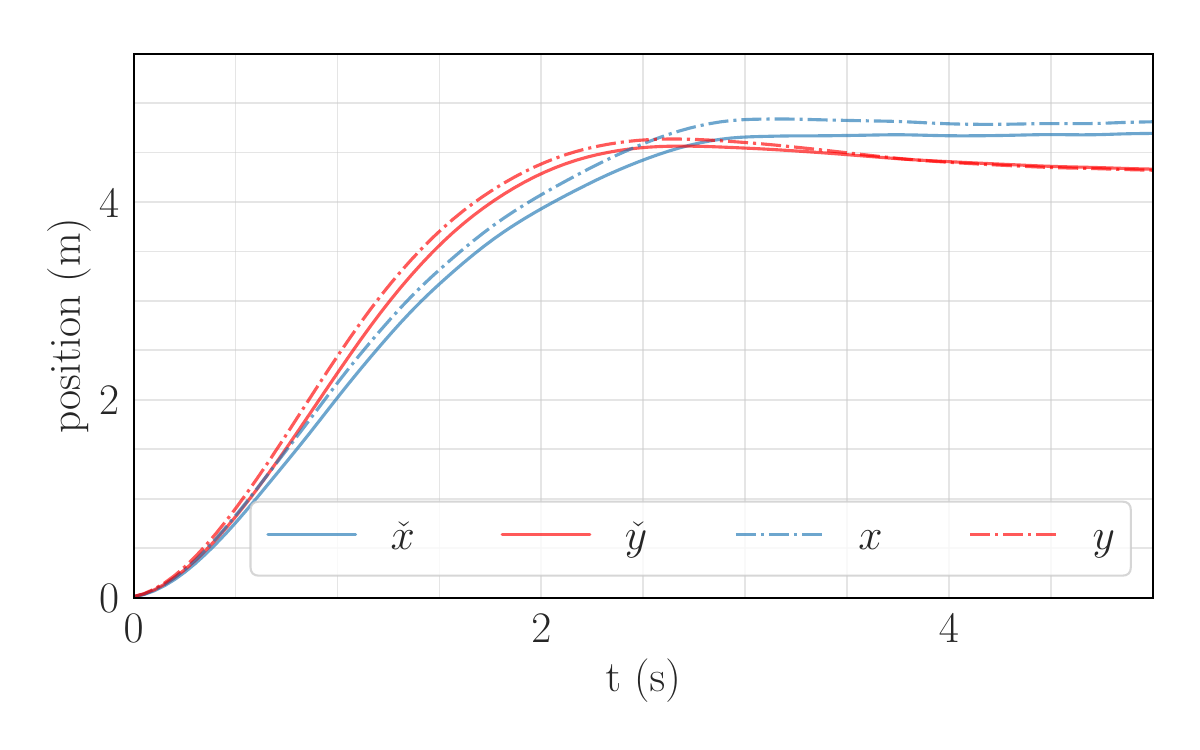}
			\caption{}
			\label{fig:16a}
		\end{subfigure}
		\hspace{2cm}
		\begin{subfigure}[b]{0.4\textwidth}
			\centering
			\includegraphics[width=\linewidth]{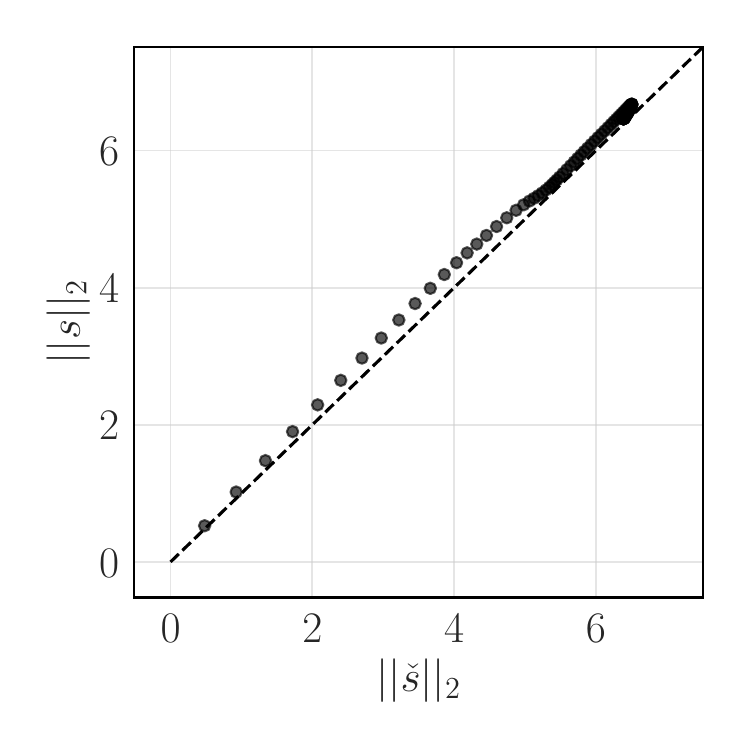}
			\caption{}
			\label{fig:16b}
		\end{subfigure}
		\caption{T2 - Comparison of the time evolution of the drone position for the real environment (solid lines) and the digital twin (dashed line). The right plot shows the high correlation between the norm of the predicted state $\bm{s}=[\dot{x},\dot{y},x,y]$ and the real state $\bm{\check{s}}$ for an episode length.}
		\label{drone_img_assimilation}   
	\end{figure*}

	In this test case, the initial weights $\bw_p^0=[a, b, c]$ (\eqref{drone_eq_Clift}  and \eqref{drone_eq_Cdrag}), are drawn from three random distributions, and a stopping criterion is defined to interrupt the $n_G$ iterations if the cost falls below a given threshold ($10^{-2}$). Figure \ref{fig:16a} shows the assimilation performance of the digital twin in predicting the real-system behavior. The two show a small deviation. This can be better appreciated in Figure \ref{fig:16b}, which compares the real and predicted states.

	% Figure \ref{drone_img_assimilation} (left) shows the evolution of this cost with the x-axis counting the iterations and the vertical lines indicating the end of the episodes.
	% On average, the first episodes require much more iterations to converge $\mathcal{L}_p$, which significantly oscillates. 
	% This behavior can be explained by the largely different trajectories $\bm{\tilde{s}}$ from one episode to another at the beginning of the training (proven by the change in $\mathcal{L}_a$), and could be smoothened by tuning the hyper-parameters of the optimizer. This was not needed since the maximum number of assimilation steps per episode was only 20 and most episodes, including the last ones, converged within one assimilation iteration.

\begin{figure*}[!ht]\center
		\begin{subfigure}[b]{0.44\textwidth}
			\centering
			\includegraphics[width=\linewidth]{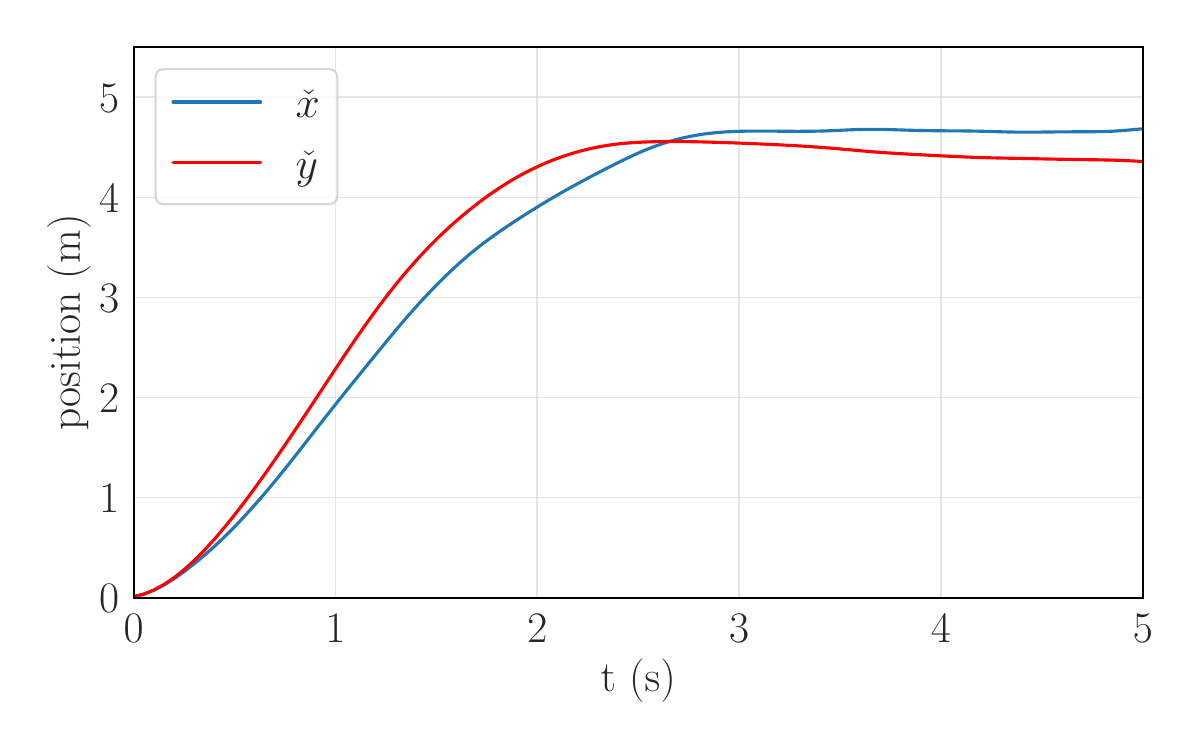}
			\caption{}
			\label{fig:17a}
		\end{subfigure}
		%\hspace{2mm}
		\begin{subfigure}[b]{0.44\textwidth}
			\centering
			\includegraphics[width=\linewidth]{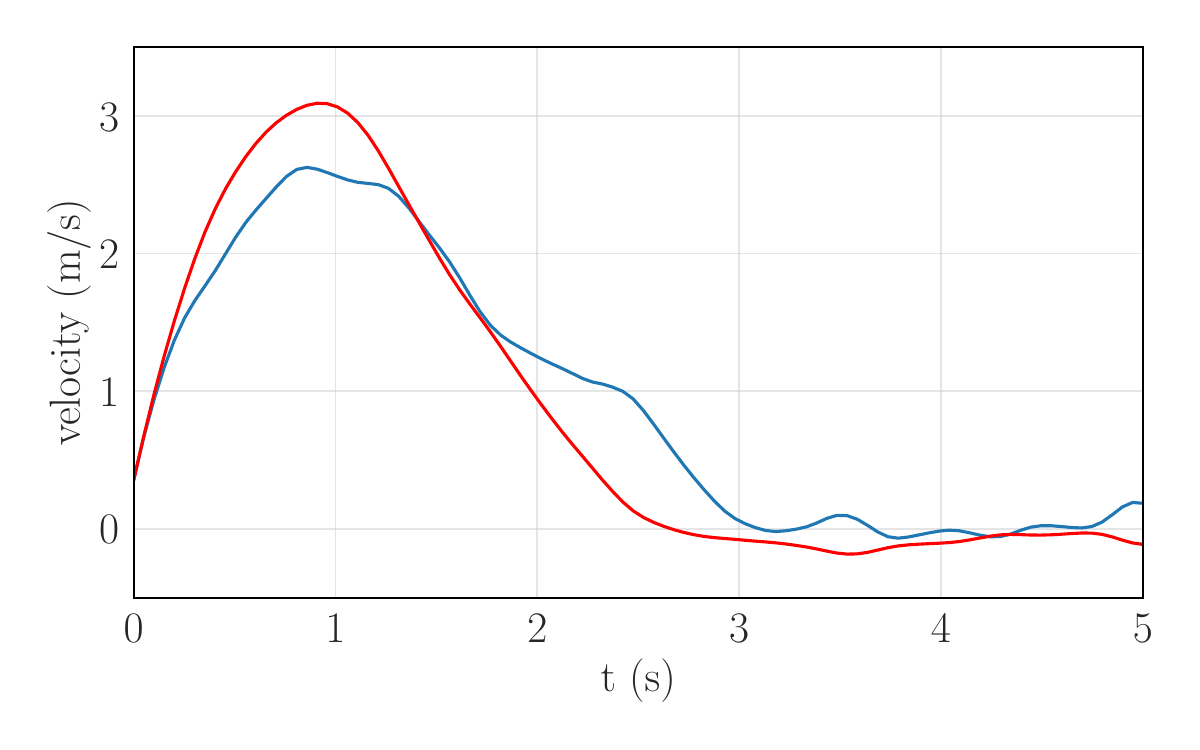}
			\caption{}
			\label{fig:17b}
		\end{subfigure}
		\begin{subfigure}[b]{0.44\textwidth}
			\centering
			\includegraphics[width=\linewidth]{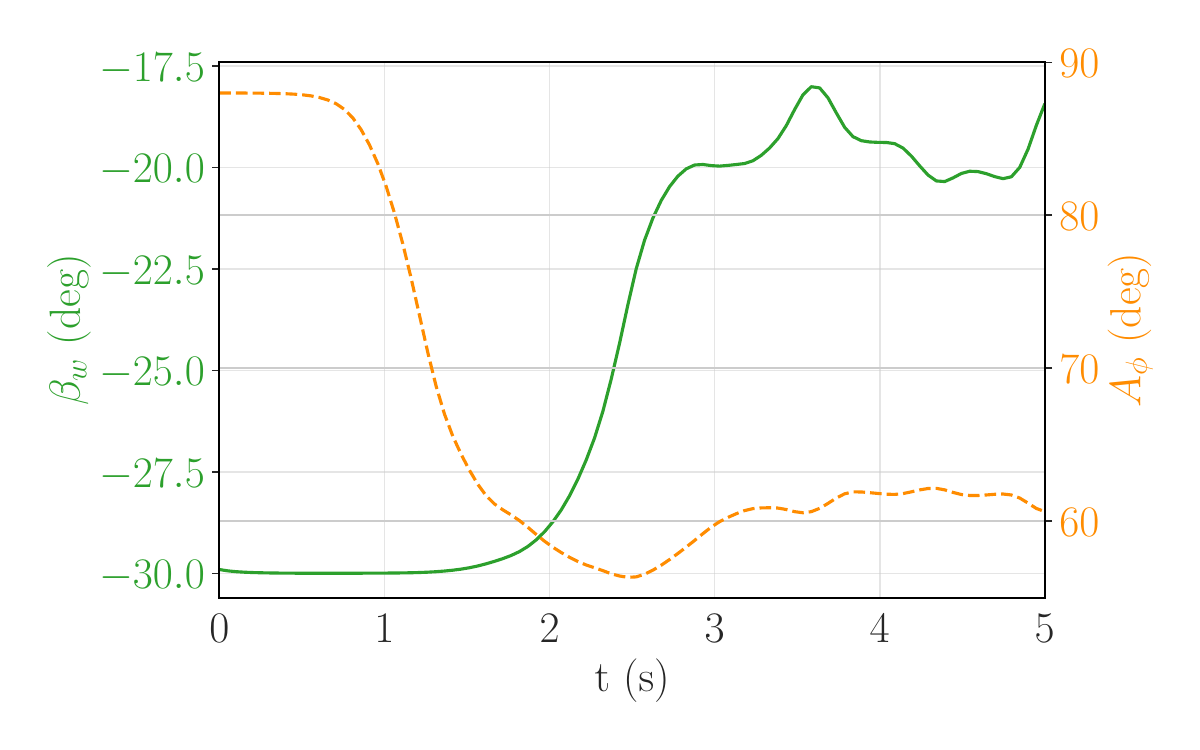}
			\caption{}
			\label{fig:17c}
		\end{subfigure}
		\caption{T2 - Time evolution of the positions $x$ and $y$ (a) and the velocities $\dot{x}$ and $\dot{y}$ (b) of the flapping wing micro air vehicle driven by the control actions $\bm{a}=[A_\phi,\beta_w]$ (c). The optimal control policy manages to lead the drone to its target $(x,y)=(5,5)$ and $(\dot{x},\dot{y})=(0,0)$ within one episode time with minor oscillations that results from the gust.}
		\label{fig:17}   
	\end{figure*}

	%\subsubsection{Optimal policy evaluation}
	Figure \ref{fig:17} characterizes the identified optimal policy, depicting state variables (\ref{fig:17a} and \ref{fig:17b}) and control actions (\ref{fig:17c}).
	% %! Transition
	The figures evidence that the drone managed to closely reach its target ($\bm{x}_f=5$ m and $\bm{\dot{x}}_f=0$, as fast as possible) thanks to the control of the stroke plane angle $\beta_w$ and the flapping amplitude $A_\phi$. The control of the latter is mainly formed of three phases in a kind of bang-bang fashion. First, the controller sets the drone's wings at their maximum angle of attack ($A_\phi = 88$ deg) to reach the target position as fast as possible. Then, it decelerates and stabilizes the drone picking a flapping angle close to the minimal one ($55$ deg), with a minimal overshoot. Finally, the wings' angle of attack is increased again, to balance the body weight with the aerodynamic lift. This force is also slightly modulated by the other control variable ($\beta_w$) which mainly drives the horizontal motion of the drone. In the same way, the controller pitches the stroke plane up to the maximum angle ($\beta_w = -30$ deg) and then stabilizes it around an angle that allows it to stay close to the desired position, $x_f$, despite the headwind. The influence of the gust is visible on $\beta_w$ even though the wind effect is still very limited, due to the selected mass and surface of the body. 
	
\begin{figure*}[!ht]\center
		% model-free
		\begin{subfigure}[b]{0.23\textwidth}
			\centering
			\includegraphics[width=\linewidth]{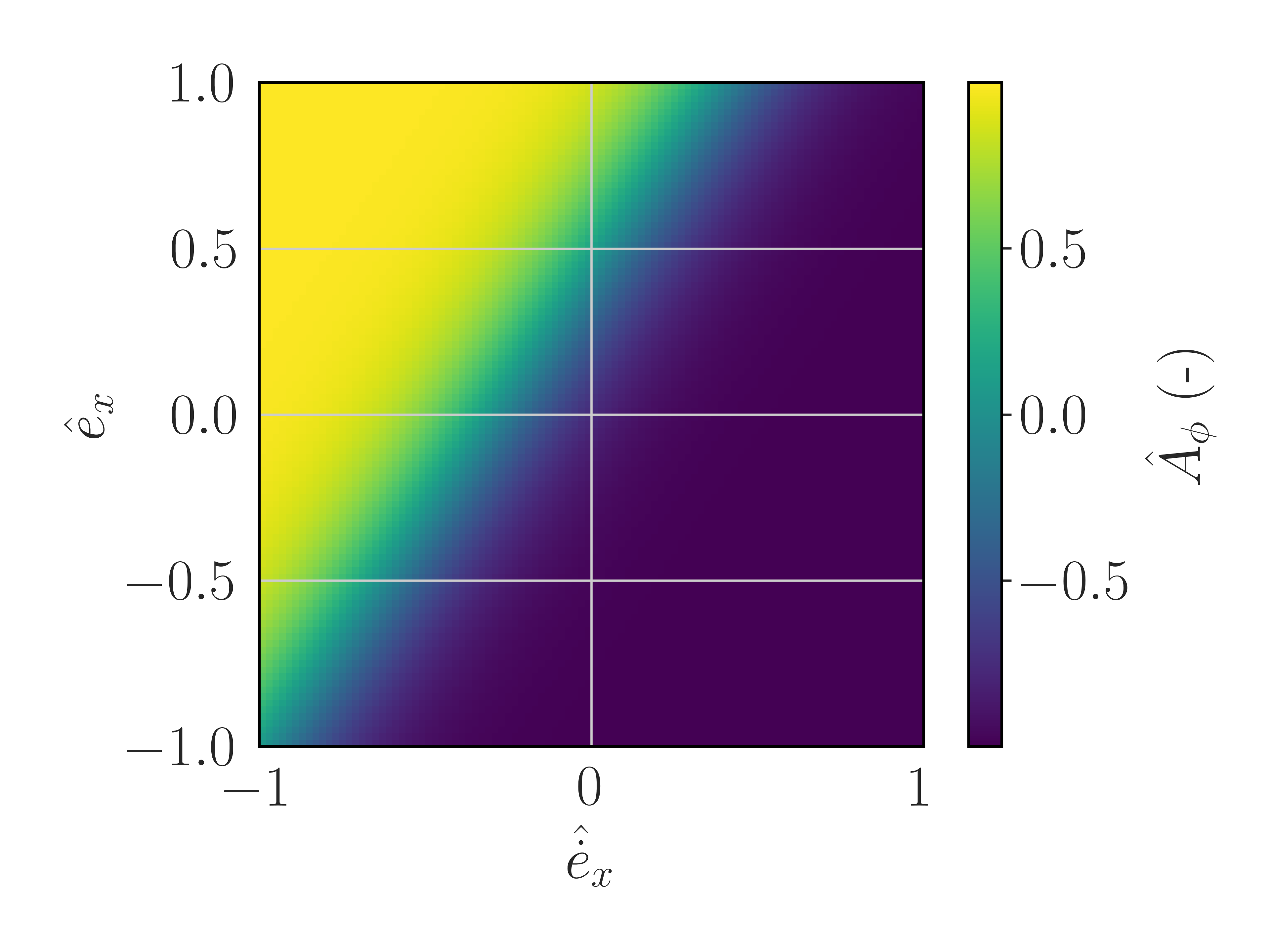}
			\caption{$\hat{A}_\phi = \pi_{mf}(\hat{e}_x, \hat{\dot{e}}_x)$}
			\label{fig:18a}
		\end{subfigure}
		% model based
		\begin{subfigure}[b]{0.23\textwidth}
			\centering
			\includegraphics[width=\linewidth]{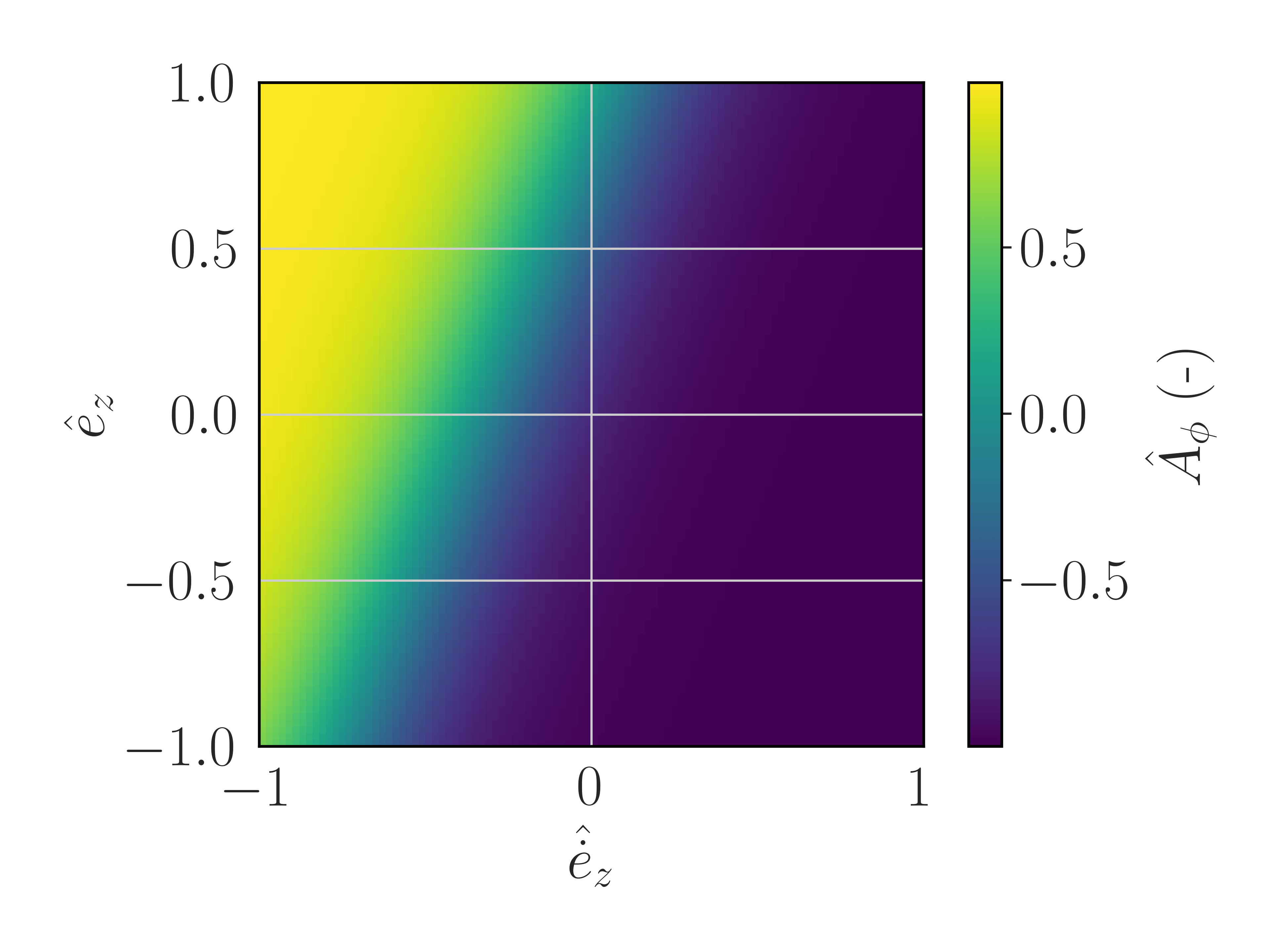}
			\caption{$\hat{A}_\phi = \pi_{mf}(\hat{e}_z, \hat{\dot{e}}_z)$}
			\label{fig:18b}
		\end{subfigure}
		% model free
		\begin{subfigure}[b]{0.23\textwidth}
			\centering
			\includegraphics[width=\linewidth]{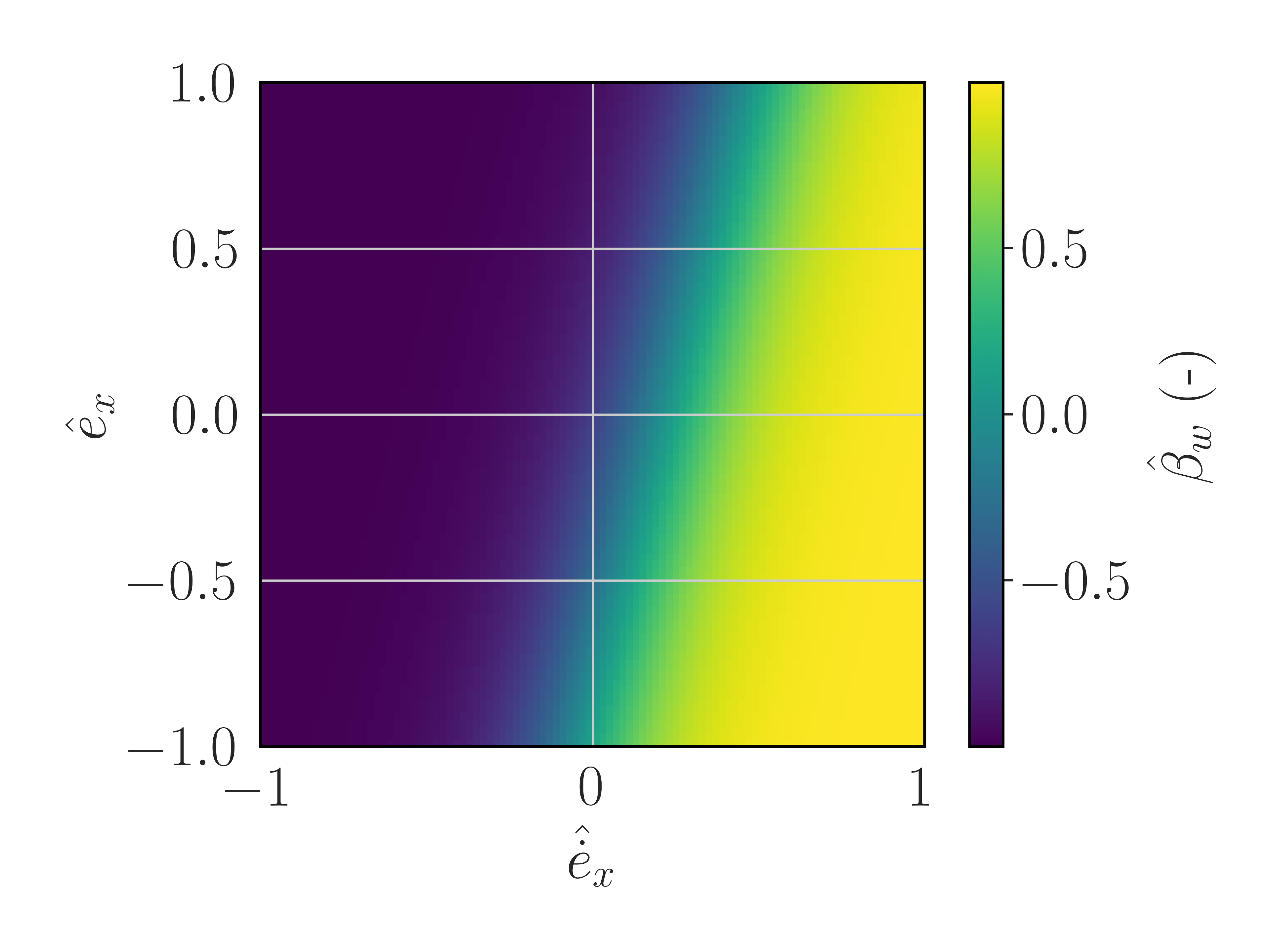}
			\caption{$\beta_w = \pi_{mf}(\hat{e}_x, \hat{\dot{e}}_x)$}
			\label{fig:18c}
		\end{subfigure}
		% model based
		\begin{subfigure}[b]{0.23\textwidth}
			\centering
			\includegraphics[width=\linewidth]{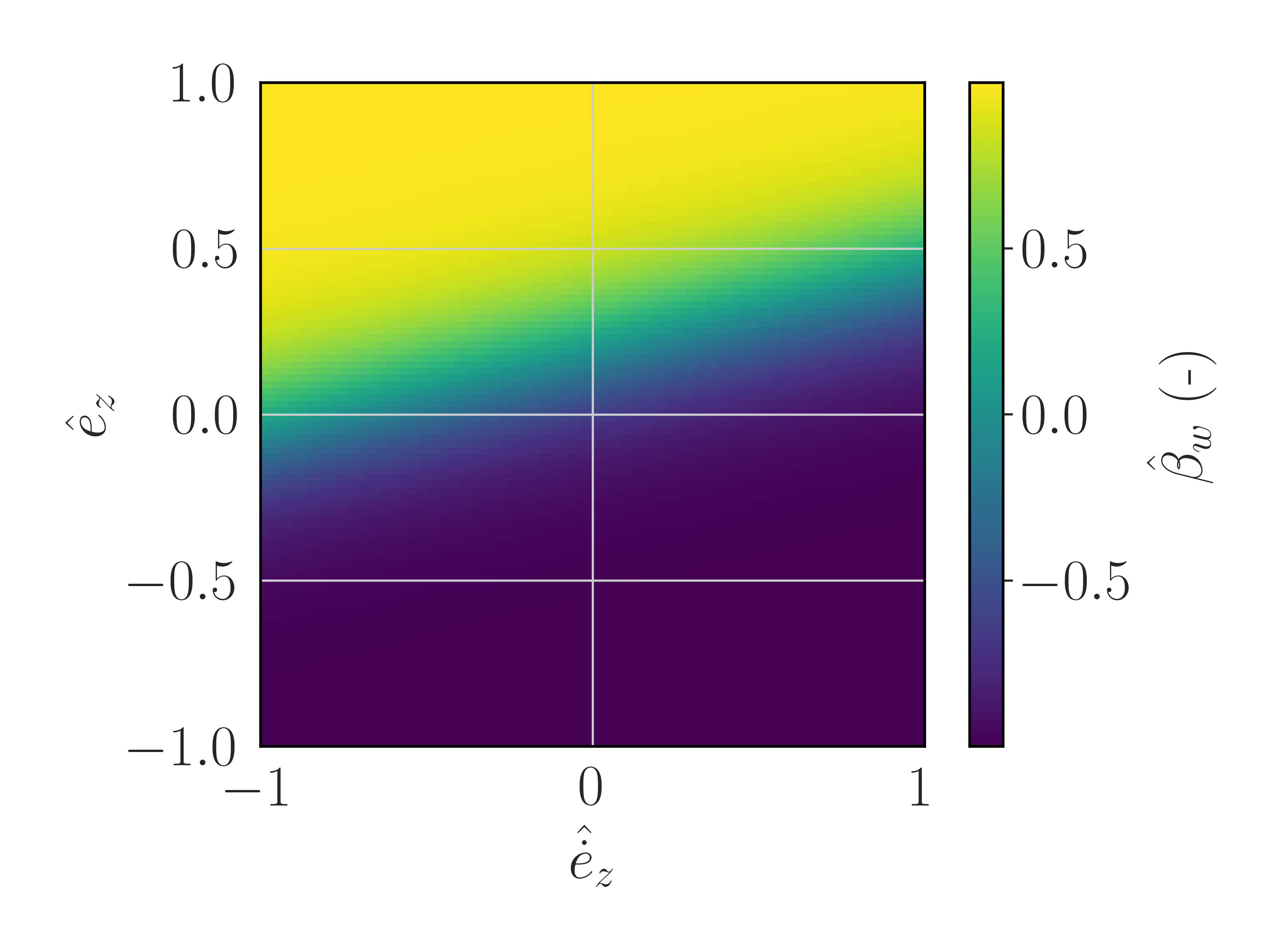}
			\caption{$\beta_w = \pi_{mf}(\hat{e}_z, \hat{\dot{e}}_z)$}
			\label{fig:18d}
		\end{subfigure}
		% model free 
		\begin{subfigure}[b]{0.23\textwidth}
			\centering
			\includegraphics[width=\linewidth]{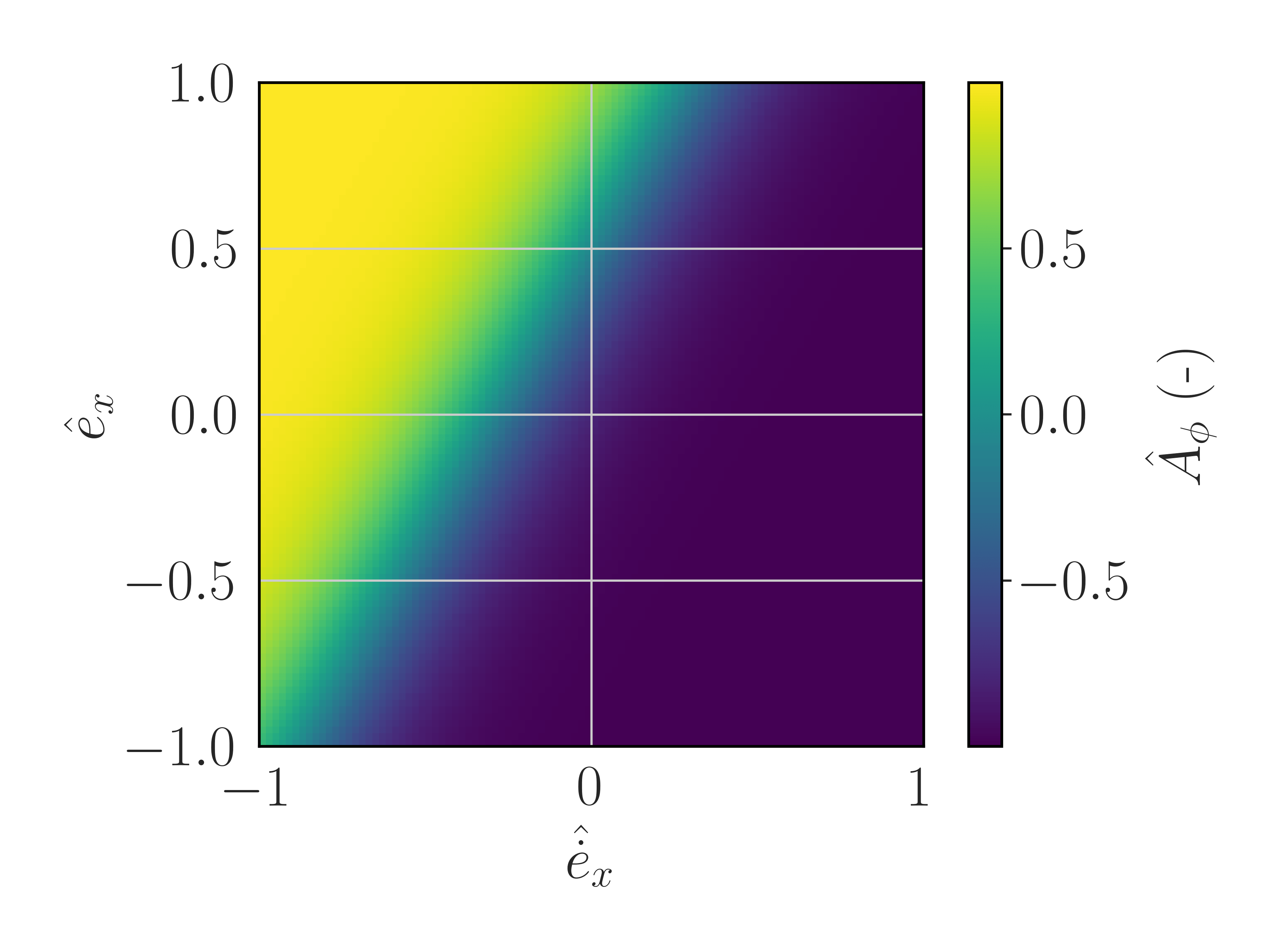}
			\caption{$\hat{A}_\phi = \pi_{mb}(\hat{e}_x, \hat{\dot{e}}_x)$}
			\label{fig:18e}
		\end{subfigure}
		% model based
		\begin{subfigure}[b]{0.23\textwidth}
			\centering
			\includegraphics[width=\linewidth]{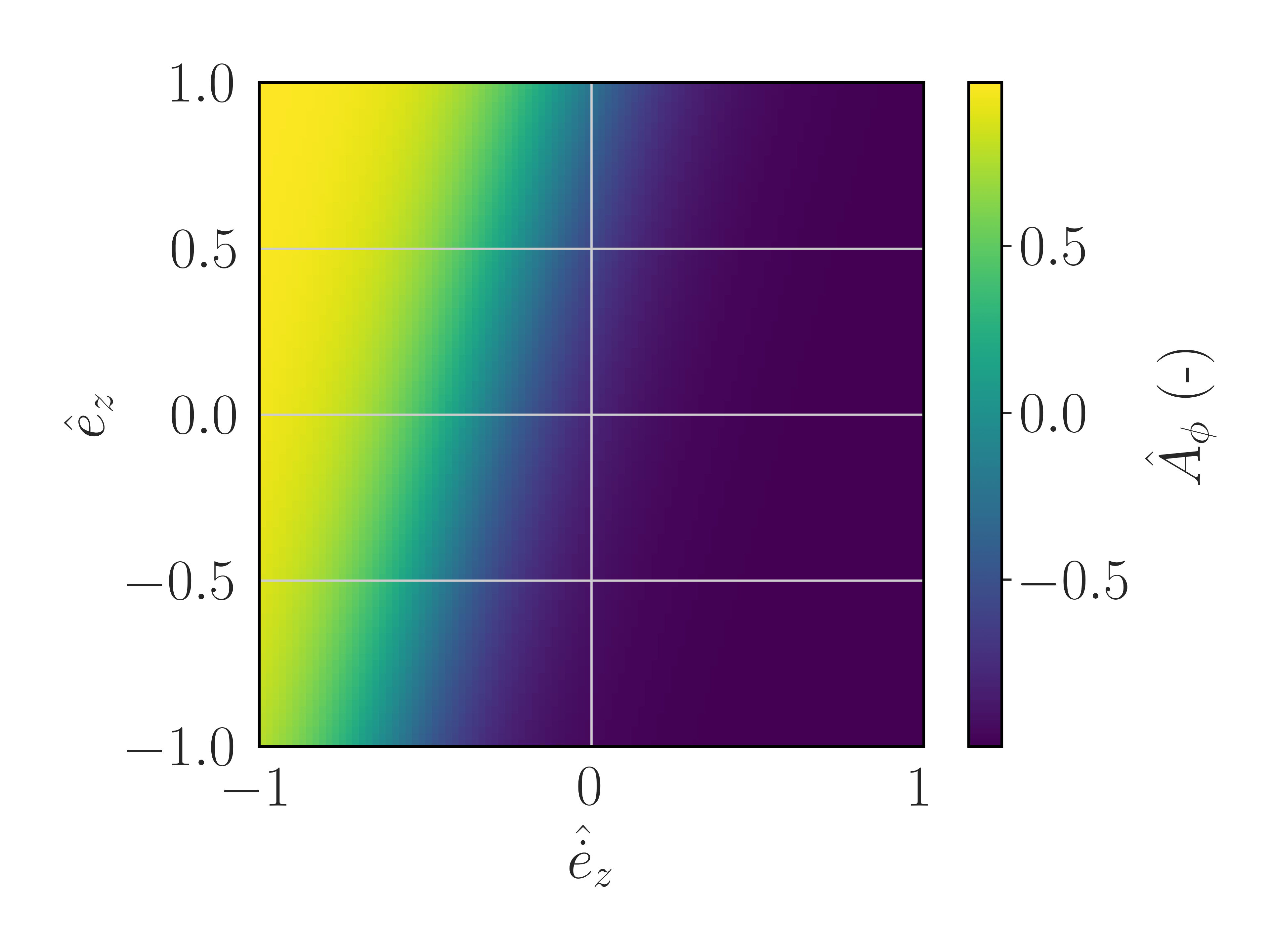}
			\caption{$\hat{A}_\phi = \pi_{mb}(\hat{e}_z, \hat{\dot{e}}_z)$}
			\label{fig:18f}
		\end{subfigure}
		\begin{subfigure}[b]{0.23\textwidth}
			\centering
			\includegraphics[width=\linewidth]{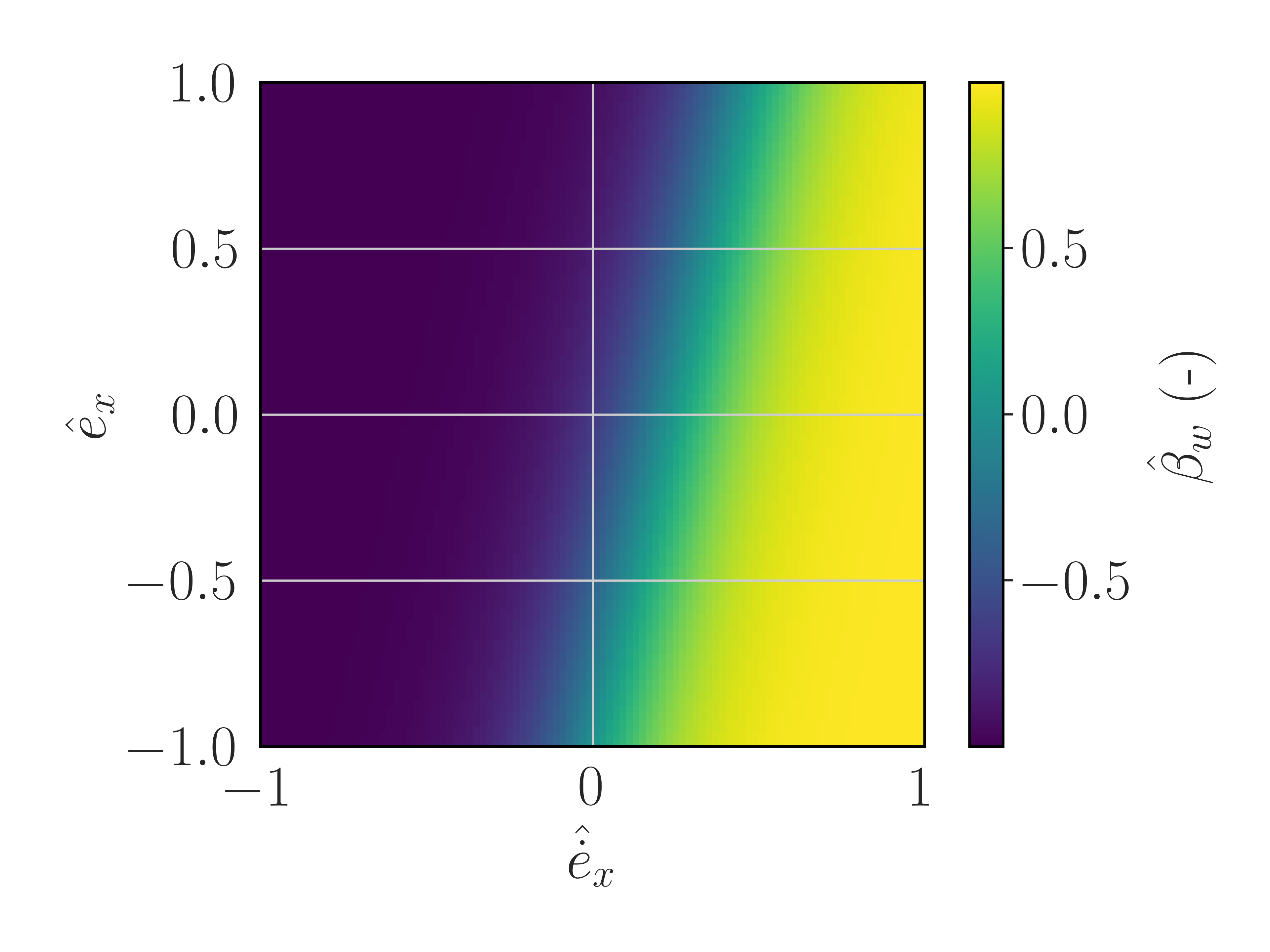}
			\caption{$\beta_w = \pi_{mb}(\hat{e}_x, \hat{\dot{e}}_x)$}
			\label{fig:18g}
		\end{subfigure}
		\begin{subfigure}[b]{0.23\textwidth}
			\centering
			\includegraphics[width=\linewidth]{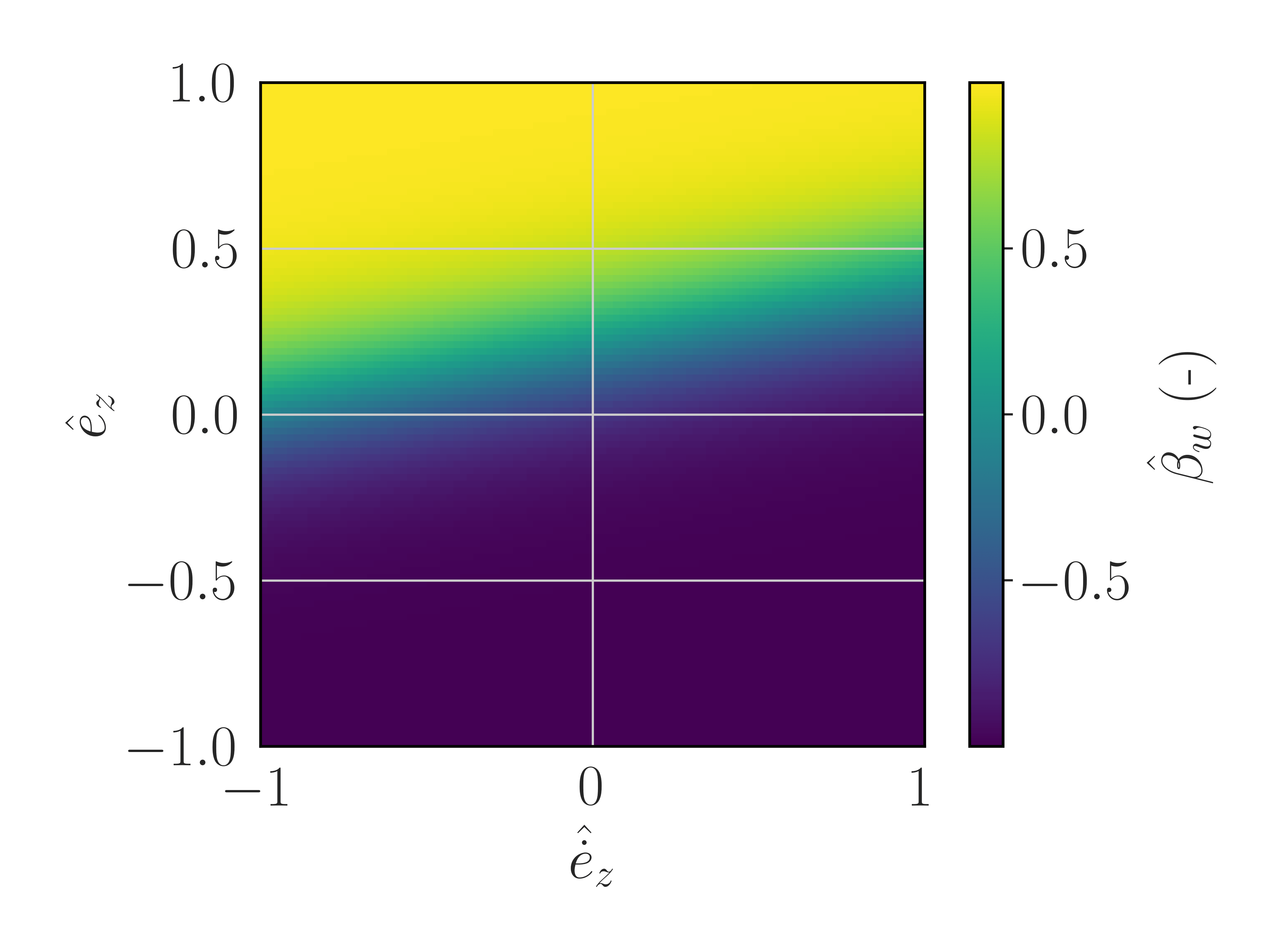}
			\caption{$\beta_w = \pi_{mb}(\hat{e}_z, \hat{\dot{e}}_z)$}
			\label{fig:18h}
		\end{subfigure}
		\caption{T2 - Comparison of model-based and model-free policies.}
		\label{fig18}   
	\end{figure*}

	Figure \ref{fig18} analyses the policies found by the model-free (Figures \ref{fig:18a} - \ref{fig:18d}) and model-based (Figures \ref{fig:18e} - \ref{fig:18h}) loops. To this end, we fix two of the four inputs to zero and span on the other two dimensions, recording the corresponding action. The two policies are remarkably similar, converging to essentially the same parametrization. The main difference between the two lies in the concavity of the hyperbolic tangent, as dictated by the biases. This affects the localization of the switching between the two bounds of the actuation space, possibly explaining the differences in their associated costs presented in Table \ref{tab:table_learning}. In both cases, for what concerns $\hat{A}_\phi$, the border between the two bounds lies on the diagonal that goes from $(\hat{\dot{e}}_x, \hat{e}_x) = (-1.0, -1.0)$ to $(\hat{\dot{e}}_x, \hat{e}_x, ) = (0.25, 1.0)$, see Figure \ref{fig:18a}. The same trend is observed in Figure \ref{fig:18b}, which analyses the dependencies from $e_z$ and its time derivative. In this case, however, the upper point of the diagonal lies around $(\hat{\dot{e}}_z, \hat{e}_z) = (0.0, 1.0)$, thus increasing $\hat{A}_\phi$ as the rate of change of $e_z$ goes negative, and decreasing it otherwise. Interestingly, the model-based policy here shows a broader decision zone, as depicted in Figure \ref{fig:10f}. Now, analysing the actuation on $\beta_w$, the threshold appears to be positioned in the diagonals $(\hat{\dot{e}}_x, \hat{e}_x) = (0.0, -1.0)$ to $(\hat{\dot{e}}_x, \hat{e}_x) = (0.75, 1.0)$. The sensitivity to the $z-$errors instead show it to lie in $(\hat{\dot{e}}_z, \hat{e}_z) = (-1.0, 0.0)$ and  $(\hat{\dot{e}}_z, \hat{e}_z) = (1, 0.5)$. Therefore, it seems that the time variation of the rate of change is the most important input on the $x-$dimension, whereas in the $z-$coordinate, it is the scalar $\hat{e}_z$. Arguably this may be caused by the headwind, oriented as opposite of the $x-$axis. Overall, there appears to be two main mechanisms at play:  (1) the errors and their rate of change seems to have the same importance when deciding how to control on $\hat{A}_\phi$ being slightly more sensitive to the scalar instantaneous errors; and (2) when deciding on how to actuate $\beta_w$, the controller balances the scalar instantaneous error in one dimension and its rate of change in the other. In fact, when $\dot{e}_x$ increases, it actuates on $\beta_w$ to get back to nominal conditions, and the same is observed on analysing the reactivity to the instantaneous error $e_z$ and $e_x$.

	%\subsubsection{Robustness of the control loops}
	
\begin{figure*}[!ht]
    \centering
		\begin{subfigure}[b]{0.45\textwidth}
			\centering
			\includegraphics[width=\textwidth]{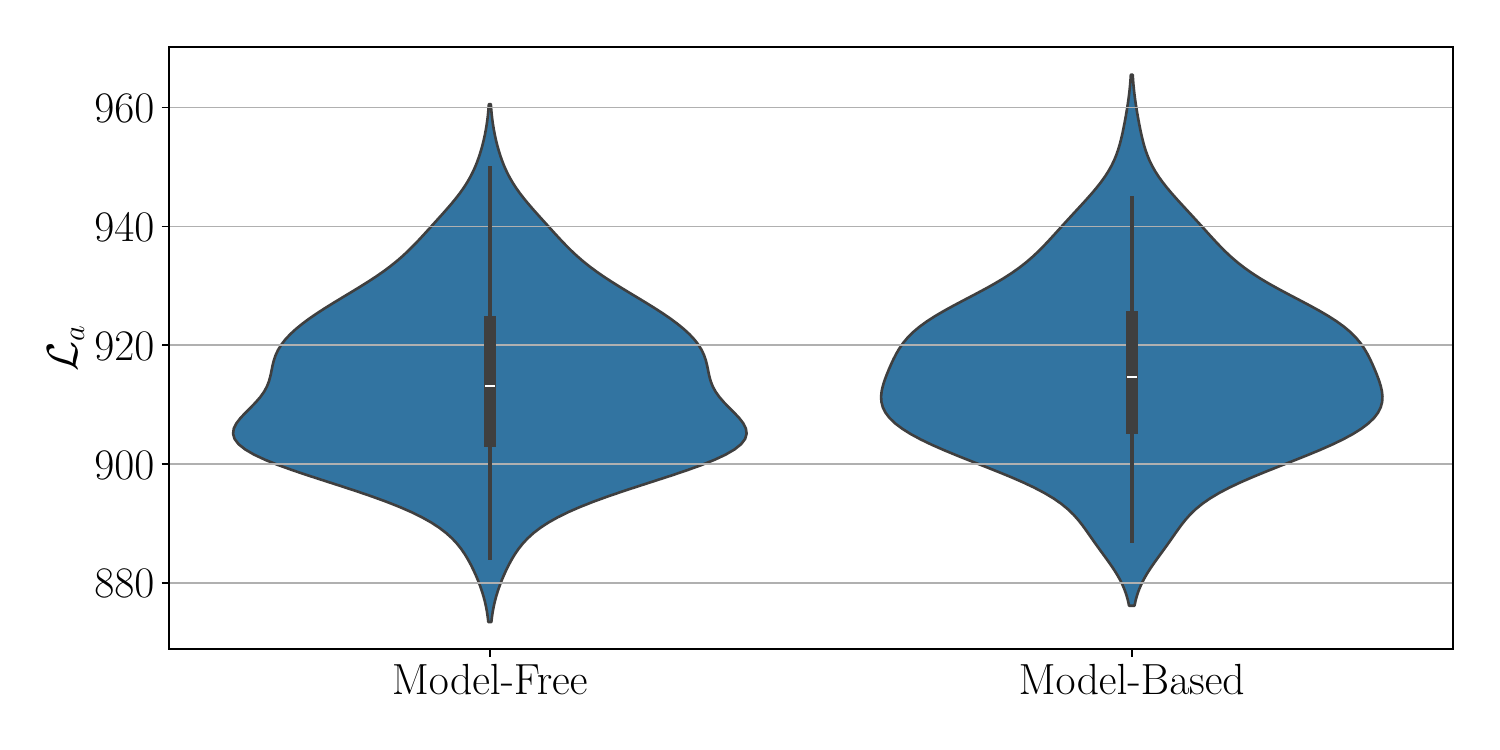}
			\caption{}
			\label{fig19a}
		\end{subfigure}
        \hspace{2cm}
		\begin{subfigure}[b]{0.23\textwidth}
			\centering
			\includegraphics[width=\textwidth]{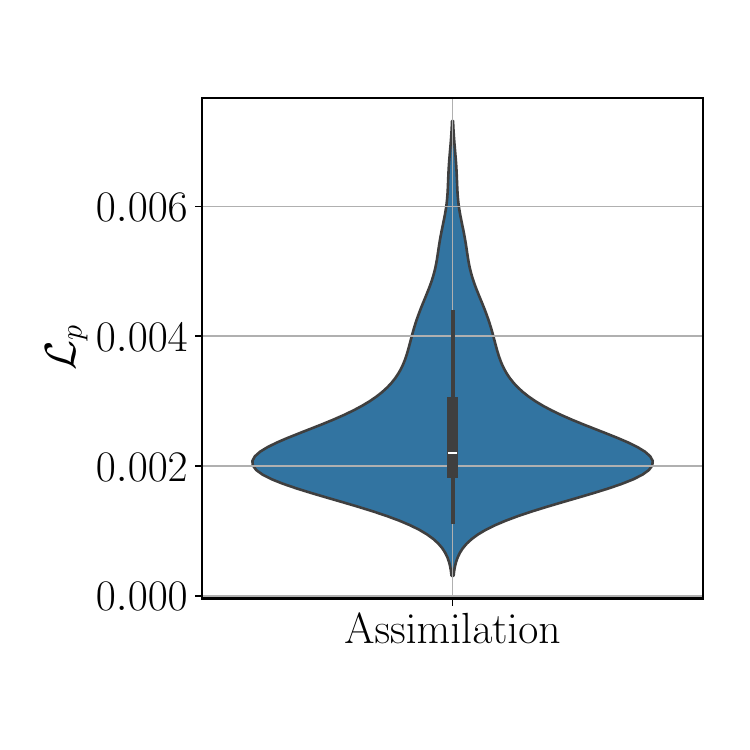}
			\caption{}
			\label{fig19b}
		\end{subfigure}
		\caption{T2 -  Robustness analysis of model-free and model-based policies (a) and assimilated closure (b), compute over 100 episodes.}% Both policies result in similar distributions with a lower average cost for the model-free policy}
		\label{fig:19}  
		
	\end{figure*}

	We conclude the analysis of the RT algorithm performances on the FWMAV presenting the robustness study of the model-free and model-based policies, shown in Figure \ref{fig:19}. One-hundred episodes were run on the real environment for each policy and their costs are compared with violin plots (Figure \ref{fig19a}). Both sets of weights result in a small cost proving their capability to handle the random gust. The model-free policy manages to reach on average a lower cost  and shows a slightly different distribution than the model-based policy, having two peaks at $\mathcal{L}_a = 905$ and $\mathcal{L}_a = 920$, whereas the model-based shows a main peak in the middle, at $\mathcal{L}_a = 910$. However, their differences are in the order $\mathcal{O}(10^{1})$, reinforcing the observation that the two have converged to a very similar policy. 
	Lastly, the same procedure is applied on the assimilation side, studying its robustness to varying operative conditions, as shown in Figure \ref{fig19b}.  It appears that the identified closure function is robust to different operative conditions, registering a mean cost of $\mathcal{L}_p \approx \mathcal{O}(10^{-3})$.
	
	\subsection{Performances in Cryogenic Storage's Thermal Management}\label{subsect:res_cryo}
	
	At last, we examine the RT results on the cryogenic thermal management test case.
\noindent
The digital twin assimilation consists of identifying the weights $\bm{w_p}=[U_s, U_{HX}, K_{JT}]$ which drive the heat transfer between the cryogenic tank and the environment, as well as the performance of the active TVS. For all simulations, this vector was initialized taking $\bm{w}_p^0=[1, 1, 10]$, with the units of the first two components being (W/m$^2$K), and the last being (GPa$\cdot$s$^2/$kg$^2$). %Furthermore, to reduce the computational cost, we terminate the optimization early if the weights $\bm{w}_p$ or the function $\mathcal{L}_p$ change within a user-defined tolerance between iterations. 

\begin{figure}[h!]\center
	\begin{subfigure}[b]{0.45\textwidth}
		\centering
		\includegraphics[width=\linewidth]{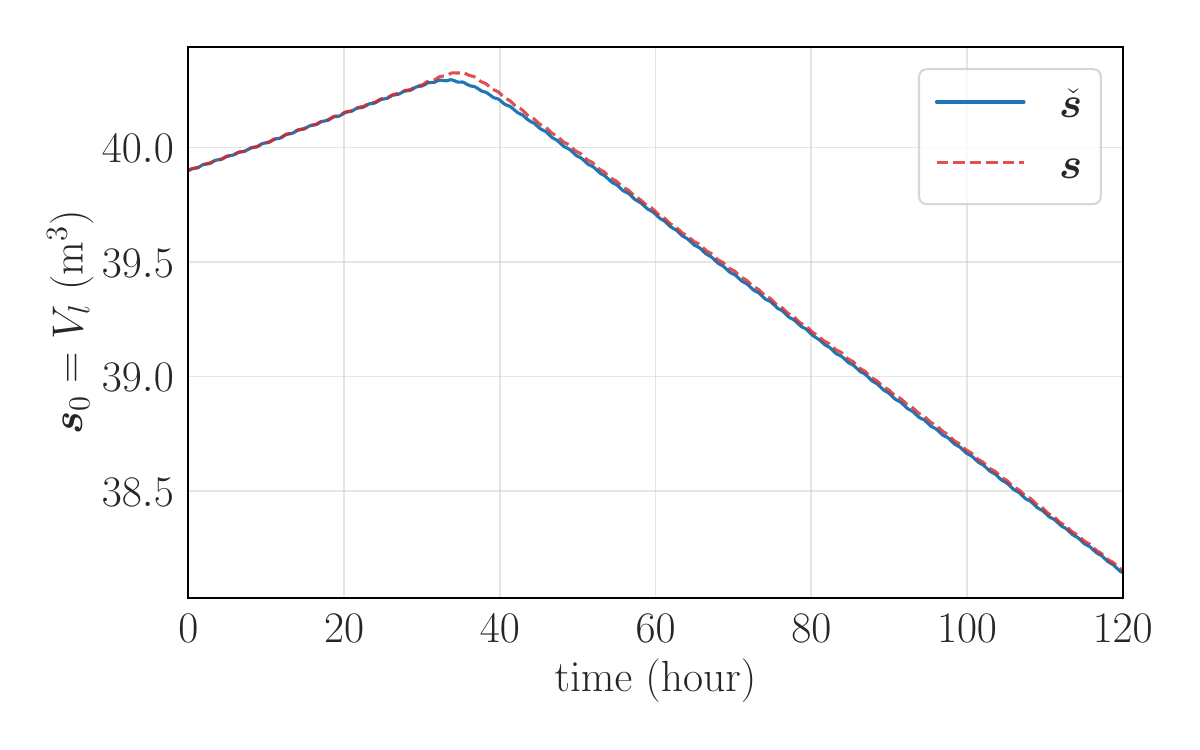}
		\caption{}
		\label{fig22a}
	\end{subfigure}
	\begin{subfigure}[b]{0.45\textwidth}
		\centering
		\includegraphics[width=\linewidth]{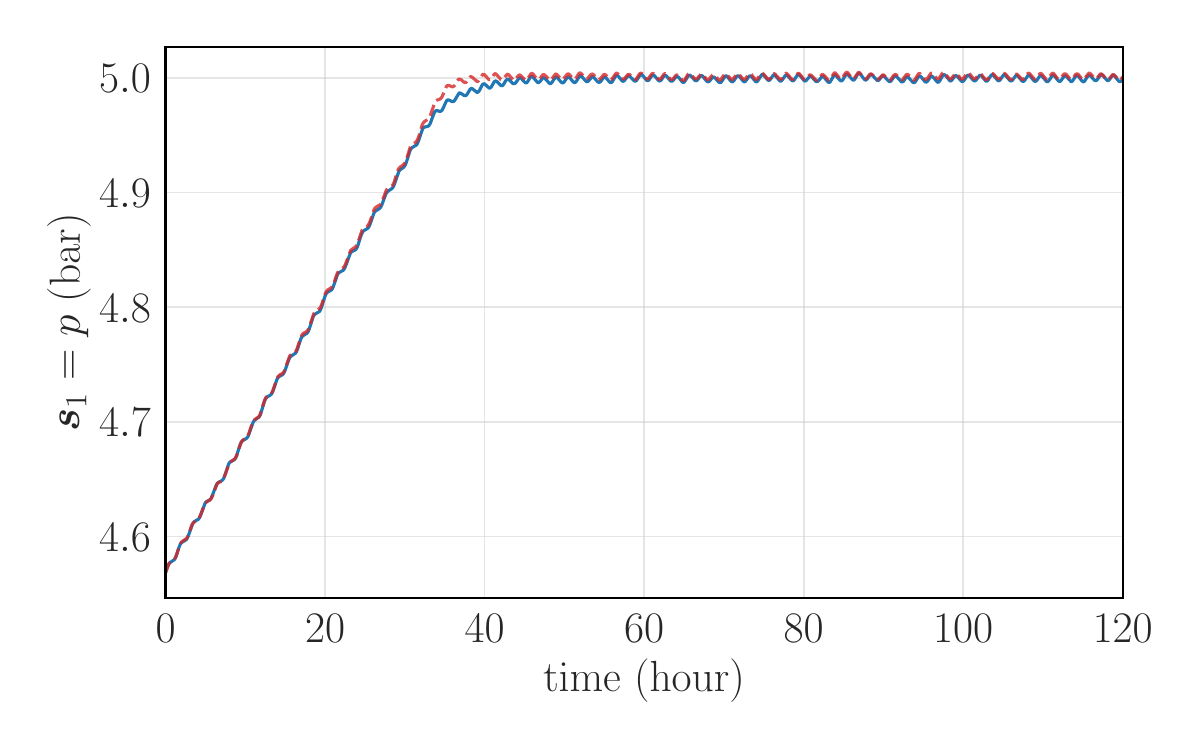}
		\caption{}
		\label{fig22b}
	\end{subfigure}
	\centering
	\begin{subfigure}[b]{0.45\textwidth}
		\centering
		\includegraphics[width=\linewidth]{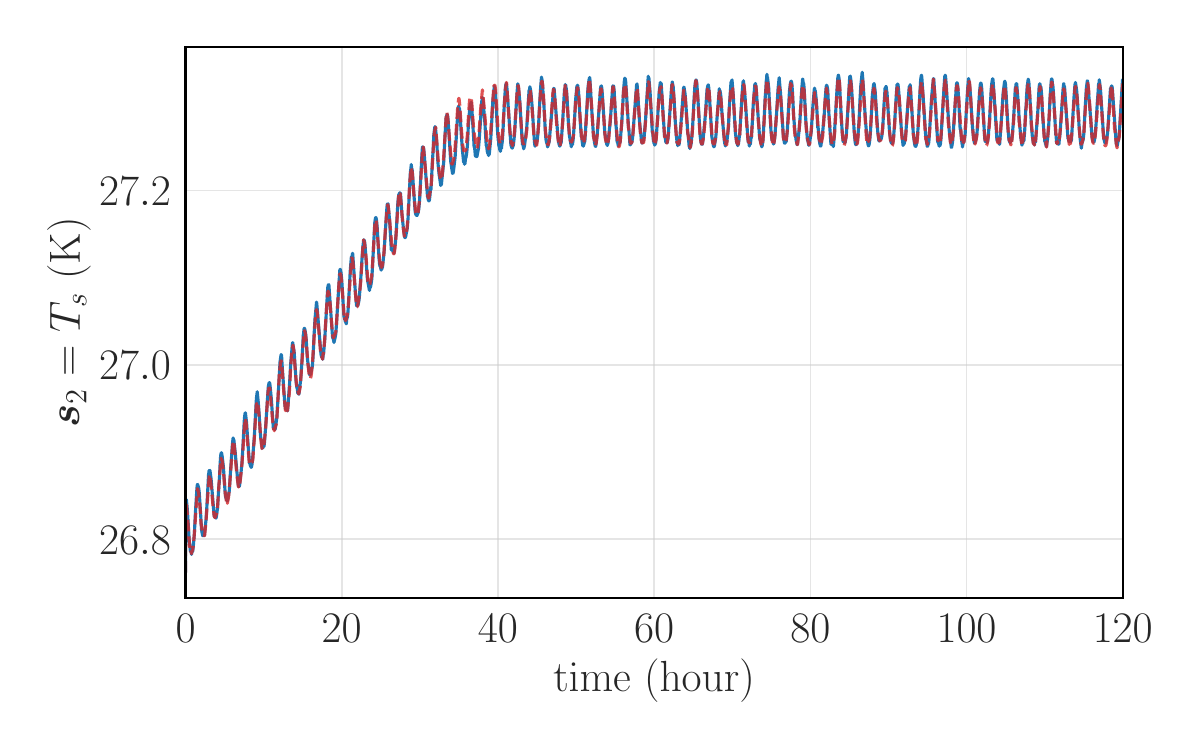}
		\caption{}
		\label{fig22c}
	\end{subfigure}
	
	\begin{subfigure}[b]{0.3\textwidth}
		\centering
		\includegraphics[width=\linewidth]{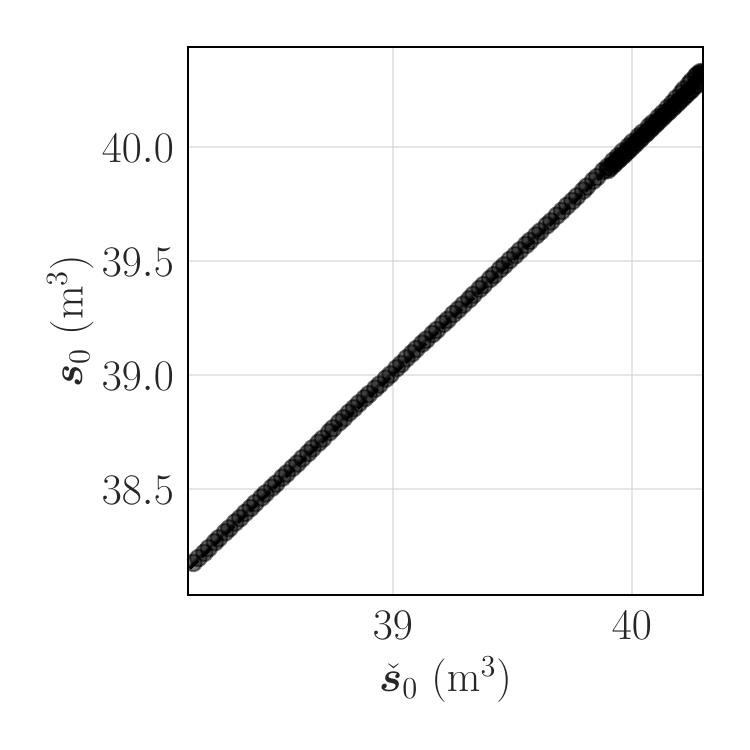}
		\caption{}
		\label{fig22d}
	\end{subfigure}
	\begin{subfigure}[b]{0.3\textwidth}
		\centering
		\includegraphics[width=\linewidth]{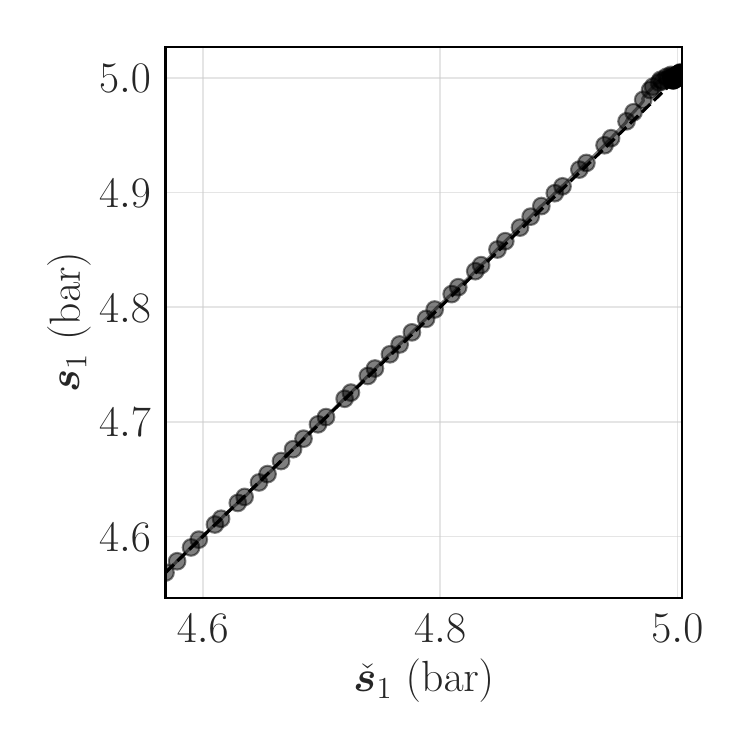}
		\caption{}
		\label{fig22e}
	\end{subfigure}
	\begin{subfigure}[b]{0.3\textwidth}
		\centering
		\includegraphics[width=\linewidth]{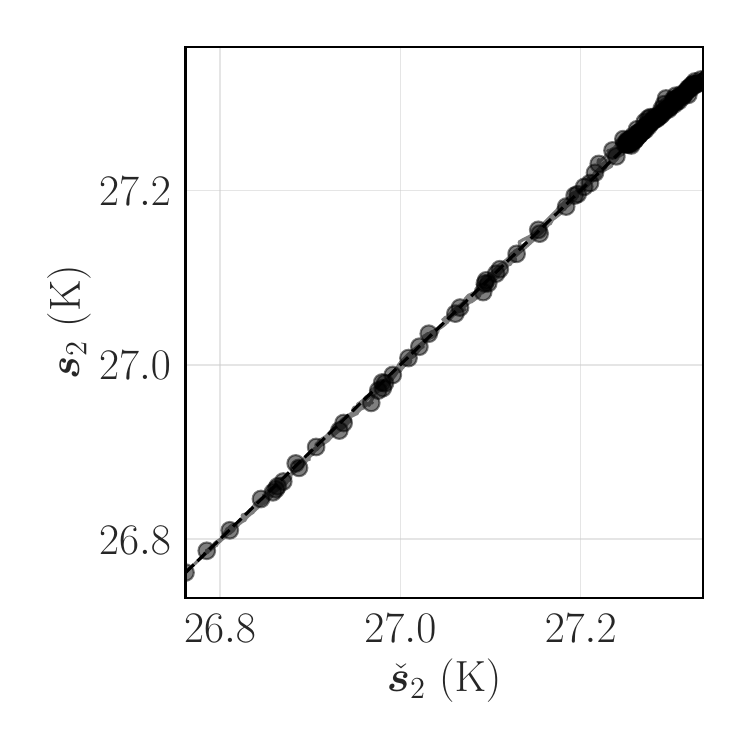}
		\caption{}
		\label{fig22f}
	\end{subfigure}
	\caption{T3 - Performance of the digital twin assimilation. (a) Cost function $\mathcal{L}_p$ normalized with its initial value $\mathcal{L}_p(\bm{w}_p^0)$ over the number of episodes. Agreement of the predicted state $\bm{s}=[V_l, p, T_s]$ with the real state $\check{\bm{s}}$ shown through (b) a scatter plot, and (c) the time series evolution.}
	\label{fig:22}
\end{figure}

% Figure \ref{fig:cryo_L_p} shows the evolution of $\mathcal{L}_p$ over the number of episodes. Similar to the previous test case, the minimization of the cost function suffers oscillations. This is linked to the different trajectories encountered in each episode due to the random initialization of the state, the different control inputs on the environment, and due to the inherent randomness of the exogenous agent. 
% While these factors produce fluctuations in the cost, they also allow for the gradient-based optimizer to avoid being trapped in local minima between the different episodes.
% Nevertheless, the adjoint-based assimilation remarkably allows for the cost to drop by two orders of magnitude within the first three episodes. After this point, while the convergence slows down, it still progresses.

The agreement between the predicted state and the real one is presented in Figure \ref{fig:22}, which shows both the time series of the real and predicted states (Figures \ref{fig22a} - \ref{fig22c}) and their direct comparison plotting one against the other (Figures \ref{fig22d} - \ref{fig22f}). All three components are successfully identified, even though slight deviations are found for $\bm{s}_1=p$ and $\bm{s}_2=T_s$.

\begin{figure*}[!ht]\center
	\begin{subfigure}[b]{0.4\textwidth}
		\centering
		\includegraphics[width=\linewidth]{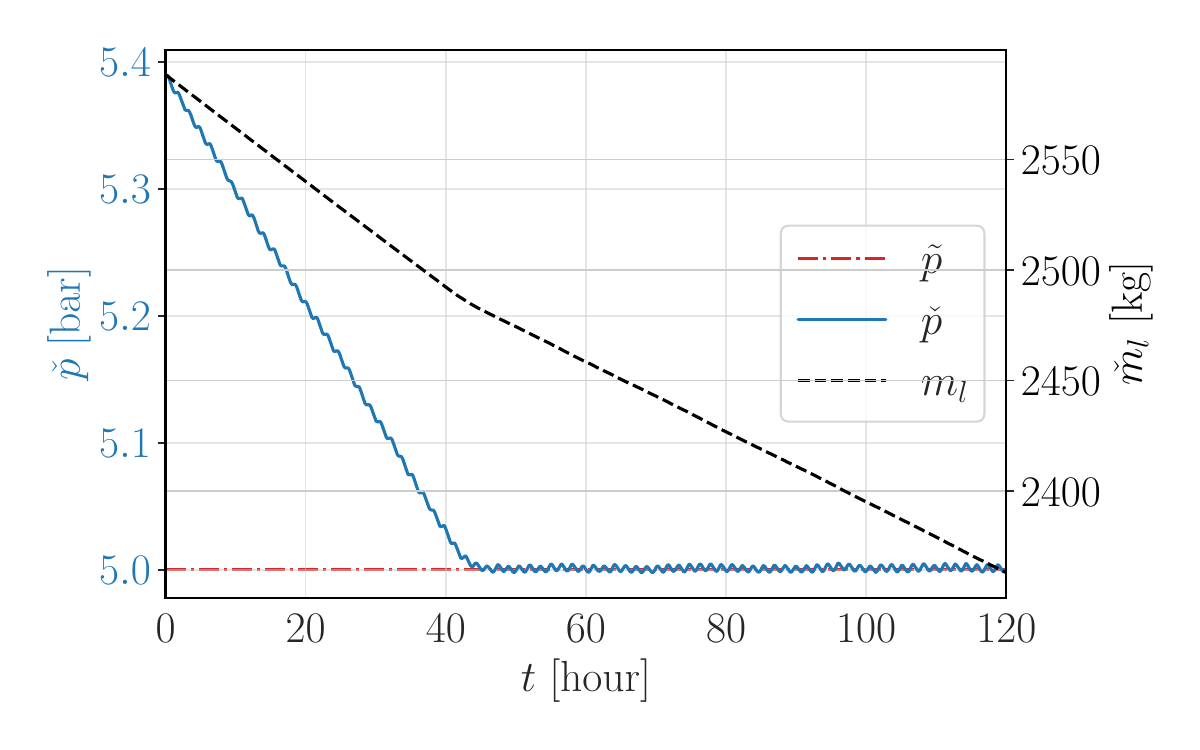}
		\caption{}
		\label{fig21a}
	\end{subfigure}
	\begin{subfigure}[b]{0.4\textwidth}
		\centering
		\includegraphics[width=\linewidth]{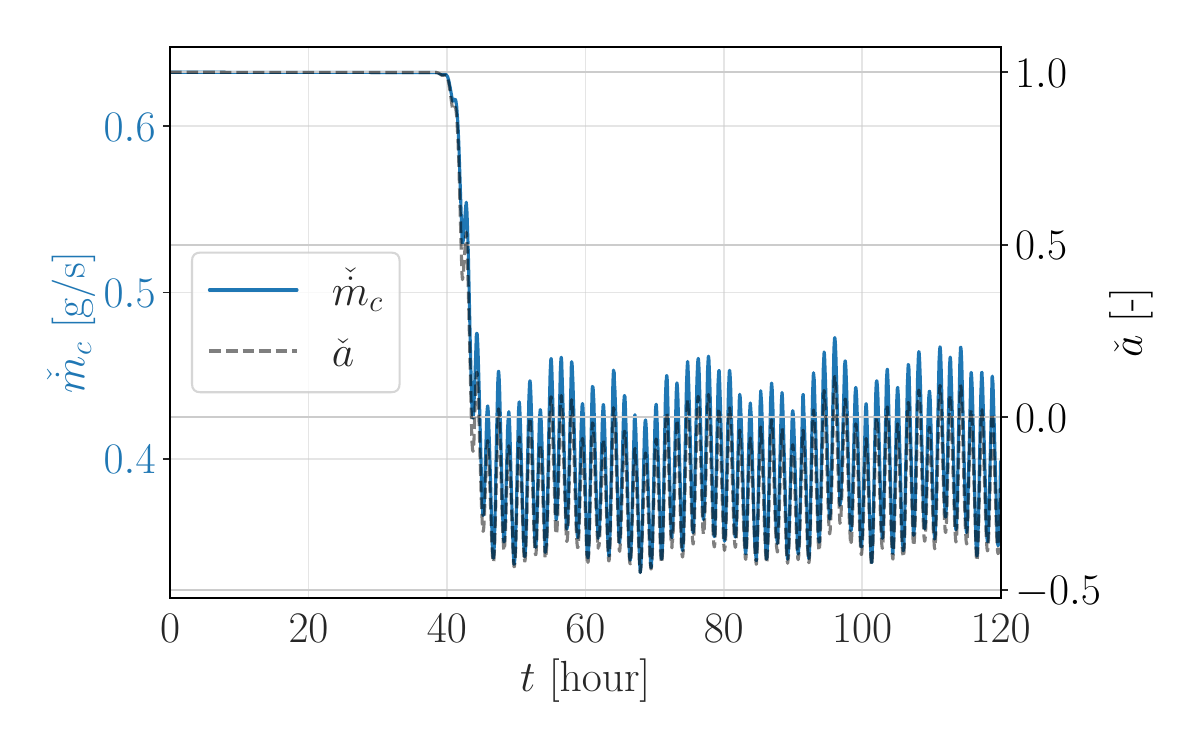}
		\caption{}
		\label{fig21b}
	\end{subfigure}
	\begin{subfigure}[b]{0.4\textwidth}
		\centering
		\includegraphics[width=\linewidth]{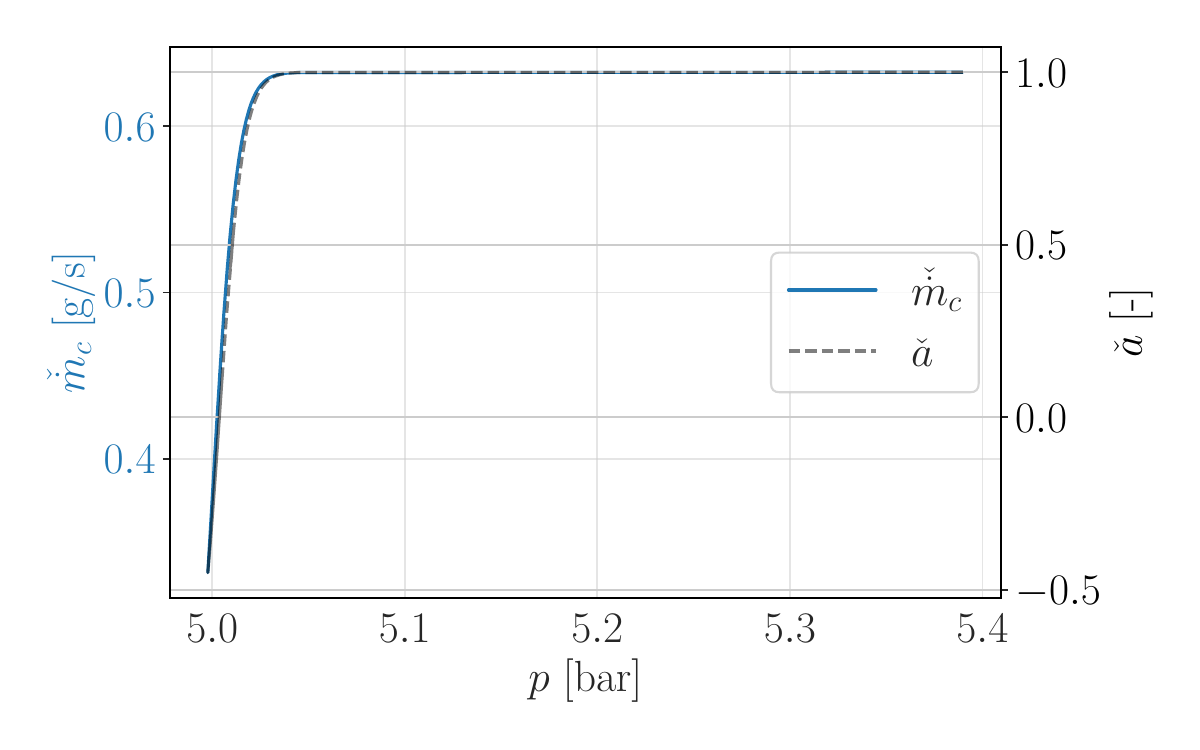}
		\caption{}
		\label{fig21c}
	\end{subfigure}
	% \begin{subfigure}[b]{0.48\textwidth}
		%     \centering
		%     \includegraphics[width=\linewidth]{fig/cryo/cryo_b_action.pdf}
		%     \caption{}
		%     \label{cryo_b_action}
		%  \end{subfigure}
	% \begin{subfigure}[b]{0.48\textwidth}
		%     \centering
		%     \includegraphics[width=\linewidth]{fig/cryo/cryo_a_state_action.pdf}
		%     \caption{}
		%     \label{cryo_a_state_action}
		% \end{subfigure}
	% \begin{subfigure}[b]{0.48\textwidth}
		%     \centering
		%     \includegraphics[width=\linewidth]{fig/cryo/cryo_b_state_action.pdf}
		%     \caption{}
		%     \label{cryo_b_state_action}
		% \end{subfigure}
	\caption{T3 - Tracking performance of the pressure $\check{p}$, and variation of propellant mass $\check{m}_l$ over the real environment, starting from an over-pressure condition. Corresponding control action $\check{a}$ and vented mass flow rate $\check{\dot{m}}_c$. Thermodynamic actuation $\check{\dot{m_v}}$ and $\check{a}$ as a function of the pressure $p$. The acting policy $\pi(\check{\bm{e}};\bm{w}_a)$ is retrieved by taking the optimal set of weights $\bm{w_a}$ from the training.}
	\label{fig:21}   
\end{figure*}

The control performances of the optimal parametrization, $\bw_a^*$, extracted from the learning curve presented in Figure \ref{fig:10d} are studied next. The current scenario involves an initial over-pressurized tank operating 0.4 bar above the reference $\Tilde{p}$. Figure \ref{fig:21} shows the time evolution of the real pressure $\check{p}$ and propellant mass $\check{m}_l$ over time, while the vented mass flow rate $\check{\dot{m}}_c$. Figures \ref{fig21b} and \ref{fig21c} show the control action over time and the corresponding pressure, respectively. Since the tank is initially in an over-pressurized state sufficiently far from $\tilde{p}$, the valve was regulated to open to its widest ($\dot{m}_c = 0.63$ g/s), allowing the pressure to rapidly reduce until the target 5 bar are reached. This pressure drop is accompanied by a consumption of 87 kg over two roughly 45 hours. Once the target is reached, the active TVS regulates the pressure around this value. To do so, $\check{\dot{m}}_c$ fluctuates between  -50\% and 25\% around its midpoint due to changes in the heat due to the disturbance, although such variations are not seen in the pressure signal.

% Figures \ref{cryo_a_state} and \ref{cryo_b_state} show the time evolution of the real pressure $\check{p}$ and propellant mass $\check{m}_l$ over time, while the vented mass flow rate $\check{\dot{m}}_c$ and normalized control action are shown in Figures \ref{cryo_a_action} and \ref{cryo_b_action}. Lastly, in Figures \ref{cryo_a_state_action} and \ref{cryo_b_state_action} we plot the control action over the corresponding pressure.

% First, we tackle scenario (1). Since the tank is initially in an over-pressurized state sufficiently far from $\tilde{p}$, the valve was regulated to open to at its widest ($\dot{m}_c = 0.63$ g/s) allowing the pressure to rapidly reduce until the target 5 bara are reached. This pressure drop is accompanied by a consumption of 87 kg over two roughly 45 hours. Once the target is reached, the active TVS regulates the pressure around this value. To do so, $\check{\dot{m}}_c$ fluctuates between  -50\% and 25\% around its midpoint due to changes in the heat ingressed from the disturbance, although such variations are not seen in the pressure signal.
% In scenario (2), since the tank is under the target value, the venting line is initially closed. However, the propellant mass still decreases during this period due to boil-off, which consequently produces a steady pressure rise. Once the target is reached, the actuation on the valve's pressure-drop is allows the pressure to stabilize while consuming roughly 125 kg of fuel during the observation time.

From these findings, we conclude that the control policy is able to completely control the tank pressure around its desired target value. However, the response time of the TVS is limited by the technical constraints assigned to the circulating pump and Joule-Thomson valve. In addition, depending on the value assigned to the $K_1$ constant in \ref{eq:cryo_J_a}, one could retrieve either faster or more fuel-efficient policies.

% \textcolor{blue}{Finish with some short remarks and outlook on applying this cool framework to more complex thermodynamic systems... account for sloshing... more complex model... simplifications of the homogeneous model... maybe do not capture the balance of phase change with the vented stuff... could lead to different optimal control policies}

\begin{figure}[h!]
	\centering 
	\begin{subfigure}[b]{0.35\textwidth}
		\centering
		\includegraphics[width=\linewidth]{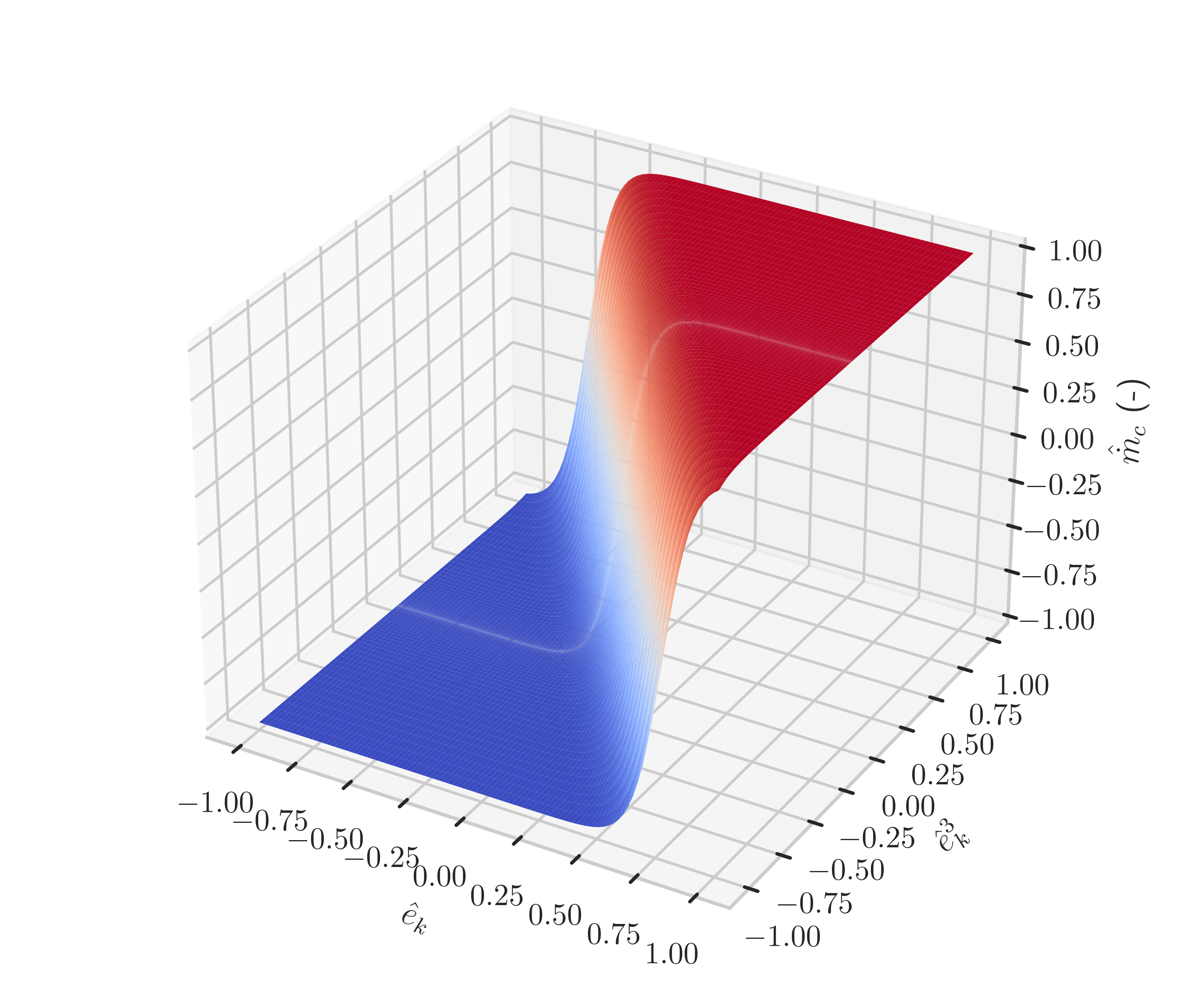}
		\caption{$\pi_{mf}$}
		\label{fig23a}
	\end{subfigure}
	\begin{subfigure}[b]{0.35\textwidth}
		\centering
		\includegraphics[width=\linewidth]{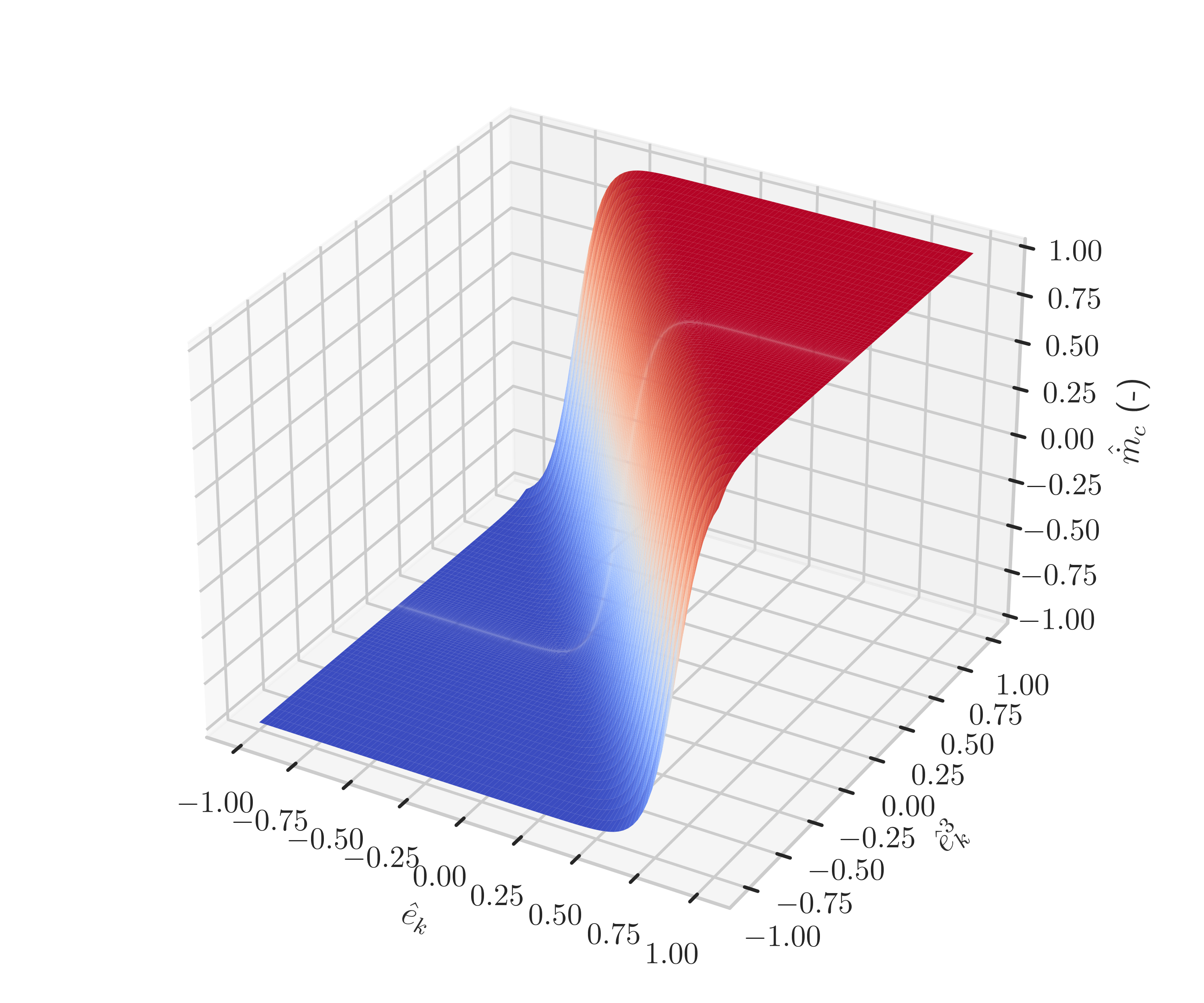}
		\caption{$\pi_{mb}$}
		\label{fig23b}
	\end{subfigure}
	\caption{T3 - Surface plot of the found policies, $\pi_{mf}$ (\ref{fig23a}) and $\pi_{mb}$ (\ref{fig23b}).}
	\label{fig:23}
\end{figure}

\begin{figure}[h!]
	\centering
	\begin{subfigure}[b]{0.45\textwidth}
		\centering
		\includegraphics[width=\textwidth]{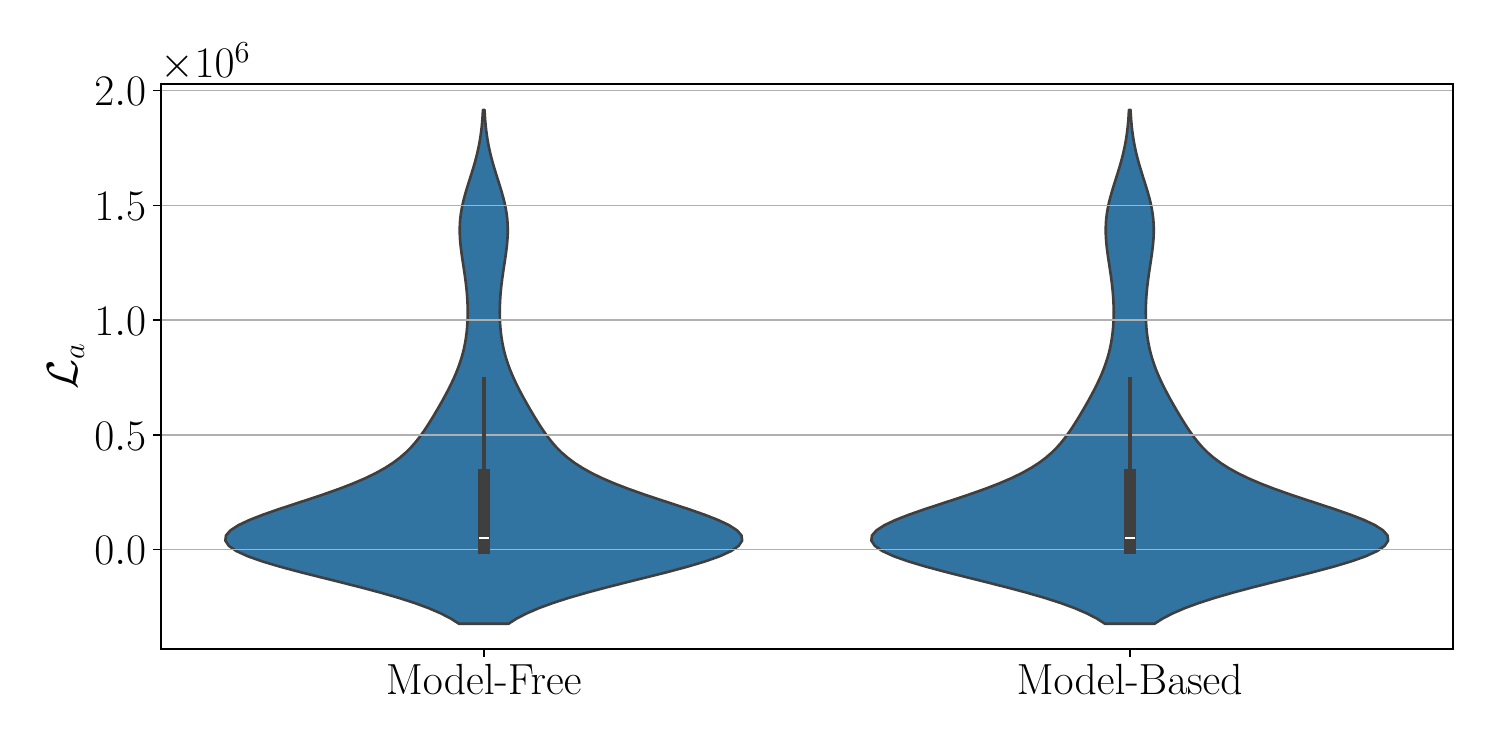}
		\caption{}
		\label{fig:25a}
  
	\end{subfigure}
    \hspace{2cm}
	\begin{subfigure}[b]{0.23\textwidth}
		\centering
		\includegraphics[width=\textwidth]{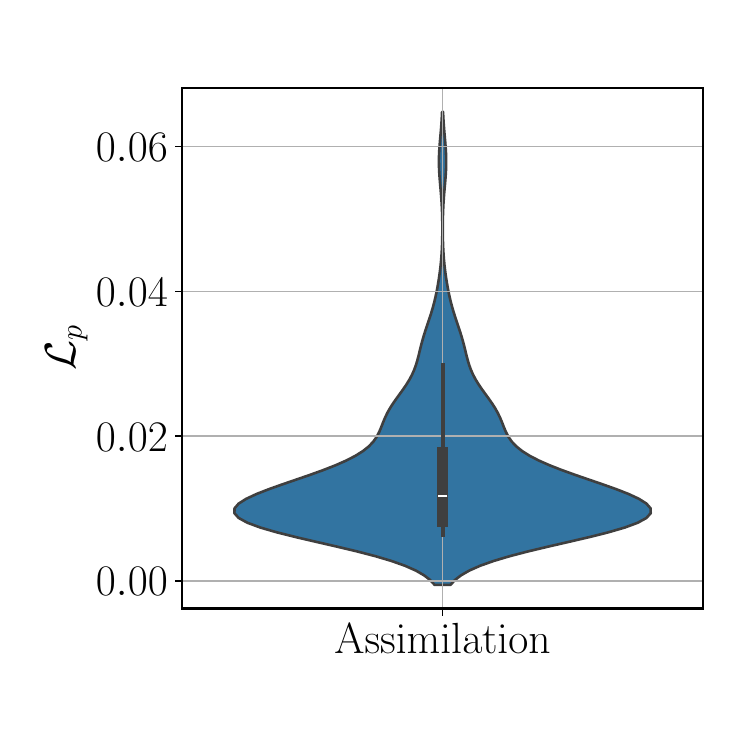}
		\caption{}
		\label{fig:25b}
	\end{subfigure}
	\caption{T3 -  Robustness analysis of of model-free and model-based policies (a) and assimilated closure (b), computed over 100 episodes.}
	\label{fig:25}
\end{figure}

We inspect the two found policies in Figure \ref{fig:23}. The two loops, model-based and model-free, converge to essentially the same parametrization. They show almost identical responses to the possible input space, with very minor discrepancies in the inclination of the hyperbolic tangent. 
Lastly, the robustness of the assimilation and control loops concludes this overview. The procedure consists in evaluating the cost of the optimal weights $\bm{w}_p$ and $\bm{w}_a$ over observations unseen during the training phase. To this end, we carried out one hundred simulations on random initial conditions and disturbances. For each of them, we computed the assimilation cost $\mathcal{L}_p$, as well as the control cost $\mathcal{L}_a$ using the model-based and model-free policies. This investigation, summarized in the violin plots presented in Figures \ref{fig:25a} and \ref{fig:25b}, proves that the weights obtained at the end of the learning phase are valid and robust even in unseen scenarios. Moreover, in contrast to what was observed in the wind turbine test case, the model-based and model-free weights are nearly identical, thus their performances are also identical as supported by Table \ref{tab:table_learning}, which shows a difference between the two control performances in the order of $\mathcal{O}(10^{-2})$. This is likely a consequence of both approaches converging in the same region of the search space. 

\section{Conclusions and Outlooks}\label{s7}

We presented a novel framework to train physics-based digital twins from real-time data while at the same time training a control agent through a combination of model-free and model-based methods. We called this framework Reinforcement Twinning (RT) and positioned it at the intersection between data assimilation, system identification, and reinforcement learning. The digital twins are formulated as physics-based models relying on closure laws expressed as parametric functions. The parameters defining these closures are identified from real-time data using an adjoint-based data assimilation strategy.

Similarly, the control agent acts on the system via a policy, also defined via parametric functions. The same adjoint-based framework is then used on the digital twin to identify the control policy and thus train the agent in the so-called ``model-based" training. At the same time, a reinforcement learning approach using the Deep Deterministic Policy Gradient (DDPG) pursues the same task of training the control agent without using explicit knowledge of the system in the so-called ``model-free" training. The ``model-based" and the ``model-free"" policies are then tested on the digital twin once this has reached satisfying predictive performances and the best-performing policy becomes the "live policy" taken on the real system. We explored a first mechanism to let both the ``model-based'' and the ``model-free'' learn from the live policy and showed that this can create a virtuous loop in which both approaches lead to a sequence of improvements over the other. 

We tested this novel framework on three widely different engineering systems: a wind turbine in a time-varying wind speed, a flapping-wing micro air vehicle traveling against gusts, and the thermal management of a cryogenic tank subject to time-varying heat loads. In these problems, the physics-based model of the digital twin is formulated as ODEs ensuring the compliance of predictions with fundamental principles (conservation of angular momentum in a wind turbine, Newton's law in the drone's trajectory, and conservation of mass/energy within the cryogenic tank) and the closure law describe unmodelled physics (aerodynamic coefficients and heat transfer correlations).

The performance of the Reinforcement Twinning framework, both in training the digital twin and in the training of the controlling agent, proves the framework's potential beyond any specific application. The merit stems from the complementary characteristics of the two approaches: the model-based is extremely sample-efficient but prone to fall stuck into local minima, while the model-free is sample-inefficient but able to better explore the parameter space and avoid model limitations.

The paths towards future extensions are many. On the one hand, we could introduce a more advanced learning feedback mechanism between the model-free and the model-based methods, which is currently limited to a ``cloning" procedure. This could allow, for example, different policy architecture between the two, unleashing the potential of reinforcement learning to handle arbitrarily complex policies in the form of artificial neural networks. On the other hand, we could introduce more sophisticated methods of integrating the digital twin in the training of the model-free loop, for instance, by using the virtual environment to boost the training of the critic network with strategies akin to traditional model-based reinforcement learning.

Nevertheless, these preliminary results open the path towards further hybridization of model-based and model-free methods and, more broadly, between traditional engineering and machine learning. Such a combination is essential to profit from the ever-growing availability of real-time data to achieve excellent predictive and control capabilities without sacrificing interpretability.

\section*{Acknowledgements}

L. Schena and R. Poletti are supported by Fonds Wetenschappelijk Onderzoek (FWO), while P. Marques and J. van Den Berghe are supported by the FRIA grant from the `Fonds de la Recherche Scientifique (F.R.S. -FNRS)'. S.Ahizi and M.A. Mendez were supported by the Flemish Agentschap Innoveren \& Ondernemen in the framework of the CSBO project ``Clean Hydrogen Propulsion for Ships (CHyPS)''.

%  with the project 1SD7823N

\bibliography{Schena_et_al_2023}

\begin{thebibliography}{193}
\expandafter\ifx\csname natexlab\endcsname\relax\def\natexlab#1{#1}\fi
\providecommand{\url}[1]{\texttt{#1}}
\providecommand{\href}[2]{#2}
\providecommand{\path}[1]{#1}
\providecommand{\DOIprefix}{doi:}
\providecommand{\ArXivprefix}{arXiv:}
\providecommand{\URLprefix}{URL: }
\providecommand{\Pubmedprefix}{pmid:}
\providecommand{\doi}[1]{\href{http://dx.doi.org/#1}{\path{#1}}}
\providecommand{\Pubmed}[1]{\href{pmid:#1}{\path{#1}}}
\providecommand{\bibinfo}[2]{#2}
\ifx\xfnm\undefined \def\xfnm[#1]{\unskip,\space#1}\fi
%Type = Article
\bibitem[{Raw(2000)}]{Rawlings2000}
\bibinfo{title}{Tutorial overview of model predictive control}.
\newblock \bibinfo{journal}{IEEE Control Syst}
  \bibinfo{year}{2000};\bibinfo{volume}{20}(\bibinfo{number}{3}):\bibinfo{pages}{38--52}.
\newblock \URLprefix \url{https://ieeexplore.ieee.org/document/845037/}.
  \DOIprefix\doi{10.1109/37.845037}.
%Type = Misc
\bibitem[{Abarbanel et~al.(2017)Abarbanel, Rozdeba and
  Shirman}]{abarbanel2017machine}
\bibinfo{author}{Abarbanel\xfnm[ H.]}, \bibinfo{author}{Rozdeba\xfnm[ P.]},
  \bibinfo{author}{Shirman\xfnm[ S.]}.
\newblock \bibinfo{title}{Machine learning, deepest learning: Statistical data
  assimilation problems}.
\newblock \bibinfo{year}{2017}.
\newblock \href{http://arxiv.org/abs/1707.01415}{\tt arXiv:1707.01415}.
%Type = Inproceedings
\bibitem[{Abbas et~al.(2020)Abbas, Wright and Pao}]{abbas2020update}
\bibinfo{author}{Abbas\xfnm[ N.J.]}, \bibinfo{author}{Wright\xfnm[ A.]},
  \bibinfo{author}{Pao\xfnm[ L.]}.
\newblock \bibinfo{title}{An update to the national renewable energy laboratory
  baseline wind turbine controller}.
\newblock In: \bibinfo{booktitle}{Journal of Physics: Conference Series}.
  \bibinfo{organization}{IOP Publishing}; volume \bibinfo{volume}{1452};
  \bibinfo{year}{2020}. p. \bibinfo{pages}{012002}.
%Type = Inproceedings
\bibitem[{Abbeel et~al.(2006)Abbeel, Quigley and Ng}]{abbeel_using_2006}
\bibinfo{author}{Abbeel\xfnm[ P.]}, \bibinfo{author}{Quigley\xfnm[ M.]},
  \bibinfo{author}{Ng\xfnm[ A.Y.]}.
\newblock \bibinfo{title}{Using inaccurate models in reinforcement learning}.
\newblock In: \bibinfo{booktitle}{Proceedings of the 23rd international
  conference on {Machine} learning - {ICML} '06}. \bibinfo{address}{Pittsburgh,
  Pennsylvania}: \bibinfo{publisher}{ACM Press}; \bibinfo{year}{2006}. p.
  \bibinfo{pages}{1--8}.
\newblock \URLprefix
  \url{http://portal.acm.org/citation.cfm?doid=1143844.1143845}.
  \DOIprefix\doi{10.1145/1143844.1143845}.
%Type = Article
\bibitem[{Ahmed et~al.(2020)Ahmed, Pawar and San}]{Ahmed2020}
\bibinfo{author}{Ahmed\xfnm[ S.E.]}, \bibinfo{author}{Pawar\xfnm[ S.]},
  \bibinfo{author}{San\xfnm[ O.]}.
\newblock \bibinfo{title}{{PyDA}: A hands-on introduction to dynamical data
  assimilation with python}.
\newblock \bibinfo{journal}{Fluids}
  \bibinfo{year}{2020};\bibinfo{volume}{5}(\bibinfo{number}{4}):\bibinfo{pages}{225}.
\newblock \DOIprefix\doi{10.3390/fluids5040225}.
%Type = Article
\bibitem[{Ammar et~al.(2022)Ammar, Nassereddine, AbdulBaky, AbouKansour,
  Tannoury, Urban and Schranz}]{ammar_digital_2022}
\bibinfo{author}{Ammar\xfnm[ A.]}, \bibinfo{author}{Nassereddine\xfnm[ H.]},
  \bibinfo{author}{AbdulBaky\xfnm[ N.]}, \bibinfo{author}{AbouKansour\xfnm[
  A.]}, \bibinfo{author}{Tannoury\xfnm[ J.]}, \bibinfo{author}{Urban\xfnm[
  H.]}, \bibinfo{author}{Schranz\xfnm[ C.]}.
\newblock \bibinfo{title}{Digital {Twins} in the {Construction} {Industry}: {A}
  {Perspective} of {Practitioners} and {Building} {Authority}}.
\newblock \bibinfo{journal}{Front Built Environ}
  \bibinfo{year}{2022};\bibinfo{volume}{8}:\bibinfo{pages}{834671}.
\newblock \URLprefix
  \url{https://www.frontiersin.org/articles/10.3389/fbuil.2022.834671/full}.
  \DOIprefix\doi{10.3389/fbuil.2022.834671}.
%Type = Inproceedings
\bibitem[{Andersson et~al.(2019)Andersson, Ribeiro, Tiels, Wahlstrom and
  Schon}]{Andersson2019}
\bibinfo{author}{Andersson\xfnm[ C.]}, \bibinfo{author}{Ribeiro\xfnm[ A.H.]},
  \bibinfo{author}{Tiels\xfnm[ K.]}, \bibinfo{author}{Wahlstrom\xfnm[ N.]},
  \bibinfo{author}{Schon\xfnm[ T.B.]}.
\newblock \bibinfo{title}{Deep convolutional networks in system
  identification}.
\newblock In: \bibinfo{booktitle}{2019 {IEEE} 58th Conference on Decision and
  Control ({CDC})}. \bibinfo{publisher}{{IEEE}}; \bibinfo{year}{2019}.
  \DOIprefix\doi{10.1109/cdc40024.2019.9030219}.
%Type = Article
\bibitem[{Arcucci et~al.(2021)Arcucci, Zhu, Hu and Guo}]{Arcucci2021}
\bibinfo{author}{Arcucci\xfnm[ R.]}, \bibinfo{author}{Zhu\xfnm[ J.]},
  \bibinfo{author}{Hu\xfnm[ S.]}, \bibinfo{author}{Guo\xfnm[ Y.K.]}.
\newblock \bibinfo{title}{Deep data assimilation: Integrating deep learning
  with data assimilation}.
\newblock \bibinfo{journal}{Applied Sciences}
  \bibinfo{year}{2021};\bibinfo{volume}{11}(\bibinfo{number}{3}):\bibinfo{pages}{1114}.
\newblock \DOIprefix\doi{10.3390/app11031114}.
%Type = Book
\bibitem[{Asch et~al.(2016)Asch, Bocquet and Nodet}]{Asch2016}
\bibinfo{author}{Asch\xfnm[ M.]}, \bibinfo{author}{Bocquet\xfnm[ M.]},
  \bibinfo{author}{Nodet\xfnm[ M.]}.
\newblock \bibinfo{title}{Data Assimilation}.
\newblock \bibinfo{publisher}{Society for Industrial and Applied Mathematics},
  \bibinfo{year}{2016}.
\newblock \DOIprefix\doi{10.1137/1.9781611974546}.
%Type = Book
\bibitem[{Astrom and Wittenmark(1994)}]{Astrom:1994}
\bibinfo{author}{Astrom\xfnm[ K.J.]}, \bibinfo{author}{Wittenmark\xfnm[ B.]}.
\newblock \bibinfo{title}{Adaptive Control}.
\newblock \bibinfo{edition}{2nd} ed.
\newblock \bibinfo{publisher}{Prentice Hall}, \bibinfo{year}{1994}.
%Type = Misc
\bibitem[{Ayed et~al.(2019)Ayed, de~Bézenac, Pajot, Brajard and
  Gallinari}]{ayed_learning_2019}
\bibinfo{author}{Ayed\xfnm[ I.]}, \bibinfo{author}{de~Bézenac\xfnm[ E.]},
  \bibinfo{author}{Pajot\xfnm[ A.]}, \bibinfo{author}{Brajard\xfnm[ J.]},
  \bibinfo{author}{Gallinari\xfnm[ P.]}.
\newblock \bibinfo{title}{Learning {Dynamical} {Systems} from {Partial}
  {Observations}}.
\newblock \bibinfo{year}{2019}.
\newblock \URLprefix \url{http://arxiv.org/abs/1902.11136};
  \bibinfo{note}{arXiv:1902.11136 [physics]}.
%Type = Article
\bibitem[{Baker et~al.(2019)Baker, Alexander, Bremer, Hagberg, Kevrekidis,
  Najm, Parashar, Patra, Sethian, Wild, Willcox and Lee}]{osti_1478744}
\bibinfo{author}{Baker\xfnm[ N.]}, \bibinfo{author}{Alexander\xfnm[ F.]},
  \bibinfo{author}{Bremer\xfnm[ T.]}, \bibinfo{author}{Hagberg\xfnm[ A.]},
  \bibinfo{author}{Kevrekidis\xfnm[ Y.]}, \bibinfo{author}{Najm\xfnm[ H.]},
  \bibinfo{author}{Parashar\xfnm[ M.]}, \bibinfo{author}{Patra\xfnm[ A.]},
  \bibinfo{author}{Sethian\xfnm[ J.]}, \bibinfo{author}{Wild\xfnm[ S.]},
  \bibinfo{author}{Willcox\xfnm[ K.]}, \bibinfo{author}{Lee\xfnm[ S.]}.
\newblock \bibinfo{title}{Workshop report on basic research needs for
  scientific machine learning: Core technologies for artificial intelligence}
  \bibinfo{year}{2019};\URLprefix \url{https://www.osti.gov/biblio/1478744}.
  \DOIprefix\doi{10.2172/1478744}.
%Type = Article
\bibitem[{Barricelli et~al.(2019)Barricelli, Casiraghi and
  Fogli}]{Barricelli2019}
\bibinfo{author}{Barricelli\xfnm[ B.R.]}, \bibinfo{author}{Casiraghi\xfnm[
  E.]}, \bibinfo{author}{Fogli\xfnm[ D.]}.
\newblock \bibinfo{title}{A survey on digital twin: Definitions,
  characteristics, applications, and design implications}.
\newblock \bibinfo{journal}{{IEEE} Access}
  \bibinfo{year}{2019};\bibinfo{volume}{7}:\bibinfo{pages}{167653--167671}.
\newblock \DOIprefix\doi{10.1109/access.2019.2953499}.
%Type = Phdthesis
\bibitem[{Barsi(2011)}]{Barsi_THESIS}
\bibinfo{author}{Barsi\xfnm[ S.]}.
\newblock \bibinfo{title}{Ventless Pressure Control of Cryogenic Storage
  Tanks}.
\newblock Ph.D. thesis; CASE Western Reserve University; \bibinfo{year}{2011}.
%Type = Article
\bibitem[{Barsi and Kassemi(2013{\natexlab{a}})}]{Barsi2013}
\bibinfo{author}{Barsi\xfnm[ S.]}, \bibinfo{author}{Kassemi\xfnm[ M.]}.
\newblock \bibinfo{title}{Investigation of tank pressurization and pressure
  control{\textemdash}part i: Experimental study}.
\newblock \bibinfo{journal}{Journal of Thermal Science and Engineering
  Applications}
  \bibinfo{year}{2013}{\natexlab{a}};\bibinfo{volume}{5}(\bibinfo{number}{4}).
\newblock \DOIprefix\doi{10.1115/1.4023891}.
%Type = Article
\bibitem[{Barsi and Kassemi(2013{\natexlab{b}})}]{Barsi2013a}
\bibinfo{author}{Barsi\xfnm[ S.]}, \bibinfo{author}{Kassemi\xfnm[ M.]}.
\newblock \bibinfo{title}{Investigation of tank pressurization and pressure
  control{\textemdash}part {II}: Numerical modeling}.
\newblock \bibinfo{journal}{Journal of Thermal Science and Engineering
  Applications}
  \bibinfo{year}{2013}{\natexlab{b}};\bibinfo{volume}{5}(\bibinfo{number}{4}).
\newblock \DOIprefix\doi{10.1115/1.4023892}.
%Type = Article
\bibitem[{van Beek et~al.(2023)van Beek, Karkaria and Chen}]{Beek2023}
\bibinfo{author}{van Beek\xfnm[ A.]}, \bibinfo{author}{Karkaria\xfnm[ V.N.]},
  \bibinfo{author}{Chen\xfnm[ W.]}.
\newblock \bibinfo{title}{Digital twins for the designs of systems: a
  perspective}.
\newblock \bibinfo{journal}{Structural and Multidisciplinary Optimization}
  \bibinfo{year}{2023};\bibinfo{volume}{66}(\bibinfo{number}{3}).
\newblock \DOIprefix\doi{10.1007/s00158-023-03488-x}.
%Type = Misc
\bibitem[{Berkenkamp et~al.(2017)Berkenkamp, Turchetta, Schoellig and
  Krause}]{berkenkamp2017safe}
\bibinfo{author}{Berkenkamp\xfnm[ F.]}, \bibinfo{author}{Turchetta\xfnm[ M.]},
  \bibinfo{author}{Schoellig\xfnm[ A.P.]}, \bibinfo{author}{Krause\xfnm[ A.]}.
\newblock \bibinfo{title}{Safe model-based reinforcement learning with
  stability guarantees}.
\newblock \bibinfo{year}{2017}.
\newblock \href{http://arxiv.org/abs/1705.08551}{\tt arXiv:1705.08551}.
%Type = Book
\bibitem[{Bertsekas(2019)}]{Bertsekas2019}
\bibinfo{author}{Bertsekas\xfnm[ D.]}.
\newblock \bibinfo{title}{Reinforcement Learning and Optimal Control}.
\newblock \bibinfo{publisher}{Athena Scientific}, \bibinfo{year}{2019}.
%Type = Article
\bibitem[{Bhatnagar et~al.(2009)Bhatnagar, Sutton, Ghavamzadeh and
  Lee}]{bhatnagar_natural_2009}
\bibinfo{author}{Bhatnagar\xfnm[ S.]}, \bibinfo{author}{Sutton\xfnm[ R.S.]},
  \bibinfo{author}{Ghavamzadeh\xfnm[ M.]}, \bibinfo{author}{Lee\xfnm[ M.]}.
\newblock \bibinfo{title}{Natural actor–critic algorithms}.
\newblock \bibinfo{journal}{Automatica}
  \bibinfo{year}{2009};\bibinfo{volume}{45}(\bibinfo{number}{11}):\bibinfo{pages}{2471--2482}.
\newblock \URLprefix
  \url{https://linkinghub.elsevier.com/retrieve/pii/S0005109809003549}.
  \DOIprefix\doi{10.1016/j.automatica.2009.07.008}.
%Type = Article
\bibitem[{Bhowmik and Spee(1999)}]{bhowmik_performance_1999}
\bibinfo{author}{Bhowmik\xfnm[ S.]}, \bibinfo{author}{Spee\xfnm[ R.]}.
\newblock \bibinfo{title}{Performance {Optimization} for {Doubly} {Fed} {Wind}
  {Power} {Generation} {Systems}}.
\newblock \bibinfo{journal}{IEEE TRANSACTIONS ON INDUSTRY APPLICATIONS}
  \bibinfo{year}{1999};\bibinfo{volume}{35}(\bibinfo{number}{4}).
%Type = Book
\bibitem[{Bianchi et~al.(2007)Bianchi, De~Battista and Mantz}]{bianchi2007wind}
\bibinfo{author}{Bianchi\xfnm[ F.D.]}, \bibinfo{author}{De~Battista\xfnm[ H.]},
  \bibinfo{author}{Mantz\xfnm[ R.J.]}.
\newblock \bibinfo{title}{Wind turbine control systems: principles, modelling
  and gain scheduling design}.
\newblock volume~\bibinfo{volume}{19}.
\newblock \bibinfo{publisher}{Springer}, \bibinfo{year}{2007}.
%Type = Article
\bibitem[{Bocquet(2011)}]{Bocquet2011}
\bibinfo{author}{Bocquet\xfnm[ M.]}.
\newblock \bibinfo{title}{Ensemble kalman filtering without the intrinsic need
  for inflation}.
\newblock \bibinfo{journal}{Nonlinear Processes in Geophysics}
  \bibinfo{year}{2011};\bibinfo{volume}{18}(\bibinfo{number}{5}):\bibinfo{pages}{735--750}.
\newblock \DOIprefix\doi{10.5194/npg-18-735-2011}.
%Type = Inproceedings
\bibitem[{Bocquet et~al.(2019)Bocquet, Brajard, Carrassi and
  Bertino}]{Bocquet2019DataAA}
\bibinfo{author}{Bocquet\xfnm[ M.]}, \bibinfo{author}{Brajard\xfnm[ J.]},
  \bibinfo{author}{Carrassi\xfnm[ A.]}, \bibinfo{author}{Bertino\xfnm[ L.]}.
\newblock \bibinfo{title}{Data assimilation as a deep learning tool to infer
  ode representations of dynamical models}.
\newblock \bibinfo{year}{2019}. .
%Type = Techreport
\bibitem[{Bocquet and Farchi(2023)}]{Bocquet2023}
\bibinfo{author}{Bocquet\xfnm[ M.]}, \bibinfo{author}{Farchi\xfnm[ A.]}.
\newblock \bibinfo{title}{Introduction to the principles and methods of data
  assimilation in the geosciences}.
\newblock \bibinfo{type}{Technical Report}; École des Ponts ParisTech;
  \bibinfo{year}{2023}.
%Type = Inproceedings
\bibitem[{Boedecker et~al.(2014{\natexlab{a}})Boedecker, Springenberg, Wülfing
  and Riedmiller}]{approx_gaussians}
\bibinfo{author}{Boedecker\xfnm[ J.]}, \bibinfo{author}{Springenberg\xfnm[
  J.T.]}, \bibinfo{author}{Wülfing\xfnm[ J.]},
  \bibinfo{author}{Riedmiller\xfnm[ M.]}.
\newblock \bibinfo{title}{Approximate real-time optimal control based on sparse
  gaussian process models}.
\newblock In: \bibinfo{booktitle}{2014 IEEE Symposium on Adaptive Dynamic
  Programming and Reinforcement Learning (ADPRL)}.
  \bibinfo{year}{2014}{\natexlab{a}}. p. \bibinfo{pages}{1--8}.
\newblock \DOIprefix\doi{10.1109/ADPRL.2014.7010608}.
%Type = Inproceedings
\bibitem[{Boedecker et~al.(2014{\natexlab{b}})Boedecker, Springenberg, Wülfing
  and Riedmiller}]{real_time_oc}
\bibinfo{author}{Boedecker\xfnm[ J.]}, \bibinfo{author}{Springenberg\xfnm[
  J.T.]}, \bibinfo{author}{Wülfing\xfnm[ J.]},
  \bibinfo{author}{Riedmiller\xfnm[ M.]}.
\newblock \bibinfo{title}{Approximate real-time optimal control based on sparse
  gaussian process models}.
\newblock In: \bibinfo{booktitle}{2014 IEEE Symposium on Adaptive Dynamic
  Programming and Reinforcement Learning (ADPRL)}.
  \bibinfo{year}{2014}{\natexlab{b}}. p. \bibinfo{pages}{1--8}.
\newblock \DOIprefix\doi{10.1109/ADPRL.2014.7010608}.
%Type = Article
\bibitem[{Bossanyi(2000)}]{bossanyi_design_2000}
\bibinfo{author}{Bossanyi\xfnm[ E.A.]}.
\newblock \bibinfo{title}{The {Design} of closed loop controllers for wind
  turbines}.
\newblock \bibinfo{journal}{Wind Energ}
  \bibinfo{year}{2000};\bibinfo{volume}{3}(\bibinfo{number}{3}):\bibinfo{pages}{149--163}.
\newblock \URLprefix \url{https://onlinelibrary.wiley.com/doi/10.1002/we.34}.
  \DOIprefix\doi{10.1002/we.34}.
%Type = Article
\bibitem[{Bradley(2019)}]{Bradley}
\bibinfo{author}{Bradley\xfnm[ A.M.]}.
\newblock \bibinfo{title}{Pde-constrained optimization and the adjoint method}
  \bibinfo{year}{2019};.
%Type = Article
\bibitem[{Brajard et~al.(2020)Brajard, Carassi, Bocquet and
  Bertino}]{brajard_combining_2020}
\bibinfo{author}{Brajard\xfnm[ J.]}, \bibinfo{author}{Carassi\xfnm[ A.]},
  \bibinfo{author}{Bocquet\xfnm[ M.]}, \bibinfo{author}{Bertino\xfnm[ L.]}.
\newblock \bibinfo{title}{Combining data assimilation and machine learning to
  emulate a dynamical model from sparse and noisy observations: a case study
  with the {Lorenz} 96 model}.
\newblock \bibinfo{journal}{Journal of Computational Science}
  \bibinfo{year}{2020};\bibinfo{volume}{44}:\bibinfo{pages}{101171}.
\newblock \URLprefix \url{http://arxiv.org/abs/2001.01520}.
  \DOIprefix\doi{10.1016/j.jocs.2020.101171}; \bibinfo{note}{arXiv:2001.01520
  [physics, stat]}.
%Type = Article
\bibitem[{Branlard et~al.(2023)Branlard, Jonkman, Brown and Zang}]{wes-2023-50}
\bibinfo{author}{Branlard\xfnm[ E.]}, \bibinfo{author}{Jonkman\xfnm[ J.]},
  \bibinfo{author}{Brown\xfnm[ C.]}, \bibinfo{author}{Zang\xfnm[ J.]}.
\newblock \bibinfo{title}{A digital-twin solution for floating offshore wind
  turbines validated using a full-scale prototype}.
\newblock \bibinfo{journal}{Wind Energy Science Discussions}
  \bibinfo{year}{2023};\bibinfo{volume}{2023}:\bibinfo{pages}{1--34}.
\newblock \URLprefix \url{https://wes.copernicus.org/preprints/wes-2023-50/}.
  \DOIprefix\doi{10.5194/wes-2023-50}.
%Type = Article
\bibitem[{Bucci et~al.(2023)Bucci, Semeraro, Allauzen, Chibbaro and
  Mathelin}]{Bucci2023}
\bibinfo{author}{Bucci\xfnm[ M.A.]}, \bibinfo{author}{Semeraro\xfnm[ O.]},
  \bibinfo{author}{Allauzen\xfnm[ A.]}, \bibinfo{author}{Chibbaro\xfnm[ S.]},
  \bibinfo{author}{Mathelin\xfnm[ L.]}.
\newblock \bibinfo{title}{Curriculum learning for data-driven modeling of
  dynamical systems}.
\newblock \bibinfo{journal}{The European Physical Journal E}
  \bibinfo{year}{2023};\bibinfo{volume}{46}(\bibinfo{number}{3}).
\newblock \DOIprefix\doi{10.1140/epje/s10189-023-00269-8}.
%Type = Article
\bibitem[{Bucci et~al.(2018)Bucci, Semeraro, Allauzen, Cordier, Wisniewski and
  Mathelin}]{bucci2018model}
\bibinfo{author}{Bucci\xfnm[ M.A.]}, \bibinfo{author}{Semeraro\xfnm[ O.]},
  \bibinfo{author}{Allauzen\xfnm[ A.]}, \bibinfo{author}{Cordier\xfnm[ L.]},
  \bibinfo{author}{Wisniewski\xfnm[ G.]}, \bibinfo{author}{Mathelin\xfnm[ L.]}.
\newblock \bibinfo{title}{Control-oriented model learning with a recurrent
  neural network}.
\newblock \bibinfo{journal}{Bulletin of the American Physical Society}
  \bibinfo{year}{2018};\bibinfo{volume}{63}.
%Type = Article
\bibitem[{Bucci et~al.(2019)Bucci, Semeraro, Allauzen, Wisniewski, Cordier and
  Mathelin}]{bucci2019control}
\bibinfo{author}{Bucci\xfnm[ M.A.]}, \bibinfo{author}{Semeraro\xfnm[ O.]},
  \bibinfo{author}{Allauzen\xfnm[ A.]}, \bibinfo{author}{Wisniewski\xfnm[ G.]},
  \bibinfo{author}{Cordier\xfnm[ L.]}, \bibinfo{author}{Mathelin\xfnm[ L.]}.
\newblock \bibinfo{title}{Control of chaotic systems by deep reinforcement
  learning}.
\newblock \bibinfo{journal}{Proceedings of the Royal Society A}
  \bibinfo{year}{2019};\bibinfo{volume}{475}(\bibinfo{number}{2231}):\bibinfo{pages}{20190351}.
%Type = Article
\bibitem[{Buizza et~al.(2022)Buizza, Casas, Nadler, Mack, Marrone, Titus,
  Cornec, Heylen, Dur, Ruiz, Heaney, Lopez, Kumar and Arcucci}]{Buizza2022}
\bibinfo{author}{Buizza\xfnm[ C.]}, \bibinfo{author}{Casas\xfnm[ C.Q.]},
  \bibinfo{author}{Nadler\xfnm[ P.]}, \bibinfo{author}{Mack\xfnm[ J.]},
  \bibinfo{author}{Marrone\xfnm[ S.]}, \bibinfo{author}{Titus\xfnm[ Z.]},
  \bibinfo{author}{Cornec\xfnm[ C.L.]}, \bibinfo{author}{Heylen\xfnm[ E.]},
  \bibinfo{author}{Dur\xfnm[ T.]}, \bibinfo{author}{Ruiz\xfnm[ L.B.]},
  \bibinfo{author}{Heaney\xfnm[ C.]}, \bibinfo{author}{Lopez\xfnm[ J.A.D.]},
  \bibinfo{author}{Kumar\xfnm[ K.S.]}, \bibinfo{author}{Arcucci\xfnm[ R.]}.
\newblock \bibinfo{title}{Data learning: Integrating data assimilation and
  machine learning}.
\newblock \bibinfo{journal}{Journal of Computational Science}
  \bibinfo{year}{2022};\bibinfo{volume}{58}:\bibinfo{pages}{101525}.
\newblock \DOIprefix\doi{10.1016/j.jocs.2021.101525}.
%Type = Article
\bibitem[{Cai et~al.(2021)Cai, Kolomenskiy, Nakata and Liu}]{Cai2021}
\bibinfo{author}{Cai\xfnm[ X.]}, \bibinfo{author}{Kolomenskiy\xfnm[ D.]},
  \bibinfo{author}{Nakata\xfnm[ T.]}, \bibinfo{author}{Liu\xfnm[ H.]}.
\newblock \bibinfo{title}{A cfd data-driven aerodynamic model for fast and
  precise prediction of flapping aerodynamics in various flight velocities}.
\newblock \bibinfo{journal}{Journal of Fluid Mechanics}
  \bibinfo{year}{2021};\bibinfo{volume}{915}:\bibinfo{pages}{A114}.
%Type = Misc
\bibitem[{Canaday et~al.(2020)Canaday, Pomerance and Gauthier}]{Canaday2020}
\bibinfo{author}{Canaday\xfnm[ D.]}, \bibinfo{author}{Pomerance\xfnm[ A.]},
  \bibinfo{author}{Gauthier\xfnm[ D.J.]}.
\newblock \bibinfo{title}{Model-free control of dynamical systems with deep
  reservoir computing}.
\newblock \bibinfo{year}{2020}.
\newblock \href{http://arxiv.org/abs/2010.02285}{\tt arXiv:2010.02285}.
%Type = Article
\bibitem[{Cao et~al.(2003)Cao, Li, Petzold and Serban}]{Cao2003}
\bibinfo{author}{Cao\xfnm[ Y.]}, \bibinfo{author}{Li\xfnm[ S.]},
  \bibinfo{author}{Petzold\xfnm[ L.]}, \bibinfo{author}{Serban\xfnm[ R.]}.
\newblock \bibinfo{title}{Adjoint sensitivity analysis for
  differential-algebraic equations: The adjoint {DAE} system and its numerical
  solution}.
\newblock \bibinfo{journal}{{SIAM} Journal on Scientific Computing}
  \bibinfo{year}{2003};\bibinfo{volume}{24}(\bibinfo{number}{3}):\bibinfo{pages}{1076--1089}.
\newblock \DOIprefix\doi{10.1137/s1064827501380630}.
%Type = Misc
\bibitem[{Carrassi et~al.(2017)Carrassi, Bocquet, Bertino and
  Evensen}]{Carrassi2017}
\bibinfo{author}{Carrassi\xfnm[ A.]}, \bibinfo{author}{Bocquet\xfnm[ M.]},
  \bibinfo{author}{Bertino\xfnm[ L.]}, \bibinfo{author}{Evensen\xfnm[ G.]}.
\newblock \bibinfo{title}{Data assimilation in the geosciences - an overview on
  methods, issues and perspectives}.
\newblock \bibinfo{year}{2017}.
\newblock \href{http://arxiv.org/abs/1709.02798}{\tt arXiv:1709.02798}.
%Type = Book
\bibitem[{Cengel and Ghajar(2019)}]{book_HT}
\bibinfo{author}{Cengel\xfnm[ Y.]}, \bibinfo{author}{Ghajar\xfnm[ A.]}.
\newblock \bibinfo{title}{Heat and Mass Transfer: Fundamentals and
  Applications}.
\newblock \bibinfo{edition}{6th} ed.
\newblock \bibinfo{publisher}{McGraw Hill}, \bibinfo{year}{2019}.
%Type = Misc
\bibitem[{Chai and Wilhite(2014)}]{chai_cryogenic_2014}
\bibinfo{author}{Chai\xfnm[ P.R.]}, \bibinfo{author}{Wilhite\xfnm[ A.W.]}.
\newblock \bibinfo{title}{Cryogenic thermal system analysis for orbital
  propellant depot}.
\newblock \bibinfo{year}{2014}.
\newblock \URLprefix
  \url{https://linkinghub.elsevier.com/retrieve/pii/S0094576514001738}.
  \DOIprefix\doi{10.1016/j.actaastro.2014.05.013}.
%Type = Article
\bibitem[{Chang and Lin(2011)}]{Chang2011}
\bibinfo{author}{Chang\xfnm[ C.C.]}, \bibinfo{author}{Lin\xfnm[ C.J.]}.
\newblock \bibinfo{title}{{LIBSVM}}.
\newblock \bibinfo{journal}{{ACM} Transactions on Intelligent Systems and
  Technology}
  \bibinfo{year}{2011};\bibinfo{volume}{2}(\bibinfo{number}{3}):\bibinfo{pages}{1--27}.
\newblock \DOIprefix\doi{10.1145/1961189.1961199}.
%Type = Misc
\bibitem[{Chatzilygeroudis et~al.(2019)Chatzilygeroudis, Vassiliades, Stulp,
  Calinon and Mouret}]{chatzilygeroudis2019survey}
\bibinfo{author}{Chatzilygeroudis\xfnm[ K.]},
  \bibinfo{author}{Vassiliades\xfnm[ V.]}, \bibinfo{author}{Stulp\xfnm[ F.]},
  \bibinfo{author}{Calinon\xfnm[ S.]}, \bibinfo{author}{Mouret\xfnm[ J.B.]}.
\newblock \bibinfo{title}{A survey on policy search algorithms for learning
  robot controllers in a handful of trials}.
\newblock \bibinfo{year}{2019}.
\newblock \href{http://arxiv.org/abs/1807.02303}{\tt arXiv:1807.02303}.
%Type = Misc
\bibitem[{Chebotar et~al.(2017)Chebotar, Hausman, Zhang, Sukhatme, Schaal and
  Levine}]{chebotar2017combining}
\bibinfo{author}{Chebotar\xfnm[ Y.]}, \bibinfo{author}{Hausman\xfnm[ K.]},
  \bibinfo{author}{Zhang\xfnm[ M.]}, \bibinfo{author}{Sukhatme\xfnm[ G.]},
  \bibinfo{author}{Schaal\xfnm[ S.]}, \bibinfo{author}{Levine\xfnm[ S.]}.
\newblock \bibinfo{title}{Combining model-based and model-free updates for
  trajectory-centric reinforcement learning}.
\newblock \bibinfo{year}{2017}.
\newblock \href{http://arxiv.org/abs/1703.03078}{\tt arXiv:1703.03078}.
%Type = Inproceedings
\bibitem[{Chen et~al.(2018)Chen, Rubanova, Bettencourt and Duvenaud}]{Chen2018}
\bibinfo{author}{Chen\xfnm[ R.T.Q.]}, \bibinfo{author}{Rubanova\xfnm[ Y.]},
  \bibinfo{author}{Bettencourt\xfnm[ J.]}, \bibinfo{author}{Duvenaud\xfnm[
  D.]}.
\newblock \bibinfo{title}{Neural ordinary differential equations}.
\newblock In: \bibinfo{booktitle}{Proceedings of the 32nd International
  Conference on Neural Information Processing Systems}. \bibinfo{address}{Red
  Hook, NY, USA}: \bibinfo{publisher}{Curran Associates Inc.}; NIPS'18;
  \bibinfo{year}{2018}. p. \bibinfo{pages}{6572–6583}.
%Type = Article
\bibitem[{Chen et~al.(1990)Chen, Billings and Grant}]{Chen1990}
\bibinfo{author}{Chen\xfnm[ S.]}, \bibinfo{author}{Billings\xfnm[ S.]},
  \bibinfo{author}{Grant\xfnm[ P.]}.
\newblock \bibinfo{title}{Non-linear system identification using neural
  networks}.
\newblock \bibinfo{journal}{International Journal of Control}
  \bibinfo{year}{1990};\bibinfo{volume}{51}(\bibinfo{number}{6}):\bibinfo{pages}{1191--1214}.
\newblock \DOIprefix\doi{10.1080/00207179008934126}.
%Type = Article
\bibitem[{Cheng et~al.(2016)Cheng, Tobalske, Powers, Hedrick, Wethington, Chiu
  and Deng}]{Cheng2016}
\bibinfo{author}{Cheng\xfnm[ B.]}, \bibinfo{author}{Tobalske\xfnm[ B.W.]},
  \bibinfo{author}{Powers\xfnm[ D.R.]}, \bibinfo{author}{Hedrick\xfnm[ T.L.]},
  \bibinfo{author}{Wethington\xfnm[ S.M.]}, \bibinfo{author}{Chiu\xfnm[ G.T.]},
  \bibinfo{author}{Deng\xfnm[ X.]}.
\newblock \bibinfo{title}{Flight mechanics and control of escape manoeuvres in
  hummingbirds. i. flight kinematics}.
\newblock \bibinfo{journal}{Journal of Experimental Biology}
  \bibinfo{year}{2016};\bibinfo{volume}{219}(\bibinfo{number}{22}):\bibinfo{pages}{3518--3531}.
%Type = Misc
\bibitem[{Cheng et~al.(2023)Cheng, Quilodran-Casas, Ouala, Farchi, Liu, Tandeo,
  Fablet, Lucor, Iooss, Brajard, Xiao, Janjic, Ding, Guo, Carrassi, Bocquet and
  Arcucci}]{cheng_machine_2023}
\bibinfo{author}{Cheng\xfnm[ S.]}, \bibinfo{author}{Quilodran-Casas\xfnm[ C.]},
  \bibinfo{author}{Ouala\xfnm[ S.]}, \bibinfo{author}{Farchi\xfnm[ A.]},
  \bibinfo{author}{Liu\xfnm[ C.]}, \bibinfo{author}{Tandeo\xfnm[ P.]},
  \bibinfo{author}{Fablet\xfnm[ R.]}, \bibinfo{author}{Lucor\xfnm[ D.]},
  \bibinfo{author}{Iooss\xfnm[ B.]}, \bibinfo{author}{Brajard\xfnm[ J.]},
  \bibinfo{author}{Xiao\xfnm[ D.]}, \bibinfo{author}{Janjic\xfnm[ T.]},
  \bibinfo{author}{Ding\xfnm[ W.]}, \bibinfo{author}{Guo\xfnm[ Y.]},
  \bibinfo{author}{Carrassi\xfnm[ A.]}, \bibinfo{author}{Bocquet\xfnm[ M.]},
  \bibinfo{author}{Arcucci\xfnm[ R.]}.
\newblock \bibinfo{title}{Machine learning with data assimilation and
  uncertainty quantification for dynamical systems: a review}.
\newblock \bibinfo{year}{2023}.
\newblock \URLprefix \url{http://arxiv.org/abs/2303.10462};
  \bibinfo{note}{arXiv:2303.10462 [cs]}.
%Type = Article
\bibitem[{Chinesta et~al.(2020)Chinesta, Cueto, Abisset-Chavanne, Duval and
  Khaldi}]{chinesta_virtual_2020}
\bibinfo{author}{Chinesta\xfnm[ F.]}, \bibinfo{author}{Cueto\xfnm[ E.]},
  \bibinfo{author}{Abisset-Chavanne\xfnm[ E.]}, \bibinfo{author}{Duval\xfnm[
  J.L.]}, \bibinfo{author}{Khaldi\xfnm[ F.E.]}.
\newblock \bibinfo{title}{Virtual, {Digital} and {Hybrid} {Twins}: {A} {New}
  {Paradigm} in {Data}-{Based} {Engineering} and {Engineered} {Data}}.
\newblock \bibinfo{journal}{Arch Computat Methods Eng}
  \bibinfo{year}{2020};\bibinfo{volume}{27}(\bibinfo{number}{1}):\bibinfo{pages}{105--134}.
\newblock \URLprefix \url{http://link.springer.com/10.1007/s11831-018-9301-4}.
  \DOIprefix\doi{10.1007/s11831-018-9301-4}.
%Type = Article
\bibitem[{Coquelet et~al.(2022)Coquelet, Bricteux, Moens and
  Chatelain}]{coquelet2022reinforcement}
\bibinfo{author}{Coquelet\xfnm[ M.]}, \bibinfo{author}{Bricteux\xfnm[ L.]},
  \bibinfo{author}{Moens\xfnm[ M.]}, \bibinfo{author}{Chatelain\xfnm[ P.]}.
\newblock \bibinfo{title}{A reinforcement-learning approach for individual
  pitch control}.
\newblock \bibinfo{journal}{Wind Energy}
  \bibinfo{year}{2022};\bibinfo{volume}{25}(\bibinfo{number}{8}):\bibinfo{pages}{1343--1362}.
%Type = Article
\bibitem[{De~Cillis et~al.(2022{\natexlab{a}})De~Cillis, Cherubini, Semeraro,
  Leonardi and De~Palma}]{de2022influenceNREL_turb}
\bibinfo{author}{De~Cillis\xfnm[ G.]}, \bibinfo{author}{Cherubini\xfnm[ S.]},
  \bibinfo{author}{Semeraro\xfnm[ O.]}, \bibinfo{author}{Leonardi\xfnm[ S.]},
  \bibinfo{author}{De~Palma\xfnm[ P.]}.
\newblock \bibinfo{title}{The influence of incoming turbulence on the dynamic
  modes of an nrel-5mw wind turbine wake}.
\newblock \bibinfo{journal}{Renewable Energy}
  \bibinfo{year}{2022}{\natexlab{a}};\bibinfo{volume}{183}:\bibinfo{pages}{601--616}.
%Type = Article
\bibitem[{De~Cillis et~al.(2022{\natexlab{b}})De~Cillis, Semeraro, Leonardi,
  De~Palma and Cherubini}]{de2022dynamicNREL}
\bibinfo{author}{De~Cillis\xfnm[ G.]}, \bibinfo{author}{Semeraro\xfnm[ O.]},
  \bibinfo{author}{Leonardi\xfnm[ S.]}, \bibinfo{author}{De~Palma\xfnm[ P.]},
  \bibinfo{author}{Cherubini\xfnm[ S.]}.
\newblock \bibinfo{title}{Dynamic-mode-decomposition of the wake of the
  nrel-5mw wind turbine impinged by a laminar inflow}.
\newblock \bibinfo{journal}{Renewable Energy}
  \bibinfo{year}{2022}{\natexlab{b}};\bibinfo{volume}{199}:\bibinfo{pages}{1--10}.
%Type = Inproceedings
\bibitem[{Deisenroth and Rasmussen(2011)}]{pilco}
\bibinfo{author}{Deisenroth\xfnm[ M.]}, \bibinfo{author}{Rasmussen\xfnm[ C.]}.
\newblock \bibinfo{title}{Pilco: A model-based and data-efficient approach to
  policy search.}
\newblock \bibinfo{year}{2011}. p. \bibinfo{pages}{465--472}.
%Type = Article
\bibitem[{Dickinson et~al.(1999)Dickinson, Lehmann and Sane}]{Dickinson1999}
\bibinfo{author}{Dickinson\xfnm[ M.H.]}, \bibinfo{author}{Lehmann\xfnm[ F.O.]},
  \bibinfo{author}{Sane\xfnm[ S.P.]}.
\newblock \bibinfo{title}{Wing rotation and the aerodynamic basis of insect
  flight}.
\newblock \bibinfo{journal}{Science}
  \bibinfo{year}{1999};\bibinfo{volume}{284}(\bibinfo{number}{5422}):\bibinfo{pages}{1954--1960}.
%Type = Incollection
\bibitem[{Dimet et~al.(2016)Dimet, Navon and
  {\c{S}}tef{\u{a}}nescu}]{Dimet2016}
\bibinfo{author}{Dimet\xfnm[ F.X.L.]}, \bibinfo{author}{Navon\xfnm[ I.M.]},
  \bibinfo{author}{{\c{S}}tef{\u{a}}nescu\xfnm[ R.]}.
\newblock \bibinfo{title}{Variational data assimilation: Optimization and
  optimal control}.
\newblock In: \bibinfo{booktitle}{Data Assimilation for Atmospheric, Oceanic
  and Hydrologic Applications (Vol. {III})}. \bibinfo{publisher}{Springer
  International Publishing}; \bibinfo{year}{2016}. p. \bibinfo{pages}{1--53}.
\newblock \DOIprefix\doi{10.1007/978-3-319-43415-5_1}.
%Type = Article
\bibitem[{Doya(2000)}]{doya2000reinforcement}
\bibinfo{author}{Doya\xfnm[ K.]}.
\newblock \bibinfo{title}{Reinforcement learning in continuous time and space}.
\newblock \bibinfo{journal}{Neural computation}
  \bibinfo{year}{2000};\bibinfo{volume}{12}(\bibinfo{number}{1}):\bibinfo{pages}{219--245}.
%Type = Misc
\bibitem[{Dulac-Arnold et~al.(2019)Dulac-Arnold, Mankowitz and
  Hester}]{dulacarnold2019challenges}
\bibinfo{author}{Dulac-Arnold\xfnm[ G.]}, \bibinfo{author}{Mankowitz\xfnm[
  D.]}, \bibinfo{author}{Hester\xfnm[ T.]}.
\newblock \bibinfo{title}{Challenges of real-world reinforcement learning}.
\newblock \bibinfo{year}{2019}.
\newblock \href{http://arxiv.org/abs/1904.12901}{\tt arXiv:1904.12901}.
%Type = Article
\bibitem[{Errico(1997)}]{Errico}
\bibinfo{author}{Errico\xfnm[ R.M.]}.
\newblock \bibinfo{title}{What is an adjoint model?}
\newblock \bibinfo{journal}{Bulletin of the American Meteorological Society}
  \bibinfo{year}{1997};\bibinfo{volume}{78}(\bibinfo{number}{11}).
%Type = Book
\bibitem[{Evensen(2009)}]{Evensen2009}
\bibinfo{author}{Evensen\xfnm[ G.]}.
\newblock \bibinfo{title}{Data Assimilation: The Ensemble Kalman Filter}.
\newblock \bibinfo{edition}{2nd} ed.
\newblock \bibinfo{publisher}{Spinger}, \bibinfo{year}{2009}.
%Type = Article
\bibitem[{Fahim et~al.(2022)Fahim, Sharma, Cao, Canberk and
  Duong}]{Fahim2022MachineLD}
\bibinfo{author}{Fahim\xfnm[ M.]}, \bibinfo{author}{Sharma\xfnm[ V.]},
  \bibinfo{author}{Cao\xfnm[ T.V.]}, \bibinfo{author}{Canberk\xfnm[ B.]},
  \bibinfo{author}{Duong\xfnm[ T.Q.]}.
\newblock \bibinfo{title}{Machine learning-based digital twin for predictive
  modeling in wind turbines}.
\newblock \bibinfo{journal}{IEEE Access}
  \bibinfo{year}{2022};\bibinfo{volume}{10}:\bibinfo{pages}{14184--14194}.
\newblock \URLprefix \url{https://api.semanticscholar.org/CorpusID:246420062}.
%Type = Inproceedings
\bibitem[{Fei et~al.(2019)Fei, Tu, Yang, Zhang and Deng}]{Fei2019}
\bibinfo{author}{Fei\xfnm[ F.]}, \bibinfo{author}{Tu\xfnm[ Z.]},
  \bibinfo{author}{Yang\xfnm[ Y.]}, \bibinfo{author}{Zhang\xfnm[ J.]},
  \bibinfo{author}{Deng\xfnm[ X.]}.
\newblock \bibinfo{title}{Flappy hummingbird: An open source dynamic simulation
  of flapping wing robots and animals}.
\newblock In: \bibinfo{booktitle}{2019 International Conference on Robotics and
  Automation (ICRA)}. \bibinfo{organization}{IEEE}; \bibinfo{year}{2019}. p.
  \bibinfo{pages}{9223--9229}.
%Type = Misc
\bibitem[{Feinberg et~al.(2018)Feinberg, Wan, Stoica, Jordan, Gonzalez and
  Levine}]{feinberg2018modelbased}
\bibinfo{author}{Feinberg\xfnm[ V.]}, \bibinfo{author}{Wan\xfnm[ A.]},
  \bibinfo{author}{Stoica\xfnm[ I.]}, \bibinfo{author}{Jordan\xfnm[ M.I.]},
  \bibinfo{author}{Gonzalez\xfnm[ J.E.]}, \bibinfo{author}{Levine\xfnm[ S.]}.
\newblock \bibinfo{title}{Model-based value estimation for efficient model-free
  reinforcement learning}.
\newblock \bibinfo{year}{2018}.
\newblock \href{http://arxiv.org/abs/1803.00101}{\tt arXiv:1803.00101}.
%Type = Misc
\bibitem[{Freed et~al.(2024)Freed, Wei, Calandra, Schneider and
  Choset}]{freed2024unifying}
\bibinfo{author}{Freed\xfnm[ B.]}, \bibinfo{author}{Wei\xfnm[ T.]},
  \bibinfo{author}{Calandra\xfnm[ R.]}, \bibinfo{author}{Schneider\xfnm[ J.]},
  \bibinfo{author}{Choset\xfnm[ H.]}.
\newblock \bibinfo{title}{Unifying model-based and model-free reinforcement
  learning with equivalent policy sets}.
\newblock \bibinfo{year}{2024}.
\newblock \URLprefix \url{https://openreview.net/forum?id=p5SurcLh24}.
%Type = Article
\bibitem[{Geer(2021)}]{Geer2021}
\bibinfo{author}{Geer\xfnm[ A.J.]}.
\newblock \bibinfo{title}{Learning earth system models from observations:
  machine learning or data assimilation?}
\newblock \bibinfo{journal}{Philosophical Transactions of the Royal Society A:
  Mathematical, Physical and Engineering Sciences}
  \bibinfo{year}{2021};\bibinfo{volume}{379}(\bibinfo{number}{2194}).
\newblock \DOIprefix\doi{10.1098/rsta.2020.0089}.
%Type = Article
\bibitem[{Gonzalez and Yu(2018)}]{Gonzalez2018}
\bibinfo{author}{Gonzalez\xfnm[ J.]}, \bibinfo{author}{Yu\xfnm[ W.]}.
\newblock \bibinfo{title}{Non-linear system modeling using {LSTM} neural
  networks}.
\newblock \bibinfo{journal}{{IFAC}-{PapersOnLine}}
  \bibinfo{year}{2018};\bibinfo{volume}{51}(\bibinfo{number}{13}):\bibinfo{pages}{485--489}.
\newblock \DOIprefix\doi{10.1016/j.ifacol.2018.07.326}.
%Type = Book
\bibitem[{Goodfellow et~al.(2016)Goodfellow, Bengio, Courville and
  Bengio}]{goodfellow2016deep}
\bibinfo{author}{Goodfellow\xfnm[ I.]}, \bibinfo{author}{Bengio\xfnm[ Y.]},
  \bibinfo{author}{Courville\xfnm[ A.]}, \bibinfo{author}{Bengio\xfnm[ Y.]}.
\newblock \bibinfo{title}{Deep learning}.
\newblock volume~\bibinfo{volume}{1}.
\newblock \bibinfo{publisher}{MIT press Cambridge}, \bibinfo{year}{2016}.
%Type = Misc
\bibitem[{Gu et~al.(2016)Gu, Lillicrap, Sutskever and
  Levine}]{gu2016continuous}
\bibinfo{author}{Gu\xfnm[ S.]}, \bibinfo{author}{Lillicrap\xfnm[ T.]},
  \bibinfo{author}{Sutskever\xfnm[ I.]}, \bibinfo{author}{Levine\xfnm[ S.]}.
\newblock \bibinfo{title}{Continuous deep q-learning with model-based
  acceleration}.
\newblock \bibinfo{year}{2016}.
\newblock \href{http://arxiv.org/abs/1603.00748}{\tt arXiv:1603.00748}.
%Type = Misc
\bibitem[{Haarnoja et~al.(2018)Haarnoja, Zhou, Abbeel and
  Levine}]{haarnoja2018soft}
\bibinfo{author}{Haarnoja\xfnm[ T.]}, \bibinfo{author}{Zhou\xfnm[ A.]},
  \bibinfo{author}{Abbeel\xfnm[ P.]}, \bibinfo{author}{Levine\xfnm[ S.]}.
\newblock \bibinfo{title}{Soft actor-critic: Off-policy maximum entropy deep
  reinforcement learning with a stochastic actor}.
\newblock \bibinfo{year}{2018}.
\newblock \href{http://arxiv.org/abs/1801.01290}{\tt arXiv:1801.01290}.
%Type = Article
\bibitem[{Haghshenas et~al.(2023)Haghshenas, Hasan, Osen and
  Mikalsen}]{haghshenas2023predictive}
\bibinfo{author}{Haghshenas\xfnm[ A.]}, \bibinfo{author}{Hasan\xfnm[ A.]},
  \bibinfo{author}{Osen\xfnm[ O.]}, \bibinfo{author}{Mikalsen\xfnm[ E.T.]}.
\newblock \bibinfo{title}{Predictive digital twin for offshore wind farms}.
\newblock \bibinfo{journal}{Energy Informatics}
  \bibinfo{year}{2023};\bibinfo{volume}{6}(\bibinfo{number}{1}):\bibinfo{pages}{1--26}.
%Type = Article
\bibitem[{Haider et~al.(2020)Haider, Shahzad, Qadri and Shah}]{Haider2020}
\bibinfo{author}{Haider\xfnm[ N.]}, \bibinfo{author}{Shahzad\xfnm[ A.]},
  \bibinfo{author}{Qadri\xfnm[ M.N.M.]}, \bibinfo{author}{Shah\xfnm[ S.I.A.]}.
\newblock \bibinfo{title}{Recent progress in flapping wings for micro aerial
  vehicle applications}.
\newblock \bibinfo{journal}{Proceedings of the Institution of Mechanical
  Engineers, Part C: Journal of Mechanical Engineering Science}
  \bibinfo{year}{2020};\bibinfo{volume}{235}(\bibinfo{number}{2}):\bibinfo{pages}{245--264}.
\newblock \DOIprefix\doi{10.1177/0954406220917426}.
%Type = Misc
\bibitem[{Han et~al.()Han, Tian, Zhang, Wang and Pan}]{hanH_InftyModelfree2020}
\bibinfo{author}{Han\xfnm[ M.]}, \bibinfo{author}{Tian\xfnm[ Y.]},
  \bibinfo{author}{Zhang\xfnm[ L.]}, \bibinfo{author}{Wang\xfnm[ J.]},
  \bibinfo{author}{Pan\xfnm[ W.]}.
\newblock \bibinfo{title}{\${{H}}\_\textbackslash infty\$ {{Model-free
  Reinforcement Learning}} with {{Robust Stability Guarantee}}}.
\newblock \URLprefix \url{http://arxiv.org/abs/1911.02875}.
  \href{http://arxiv.org/abs/1911.02875}{\tt arXiv:1911.02875}.
%Type = Inproceedings
\bibitem[{Hastings et~al.(2005)Hastings, Tucker, Flachbart, Hedayat and
  Nelson}]{Hastings2005}
\bibinfo{author}{Hastings\xfnm[ L.]}, \bibinfo{author}{Tucker\xfnm[ S.]},
  \bibinfo{author}{Flachbart\xfnm[ R.]}, \bibinfo{author}{Hedayat\xfnm[ A.]},
  \bibinfo{author}{Nelson\xfnm[ S.]}.
\newblock \bibinfo{title}{Marshall space flight center in-space cryogenic fluid
  management program overview}.
\newblock In: \bibinfo{booktitle}{41st {AIAA}/{ASME}/{SAE}/{ASEE} Joint
  Propulsion Conference and Exhibit}. \bibinfo{publisher}{American Institute of
  Aeronautics and Astronautics}; \bibinfo{year}{2005}.
  \DOIprefix\doi{10.2514/6.2005-3561}.
%Type = Techreport
\bibitem[{Hastings et~al.(2003)Hastings, Flachbart, Martin, Hedayat, Fazah,
  Lak, Nguyen and Bailey}]{Hastings2003_TVS}
\bibinfo{author}{Hastings\xfnm[ L.J.]}, \bibinfo{author}{Flachbart\xfnm[ R.]},
  \bibinfo{author}{Martin\xfnm[ J.]}, \bibinfo{author}{Hedayat\xfnm[ A.]},
  \bibinfo{author}{Fazah\xfnm[ M.]}, \bibinfo{author}{Lak\xfnm[ T.]},
  \bibinfo{author}{Nguyen\xfnm[ H.]}, \bibinfo{author}{Bailey\xfnm[ J.]}.
\newblock \bibinfo{title}{Spray {Bar} {Zero}-{Gravity} {Vent} {System} for
  {On}-{Orbit} {Liquid} {Hydrogen} {Storage}}.
\newblock \bibinfo{type}{Technical {Memorandum} ({TM})}
  \bibinfo{number}{NASA/TM-2003-212926}; National Aeronautics and Space
  Administration, Marshall Space Flight Center; \bibinfo{address}{Alabama
  35812}; \bibinfo{year}{2003}.
\newblock \URLprefix
  \url{https://ntrs.nasa.gov/api/citations/20040000092/downloads/20040000092.pdf}.
%Type = Article
\bibitem[{Hedengren et~al.(2014)Hedengren, Shishavan, Powell and
  Edgar}]{hedengren_nonlinear_2014}
\bibinfo{author}{Hedengren\xfnm[ J.D.]}, \bibinfo{author}{Shishavan\xfnm[
  R.A.]}, \bibinfo{author}{Powell\xfnm[ K.M.]}, \bibinfo{author}{Edgar\xfnm[
  T.F.]}.
\newblock \bibinfo{title}{Nonlinear modeling, estimation and predictive control
  in {APMonitor}}.
\newblock \bibinfo{journal}{Computers \& Chemical Engineering}
  \bibinfo{year}{2014};\bibinfo{volume}{70}:\bibinfo{pages}{133--148}.
\newblock \URLprefix
  \url{https://linkinghub.elsevier.com/retrieve/pii/S0098135414001306}.
  \DOIprefix\doi{10.1016/j.compchemeng.2014.04.013}.
%Type = Misc
\bibitem[{Heess et~al.(2015)Heess, Wayne, Silver, Lillicrap, Tassa and
  Erez}]{heess2015learning}
\bibinfo{author}{Heess\xfnm[ N.]}, \bibinfo{author}{Wayne\xfnm[ G.]},
  \bibinfo{author}{Silver\xfnm[ D.]}, \bibinfo{author}{Lillicrap\xfnm[ T.]},
  \bibinfo{author}{Tassa\xfnm[ Y.]}, \bibinfo{author}{Erez\xfnm[ T.]}.
\newblock \bibinfo{title}{Learning continuous control policies by stochastic
  value gradients}.
\newblock \bibinfo{year}{2015}.
\newblock \href{http://arxiv.org/abs/1510.09142}{\tt arXiv:1510.09142}.
%Type = Article
\bibitem[{Hochreiter and Schmidhuber(1997)}]{Hochreiter1997}
\bibinfo{author}{Hochreiter\xfnm[ S.]}, \bibinfo{author}{Schmidhuber\xfnm[
  J.]}.
\newblock \bibinfo{title}{Long short-term memory}.
\newblock \bibinfo{journal}{Neural Computation}
  \bibinfo{year}{1997};\bibinfo{volume}{9}(\bibinfo{number}{8}):\bibinfo{pages}{1735--1780}.
\newblock \DOIprefix\doi{10.1162/neco.1997.9.8.1735}.
%Type = Article
\bibitem[{Howlader et~al.(2010)Howlader, Urasaki, Uchida, Yona, Senjyu, Kim and
  Saber}]{howlader_parameter_2010}
\bibinfo{author}{Howlader\xfnm[ A.M.]}, \bibinfo{author}{Urasaki\xfnm[ N.]},
  \bibinfo{author}{Uchida\xfnm[ K.]}, \bibinfo{author}{Yona\xfnm[ A.]},
  \bibinfo{author}{Senjyu\xfnm[ T.]}, \bibinfo{author}{Kim\xfnm[ C.H.]},
  \bibinfo{author}{Saber\xfnm[ A.Y.]}.
\newblock \bibinfo{title}{Parameter {Identification} of {Wind} {Turbine} for
  {Maximum} {Power}-point {Tracking} {Control}}.
\newblock \bibinfo{journal}{Electric Power Components and Systems}
  \bibinfo{year}{2010};\bibinfo{volume}{38}(\bibinfo{number}{5}):\bibinfo{pages}{603--614}.
\newblock \URLprefix
  \url{http://www.tandfonline.com/doi/abs/10.1080/15325000903376974}.
  \DOIprefix\doi{10.1080/15325000903376974}.
%Type = Article
\bibitem[{Hunt et~al.(1992)Hunt, Sbarbaro, Żbikowski and
  Gawthrop}]{HUNT19921083}
\bibinfo{author}{Hunt\xfnm[ K.]}, \bibinfo{author}{Sbarbaro\xfnm[ D.]},
  \bibinfo{author}{Żbikowski\xfnm[ R.]}, \bibinfo{author}{Gawthrop\xfnm[ P.]}.
\newblock \bibinfo{title}{Neural networks for control systems—a survey}.
\newblock \bibinfo{journal}{Automatica}
  \bibinfo{year}{1992};\bibinfo{volume}{28}(\bibinfo{number}{6}):\bibinfo{pages}{1083--1112}.
\newblock \URLprefix
  \url{https://www.sciencedirect.com/science/article/pii/000510989290053I}.
  \DOIprefix\doi{https://doi.org/10.1016/0005-1098(92)90053-I}.
%Type = Article
\bibitem[{Imai et~al.(2020)Imai, Nishida, Kawanami, Umemura and
  Himeno}]{Imai2020}
\bibinfo{author}{Imai\xfnm[ R.]}, \bibinfo{author}{Nishida\xfnm[ K.]},
  \bibinfo{author}{Kawanami\xfnm[ O.]}, \bibinfo{author}{Umemura\xfnm[ Y.]},
  \bibinfo{author}{Himeno\xfnm[ T.]}.
\newblock \bibinfo{title}{Ground based experiment and numerical calculation on
  thermodynamic vent system in propellant tank for future cryogenic propulsion
  system}.
\newblock \bibinfo{journal}{Cryogenics}
  \bibinfo{year}{2020};\bibinfo{volume}{109}:\bibinfo{pages}{103095}.
\newblock \DOIprefix\doi{10.1016/j.cryogenics.2020.103095}.
%Type = Article
\bibitem[{Jaeger and Haas(2004)}]{Jaeger2004}
\bibinfo{author}{Jaeger\xfnm[ H.]}, \bibinfo{author}{Haas\xfnm[ H.]}.
\newblock \bibinfo{title}{Harnessing nonlinearity: Predicting chaotic systems
  and saving energy in wireless communication}.
\newblock \bibinfo{journal}{Science}
  \bibinfo{year}{2004};\bibinfo{volume}{304}(\bibinfo{number}{5667}):\bibinfo{pages}{78--80}.
\newblock \DOIprefix\doi{10.1126/science.1091277}.
%Type = Misc
\bibitem[{Janner et~al.(2021)Janner, Fu, Zhang and Levine}]{janner2021trust}
\bibinfo{author}{Janner\xfnm[ M.]}, \bibinfo{author}{Fu\xfnm[ J.]},
  \bibinfo{author}{Zhang\xfnm[ M.]}, \bibinfo{author}{Levine\xfnm[ S.]}.
\newblock \bibinfo{title}{When to trust your model: Model-based policy
  optimization}.
\newblock \bibinfo{year}{2021}.
\newblock \href{http://arxiv.org/abs/1906.08253}{\tt arXiv:1906.08253}.
%Type = Article
\bibitem[{Jiang et~al.(2021)Jiang, Sun, Li, Zuo and Huang}]{Jiang2021}
\bibinfo{author}{Jiang\xfnm[ W.]}, \bibinfo{author}{Sun\xfnm[ P.]},
  \bibinfo{author}{Li\xfnm[ P.]}, \bibinfo{author}{Zuo\xfnm[ Z.]},
  \bibinfo{author}{Huang\xfnm[ Y.]}.
\newblock \bibinfo{title}{Transient thermal behavior of multi-layer insulation
  coupled with vapor cooled shield used for liquid hydrogen storage tank}.
\newblock \bibinfo{journal}{Energy}
  \bibinfo{year}{2021};\bibinfo{volume}{231}:\bibinfo{pages}{120859}.
\newblock \DOIprefix\doi{10.1016/j.energy.2021.120859}.
%Type = Book
\bibitem[{Johnson(2004)}]{johnson2004}
\bibinfo{author}{Johnson\xfnm[ K.E.]}.
\newblock \bibinfo{title}{Adaptive torque control of variable speed wind
  turbines}.
\newblock \bibinfo{publisher}{University of Colorado at Boulder},
  \bibinfo{year}{2004}.
%Type = Article
\bibitem[{Johnson(2006)}]{johnson2006}
\bibinfo{author}{Johnson\xfnm[ K.E.]}.
\newblock \bibinfo{title}{Control of variable-speed wind turbines: standard and
  adaptive techniques for maximizing energy capture}.
\newblock \bibinfo{journal}{IEEE Control Systems Magazine}
  \bibinfo{year}{2006};\bibinfo{volume}{26}(\bibinfo{number}{3}):\bibinfo{pages}{70--81}.
%Type = Techreport
\bibitem[{Jonkman(2006)}]{jonkman2006turbsim}
\bibinfo{author}{Jonkman\xfnm[ B.J.]}.
\newblock \bibinfo{title}{TurbSim user's guide}.
\newblock \bibinfo{type}{Technical Report}; National Renewable Energy
  Lab.(NREL), Golden, CO (United States); \bibinfo{year}{2006}.
%Type = Techreport
\bibitem[{Jonkman et~al.(2009)Jonkman, Butterfield, Musial and
  Scott}]{jonkman2009definition}
\bibinfo{author}{Jonkman\xfnm[ J.]}, \bibinfo{author}{Butterfield\xfnm[ S.]},
  \bibinfo{author}{Musial\xfnm[ W.]}, \bibinfo{author}{Scott\xfnm[ G.]}.
\newblock \bibinfo{title}{Definition of a 5-MW reference wind turbine for
  offshore system development}.
\newblock \bibinfo{type}{Technical Report}; National Renewable Energy
  Lab.(NREL), Golden, CO (United States); \bibinfo{year}{2009}.
%Type = Article
\bibitem[{Jordan and Jacobs(1994)}]{jordanmixture1994}
\bibinfo{author}{Jordan\xfnm[ M.]}, \bibinfo{author}{Jacobs\xfnm[ R.]}.
\newblock \bibinfo{title}{Hierarchical mixtures of experts and the}.
\newblock \bibinfo{journal}{Neural computation}
  \bibinfo{year}{1994};\bibinfo{volume}{6}:\bibinfo{pages}{181--}.
%Type = Inproceedings
\bibitem[{Kahn et~al.(2017)Kahn, Zhang, Levine and Abbeel}]{kahn2017plato}
\bibinfo{author}{Kahn\xfnm[ G.]}, \bibinfo{author}{Zhang\xfnm[ T.]},
  \bibinfo{author}{Levine\xfnm[ S.]}, \bibinfo{author}{Abbeel\xfnm[ P.]}.
\newblock \bibinfo{title}{Plato: Policy learning using adaptive trajectory
  optimization}.
\newblock In: \bibinfo{booktitle}{2017 IEEE International Conference on
  Robotics and Automation (ICRA)}. \bibinfo{organization}{IEEE};
  \bibinfo{year}{2017}. p. \bibinfo{pages}{3342--3349}.
%Type = Article
\bibitem[{Kalnay et~al.(2007)Kalnay, Li, Miyoshi, Yang and
  Ballabrera-Poy}]{Kalnay2007}
\bibinfo{author}{Kalnay\xfnm[ E.]}, \bibinfo{author}{Li\xfnm[ H.]},
  \bibinfo{author}{Miyoshi\xfnm[ T.]}, \bibinfo{author}{Yang\xfnm[ S.C.]},
  \bibinfo{author}{Ballabrera-Poy\xfnm[ J.]}.
\newblock \bibinfo{title}{4-d-var or ensemble kalman filter?}
\newblock \bibinfo{journal}{Tellus A: Dynamic Meteorology and Oceanography}
  \bibinfo{year}{2007};\bibinfo{volume}{59}(\bibinfo{number}{5}):\bibinfo{pages}{758}.
\newblock \DOIprefix\doi{10.1111/j.1600-0870.2007.00261.x}.
%Type = Misc
\bibitem[{Khandelwal et~al.(2021)Khandelwal, Nadler, Arcucci, Knottenbelt and
  Guo}]{Khandelwal2021}
\bibinfo{author}{Khandelwal\xfnm[ P.]}, \bibinfo{author}{Nadler\xfnm[ P.]},
  \bibinfo{author}{Arcucci\xfnm[ R.]}, \bibinfo{author}{Knottenbelt\xfnm[ W.]},
  \bibinfo{author}{Guo\xfnm[ Y.K.]}.
\newblock \bibinfo{title}{A scalable inference method for large dynamic
  economic systems}.
\newblock \bibinfo{year}{2021}.
\newblock \href{http://arxiv.org/abs/2110.14346}{\tt arXiv:2110.14346}.
%Type = Article
\bibitem[{Kruyt et~al.(2014)Kruyt, Quicaz{\'a}n-Rubio, Van~Heijst, Altshuler
  and Lentink}]{Kruyt2014}
\bibinfo{author}{Kruyt\xfnm[ J.W.]}, \bibinfo{author}{Quicaz{\'a}n-Rubio\xfnm[
  E.M.]}, \bibinfo{author}{Van~Heijst\xfnm[ G.F.]},
  \bibinfo{author}{Altshuler\xfnm[ D.L.]}, \bibinfo{author}{Lentink\xfnm[ D.]}.
\newblock \bibinfo{title}{Hummingbird wing efficacy depends on aspect ratio and
  compares with helicopter rotors}.
\newblock \bibinfo{journal}{Journal of the royal society interface}
  \bibinfo{year}{2014};\bibinfo{volume}{11}(\bibinfo{number}{99}):\bibinfo{pages}{20140585}.
%Type = Article
\bibitem[{Kurutach et~al.(2018)Kurutach, Clavera, Duan, Tamar and
  Abbeel}]{kurutach2018model}
\bibinfo{author}{Kurutach\xfnm[ T.]}, \bibinfo{author}{Clavera\xfnm[ I.]},
  \bibinfo{author}{Duan\xfnm[ Y.]}, \bibinfo{author}{Tamar\xfnm[ A.]},
  \bibinfo{author}{Abbeel\xfnm[ P.]}.
\newblock \bibinfo{title}{Model-ensemble trust-region policy optimization}.
\newblock \bibinfo{journal}{arXiv preprint arXiv:180210592}
  \bibinfo{year}{2018};.
%Type = Book
\bibitem[{Lahoz et~al.(2010)Lahoz, Khattatov and Menard}]{Lahoz2010}
\bibinfo{editor}{Lahoz\xfnm[ W.]}, \bibinfo{editor}{Khattatov\xfnm[ B.]},
  \bibinfo{editor}{Menard\xfnm[ R.]}, editors.
\newblock \bibinfo{title}{Data Assimilation}.
\newblock \bibinfo{publisher}{Springer Berlin Heidelberg},
  \bibinfo{year}{2010}.
\newblock \DOIprefix\doi{10.1007/978-3-540-74703-1}.
%Type = Inproceedings
\bibitem[{Laks et~al.(2009)Laks, Pao and Wright}]{Laks2009}
\bibinfo{author}{Laks\xfnm[ J.H.]}, \bibinfo{author}{Pao\xfnm[ L.Y.]},
  \bibinfo{author}{Wright\xfnm[ A.D.]}.
\newblock \bibinfo{title}{Control of wind turbines: Past, present, and future}.
\newblock In: \bibinfo{booktitle}{2009 American Control Conference}.
  \bibinfo{publisher}{{IEEE}}; \bibinfo{year}{2009}.
  \DOIprefix\doi{10.1109/acc.2009.5160590}.
%Type = Article
\bibitem[{LeCun et~al.(1989)LeCun, Boser, Denker, Henderson, Howard, Hubbard
  and Jackel}]{LeCun1989}
\bibinfo{author}{LeCun\xfnm[ Y.]}, \bibinfo{author}{Boser\xfnm[ B.]},
  \bibinfo{author}{Denker\xfnm[ J.S.]}, \bibinfo{author}{Henderson\xfnm[ D.]},
  \bibinfo{author}{Howard\xfnm[ R.E.]}, \bibinfo{author}{Hubbard\xfnm[ W.]},
  \bibinfo{author}{Jackel\xfnm[ L.D.]}.
\newblock \bibinfo{title}{Backpropagation applied to handwritten zip code
  recognition}.
\newblock \bibinfo{journal}{Neural Computation}
  \bibinfo{year}{1989};\bibinfo{volume}{1}(\bibinfo{number}{4}):\bibinfo{pages}{541--551}.
\newblock \DOIprefix\doi{10.1162/neco.1989.1.4.541}.
%Type = Article
\bibitem[{Lee et~al.(2016)Lee, Lua, Lim and Yeo}]{Lee2016}
\bibinfo{author}{Lee\xfnm[ Y.]}, \bibinfo{author}{Lua\xfnm[ K.B.]},
  \bibinfo{author}{Lim\xfnm[ T.]}, \bibinfo{author}{Yeo\xfnm[ K.]}.
\newblock \bibinfo{title}{A quasi-steady aerodynamic model for flapping flight
  with improved adaptability}.
\newblock \bibinfo{journal}{Bioinspiration \& biomimetics}
  \bibinfo{year}{2016};\bibinfo{volume}{11}(\bibinfo{number}{3}):\bibinfo{pages}{036005}.
%Type = Article
\bibitem[{Leishman(2002)}]{leishman2002challenges}
\bibinfo{author}{Leishman\xfnm[ J.G.]}.
\newblock \bibinfo{title}{Challenges in modelling the unsteady aerodynamics of
  wind turbines}.
\newblock \bibinfo{journal}{Wind Energy: An International Journal for Progress
  and Applications in Wind Power Conversion Technology}
  \bibinfo{year}{2002};\bibinfo{volume}{5}(\bibinfo{number}{2-3}):\bibinfo{pages}{85--132}.
%Type = Misc
\bibitem[{Lemmon et~al.(2018)Lemmon, Bell, Huber and McLinden}]{REFPROP10}
\bibinfo{author}{Lemmon\xfnm[ E.W.]}, \bibinfo{author}{Bell\xfnm[ I.H.]},
  \bibinfo{author}{Huber\xfnm[ M.L.]}, \bibinfo{author}{McLinden\xfnm[ M.O.]}.
\newblock \bibinfo{title}{{NIST Standard Reference Database 23: Reference Fluid
  Thermodynamic and Transport Properties-REFPROP, Version 10.0, National
  Institute of Standards and Technology}}.
\newblock \bibinfo{year}{2018}.
\newblock \URLprefix \url{https://www.nist.gov/srd/refprop}.
  \DOIprefix\doi{https://doi.org/10.18434/T4/1502528}.
%Type = Article
\bibitem[{Levine and Abbeel(2014)}]{levine2014learning}
\bibinfo{author}{Levine\xfnm[ S.]}, \bibinfo{author}{Abbeel\xfnm[ P.]}.
\newblock \bibinfo{title}{Learning neural network policies with guided policy
  search under unknown dynamics}.
\newblock \bibinfo{journal}{Advances in neural information processing systems}
  \bibinfo{year}{2014};\bibinfo{volume}{27}.
%Type = Inproceedings
\bibitem[{Levine and Koltun(2013)}]{levine2013guided}
\bibinfo{author}{Levine\xfnm[ S.]}, \bibinfo{author}{Koltun\xfnm[ V.]}.
\newblock \bibinfo{title}{Guided policy search}.
\newblock In: \bibinfo{booktitle}{International conference on machine
  learning}. \bibinfo{organization}{PMLR}; \bibinfo{year}{2013}. p.
  \bibinfo{pages}{1--9}.
%Type = Misc
\bibitem[{Lillicrap et~al.(2019)Lillicrap, Hunt, Pritzel, Heess, Erez, Tassa,
  Silver and Wierstra}]{lillicrap2019continuous}
\bibinfo{author}{Lillicrap\xfnm[ T.P.]}, \bibinfo{author}{Hunt\xfnm[ J.J.]},
  \bibinfo{author}{Pritzel\xfnm[ A.]}, \bibinfo{author}{Heess\xfnm[ N.]},
  \bibinfo{author}{Erez\xfnm[ T.]}, \bibinfo{author}{Tassa\xfnm[ Y.]},
  \bibinfo{author}{Silver\xfnm[ D.]}, \bibinfo{author}{Wierstra\xfnm[ D.]}.
\newblock \bibinfo{title}{Continuous control with deep reinforcement learning}.
\newblock \bibinfo{year}{2019}.
\newblock \href{http://arxiv.org/abs/1509.02971}{\tt arXiv:1509.02971}.
%Type = Article
\bibitem[{Lin et~al.(1991)Lin, Dresar and Hasan}]{Lin1991_TVS}
\bibinfo{author}{Lin\xfnm[ C.]}, \bibinfo{author}{Dresar\xfnm[ N.]},
  \bibinfo{author}{Hasan\xfnm[ M.]}.
\newblock \bibinfo{title}{Pressure control analysis of cryogenic storage
  systems}.
\newblock \bibinfo{journal}{Journal of Propulsion and Power}
  \bibinfo{year}{1991};\bibinfo{volume}{20}.
\newblock \DOIprefix\doi{10.2514/1.10387}.
%Type = Article
\bibitem[{Lin et~al.(1996)Lin, Horne, Tino and Giles}]{narx}
\bibinfo{author}{Lin\xfnm[ T.]}, \bibinfo{author}{Horne\xfnm[ B.]},
  \bibinfo{author}{Tino\xfnm[ P.]}, \bibinfo{author}{Giles\xfnm[ C.]}.
\newblock \bibinfo{title}{Learning long-term dependencies in narx recurrent
  neural networks}.
\newblock \bibinfo{journal}{IEEE Transactions on Neural Networks}
  \bibinfo{year}{1996};\bibinfo{volume}{7}(\bibinfo{number}{6}):\bibinfo{pages}{1329--1338}.
\newblock \DOIprefix\doi{10.1109/72.548162}.
%Type = Misc
\bibitem[{Liu and MacArt(2023)}]{liu2023adjointbased}
\bibinfo{author}{Liu\xfnm[ X.]}, \bibinfo{author}{MacArt\xfnm[ J.F.]}.
\newblock \bibinfo{title}{Adjoint-based machine learning for active flow
  control}.
\newblock \bibinfo{year}{2023}.
\newblock \href{http://arxiv.org/abs/2307.09980}{\tt arXiv:2307.09980}.
%Type = Misc
\bibitem[{Liu and Wang(2021)}]{Liu2021}
\bibinfo{author}{Liu\xfnm[ X.Y.]}, \bibinfo{author}{Wang\xfnm[ J.X.]}.
\newblock \bibinfo{title}{Physics-informed dyna-style model-based deep
  reinforcement learning for dynamic control}.
\newblock \bibinfo{year}{2021}.
\newblock \DOIprefix\doi{10.1098/rspa.2021.0618}.
  \href{http://arxiv.org/abs/2108.00128}{\tt arXiv:2108.00128}.
%Type = Article
\bibitem[{Ljung(2008)}]{Ljung2008}
\bibinfo{author}{Ljung\xfnm[ L.]}.
\newblock \bibinfo{title}{Perspectives on system identification}.
\newblock \bibinfo{journal}{{IFAC} Proceedings Volumes}
  \bibinfo{year}{2008};\bibinfo{volume}{41}(\bibinfo{number}{2}):\bibinfo{pages}{7172--7184}.
\newblock \DOIprefix\doi{10.3182/20080706-5-kr-1001.01215}.
%Type = Article
\bibitem[{Ljung et~al.(2020)Ljung, Andersson, Tiels and
  Sch{\"o}n}]{Ljung2020DeepLA}
\bibinfo{author}{Ljung\xfnm[ L.]}, \bibinfo{author}{Andersson\xfnm[ C.R.]},
  \bibinfo{author}{Tiels\xfnm[ K.]}, \bibinfo{author}{Sch{\"o}n\xfnm[ T.B.]}.
\newblock \bibinfo{title}{Deep learning and system identification}.
\newblock \bibinfo{journal}{IFAC-PapersOnLine} \bibinfo{year}{2020};\URLprefix
  \url{https://api.semanticscholar.org/CorpusID:226118683}.
%Type = Article
\bibitem[{Lorenc(1986)}]{Lorenc1986}
\bibinfo{author}{Lorenc\xfnm[ A.C.]}.
\newblock \bibinfo{title}{Analysis methods for numerical weather prediction}.
\newblock \bibinfo{journal}{Quarterly Journal of the Royal Meteorological
  Society}
  \bibinfo{year}{1986};\bibinfo{volume}{112}(\bibinfo{number}{474}):\bibinfo{pages}{1177--1194}.
\newblock \DOIprefix\doi{10.1002/qj.49711247414}.
%Type = Article
\bibitem[{Lorenc et~al.(2015)Lorenc, Bowler, Clayton, Pring and
  Fairbairn}]{Lorenc2015}
\bibinfo{author}{Lorenc\xfnm[ A.C.]}, \bibinfo{author}{Bowler\xfnm[ N.E.]},
  \bibinfo{author}{Clayton\xfnm[ A.M.]}, \bibinfo{author}{Pring\xfnm[ S.R.]},
  \bibinfo{author}{Fairbairn\xfnm[ D.]}.
\newblock \bibinfo{title}{Comparison of hybrid-4denvar and hybrid-4dvar data
  assimilation methods for global {NWP}}.
\newblock \bibinfo{journal}{Monthly Weather Review}
  \bibinfo{year}{2015};\bibinfo{volume}{143}(\bibinfo{number}{1}):\bibinfo{pages}{212--229}.
\newblock \DOIprefix\doi{10.1175/mwr-d-14-00195.1}.
%Type = Misc
\bibitem[{Luo et~al.(2022{\natexlab{a}})Luo, Xu, Lai, Chen, Zhang and
  Yu}]{luo_survey_2022}
\bibinfo{author}{Luo\xfnm[ F.M.]}, \bibinfo{author}{Xu\xfnm[ T.]},
  \bibinfo{author}{Lai\xfnm[ H.]}, \bibinfo{author}{Chen\xfnm[ X.H.]},
  \bibinfo{author}{Zhang\xfnm[ W.]}, \bibinfo{author}{Yu\xfnm[ Y.]}.
\newblock \bibinfo{title}{A survey on model-based reinforcement learning}.
\newblock \bibinfo{year}{2022}{\natexlab{a}}.
\newblock \URLprefix \url{http://arxiv.org/abs/2206.09328}.
  \DOIprefix\doi{10.48550/arXiv.2206.09328}.
  \href{http://arxiv.org/abs/2206.09328 [cs]}{\tt arXiv:2206.09328 [cs]}.
%Type = Misc
\bibitem[{Luo et~al.(2022{\natexlab{b}})Luo, Xu, Lai, Chen, Zhang and
  Yu}]{luo2022survey}
\bibinfo{author}{Luo\xfnm[ F.M.]}, \bibinfo{author}{Xu\xfnm[ T.]},
  \bibinfo{author}{Lai\xfnm[ H.]}, \bibinfo{author}{Chen\xfnm[ X.H.]},
  \bibinfo{author}{Zhang\xfnm[ W.]}, \bibinfo{author}{Yu\xfnm[ Y.]}.
\newblock \bibinfo{title}{A survey on model-based reinforcement learning}.
\newblock \bibinfo{year}{2022}{\natexlab{b}}.
\newblock \href{http://arxiv.org/abs/2206.09328}{\tt arXiv:2206.09328}.
%Type = Misc
\bibitem[{Lutter et~al.(2021)Lutter, Belousov, Mannor, Fox, Garg and
  Peters}]{lutter2021continuoustime}
\bibinfo{author}{Lutter\xfnm[ M.]}, \bibinfo{author}{Belousov\xfnm[ B.]},
  \bibinfo{author}{Mannor\xfnm[ S.]}, \bibinfo{author}{Fox\xfnm[ D.]},
  \bibinfo{author}{Garg\xfnm[ A.]}, \bibinfo{author}{Peters\xfnm[ J.]}.
\newblock \bibinfo{title}{Continuous-time fitted value iteration for robust
  policies}.
\newblock \bibinfo{year}{2021}.
\newblock \href{http://arxiv.org/abs/2110.01954}{\tt arXiv:2110.01954}.
%Type = Misc
\bibitem[{Lutter et~al.(2019)Lutter, Ritter and Peters}]{lutter2019deep}
\bibinfo{author}{Lutter\xfnm[ M.]}, \bibinfo{author}{Ritter\xfnm[ C.]},
  \bibinfo{author}{Peters\xfnm[ J.]}.
\newblock \bibinfo{title}{Deep lagrangian networks: Using physics as model
  prior for deep learning}.
\newblock \bibinfo{year}{2019}.
\newblock \href{http://arxiv.org/abs/1907.04490}{\tt arXiv:1907.04490}.
%Type = Misc
\bibitem[{Lutter et~al.(2020)Lutter, Silberbauer, Watson and
  Peters}]{lutter2020differentiable}
\bibinfo{author}{Lutter\xfnm[ M.]}, \bibinfo{author}{Silberbauer\xfnm[ J.]},
  \bibinfo{author}{Watson\xfnm[ J.]}, \bibinfo{author}{Peters\xfnm[ J.]}.
\newblock \bibinfo{title}{Differentiable physics models for real-world offline
  model-based reinforcement learning}.
\newblock \bibinfo{year}{2020}.
\newblock \href{http://arxiv.org/abs/2011.01734}{\tt arXiv:2011.01734}.
%Type = Inproceedings
\bibitem[{Madhavan(1993)}]{Madhavana}
\bibinfo{author}{Madhavan\xfnm[ P.]}.
\newblock \bibinfo{title}{Recurrent neural network for time series prediction}.
\newblock In: \bibinfo{booktitle}{Proceedings of the 15th Annual International
  Conference of the {IEEE} Engineering in Medicine and Biology Society}.
  \bibinfo{publisher}{{IEEE}}; \bibinfo{year}{1993}.
  \DOIprefix\doi{10.1109/iembs.1993.978527}.
%Type = Inproceedings
\bibitem[{Marques et~al.(2023)Marques, Ahizi and Mendez}]{Marques2023}
\bibinfo{author}{Marques\xfnm[ P.]}, \bibinfo{author}{Ahizi\xfnm[ S.]},
  \bibinfo{author}{Mendez\xfnm[ M.A.]}.
\newblock \bibinfo{title}{Real {Time} {Data} {Assimilation} for the
  {Thermodynamic} {Modeling} of a {Cryogenic} {Fuel} {Tank}}.
\newblock In: \bibinfo{booktitle}{36th {International} {Conference} on
  {Efficiency}, {Cost}, {Optimization}, {Simulation} and {Environmental}
  {Impact} of {Energy} {Systems} ({ECOS} 2023)}. \bibinfo{address}{Las Palmas
  De Gran Canaria, Spain}: \bibinfo{publisher}{ECOS 2023};
  \bibinfo{year}{2023}. p. \bibinfo{pages}{1041--1052}.
\newblock \URLprefix \url{http://www.proceedings.com/069564-0095.html}.
  \DOIprefix\doi{10.52202/069564-0095}.
%Type = Article
\bibitem[{Mer et~al.(2016{\natexlab{a}})Mer, Fernandez, Thibault and
  Corre}]{Mer2016a}
\bibinfo{author}{Mer\xfnm[ S.]}, \bibinfo{author}{Fernandez\xfnm[ D.]},
  \bibinfo{author}{Thibault\xfnm[ J.P.]}, \bibinfo{author}{Corre\xfnm[ C.]}.
\newblock \bibinfo{title}{Optimal design of a thermodynamic vent system for
  cryogenic propellant storage}.
\newblock \bibinfo{journal}{Cryogenics}
  \bibinfo{year}{2016}{\natexlab{a}};\bibinfo{volume}{80}:\bibinfo{pages}{127--137}.
\newblock \DOIprefix\doi{10.1016/j.cryogenics.2016.09.012}.
%Type = Article
\bibitem[{Mer et~al.(2016{\natexlab{b}})Mer, Thibault and Corre}]{Mer2016}
\bibinfo{author}{Mer\xfnm[ S.]}, \bibinfo{author}{Thibault\xfnm[ J.P.]},
  \bibinfo{author}{Corre\xfnm[ C.]}.
\newblock \bibinfo{title}{Active insulation technique applied to the
  experimental analysis of a thermodynamic control system for cryogenic
  propellant storage}.
\newblock \bibinfo{journal}{Journal of Thermal Science and Engineering
  Applications}
  \bibinfo{year}{2016}{\natexlab{b}};\bibinfo{volume}{8}(\bibinfo{number}{2}).
\newblock \DOIprefix\doi{10.1115/1.4032761}.
%Type = Misc
\bibitem[{Mnih et~al.(2013{\natexlab{a}})Mnih, Kavukcuoglu, Silver, Graves,
  Antonoglou, Wierstra and Riedmiller}]{mnih2013playing}
\bibinfo{author}{Mnih\xfnm[ V.]}, \bibinfo{author}{Kavukcuoglu\xfnm[ K.]},
  \bibinfo{author}{Silver\xfnm[ D.]}, \bibinfo{author}{Graves\xfnm[ A.]},
  \bibinfo{author}{Antonoglou\xfnm[ I.]}, \bibinfo{author}{Wierstra\xfnm[ D.]},
  \bibinfo{author}{Riedmiller\xfnm[ M.]}.
\newblock \bibinfo{title}{Playing atari with deep reinforcement learning}.
\newblock \bibinfo{year}{2013}{\natexlab{a}}.
\newblock \href{http://arxiv.org/abs/1312.5602}{\tt arXiv:1312.5602}.
%Type = Misc
\bibitem[{Mnih et~al.(2013{\natexlab{b}})Mnih, Kavukcuoglu, Silver, Graves,
  Antonoglou, Wierstra and Riedmiller}]{Mnih2013}
\bibinfo{author}{Mnih\xfnm[ V.]}, \bibinfo{author}{Kavukcuoglu\xfnm[ K.]},
  \bibinfo{author}{Silver\xfnm[ D.]}, \bibinfo{author}{Graves\xfnm[ A.]},
  \bibinfo{author}{Antonoglou\xfnm[ I.]}, \bibinfo{author}{Wierstra\xfnm[ D.]},
  \bibinfo{author}{Riedmiller\xfnm[ M.]}.
\newblock \bibinfo{title}{Playing atari with deep reinforcement learning}.
\newblock \bibinfo{year}{2013}{\natexlab{b}}.
\newblock \href{http://arxiv.org/abs/http://arxiv.org/abs/1312.5602v1}{\tt
  arXiv:http://arxiv.org/abs/1312.5602v1}.
%Type = Article
\bibitem[{Mnih et~al.(2015)Mnih, Kavukcuoglu, Silver, Rusu, Veness, Bellemare,
  Graves, Riedmiller, Fidjeland, Ostrovski, Petersen, Beattie, Sadik,
  Antonoglou, King, Kumaran, Wierstra, Legg and Hassabis}]{Mnih2015}
\bibinfo{author}{Mnih\xfnm[ V.]}, \bibinfo{author}{Kavukcuoglu\xfnm[ K.]},
  \bibinfo{author}{Silver\xfnm[ D.]}, \bibinfo{author}{Rusu\xfnm[ A.A.]},
  \bibinfo{author}{Veness\xfnm[ J.]}, \bibinfo{author}{Bellemare\xfnm[ M.G.]},
  \bibinfo{author}{Graves\xfnm[ A.]}, \bibinfo{author}{Riedmiller\xfnm[ M.]},
  \bibinfo{author}{Fidjeland\xfnm[ A.K.]}, \bibinfo{author}{Ostrovski\xfnm[
  G.]}, \bibinfo{author}{Petersen\xfnm[ S.]}, \bibinfo{author}{Beattie\xfnm[
  C.]}, \bibinfo{author}{Sadik\xfnm[ A.]}, \bibinfo{author}{Antonoglou\xfnm[
  I.]}, \bibinfo{author}{King\xfnm[ H.]}, \bibinfo{author}{Kumaran\xfnm[ D.]},
  \bibinfo{author}{Wierstra\xfnm[ D.]}, \bibinfo{author}{Legg\xfnm[ S.]},
  \bibinfo{author}{Hassabis\xfnm[ D.]}.
\newblock \bibinfo{title}{Human-level control through deep reinforcement
  learning}.
\newblock \bibinfo{journal}{Nature}
  \bibinfo{year}{2015};\bibinfo{volume}{518}(\bibinfo{number}{7540}):\bibinfo{pages}{529--533}.
\newblock \DOIprefix\doi{10.1038/nature14236}.
%Type = Misc
\bibitem[{Moerland et~al.(2022{\natexlab{a}})Moerland, Broekens, Plaat and
  Jonker}]{moerland_model-based_2022}
\bibinfo{author}{Moerland\xfnm[ T.M.]}, \bibinfo{author}{Broekens\xfnm[ J.]},
  \bibinfo{author}{Plaat\xfnm[ A.]}, \bibinfo{author}{Jonker\xfnm[ C.M.]}.
\newblock \bibinfo{title}{Model-based reinforcement learning: A survey}.
\newblock \bibinfo{year}{2022}{\natexlab{a}}.
\newblock \URLprefix \url{http://arxiv.org/abs/2006.16712}.
  \DOIprefix\doi{10.48550/arXiv.2006.16712}.
  \href{http://arxiv.org/abs/2006.16712 [cs, stat]}{\tt arXiv:2006.16712 [cs,
  stat]}.
%Type = Misc
\bibitem[{Moerland et~al.(2022{\natexlab{b}})Moerland, Broekens, Plaat and
  Jonker}]{moerland2022modelbased}
\bibinfo{author}{Moerland\xfnm[ T.M.]}, \bibinfo{author}{Broekens\xfnm[ J.]},
  \bibinfo{author}{Plaat\xfnm[ A.]}, \bibinfo{author}{Jonker\xfnm[ C.M.]}.
\newblock \bibinfo{title}{Model-based reinforcement learning: A survey}.
\newblock \bibinfo{year}{2022}{\natexlab{b}}.
\newblock \href{http://arxiv.org/abs/2006.16712}{\tt arXiv:2006.16712}.
%Type = Techreport
\bibitem[{Moriarty and Hansen(2005)}]{moriarty2005aerodyn}
\bibinfo{author}{Moriarty\xfnm[ P.J.]}, \bibinfo{author}{Hansen\xfnm[ A.C.]}.
\newblock \bibinfo{title}{AeroDyn theory manual}.
\newblock \bibinfo{type}{Technical Report}; National Renewable Energy Lab.,
  Golden, CO (US); \bibinfo{year}{2005}.
%Type = Inproceedings
\bibitem[{Motil et~al.(2007)Motil, Meyer and Tucker}]{Motil2007}
\bibinfo{author}{Motil\xfnm[ S.]}, \bibinfo{author}{Meyer\xfnm[ M.]},
  \bibinfo{author}{Tucker\xfnm[ S.]}.
\newblock \bibinfo{title}{Cryogenic fluid management technologies for advanced
  green propulsion systems}.
\newblock In: \bibinfo{booktitle}{45th {AIAA} Aerospace Sciences Meeting and
  Exhibit}. \bibinfo{publisher}{American Institute of Aeronautics and
  Astronautics}; \bibinfo{year}{2007}. \DOIprefix\doi{10.2514/6.2007-343}.
%Type = Inproceedings
\bibitem[{Nadler et~al.(2020)Nadler, Arcucci and Guo}]{Nadler2020}
\bibinfo{author}{Nadler\xfnm[ P.]}, \bibinfo{author}{Arcucci\xfnm[ R.]},
  \bibinfo{author}{Guo\xfnm[ Y.]}.
\newblock \bibinfo{title}{A neural sir model for global forecasting}.
\newblock In: \bibinfo{editor}{Alsentzer\xfnm[ E.]},
  \bibinfo{editor}{McDermott\xfnm[ M.B.A.]}, \bibinfo{editor}{Falck\xfnm[ F.]},
  \bibinfo{editor}{Sarkar\xfnm[ S.K.]}, \bibinfo{editor}{Roy\xfnm[ S.]},
  \bibinfo{editor}{Hyland\xfnm[ S.L.]}, editors.
  \bibinfo{booktitle}{Proceedings of the Machine Learning for Health NeurIPS
  Workshop}. \bibinfo{publisher}{PMLR}; volume \bibinfo{volume}{136} of
  \textit{\bibinfo{series}{Proceedings of Machine Learning Research}};
  \bibinfo{year}{2020}. p. \bibinfo{pages}{254--266}.
\newblock \URLprefix \url{https://proceedings.mlr.press/v136/nadler20a.html}.
%Type = Misc
\bibitem[{Nagabandi et~al.(2017)Nagabandi, Kahn, Fearing and
  Levine}]{nagabandi2017neural}
\bibinfo{author}{Nagabandi\xfnm[ A.]}, \bibinfo{author}{Kahn\xfnm[ G.]},
  \bibinfo{author}{Fearing\xfnm[ R.S.]}, \bibinfo{author}{Levine\xfnm[ S.]}.
\newblock \bibinfo{title}{Neural network dynamics for model-based deep
  reinforcement learning with model-free fine-tuning}.
\newblock \bibinfo{year}{2017}.
\newblock \href{http://arxiv.org/abs/1708.02596}{\tt arXiv:1708.02596}.
%Type = Book
\bibitem[{Nelles(2001)}]{Nelles2001}
\bibinfo{author}{Nelles\xfnm[ O.]}.
\newblock \bibinfo{title}{Nonlinear System Identification}.
\newblock \bibinfo{publisher}{Springer Berlin Heidelberg},
  \bibinfo{year}{2001}.
\newblock \DOIprefix\doi{10.1007/978-3-662-04323-3}.
%Type = Inproceedings
\bibitem[{Nicolao(2003)}]{Nicolao2003SystemI}
\bibinfo{author}{Nicolao\xfnm[ G.D.]}.
\newblock \bibinfo{title}{System identification : Problems and perspectives}.
\newblock In: \bibinfo{booktitle}{12th Workshop on Qualitative Reasoning}.
  \bibinfo{year}{2003}. \URLprefix
  \url{https://api.semanticscholar.org/CorpusID:26963940}.
%Type = Book
\bibitem[{Norgaard et~al.(2000)Norgaard, Ravn, Poulsen and
  Hansen}]{Norgaard2000}
\bibinfo{author}{Norgaard\xfnm[ M.]}, \bibinfo{author}{Ravn\xfnm[ O.]},
  \bibinfo{author}{Poulsen\xfnm[ N.]}, \bibinfo{author}{Hansen\xfnm[ L.]}.
\newblock \bibinfo{title}{Neural Networks for Modelling and Control of Dynamic
  Systems}.
\newblock Advanced Textbooks in Control and Signal Processing.
  \bibinfo{publisher}{Springer London}, \bibinfo{year}{2000}.
%Type = Article
\bibitem[{Olatunji et~al.(2021)Olatunji, Adedeji, Madushele and
  Jen}]{Olatunji2021OverviewOD}
\bibinfo{author}{Olatunji\xfnm[ O.O.]}, \bibinfo{author}{Adedeji\xfnm[ P.A.]},
  \bibinfo{author}{Madushele\xfnm[ N.]}, \bibinfo{author}{Jen\xfnm[ T.C.]}.
\newblock \bibinfo{title}{Overview of digital twin technology in wind turbine
  fault diagnosis and condition monitoring}.
\newblock \bibinfo{journal}{2021 IEEE 12th International Conference on
  Mechanical and Intelligent Manufacturing Technologies (ICMIMT)}
  \bibinfo{year}{2021};:\bibinfo{pages}{201--207}\URLprefix
  \url{https://api.semanticscholar.org/CorpusID:236190314}.
%Type = Article
\bibitem[{Ortega-Jim{\'e}nez and Dudley(2018)}]{Ortega2018}
\bibinfo{author}{Ortega-Jim{\'e}nez\xfnm[ V.M.]}, \bibinfo{author}{Dudley\xfnm[
  R.]}.
\newblock \bibinfo{title}{Ascending flight and decelerating vertical glides in
  anna's hummingbirds}.
\newblock \bibinfo{journal}{Journal of Experimental Biology}
  \bibinfo{year}{2018};\bibinfo{volume}{221}(\bibinfo{number}{24}):\bibinfo{pages}{jeb191171}.
%Type = Misc
\bibitem[{Panzarella et~al.(2004)Panzarella, Plachta and
  Kassemi}]{panzarella_pressure_2004}
\bibinfo{author}{Panzarella\xfnm[ C.]}, \bibinfo{author}{Plachta\xfnm[ D.]},
  \bibinfo{author}{Kassemi\xfnm[ M.]}.
\newblock \bibinfo{title}{Pressure control of large cryogenic tanks in
  microgravity}.
\newblock \bibinfo{year}{2004}.
\newblock \URLprefix
  \url{https://linkinghub.elsevier.com/retrieve/pii/S0011227504000633}.
  \DOIprefix\doi{10.1016/j.cryogenics.2004.03.009}.
%Type = Misc
\bibitem[{Panzarella and Kassemi(2003)}]{panzarella_validity_2003}
\bibinfo{author}{Panzarella\xfnm[ C.H.]}, \bibinfo{author}{Kassemi\xfnm[ M.]}.
\newblock \bibinfo{title}{On the validity of purely thermodynamic descriptions
  of two-phase cryogenic fluid storage}.
\newblock \bibinfo{year}{2003}.
\newblock \URLprefix
  \url{http://www.journals.cambridge.org/abstract_S0022112003004002}.
  \DOIprefix\doi{10.1017/S0022112003004002}.
%Type = Inproceedings
\bibitem[{Pao and Johnson(2009)}]{Pao2009}
\bibinfo{author}{Pao\xfnm[ L.Y.]}, \bibinfo{author}{Johnson\xfnm[ K.E.]}.
\newblock \bibinfo{title}{A tutorial on the dynamics and control of wind
  turbines and wind farms}.
\newblock In: \bibinfo{booktitle}{2009 American Control Conference}.
  \bibinfo{publisher}{{IEEE}}; \bibinfo{year}{2009}.
  \DOIprefix\doi{10.1109/acc.2009.5160195}.
%Type = Misc
\bibitem[{Pillonetto et~al.(2023)Pillonetto, Aravkin, Gedon, Ljung, Ribeiro and
  Schön}]{Pillonetto2023}
\bibinfo{author}{Pillonetto\xfnm[ G.]}, \bibinfo{author}{Aravkin\xfnm[ A.]},
  \bibinfo{author}{Gedon\xfnm[ D.]}, \bibinfo{author}{Ljung\xfnm[ L.]},
  \bibinfo{author}{Ribeiro\xfnm[ A.H.]}, \bibinfo{author}{Schön\xfnm[ T.B.]}.
\newblock \bibinfo{title}{Deep networks for system identification: a survey}.
\newblock \bibinfo{year}{2023}.
\newblock \href{http://arxiv.org/abs/2301.12832}{\tt arXiv:2301.12832}.
%Type = Article
\bibitem[{Pimenta et~al.(2020)Pimenta, Pacheco, Branco, Teixeira and
  Magalhães}]{Pimenta_2020}
\bibinfo{author}{Pimenta\xfnm[ F.]}, \bibinfo{author}{Pacheco\xfnm[ J.]},
  \bibinfo{author}{Branco\xfnm[ C.M.]}, \bibinfo{author}{Teixeira\xfnm[ C.M.]},
  \bibinfo{author}{Magalhães\xfnm[ F.]}.
\newblock \bibinfo{title}{Development of a digital twin of an onshore wind
  turbine using monitoring data}.
\newblock \bibinfo{journal}{Journal of Physics: Conference Series}
  \bibinfo{year}{2020};\bibinfo{volume}{1618}(\bibinfo{number}{2}):\bibinfo{pages}{022065}.
\newblock \URLprefix \url{https://dx.doi.org/10.1088/1742-6596/1618/2/022065}.
  \DOIprefix\doi{10.1088/1742-6596/1618/2/022065}.
%Type = Article
\bibitem[{Pino et~al.(2023)Pino, Schena, Rabault and Mendez}]{Pino2023}
\bibinfo{author}{Pino\xfnm[ F.]}, \bibinfo{author}{Schena\xfnm[ L.]},
  \bibinfo{author}{Rabault\xfnm[ J.]}, \bibinfo{author}{Mendez\xfnm[ M.A.]}.
\newblock \bibinfo{title}{Comparative analysis of machine learning methods for
  active flow control}.
\newblock \bibinfo{journal}{Journal of Fluid Mechanics}
  \bibinfo{year}{2023};\bibinfo{volume}{958}.
\newblock \DOIprefix\doi{10.1017/jfm.2023.76}.
%Type = Article
\bibitem[{Pinosky et~al.(2023)Pinosky, Abraham, Broad, Argall and
  Murphey}]{pinosky2022}
\bibinfo{author}{Pinosky\xfnm[ A.]}, \bibinfo{author}{Abraham\xfnm[ I.]},
  \bibinfo{author}{Broad\xfnm[ A.]}, \bibinfo{author}{Argall\xfnm[ B.]},
  \bibinfo{author}{Murphey\xfnm[ T.D.]}.
\newblock \bibinfo{title}{Hybrid control for combining model-based and
  model-free reinforcement learning}.
\newblock \bibinfo{journal}{The International Journal of Robotics Research}
  \bibinfo{year}{2023};\bibinfo{volume}{42}(\bibinfo{number}{6}):\bibinfo{pages}{337--355}.
\newblock \URLprefix \url{https://doi.org/10.1177/02783649221083331}.
  \DOIprefix\doi{10.1177/02783649221083331}.
  \href{http://arxiv.org/abs/https://doi.org/10.1177/02783649221083331}{\tt
  arXiv:https://doi.org/10.1177/02783649221083331}.
%Type = Misc
\bibitem[{Pong et~al.(2020)Pong, Gu, Dalal and Levine}]{pong2020temporal}
\bibinfo{author}{Pong\xfnm[ V.]}, \bibinfo{author}{Gu\xfnm[ S.]},
  \bibinfo{author}{Dalal\xfnm[ M.]}, \bibinfo{author}{Levine\xfnm[ S.]}.
\newblock \bibinfo{title}{Temporal difference models: Model-free deep rl for
  model-based control}.
\newblock \bibinfo{year}{2020}.
\newblock \href{http://arxiv.org/abs/1802.09081}{\tt arXiv:1802.09081}.
%Type = Incollection
\bibitem[{Pu and Kalnay(2018)}]{Pu2018}
\bibinfo{author}{Pu\xfnm[ Z.]}, \bibinfo{author}{Kalnay\xfnm[ E.]}.
\newblock \bibinfo{title}{Numerical weather prediction basics: Models,
  numerical methods, and data assimilation}.
\newblock In: \bibinfo{booktitle}{Handbook of Hydrometeorological Ensemble
  Forecasting}. \bibinfo{publisher}{Springer Berlin Heidelberg};
  \bibinfo{year}{2018}. p. \bibinfo{pages}{1--31}.
\newblock \DOIprefix\doi{10.1007/978-3-642-40457-3_11-1}.
%Type = Book
\bibitem[{Puterman(1994)}]{Puterman1994}
\bibinfo{author}{Puterman\xfnm[ M.L.]}.
\newblock \bibinfo{title}{Markov Decision Processes}.
\newblock \bibinfo{publisher}{Wiley}, \bibinfo{year}{1994}.
\newblock \DOIprefix\doi{10.1002/9780470316887}.
%Type = Article
\bibitem[{Qin et~al.(2021)Qin, Li, Sun and Huang}]{Qin2021}
\bibinfo{author}{Qin\xfnm[ X.]}, \bibinfo{author}{Li\xfnm[ P.]},
  \bibinfo{author}{Sun\xfnm[ P.]}, \bibinfo{author}{Huang\xfnm[ Y.]}.
\newblock \bibinfo{title}{Testing and comparison of a thermodynamic vent system
  operating in different modes in a liquid nitrogen tank}.
\newblock \bibinfo{journal}{Applied Thermal Engineering}
  \bibinfo{year}{2021};\bibinfo{volume}{197}:\bibinfo{pages}{117393}.
\newblock \DOIprefix\doi{10.1016/j.applthermaleng.2021.117393}.
%Type = Misc
\bibitem[{Qu et~al.(2020)Qu, Yu, Low and Wierman}]{qu2020combining}
\bibinfo{author}{Qu\xfnm[ G.]}, \bibinfo{author}{Yu\xfnm[ C.]},
  \bibinfo{author}{Low\xfnm[ S.]}, \bibinfo{author}{Wierman\xfnm[ A.]}.
\newblock \bibinfo{title}{Combining model-based and model-free methods for
  nonlinear control: A provably convergent policy gradient approach}.
\newblock \bibinfo{year}{2020}.
\newblock \href{http://arxiv.org/abs/2006.07476}{\tt arXiv:2006.07476}.
%Type = Misc
\bibitem[{Rahman et~al.(2022)Rahman, Drgoňa, Tuor and Strube}]{Rahman2022}
\bibinfo{author}{Rahman\xfnm[ A.]}, \bibinfo{author}{Drgoňa\xfnm[ J.]},
  \bibinfo{author}{Tuor\xfnm[ A.]}, \bibinfo{author}{Strube\xfnm[ J.]}.
\newblock \bibinfo{title}{Neural ordinary differential equations for nonlinear
  system identification}.
\newblock \bibinfo{year}{2022}.
\newblock \href{http://arxiv.org/abs/2203.00120}{\tt arXiv:2203.00120}.
%Type = Article
\bibitem[{Raissi et~al.(2019)Raissi, Perdikaris and
  Karniadakis}]{raissi2019physics}
\bibinfo{author}{Raissi\xfnm[ M.]}, \bibinfo{author}{Perdikaris\xfnm[ P.]},
  \bibinfo{author}{Karniadakis\xfnm[ G.E.]}.
\newblock \bibinfo{title}{Physics-informed neural networks: A deep learning
  framework for solving forward and inverse problems involving nonlinear
  partial differential equations}.
\newblock \bibinfo{journal}{Journal of Computational physics}
  \bibinfo{year}{2019};\bibinfo{volume}{378}:\bibinfo{pages}{686--707}.
%Type = Misc
\bibitem[{Ramesh and Ravindran(2023)}]{ramesh2023physicsinformed}
\bibinfo{author}{Ramesh\xfnm[ A.]}, \bibinfo{author}{Ravindran\xfnm[ B.]}.
\newblock \bibinfo{title}{Physics-informed model-based reinforcement learning}.
\newblock \bibinfo{year}{2023}.
\newblock \href{http://arxiv.org/abs/2212.02179}{\tt arXiv:2212.02179}.
%Type = Misc
\bibitem[{Rasheed et~al.(2019)Rasheed, San and Kvamsdal}]{rasheed_digital_2019}
\bibinfo{author}{Rasheed\xfnm[ A.]}, \bibinfo{author}{San\xfnm[ O.]},
  \bibinfo{author}{Kvamsdal\xfnm[ T.]}.
\newblock \bibinfo{title}{Digital {Twin}: {Values}, {Challenges} and
  {Enablers}}.
\newblock \bibinfo{year}{2019}.
\newblock \URLprefix \url{http://arxiv.org/abs/1910.01719};
  \bibinfo{note}{arXiv:1910.01719 [eess]}.
%Type = Book
\bibitem[{Rasmussen and Williams(2005)}]{CarlEdward}
\bibinfo{author}{Rasmussen\xfnm[ C.E.]}, \bibinfo{author}{Williams\xfnm[
  C.K.I.]}.
\newblock \bibinfo{title}{Gaussian Processes for Machine Learning}.
\newblock \bibinfo{publisher}{MIT Press Ltd}, \bibinfo{year}{2005}.
%Type = Inproceedings
\bibitem[{Richards et~al.(2018)Richards, Berkenkamp and
  Krause}]{richards2018lyapunov}
\bibinfo{author}{Richards\xfnm[ S.M.]}, \bibinfo{author}{Berkenkamp\xfnm[ F.]},
  \bibinfo{author}{Krause\xfnm[ A.]}.
\newblock \bibinfo{title}{The lyapunov neural network: Adaptive stability
  certification for safe learning of dynamical systems}.
\newblock In: \bibinfo{booktitle}{Conference on Robot Learning}.
  \bibinfo{organization}{PMLR}; \bibinfo{year}{2018}. p.
  \bibinfo{pages}{466--476}.
%Type = Incollection
\bibitem[{Routray et~al.(2016)Routray, Osuri, Pattanayak and
  Mohanty}]{Routray2016}
\bibinfo{author}{Routray\xfnm[ A.]}, \bibinfo{author}{Osuri\xfnm[ K.K.]},
  \bibinfo{author}{Pattanayak\xfnm[ S.]}, \bibinfo{author}{Mohanty\xfnm[
  U.C.]}.
\newblock \bibinfo{title}{Introduction to data assimilation techniques and
  ensemble kalman filter}.
\newblock In: \bibinfo{booktitle}{Advanced Numerical Modeling and Data
  Assimilation Techniques for Tropical Cyclone Prediction}.
  \bibinfo{publisher}{Springer Netherlands}; \bibinfo{year}{2016}. p.
  \bibinfo{pages}{307--330}.
\newblock \DOIprefix\doi{10.5822/978-94-024-0896-6_11}.
%Type = Article
\bibitem[{Saint-Drenan et~al.(2020)Saint-Drenan, Besseau, Jansen, Staffell,
  Troccoli, Dubus, Schmidt, Gruber, Simões and
  Heier}]{saint-drenan_parametric_2020}
\bibinfo{author}{Saint-Drenan\xfnm[ Y.M.]}, \bibinfo{author}{Besseau\xfnm[
  R.]}, \bibinfo{author}{Jansen\xfnm[ M.]}, \bibinfo{author}{Staffell\xfnm[
  I.]}, \bibinfo{author}{Troccoli\xfnm[ A.]}, \bibinfo{author}{Dubus\xfnm[
  L.]}, \bibinfo{author}{Schmidt\xfnm[ J.]}, \bibinfo{author}{Gruber\xfnm[
  K.]}, \bibinfo{author}{Simões\xfnm[ S.G.]}, \bibinfo{author}{Heier\xfnm[
  S.]}.
\newblock \bibinfo{title}{A parametric model for wind turbine power curves
  incorporating environmental conditions}.
\newblock \bibinfo{journal}{Renewable Energy}
  \bibinfo{year}{2020};\bibinfo{volume}{157}:\bibinfo{pages}{754--768}.
\newblock \URLprefix
  \url{https://www.sciencedirect.com/science/article/pii/S0960148120306613}.
  \DOIprefix\doi{https://doi.org/10.1016/j.renene.2020.04.123}.
%Type = Inproceedings
\bibitem[{Salzman(1996)}]{Salzman1996}
\bibinfo{author}{Salzman\xfnm[ J.A.]}.
\newblock \bibinfo{title}{Fluid management in space-based systems}.
\newblock In: \bibinfo{booktitle}{Engineering, Construction, and Operations in
  Space V}. \bibinfo{publisher}{American Society of Civil Engineers};
  \bibinfo{year}{1996}. \DOIprefix\doi{10.1061/40177(207)71}.
%Type = Article
\bibitem[{Sane(2003)}]{Sane2003}
\bibinfo{author}{Sane\xfnm[ S.P.]}.
\newblock \bibinfo{title}{The aerodynamics of insect flight}.
\newblock \bibinfo{journal}{Journal of experimental biology}
  \bibinfo{year}{2003};\bibinfo{volume}{206}(\bibinfo{number}{23}):\bibinfo{pages}{4191--4208}.
%Type = Article
\bibitem[{Sane and Dickinson(2001)}]{Sane2001}
\bibinfo{author}{Sane\xfnm[ S.P.]}, \bibinfo{author}{Dickinson\xfnm[ M.H.]}.
\newblock \bibinfo{title}{The control of flight force by a flapping wing: lift
  and drag production}.
\newblock \bibinfo{journal}{Journal of experimental biology}
  \bibinfo{year}{2001};\bibinfo{volume}{204}(\bibinfo{number}{15}):\bibinfo{pages}{2607--2626}.
%Type = Article
\bibitem[{Sastry and Isidori(1989)}]{sastry1989adaptive}
\bibinfo{author}{Sastry\xfnm[ S.S.]}, \bibinfo{author}{Isidori\xfnm[ A.]}.
\newblock \bibinfo{title}{Adaptive control of linearizable systems}.
\newblock \bibinfo{journal}{IEEE Transactions on Automatic Control}
  \bibinfo{year}{1989};\bibinfo{volume}{34}(\bibinfo{number}{11}):\bibinfo{pages}{1123--1131}.
%Type = Misc
\bibitem[{Schaul et~al.(2015)Schaul, Quan, Antonoglou and Silver}]{Schaul2015}
\bibinfo{author}{Schaul\xfnm[ T.]}, \bibinfo{author}{Quan\xfnm[ J.]},
  \bibinfo{author}{Antonoglou\xfnm[ I.]}, \bibinfo{author}{Silver\xfnm[ D.]}.
\newblock \bibinfo{title}{Prioritized experience replay}.
\newblock \bibinfo{year}{2015}.
\newblock \DOIprefix\doi{10.48550/ARXIV.1511.05952}.
%Type = Misc
\bibitem[{Schaul et~al.(2016)Schaul, Quan, Antonoglou and
  Silver}]{schaul2016prioritized}
\bibinfo{author}{Schaul\xfnm[ T.]}, \bibinfo{author}{Quan\xfnm[ J.]},
  \bibinfo{author}{Antonoglou\xfnm[ I.]}, \bibinfo{author}{Silver\xfnm[ D.]}.
\newblock \bibinfo{title}{Prioritized experience replay}.
\newblock \bibinfo{year}{2016}.
\newblock \href{http://arxiv.org/abs/1511.05952}{\tt arXiv:1511.05952}.
%Type = Article
\bibitem[{Schoukens and Ljung(2019)}]{Schoukens2019}
\bibinfo{author}{Schoukens\xfnm[ J.]}, \bibinfo{author}{Ljung\xfnm[ L.]}.
\newblock \bibinfo{title}{Nonlinear system identification: A user-oriented road
  map}.
\newblock \bibinfo{journal}{{IEEE} Control Systems}
  \bibinfo{year}{2019};\bibinfo{volume}{39}(\bibinfo{number}{6}):\bibinfo{pages}{28--99}.
\newblock \DOIprefix\doi{10.1109/mcs.2019.2938121}.
%Type = Article
\bibitem[{Schwenzer et~al.(2021)Schwenzer, Ay, Bergs and
  Abel}]{schwenzer_review_2021}
\bibinfo{author}{Schwenzer\xfnm[ M.]}, \bibinfo{author}{Ay\xfnm[ M.]},
  \bibinfo{author}{Bergs\xfnm[ T.]}, \bibinfo{author}{Abel\xfnm[ D.]}.
\newblock \bibinfo{title}{Review on model predictive control: an engineering
  perspective}.
\newblock \bibinfo{journal}{Int J Adv Manuf Technol}
  \bibinfo{year}{2021};\bibinfo{volume}{117}(\bibinfo{number}{5-6}):\bibinfo{pages}{1327--1349}.
\newblock \URLprefix
  \url{https://link.springer.com/10.1007/s00170-021-07682-3}.
  \DOIprefix\doi{10.1007/s00170-021-07682-3}.
%Type = Article
\bibitem[{Silver et~al.(2016)Silver, Huang, Maddison, Guez, Sifre, van~den
  Driessche, Schrittwieser, Antonoglou, Panneershelvam, Lanctot, Dieleman,
  Grewe, Nham, Kalchbrenner, Sutskever, Lillicrap, Leach, Kavukcuoglu, Graepel
  and Hassabis}]{Silver2016}
\bibinfo{author}{Silver\xfnm[ D.]}, \bibinfo{author}{Huang\xfnm[ A.]},
  \bibinfo{author}{Maddison\xfnm[ C.J.]}, \bibinfo{author}{Guez\xfnm[ A.]},
  \bibinfo{author}{Sifre\xfnm[ L.]}, \bibinfo{author}{van~den Driessche\xfnm[
  G.]}, \bibinfo{author}{Schrittwieser\xfnm[ J.]},
  \bibinfo{author}{Antonoglou\xfnm[ I.]}, \bibinfo{author}{Panneershelvam\xfnm[
  V.]}, \bibinfo{author}{Lanctot\xfnm[ M.]}, \bibinfo{author}{Dieleman\xfnm[
  S.]}, \bibinfo{author}{Grewe\xfnm[ D.]}, \bibinfo{author}{Nham\xfnm[ J.]},
  \bibinfo{author}{Kalchbrenner\xfnm[ N.]}, \bibinfo{author}{Sutskever\xfnm[
  I.]}, \bibinfo{author}{Lillicrap\xfnm[ T.]}, \bibinfo{author}{Leach\xfnm[
  M.]}, \bibinfo{author}{Kavukcuoglu\xfnm[ K.]}, \bibinfo{author}{Graepel\xfnm[
  T.]}, \bibinfo{author}{Hassabis\xfnm[ D.]}.
\newblock \bibinfo{title}{Mastering the game of go with deep neural networks
  and tree search}.
\newblock \bibinfo{journal}{Nature}
  \bibinfo{year}{2016};\bibinfo{volume}{529}(\bibinfo{number}{7587}):\bibinfo{pages}{484--489}.
%Type = Article
\bibitem[{Silver et~al.(2018)Silver, Hubert, Schrittwieser, Antonoglou, Lai,
  Guez, Lanctot, Sifre, Kumaran, Graepel, Lillicrap, Simonyan and
  Hassabis}]{Silver2018}
\bibinfo{author}{Silver\xfnm[ D.]}, \bibinfo{author}{Hubert\xfnm[ T.]},
  \bibinfo{author}{Schrittwieser\xfnm[ J.]}, \bibinfo{author}{Antonoglou\xfnm[
  I.]}, \bibinfo{author}{Lai\xfnm[ M.]}, \bibinfo{author}{Guez\xfnm[ A.]},
  \bibinfo{author}{Lanctot\xfnm[ M.]}, \bibinfo{author}{Sifre\xfnm[ L.]},
  \bibinfo{author}{Kumaran\xfnm[ D.]}, \bibinfo{author}{Graepel\xfnm[ T.]},
  \bibinfo{author}{Lillicrap\xfnm[ T.]}, \bibinfo{author}{Simonyan\xfnm[ K.]},
  \bibinfo{author}{Hassabis\xfnm[ D.]}.
\newblock \bibinfo{title}{A general reinforcement learning algorithm that
  masters chess, shogi, and go through self-play}.
\newblock \bibinfo{journal}{Science}
  \bibinfo{year}{2018};\bibinfo{volume}{362}(\bibinfo{number}{6419}):\bibinfo{pages}{1140--1144}.
%Type = Article
\bibitem[{Silver et~al.(2014)Silver, Lever, Heess, Degris, Wierstra and
  Riedmiller}]{silver_deterministic_nodate}
\bibinfo{author}{Silver\xfnm[ D.]}, \bibinfo{author}{Lever\xfnm[ G.]},
  \bibinfo{author}{Heess\xfnm[ N.]}, \bibinfo{author}{Degris\xfnm[ T.]},
  \bibinfo{author}{Wierstra\xfnm[ D.]}, \bibinfo{author}{Riedmiller\xfnm[ M.]}.
\newblock \bibinfo{title}{Deterministic {Policy} {Gradient} {Algorithms}}
  \bibinfo{year}{2014};.
%Type = Article
\bibitem[{Sjöberg et~al.(1994)Sjöberg, Hjalmarsson and Ljung}]{Sjoeberg1994}
\bibinfo{author}{Sjöberg\xfnm[ J.]}, \bibinfo{author}{Hjalmarsson\xfnm[ H.]},
  \bibinfo{author}{Ljung\xfnm[ L.]}.
\newblock \bibinfo{title}{Neural networks in system identification}.
\newblock \bibinfo{journal}{{IFAC} Proceedings Volumes}
  \bibinfo{year}{1994};\bibinfo{volume}{27}(\bibinfo{number}{8}):\bibinfo{pages}{359--382}.
\newblock \DOIprefix\doi{10.1016/s1474-6670(17)47737-8}.
%Type = Article
\bibitem[{Smola and Schölkopf(2004)}]{Smola2004}
\bibinfo{author}{Smola\xfnm[ A.J.]}, \bibinfo{author}{Schölkopf\xfnm[ B.]}.
\newblock \bibinfo{title}{A tutorial on support vector regression}.
\newblock \bibinfo{journal}{Statistics and Computing}
  \bibinfo{year}{2004};\bibinfo{volume}{14}(\bibinfo{number}{3}):\bibinfo{pages}{199--222}.
\newblock \DOIprefix\doi{10.1023/b:stco.0000035301.49549.88}.
%Type = Article
\bibitem[{Staffell and Green(2014)}]{STAFFELL2014775}
\bibinfo{author}{Staffell\xfnm[ I.]}, \bibinfo{author}{Green\xfnm[ R.]}.
\newblock \bibinfo{title}{How does wind farm performance decline with age?}
\newblock \bibinfo{journal}{Renewable Energy}
  \bibinfo{year}{2014};\bibinfo{volume}{66}:\bibinfo{pages}{775--786}.
\newblock \URLprefix
  \url{https://www.sciencedirect.com/science/article/pii/S0960148113005727}.
  \DOIprefix\doi{https://doi.org/10.1016/j.renene.2013.10.041}.
%Type = Book
\bibitem[{Stengel(1994)}]{stengel1994optimal}
\bibinfo{author}{Stengel\xfnm[ R.F.]}.
\newblock \bibinfo{title}{Optimal control and estimation}.
\newblock \bibinfo{publisher}{Courier Corporation}, \bibinfo{year}{1994}.
%Type = Article
\bibitem[{Sutton(1991)}]{sutton_dyna}
\bibinfo{author}{Sutton\xfnm[ R.S.]}.
\newblock \bibinfo{title}{Dyna, an integrated architecture for learning,
  planning, and reacting}.
\newblock \bibinfo{journal}{SIGART Bull}
  \bibinfo{year}{1991};\bibinfo{volume}{2}(\bibinfo{number}{4}):\bibinfo{pages}{160–163}.
\newblock \URLprefix \url{https://doi.org/10.1145/122344.122377}.
  \DOIprefix\doi{10.1145/122344.122377}.
%Type = Book
\bibitem[{Sutton and Barto(2018)}]{sutton2018reinforcement}
\bibinfo{author}{Sutton\xfnm[ R.S.]}, \bibinfo{author}{Barto\xfnm[ A.G.]}.
\newblock \bibinfo{title}{Reinforcement learning: An introduction}.
\newblock \bibinfo{publisher}{MIT press}, \bibinfo{year}{2018}.
%Type = Book
\bibitem[{Suykens et~al.(1996)Suykens, Vandewalle and Moor}]{Suykens1996}
\bibinfo{author}{Suykens\xfnm[ J.A.K.]}, \bibinfo{author}{Vandewalle\xfnm[
  J.P.L.]}, \bibinfo{author}{Moor\xfnm[ B.L.R.D.]}.
\newblock \bibinfo{title}{Artificial Neural Networks for Modelling and Control
  of Non-Linear Systems}.
\newblock \bibinfo{publisher}{Springer {US}}, \bibinfo{year}{1996}.
\newblock \DOIprefix\doi{10.1007/978-1-4757-2493-6}.
%Type = Incollection
\bibitem[{Szita(2012)}]{Szita2012}
\bibinfo{author}{Szita\xfnm[ I.]}.
\newblock \bibinfo{title}{Reinforcement learning in games}.
\newblock In: \bibinfo{booktitle}{Adaptation, Learning, and Optimization}.
  \bibinfo{publisher}{Springer Berlin Heidelberg}; \bibinfo{year}{2012}. p.
  \bibinfo{pages}{539--577}.
%Type = Article
\bibitem[{Taha et~al.(2012)Taha, Hajj and Nayfeh}]{Taha2012}
\bibinfo{author}{Taha\xfnm[ H.E.]}, \bibinfo{author}{Hajj\xfnm[ M.R.]},
  \bibinfo{author}{Nayfeh\xfnm[ A.H.]}.
\newblock \bibinfo{title}{Flight dynamics and control of flapping-wing mavs: a
  review}.
\newblock \bibinfo{journal}{Nonlinear Dynamics}
  \bibinfo{year}{2012};\bibinfo{volume}{70}:\bibinfo{pages}{907--939}.
%Type = Article
\bibitem[{Talagrand and Courtier(1987)}]{Talagrand1987}
\bibinfo{author}{Talagrand\xfnm[ O.]}, \bibinfo{author}{Courtier\xfnm[ P.]}.
\newblock \bibinfo{title}{Variational assimilation of meteorological
  observations with the adjoint vorticity equation. i: Theory}.
\newblock \bibinfo{journal}{Quarterly Journal of the Royal Meteorological
  Society}
  \bibinfo{year}{1987};\bibinfo{volume}{113}(\bibinfo{number}{478}):\bibinfo{pages}{1311--1328}.
\newblock \DOIprefix\doi{10.1002/qj.49711347812}.
%Type = Article
\bibitem[{Tang and Hsieh(2001)}]{Tang2001}
\bibinfo{author}{Tang\xfnm[ Y.]}, \bibinfo{author}{Hsieh\xfnm[ W.W.]}.
\newblock \bibinfo{title}{Coupling neural networks to incomplete dynamical
  systems via variational data assimilation}.
\newblock \bibinfo{journal}{Monthly Weather Review}
  \bibinfo{year}{2001};\bibinfo{volume}{129}(\bibinfo{number}{4}):\bibinfo{pages}{818--834}.
\newblock \DOIprefix\doi{10.1175/1520-0493(2001)129<0818:cnntid>2.0.co;2}.
%Type = Inproceedings
\bibitem[{Tassa et~al.()Tassa, Erez and
  Todorov}]{tassaSynthesisStabilizationComplex2012}
\bibinfo{author}{Tassa\xfnm[ Y.]}, \bibinfo{author}{Erez\xfnm[ T.]},
  \bibinfo{author}{Todorov\xfnm[ E.]}.
\newblock \bibinfo{title}{Synthesis and stabilization of complex behaviors
  through online trajectory optimization}.
\newblock In: \bibinfo{booktitle}{2012 {{IEEE}}/{{RSJ International
  Conference}} on {{Intelligent Robots}} and {{Systems}}}.
  \bibinfo{publisher}{IEEE}. p. \bibinfo{pages}{4906--4913}.
\newblock \URLprefix \url{http://ieeexplore.ieee.org/document/6386025/}.
  \DOIprefix\doi{10.1109/IROS.2012.6386025}.
%Type = Misc
\bibitem[{Tekinerdogan(2022)}]{tekinerdogan_notion_2022}
\bibinfo{author}{Tekinerdogan\xfnm[ B.]}.
\newblock \bibinfo{title}{On the notion of digital twins: A modeling
  perspective}.
\newblock \bibinfo{year}{2022}.
\newblock \URLprefix \url{https://www.mdpi.com/2079-8954/11/1/15}.
  \DOIprefix\doi{10.3390/systems11010015}.
%Type = Article
\bibitem[{Tham(1995)}]{tham1995}
\bibinfo{author}{Tham\xfnm[ C.K.]}.
\newblock \bibinfo{title}{Reinforcement learning of multiple tasks using a
  hierarchical cmac architecture}.
\newblock \bibinfo{journal}{Robotics and Autonomous Systems}
  \bibinfo{year}{1995};\bibinfo{volume}{15}(\bibinfo{number}{4}):\bibinfo{pages}{247--274}.
\newblock \URLprefix
  \url{https://www.sciencedirect.com/science/article/pii/092188909500005Z}.
  \DOIprefix\doi{https://doi.org/10.1016/0921-8890(95)00005-Z};
  \bibinfo{note}{reinforcement Learning and Robotics}.
%Type = Article
\bibitem[{Uc-Cetina et~al.(2022)Uc-Cetina, Navarro-Guerrero, Martin-Gonzalez,
  Weber and Wermter}]{UcCetina2022}
\bibinfo{author}{Uc-Cetina\xfnm[ V.]}, \bibinfo{author}{Navarro-Guerrero\xfnm[
  N.]}, \bibinfo{author}{Martin-Gonzalez\xfnm[ A.]},
  \bibinfo{author}{Weber\xfnm[ C.]}, \bibinfo{author}{Wermter\xfnm[ S.]}.
\newblock \bibinfo{title}{Survey on reinforcement learning for language
  processing}.
\newblock \bibinfo{journal}{Artificial Intelligence Review}
  \bibinfo{year}{2022};\bibinfo{volume}{56}(\bibinfo{number}{2}):\bibinfo{pages}{1543--1575}.
\newblock \DOIprefix\doi{10.1007/s10462-022-10205-5}.
%Type = Inproceedings
\bibitem[{Van Den~Berg()}]{vandenbergIteratedLQRSmoothing2014}
\bibinfo{author}{Van Den~Berg\xfnm[ J.]}.
\newblock \bibinfo{title}{Iterated {{LQR}} smoothing for locally-optimal
  feedback control of systems with non-linear dynamics and non-quadratic cost}.
\newblock In: \bibinfo{booktitle}{2014 {{American Control Conference}}}.
  \bibinfo{publisher}{IEEE}. p. \bibinfo{pages}{1912--1918}.
\newblock \URLprefix \url{http://ieeexplore.ieee.org/document/6859404/}.
  \DOIprefix\doi{10.1109/ACC.2014.6859404}.
%Type = Article
\bibitem[{Wagner et~al.(2019)Wagner, Schleich, Haefner, Kuhnle, Wartzack and
  Lanza}]{Wagner2019}
\bibinfo{author}{Wagner\xfnm[ R.]}, \bibinfo{author}{Schleich\xfnm[ B.]},
  \bibinfo{author}{Haefner\xfnm[ B.]}, \bibinfo{author}{Kuhnle\xfnm[ A.]},
  \bibinfo{author}{Wartzack\xfnm[ S.]}, \bibinfo{author}{Lanza\xfnm[ G.]}.
\newblock \bibinfo{title}{Challenges and potentials of digital twins and
  industry 4.0 in product design and production for high performance products}.
\newblock \bibinfo{journal}{Procedia {CIRP}}
  \bibinfo{year}{2019};\bibinfo{volume}{84}:\bibinfo{pages}{88--93}.
\newblock \DOIprefix\doi{10.1016/j.procir.2019.04.219}.
%Type = Article
\bibitem[{Wang et~al.(2017)Wang, Huang, Wu, Wang and Lei}]{Wang2017_TVS}
\bibinfo{author}{Wang\xfnm[ B.]}, \bibinfo{author}{Huang\xfnm[ Y.]},
  \bibinfo{author}{Wu\xfnm[ J.]}, \bibinfo{author}{Wang\xfnm[ T.]},
  \bibinfo{author}{Lei\xfnm[ G.]}.
\newblock \bibinfo{title}{Experimental study on pressure control of liquid
  nitrogen tank by thermodynamic vent system}.
\newblock \bibinfo{journal}{Applied Thermal Engineering}
  \bibinfo{year}{2017};\bibinfo{volume}{125}:\bibinfo{pages}{1037--1046}.
\newblock \URLprefix
  \url{https://www.sciencedirect.com/science/article/pii/S1359431117315491}.
  \DOIprefix\doi{https://doi.org/10.1016/j.applthermaleng.2017.07.067}.
%Type = Article
\bibitem[{Wang et~al.(2000)Wang, Zou and Zhu}]{Wang2000}
\bibinfo{author}{Wang\xfnm[ B.]}, \bibinfo{author}{Zou\xfnm[ X.]},
  \bibinfo{author}{Zhu\xfnm[ J.]}.
\newblock \bibinfo{title}{Data assimilation and its applications}.
\newblock \bibinfo{journal}{Proceedings of the National Academy of Sciences}
  \bibinfo{year}{2000};\bibinfo{volume}{97}(\bibinfo{number}{21}):\bibinfo{pages}{11143--11144}.
\newblock \DOIprefix\doi{10.1073/pnas.97.21.11143}.
%Type = Misc
\bibitem[{Weber et~al.(2017)Weber, Racanière, Reichert, Buesing, Guez,
  Rezende, Badia, Vinyals, Heess, Li, Pascanu, Battaglia, Hassabis, Silver and
  Wierstra}]{Weber2017}
\bibinfo{author}{Weber\xfnm[ T.]}, \bibinfo{author}{Racanière\xfnm[ S.]},
  \bibinfo{author}{Reichert\xfnm[ D.P.]}, \bibinfo{author}{Buesing\xfnm[ L.]},
  \bibinfo{author}{Guez\xfnm[ A.]}, \bibinfo{author}{Rezende\xfnm[ D.J.]},
  \bibinfo{author}{Badia\xfnm[ A.P.]}, \bibinfo{author}{Vinyals\xfnm[ O.]},
  \bibinfo{author}{Heess\xfnm[ N.]}, \bibinfo{author}{Li\xfnm[ Y.]},
  \bibinfo{author}{Pascanu\xfnm[ R.]}, \bibinfo{author}{Battaglia\xfnm[ P.]},
  \bibinfo{author}{Hassabis\xfnm[ D.]}, \bibinfo{author}{Silver\xfnm[ D.]},
  \bibinfo{author}{Wierstra\xfnm[ D.]}.
\newblock \bibinfo{title}{Imagination-augmented agents for deep reinforcement
  learning}.
\newblock \bibinfo{year}{2017}.
\newblock \href{http://arxiv.org/abs/1707.06203}{\tt arXiv:1707.06203}.
%Type = Misc
\bibitem[{Werner and Peitz(2023)}]{Werner2023}
\bibinfo{author}{Werner\xfnm[ S.]}, \bibinfo{author}{Peitz\xfnm[ S.]}.
\newblock \bibinfo{title}{Learning a model is paramount for sample efficiency
  in reinforcement learning control of pdes}.
\newblock \bibinfo{year}{2023}.
\newblock \href{http://arxiv.org/abs/2302.07160}{\tt arXiv:2302.07160}.
%Type = Article
\bibitem[{Whitney and Wood(2010)}]{Whitney2010}
\bibinfo{author}{Whitney\xfnm[ J.P.]}, \bibinfo{author}{Wood\xfnm[ R.J.]}.
\newblock \bibinfo{title}{Aeromechanics of passive rotation in flapping
  flight}.
\newblock \bibinfo{journal}{Journal of fluid mechanics}
  \bibinfo{year}{2010};\bibinfo{volume}{660}:\bibinfo{pages}{197--220}.
%Type = Misc
\bibitem[{Willard et~al.(2020)Willard, Jia, Xu, Steinbach and
  Kumar}]{Willard2020}
\bibinfo{author}{Willard\xfnm[ J.]}, \bibinfo{author}{Jia\xfnm[ X.]},
  \bibinfo{author}{Xu\xfnm[ S.]}, \bibinfo{author}{Steinbach\xfnm[ M.]},
  \bibinfo{author}{Kumar\xfnm[ V.]}.
\newblock \bibinfo{title}{Integrating scientific knowledge with machine
  learning for engineering and environmental systems}.
\newblock \bibinfo{year}{2020}.
\newblock \DOIprefix\doi{10.48550/ARXIV.2003.04919}.
%Type = Article
\bibitem[{Wright and Davidson(2020)}]{Wright2020}
\bibinfo{author}{Wright\xfnm[ L.]}, \bibinfo{author}{Davidson\xfnm[ S.]}.
\newblock \bibinfo{title}{How to tell the difference between a model and a
  digital twin}.
\newblock \bibinfo{journal}{Advanced Modeling and Simulation in Engineering
  Sciences} \bibinfo{year}{2020};\bibinfo{volume}{7}(\bibinfo{number}{1}).
\newblock \DOIprefix\doi{10.1186/s40323-020-00147-4}.
%Type = Article
\bibitem[{Xue et~al.(2023)Xue, Cai, Xu and Liu}]{Xue2023}
\bibinfo{author}{Xue\xfnm[ Y.]}, \bibinfo{author}{Cai\xfnm[ X.]},
  \bibinfo{author}{Xu\xfnm[ R.]}, \bibinfo{author}{Liu\xfnm[ H.]}.
\newblock \bibinfo{title}{Wing kinematics-based flight control strategy in
  insect-inspired flight systems: Deep reinforcement learning gives solutions
  and inspires controller design in flapping mavs}.
\newblock \bibinfo{journal}{Biomimetics}
  \bibinfo{year}{2023};\bibinfo{volume}{8}(\bibinfo{number}{3}):\bibinfo{pages}{295}.
%Type = Inproceedings
\bibitem[{Yamada et~al.(1997)Yamada, Watanabe and Nakashima}]{hybrid1997}
\bibinfo{author}{Yamada\xfnm[ S.]}, \bibinfo{author}{Watanabe\xfnm[ A.]},
  \bibinfo{author}{Nakashima\xfnm[ M.]}.
\newblock \bibinfo{title}{Hybrid reinforcement learning and its application to
  biped robot control}.
\newblock In: \bibinfo{editor}{Jordan\xfnm[ M.]}, \bibinfo{editor}{Kearns\xfnm[
  M.]}, \bibinfo{editor}{Solla\xfnm[ S.]}, editors.
  \bibinfo{booktitle}{Advances in Neural Information Processing Systems}.
  \bibinfo{publisher}{MIT Press}; volume~\bibinfo{volume}{10};
  \bibinfo{year}{1997}. \URLprefix
  \url{https://proceedings.neurips.cc/paper_files/paper/1997/file/7895fc13088ee37f511913bac71fa66f-Paper.pdf}.
%Type = Misc
\bibitem[{Çağatay Yıldız et~al.(2021)Çağatay Yıldız, Heinonen and
  Lähdesmäki}]{yıldız2021continuoustime}
\bibinfo{author}{Çağatay Yıldız\xfnm[]}, \bibinfo{author}{Heinonen\xfnm[
  M.]}, \bibinfo{author}{Lähdesmäki\xfnm[ H.]}.
\newblock \bibinfo{title}{Continuous-time model-based reinforcement learning}.
\newblock \bibinfo{year}{2021}.
\newblock \href{http://arxiv.org/abs/2102.04764}{\tt arXiv:2102.04764}.
%Type = Article
\bibitem[{Zhang and Moore(1991)}]{Zhang1991SystemIU}
\bibinfo{author}{Zhang\xfnm[ C.]}, \bibinfo{author}{Moore\xfnm[ K.L.]}.
\newblock \bibinfo{title}{System identification using neural networks}.
\newblock \bibinfo{journal}{[1991] Proceedings of the 30th IEEE Conference on
  Decision and Control} \bibinfo{year}{1991};:\bibinfo{pages}{873--874
  vol.1}\URLprefix \url{https://api.semanticscholar.org/CorpusID:60798054}.
%Type = Article
\bibitem[{Zhang and Constantinescu(2023)}]{Zhang2023}
\bibinfo{author}{Zhang\xfnm[ H.]}, \bibinfo{author}{Constantinescu\xfnm[
  E.M.]}.
\newblock \bibinfo{title}{Optimal checkpointing for adjoint multistage
  time-stepping schemes}.
\newblock \bibinfo{journal}{Journal of Computational Science}
  \bibinfo{year}{2023};\bibinfo{volume}{66}:\bibinfo{pages}{101913}.
\newblock \DOIprefix\doi{10.1016/j.jocs.2022.101913}.
%Type = Article
\bibitem[{Zheng and Jin(2022)}]{Zheng2022}
\bibinfo{author}{Zheng\xfnm[ X.]}, \bibinfo{author}{Jin\xfnm[ T.]}.
\newblock \bibinfo{title}{A reliable method of wind power fluctuation smoothing
  strategy based on multidimensional non-linear exponential smoothing
  short-term forecasting}.
\newblock \bibinfo{journal}{{IET} Renewable Power Generation}
  \bibinfo{year}{2022};\bibinfo{volume}{16}(\bibinfo{number}{16}):\bibinfo{pages}{3573--3586}.
\newblock \DOIprefix\doi{10.1049/rpg2.12395}.

\end{thebibliography}

\end{document}